\documentclass[preprint]{jpsj3}

\usepackage{txfonts}
\usepackage{graphicx}
\usepackage{dcolumn}
\usepackage{bm}
\usepackage[usenames,dvipsnames]{xcolor}
\usepackage{ulem}
\usepackage{amsmath}
\usepackage{braket}
\usepackage{enumerate}
\usepackage{multirow}

\newcommand{\red}[1]{\textcolor{black}{#1}}

\newcommand{\SH}[1]{\textcolor{black}{#1}}

\newcommand{\green}[1]{\textcolor{black}{#1}}

\title{
Theory of Many-Body Multipole Operators in Single-Centered Electron Systems: \\ 
Two-Body Toroidal Monopoles in Spinless Orbitals
}

\author{Shingo Kuniyoshi, Rikuto Oiwa, and Satoru Hayami}
\inst{Graduate School of Science, Hokkaido University, Sapporo 060-0810, Japan}

\abst{
One-body multipole operators are defined as irreducible representations of rotational symmetry together with spatial-inversion \red{$\mathcal{P}$} and time-reversal \red{$\mathcal{T}$} \red{parities}, providing a systematic framework for classifying electronic internal degrees of freedom and for describing a wide variety of composite order parameters.
While this formalism has been successfully established for the one-body operator space
\red{and the rotational-symmetry-based classification of antisymmetrized many-electron spaces has long been developed in atomic and nuclear shell theory, 
its 
reformulation within the framework
of electric, magnetic, electric toroidal, and magnetic toroidal multipoles characterized by $\mathcal{P}$ and $\mathcal{T}$ parities \red{has not been systematically developed}.
}
In this paper, we extend the multipole formalism, \red{including  $\mathcal{P}$ and $\mathcal{T}$ parities}, in the one-body operator space to the many-body operator space.
By formulating fermionic creation and annihilation operators as spherical tensors and employing Clebsch--Gordan coupling combined with the exterior (Grassmann) algebra, we construct an irreducible decomposition of many-body operators that fully incorporates fermionic antisymmetrization.
As a concrete application, we classify monopoles appearing in spinless many-body operators.
In particular, we show that the electric toroidal monopole, a pseudoscalar breaking spatial-inversion symmetry, and the magnetic toroidal monopole, a time-reversal-odd scalar, become active in spinless interacting many-body systems, although they are absent in the spinless one-body hybrid orbital space.
}

\begin{document}
\maketitle

A central task in the theory of ordering is to identify the active low-energy degrees of freedom and their symmetry constraints.
While local-moment magnetism is a canonical example of ordering, $f$-electron compounds often exhibit orbital multipoles as  relevant order parameters, giving rise to a rich variety of phases and responses beyond conventional dipolar magnetism.
This has motivated the development of one-body electronic multipole representation theory
, which systematically organizes spin and orbital degrees of freedom into irreducible components under rotations and  has become a standard  tool in condensed-matter physics~\cite{Kuramoto_JPSJ_2008,Kusunose_JPSJ_2008,Santini_RMP_2009,Kuramoto_JPSJ_2009}.

According to their spatial-inversion $\mathcal{P}$ and time-reversal $\mathcal{T}$ parities, 
one-body multipole operators in the particle-number--conserving space $\{c^{\dagger} c, c c^{\dagger}\}$ are classified into four types: electric (E), magnetic (M), magnetic toroidal (MT), and electric toroidal (ET) multipoles~\cite{dubovik1990toroid, Hayami_JPSJ_2018, SH_MY_YY_HK_mul_2018, HW_YY_Mul_2018, Kusunose_JPSJ_2020,Kusunose_IOP_2022,Hayami_JPSJ_2024}.
These multipoles are  labeled by  irreducible representations of the rotation group,  specified by the rank $
j$ and component $m$ as $X_{{j}m}$ ($X=Q,M,T,G$), with $\mathcal{P}$ and $\mathcal{T}$ parities given by $Q_{{jm}}: [(-1)^{{j}},+1]$, $M_{{jm}}: [(-1)^{{j}+1},-1]$, $T_{{jm}}: [(-1)^{{j}},-1]$, and $G_{{jm}}: [(-1)^{j+1},+1]$.
Such multipole theory has also recently been extended to the particle-number--nonconserving one-body space $\{c^{\dagger}c^{\dagger},\, cc\}$, providing a systematic 
 classification of multi-orbital superconducting order parameters~\cite{Kirikoshi_PRB_2024}.

Since one-body multipole operators form a complete and orthonormal basis in the  one-body operator space,  they can describe  any parity-specific anisotropic distributions of electronic degrees of freedom~\cite{Kusunose_PRB_2023}. 
Systematic first-principles schemes for evaluating these multipoles
have also been estabilished~\cite{Ederer_PRB_2007, Spaldin_PRB_2013, Florian_PRB_2016, suzuki2018first,Oiwa_PRL_2022,Hoshino_PRL_2023,Inda_JCP_2024,Oiwa_PRB_2025,Oiwa_PRR_2025,Miki_PRL_2025,Xie_arxiv_2025}. 
A representative example is the ET monopole $G_0$,
a $\mathcal{T}$-even and $\mathcal{P}$-odd pseudoscalar that serves as 
a descriptor of electronic chirality 
and underlies various chirality-induced phenomena~\cite{SH_MY_YY_HK_mul_2018, Hayami_PhysRevLett.122.147602, Oiwa_PRL_2022, Kishine_IJC_2022, hayami2023chiral, Kusunose_APL_2024,Hoshino_PRL_2023,Oiwa_PRR_2025, Miki_PRL_2025, ishitobi2026purely}, such as optical activity~\cite{1979JETPL} and chirality-induced spin selectivity.\red{\cite{R.Neeman_ACR_2020_CISS}}
% ~\cite{B.Gohler_nat_2011_CISS, O.Ben_nat_2017_CISS, K.Michaeli_PNAS_2019_CISS, Suda_natcom_2019_CISS, A.Inui2020prl_CrNb3S6, R.Neeman_ACR_2020_CISS, shitade2020geometric, Waldeck2021aplmat_CISS, Bloom_2024_chemrev_CISS}.
Another example is the MT monopole $T_0$, a $\mathcal{T}$-odd  scalar, which has recently attracted  interest as an unconventional  order parameter in $\mathcal{T}$-breaking magnets.
% {\cyan{~\cite{Hayami_T0_PRB_2023}}}
{\red{~\cite{Hayami_T0_PRB_2023}}}

The developments above, however, are still largely restricted to the one-body operator spaces, namely the particle-number--conserving sector $\{c^{\dagger}c,\, cc^{\dagger}\}$ and the particle-number--nonconserving sector $\{c^{\dagger}c^{\dagger},\, cc\}$.
\red{ For interacting electron systems, the group-theoretical construction and classification of antisymmetrized many-electron spaces and tensor operators based on rotational symmetry have long been established in atomic and nuclear shell theory~\cite{Weyl1939,Judd19631967,deShalitTalmi1963,Racah19431949}. 
However, this conventional classification has not been systematically reorganized from the viewpoint of the four multipole types $(Q,M,T,G)$ 
% \cyan{distinguished by $\mathcal{P}$ and $\mathcal{T}$ parities}.
distinguished by $\mathcal{P}$ and $\mathcal{T}$ parities.}
This limitation is particularly crucial in spinless systems, where the toroidal monopoles $G_0$ and $T_0$ are inactive at the one-body level in single-centered electron systems: $G_0$ becomes active only for spinful electrons, whereas $T_0$ remains inactive irrespective of spin.
Their emergence therefore necessarily requires higher-body operator sectors.

In this Letter, we extend the concept of one-body multipole operators to the many-body operator space and develop a systematic scheme for constructing, classifying, and labeling many-body multipole operators \red{using standard methods of group theory.~\cite{Weyl1939,Judd19631967,deShalitTalmi1963,Racah19431949}}
We construct an irreducible operator basis by representing fermionic creation and annihilation operators as spherical tensors and performing Clebsch--Gordan couplings under full fermionic antisymmetrization.
Technically, this is achieved through a character-theoretic formulation of the exterior (Grassmann) algebra, together with a generating function expressed as a Fredholm determinant.
The resulting many-body operators are further organized according to the particle-number $U(1)$ gauge symmetry.
As a concrete application, we classify monopoles in spinless many-body operators and demonstrate that the ET and MT monopoles, which are absent in the spinless one-body operator space, become active in spinless interacting electron systems.

\begin{table*}[t]
  \centering
  \caption{
  Classification of multipole operators constructed from products of spinless $p$-orbital creation and annihilation operators. 
  The subscript $(\Delta N)$ denotes the particle-number change; for example, $\mathrm{ET}_{(2)}$ is a Hermitian electric-toroidal multipole with $\Delta N=2$.
  \red{The blank cells indicate that the corresponding representation is absent.}
  }
  \vspace{+0.1ex}
  \label{tab:p}
  \setlength{\tabcolsep}{4pt}
  \setlength{\dbltextfloatsep}{6pt plus 2pt minus 2pt}
  \setlength{\dblfloatsep}{6pt plus 2pt minus 2pt}
  \renewcommand{\arraystretch}{1.1}
  \begin{tabular}{|l||c|c|c||c|c|c|}
    \hline
      & \multicolumn{3}{c||}{$\Delta N = 0$}
      & \multicolumn{3}{c|}{$\Delta N \ne 0$} \\
    \hline
      & \textbf{Monopole} & \textbf{Dipole} & \textbf{Quadrupole}
      & \textbf{Monopole} & \textbf{Dipole} & \textbf{Quadrupole} \\
    \hline
    $\Lambda^0, \Lambda^6$ & E  &   &   &   &   &   \\
    \hline
    $\Lambda^1, \Lambda^5$ &   &   &   &   & $\rm E_{(1)}$, $\rm MT_{(1)}$  &   \\
    \hline
    $\Lambda^2, \Lambda^4$ & E  &  M & E  &   & $\rm M_{(2)}$, $\rm ET_{(2)}$  &   \\
    \hline
    $\Lambda^3$            &   &   &   & $\rm M_{(1)}$, $\rm ET_{(1)}$, $\rm M_{(3)}$, $\rm ET_{(3)}$  & $\rm E_{(1)}$, $\rm MT_{(1)}$  & $\rm M_{(1)}$, $\rm ET_{(1)}$  \\
    \hline
  \end{tabular}
    \vspace{-6ex}
\end{table*}

To formulate a systematic construction of many-body multipole operators, we first summarize the rotational and $U(1)$ transformation properties of fermionic creation and annihilation operators.
The creation operators $c^\dagger_{jm}$ transform under rotations as the $m$ component of a rank-$j$ irreducible spherical tensor~\cite{Weyl1939,Judd19631967,deShalitTalmi1963,Racah19431949,Gaigalas_1996,Gaigalas_2004,Iwahara_PRB_2015}:
\begin{equation}
 R\,c^\dagger_{jm}\,R^{-1}=\sum_{m'} D^{(j)}_{m m'}(R)\,c^\dagger_{jm'} ,
\end{equation}
where $R$ denotes a rotation and $D^{(j)}(R)$ is the corresponding Wigner $D$ matrix.
\red{Since the annihilation operators $c_{jm}$ transform according to the conjugate representation, we introduce the spherical-tensor conjugate}
\begin{equation}
\tilde c_{jm}:=(-1)^{j-m}c_{j,-m},
\end{equation}
which transforms in the same way as $c^\dagger_{jm}$.
With this choice, irreducible many-body multipole operators can be constructed from products of creation and annihilation operators through successive Clebsch--Gordan couplings.
In addition, considering that $c^\dagger_{jm}$ and $c_{jm}$ change the particle number by $+1$ and $-1$, respectively, 
they can be assigned a $U(1)$ charge (weight) under the particle-number symmetry.
The symmetry content of the operator space generated by products of $c^\dagger$ and $c$ is then determined by its irreducible decomposition under $SU(2)\times U(1)$.
Moreover, for higher-body operators, fermionic statistics enforces antisymmetrization, under which some irreducible components suggested by a formal tensor-product construction vanish identically.
To incorporate this constraint systematically, we formulate the construction in the exterior (wedge) algebra of fermionic modes, namely the Grassmann algebra.

We then consider the operator space generated solely by creation operators {$\{c^\dagger_{jm}\}$ or  \red{tensor-conjugated} annihilation operators $\{\tilde{c}_{jm}\}$}.
The corresponding one-particle representation space is denoted by
${v}_{j, \pm1}:={v}_j \otimes \mathcal{N}_{ \pm1}$,
where ${v}_j$ is the rank{-}$j$ irreducible representation of $SU(2)$ and $\mathcal{N}_{\pm1}$ is the one-dimensional representation carrying $U(1)$ charge $\pm1$.
Under the particle-number rotation $U(\phi)=e^{-i\phi \hat N}$, this convention reads $U(\phi)\,{(}c^\dagger_{jm}{, \tilde{c}_{jm})}\,U(\phi)^{-1}={(}e^{-i\phi {(+1)}}c^\dagger_{jm}{, e^{-i\phi {(-1)}} \tilde{c}_{jm})}$.
The representation generated by $n$ creation {(annihilation)} operators is formally defined on the tensor-product space $({v}_{j,\pm1})^{\otimes n}$, whose irreducible decomposition yields the candidate $SU(2)$ ranks $k$.
Fermionic statistics, however, restricts the physical subspace to the antisymmetrized exterior-power space $\Lambda^n {v}_{j,\pm1} \subset {({v}_{j,\pm1})}^{\otimes n}$, so that the surviving irreducible content is obtained by decomposing $\Lambda^n {v}_{j,\pm1}$.

Let $\rho(g)$ {denote} the representation matrix of a group element $g\in SU(2)\times U(1)$ on a finite {$d$}-dimensional representation space $\mathcal{W}$.
The character of the exterior-power representation $\Lambda^n \mathcal{W}$ is then defined as 
\begin{equation}
 \chi_{\Lambda^n\mathcal{W}}(g) := \Tr \!\left[\Lambda^n \rho(g)\right].
\end{equation}
It is generated by the Fredholm determinant:\red{~\cite{Weyl1939}}
\begin{equation}
 \det \!\left(I+z\,\rho(g)\right)
 =\sum_{n=0}^{{d}} z^n \chi_{\Lambda^n\mathcal{W}}(g).
\label{eq:fredholm}
\end{equation}
By extracting the coefficient of $z^n$ in Eq.~(\ref{eq:fredholm}), one obtains the character of the grade-$n$ sector, namely the space spanned by antisymmetrized $n$-fold products of creation (or annihilation) operators.
Since the inversion parity of an operator product is given by the product of the individual parities,
the $\mathcal P$ parity of the grade-$n$ sector follows immediately.
When $c^\dagger_{jm}$ is $\mathcal P$-odd (and likewise $c_{jm}$), any operator of odd grade is $\mathcal P$-odd.

As an example,\red{\cite{Racah19431949}} consider the spinless $p$ orbital  ${j = 1}$ {creation operator space $\{c^\dagger_{1,+1}, c^\dagger_{1,0}, c^\dagger_{1,-1}\}$}.
For a  rotation {by an angle $\theta$} about the $z$ axis,  the representation matrix is
$D^{(1)}_z(\theta)=\mathrm{diag}(e^{-i\theta},\,1,\,e^{+i\theta})$.
Including the particle-number $U(1)$ phase factor $e^{-i\phi}$, the full representation matrix is $\rho{(\theta , \phi)} = e^{-i\phi} D^{(1)}_z(\theta)$.
Then, the characters are generated by the Fredholm determinant
\begin{align}
 \det &\left(I + z \rho{(\theta {, \phi})} \right) \notag\\
  &= (1+ze^{-i\phi}e^{-i\theta})(1+\SH{z} e^{-i\phi}%z
  )(1+ze^{-i\phi}e^{+i\theta}),
\end{align}
so that the coefficients of $z^n$ yield
\begin{align}
  \chi_{\Lambda^0}{(\theta {, \phi})} &= 1,\\
  \chi_{\Lambda^1} {(\theta {, \phi})} &= e^{-i\phi}(1 + e^{-i\theta} + e^{i\theta})
  = e^{-i\phi}\Tr\!\left[D^{(1)}_z\left(\theta\right)\right],\\
  \chi_{\Lambda^2} {(\theta {, \phi})} &= e^{-i2\phi}(1 + e^{-i\theta} + e^{i\theta})
  = e^{-i2\phi}\Tr\!\left[D^{(1)}_z\left(\theta\right)\right],\\
  \chi_{\Lambda^3} {(\theta {, \phi})} &= e^{-i3\phi}.
\end{align}
Here $\chi_{\Lambda^n}$ is the character of the grade-$n$ sector $\Lambda^n v_{1,+1}$ generated by $n$-th exterior-power of the $p$-orbital creation operators $\{c^\dagger_{l=1,m}\}$.
Since spinless $p$ orbitals are $\mathcal{P}$-odd, the grade-$n$ sector has parity $(-1)^n$: even (odd) grades are $\mathcal{P}$-even (odd).
For instance, the grade-$0$ sector corresponds to constants and thus contains only a rank-$0$ monopole ($\mathcal{P}$-even).
The grade-$1$ sector consists of the creation operators themselves and carries rank $1$ ($\mathcal{P}$-odd), the grade-$2$ sector corresponds to antisymmetrized two-fermion states; again only rank $1$ appears ($\mathcal{P}$-even), and the grade-$3$ sector is one-dimensional  monopole $c^\dagger_{1}c^\dagger_{0}c^\dagger_{-1}$ ($\mathcal{P}$-odd). 
For a complicated character, it is often useful to decompose it into irreducible characters by using the following orthogonality relation:
\begin{align}
 &\int_{0}^{2\pi}\frac{d\phi}{2\pi}
   \int_{0}^{2\pi}\frac{d\theta}{\pi}
   \sin^2 \left(\frac{\theta}{2}\right)
   \bigl(\chi_{j,n}\bigr)^*
   \chi_{j',n'}
   = \delta_{jj'}\,\delta_{nn'}.
\end{align}
Here $\chi_{j,n} := \chi_n(\phi)\,\chi_j(\theta)$, where
$\chi_n(\phi)=e^{\pm i n\phi}$ is the $U(1)$ character and
$\chi_j(\theta)=\sin \bigl((2j+1)\theta/2\bigr)/\sin(\theta/2)$ is the character of the rank $j$ irreducible representation of $SU(2)$.
The characters for the space spanned by the \red{tensor-conjugated} annihilation operators $\{\tilde{c}_{j,m}\}$ are obtained in the same manner.

We now consider the single-orbital representation space including both creation and \red{tensor-conjugated} annihilation operators,
$\mathcal{V}_{j}:=v_{j,+1}\oplus v_{j,-1}$.
The corresponding representation matrix is block diagonal and can be written as
$\rho{(\theta, \phi)} = e^{-i\phi} D^{(j)}({\theta})\oplus e^{+i\phi} D^{(j)}({\theta})$.
Applying the Fredholm determinant to 
$\rho{({\theta, \phi})}$ yields the character of each grade{-}$n$ sector $\Lambda^n\mathcal{V}_j$.
In the particle-number-conserving sector with $\Delta N = 0$, grade{-}${(}n=2k{)}$ corresponds to $k$-body operators containing $k$ creation and $k$ annihilation operators.

Hermitianization of the operators enables a further classification into $\mathcal{T}$-even and $\mathcal{T}$-odd  sectors.
For instance, the grade-2 one-body operators, $c^{\dagger} c$, $c c^{\dagger}$ ($\Delta N = 0$){, as well as}  $c^{\dagger} c^{\dagger}$, $c^{} c^{}$ ($\Delta N = 2$), can be decomposed into $\mathcal{T}$-even and $\mathcal{T}$-odd  combinations as follows.
\begin{align}
&\mathcal{T}\text{-even}: \qquad (c^{\dagger}_{1} c^{}_{2} + c^{\dagger}_{2} c^{}_{1}), \qquad (c^{\dagger}_{1} c^{\dagger}_{2} + c^{}_{2} c^{}_{1}), \qquad\\
&\mathcal{T}\text{-odd}: \qquad i(c^{\dagger}_{1} c^{}_{2} - c^{{\dagger}}_{2} c^{ }_{1}),  \qquad i(c^{\dagger}_{1} c^{\dagger}_{2} - c^{}_{2} c^{}_{1}).
\end{align}
As a result,  one-body multipole operators fall into four symmetry-distinct classes as 
% $Q_{ {jm}}$ $[(\mathcal{P}, \mathcal{T}) = ((-1)^{ {j}}, +1)]$, $M_{ {jm}}$ $[(-1)^{ {j}+1}, -1]$, $T_{ {jm}}$ $[(-1)^{ {j}}, -1]$, and $G_{ {jm}}$ $[(-1)^{ {j}+1}, +1]$.
$Q_{jm}$, $M_{jm}$, $T_{jm}$, and $G_{jm}$.
Operators of higher grade are classified in the same manner.
Note that for spinful electrons, the $\mathcal{T}$ parity of  odd-grade operators {($n=2k+1$)} is not uniquely determined, since their spinor nature prevents them from being  eigenoperators of the $\mathcal{T}$ parity.

Table~\ref{tab:p} summarizes the classification of many-body spinless multipoles for $p$ orbitals in single-centered electron systems.
One finds that the grade-$n$ sector ($\Lambda^{n}$) has the same representation content as the grade-$(6-n)$ sector ($\Lambda^{6-n}$),
\red{reflecting Hodge duality in the exterior algebra; the corresponding sector pairing can also be viewed as particle--hole conjugation in the finite local Fock space.}
For example, in the particle-number--conserving two-body sector $\Lambda^{4}$, the electric monopole, magnetic dipoles, and electric quadrupoles are active, as in the conventional one-body sector $\Lambda^{2}$.
In particle-number--nonconserving sectors ($\Delta N \ne 0$), the subscript $(\Delta N)$ denotes the particle-number change. 
Notably, spinless ET dipoles become allowed in $\Lambda^{2}$ and the two-body sector $\Lambda^{4}$ with $\Delta N=2$, which are forbidden in particle-number--conserving sectors ($\Delta N = 0$)~\cite{Hayami_doi:10.7566/JPSJ.91.113702}.

Next, let us discuss the hybrid systems involving multiple orbital sectors. 
One proceeds in the same way by defining the one-particle representation space as a direct sum, $\mathcal{V}=\mathcal{V}_{j}\oplus \mathcal{V}_{j'}\oplus \cdots$, and analyzing the characters of the exterior powers $\Lambda^n\mathcal{V}$.
To keep track of how many operators are taken from each orbital sector, we introduce orbital-dependent auxiliary variables. As an illustrative example, consider an $s$--$p$ hybrid system ($\mathcal{V}=\mathcal{V}_{s}\oplus \mathcal{V}_{p}$).
The generating function for a rotation $R_z(\theta)$ is defined as
\begin{align}
\det \!\left(I+z\, \rho^{(s)}(\theta,\phi_s)\right)\,
 \det \!\left(I+w\, \rho^{(p)}(\theta,\phi_p)\right),
\end{align}
with
$\rho^{(s)}(\theta,\phi_s)= e^{-i\phi_s} D^{(0)}(R_z(\theta))\oplus e^{+i\phi_s} D^{(0)}(R_z(\theta))$ and
$\rho^{(p)}(\theta,\phi_p)= e^{-i\phi_p} D^{(1)}(R_z(\theta))\oplus e^{+i\phi_p} D^{(1)}(R_z(\theta))$.
Expanding it in $z$ and $w$, the coefficient of $z^{a}w^{b}$ gives the character of the $(a,b)$ sector in the standard decomposition $\Lambda^{a+b}(\mathcal{V}_{s}\oplus \mathcal{V}_{p})=\bigoplus_{a+b=n}\Lambda^{a}\mathcal{V}_{s}\otimes \Lambda^{b}\mathcal{V}_{p}$.
In particular, the $z^{2}$ and $w^2$ terms correspond to the one-body $s$ and $p$ sectors, respectively, while the $zw$ terms yield the one-body multipole contents in the $s$--$p$ hybrid sector~\cite{Hayami_JPSJ_2018}.
Within this hybrid sector, the contributions are further separated into four $U(1)$-weight factors,
$e^{-i\phi_s}e^{-i\phi_p}$, $e^{+i\phi_s}e^{+i\phi_p}$, $e^{+i\phi_s}e^{-i\phi_p}$, and $e^{-i\phi_s}e^{+i\phi_p}$,
corresponding to $c^\dagger_s c^\dagger_p$, $c_s c_p$, $c^\dagger_p c_s$, and $c^\dagger_s c_p$, respectively.
Imposing Hermiticity, the pair{s} {($c^\dagger_s c_p$}, $c^\dagger_p c_s${)} {and ($c_{s}^{\dagger} c_{p}^{\dagger}$, $c_{s}^{} c_{p}^{}$)} yield four Hermitian combinations with definite $\mathcal{T}$ parity as follows:
\begin{align}
&\mathcal{T}\text{-even}: \qquad (c^{\dagger}_{s} c^{}_{p} + c^{\dagger}_{p} c^{}_{s}), \qquad (c^{\dagger}_{s} c^{\dagger}_{p} + c^{}_{p} c^{}_{s}), \qquad\\
&\mathcal{T}\text{-odd}: \qquad i(c^{\dagger}_{s} c^{}_{p} - c^{{\dagger}}_{p} c^{
}_{s}),  \qquad i(c^{\dagger}_{s} c^{\dagger}_{p} - c^{}_{p} c^{}_{s}).
\end{align}
Moreover, since the hybrid operators involve one $p$ orbital ($\mathcal{P}$-odd) and one $s$ orbital ($\mathcal{P}$-even), they are overall $\mathcal{P}$-odd. 
Consequently, the $s$--$p$ hybrid one-body multipoles are rank $1$ and $\mathcal{P}$-odd, with both $\mathcal{T}$-even and $\mathcal{T}$-odd components; they are therefore  classified as the E and MT dipoles~\cite{Hayami_JPSJ_2018,Hayami_JPSJ_2024}. Higher-body hybrid multipoles are obtained in the same manner from the 
coefficients {$z^{a}w^{b}$} with $a+b\ge 4$.

In Table~\ref{tab:two-body}, we summarize the numbers of particle-number--conserving ($\Delta N = 0$) hybrid monopoles in spinless hybrid systems, namely E monopole ($Q_{0}$), M monopole ($M_{0}$), MT monopole ($T_{0}$), and ET monopole ($G_{0}$).
Here the superscript $(2)$ denotes a two-body operator; one-body operators are written without a superscript.
The most important point in Table~\ref{tab:two-body} is that the monopoles $M_0$, $T_0$, and $G_0$, which are absent at the one-body level, become allowed in the two-body  operator space~\cite{Hayami_JPSJ_2018,Hayami_JPSJ_2024}.
The equality in the active numbers of $Q_0$ ($M_0$) and $T_0$ ($G_0$) reflects the time-reversal pairing obtained by forming Hermitian combinations of hybrid operators.

\begin{table}[t]
  \centering
  \caption{
    Numbers of particle-number-conserving ($\Delta N = 0$)  hybrid E monopole ($Q_{0}$), M monopole ($M_{0}$), MT monopole ($T_{0}$), and ET monopole ($G_{0}$) in the spinless hybrid systems; $X= Q, M, T, G$.
     The superscripts (2) denotes two-body operators; $X_0$ without a superscript stands for one-body operators.    
  }
  \vspace{+0.1ex}
  \label{tab:two-body}
   \begin{tabular}{lc|cc}
    \hline\hline    
       & $X_{0}$ & $M_{0}^{(2)}$, $G_{0}^{(2)}$ & $Q_{0}^{(2)}$, $T_{0}^{(2)}$\\
    \hline
    $sp$   & %$\times$
    {--} & \textbf{1} & %$\times$ & 
    %$\times$
    {--} \\ \hline
    $sd$   & %$\times$
    {--} & %$\times$
    {--} & %$\times$ & 
    %$\times$
    {--} \\ \hline
    $sf$   & %$\times$
    {--} & \textbf{1} & %$\times$ & 
    %$\times$
    {--} \\ \hline
    $pd$   & %$\times$
    {--} & \textbf{3} & %$\times$ & 
    \textbf{1} \\ \hline
    $pf$   & %$\times$
    {--} & %$\times$
    {--} & %$\times$ & 
    \textbf{2} \\ \hline
    $df$   & %$\times$
    {--} & \textbf{5} & %$\times$ &
     \textbf{2} \\ \hline
    $spd$  & %-
    %{$\times$}
    {--} & \textbf{2} & %- & 
    \textbf{1} \\ \hline
    $spf$  & %-
    %{$\times$}
    {--} & \textbf{2} & %- & 
    %$\times$
    {--} \\ \hline
    $sdf$  & %-
    %{$\times$}
    {--} & \textbf{2} & %- & 
    \textbf{1} \\ \hline
    $pdf$  & %-
    %{$\times$}
    {--} & \textbf{8} & %- & 
    \textbf{4} \\ \hline
    $spdf$ & %-
    %{$\times$}
    {--} & %$\times$
    {--}      &% - & 
    \textbf{3} \\ 
    \hline\hline
    \end{tabular}
  % \end{ruledtabular}
  \vspace{-6ex}
\end{table}

Among two-body monopole operators, we focus on the ET monopole, $G_0^{(2)}$, and the MT monopole, $T_0^{(2)}$.
Below, we present their explicit constructions for spinless two-orbital hybrid systems
($sp$, $pd$, and $pf$ orbitals).
The expressions for other two-body monopole operators and/or hybrid orbitals are presented in Supplemental Material~\cite{SM}.

In the $sp$-hybrid systems, $G_0^{(2)}$ is interpreted as the scalar coupling of the $sp$-hybrid MT dipole $\bm{T}_1^{sp}$ with the single $p$-orbital M dipole $\bm{M}_1^{p}$:
\begin{align}
G_0^{(2)} &=
\frac{1}{2} \left(\left[[c^\dagger_s \times c^\dagger_p]_{1} \times[\tilde c_p \times \tilde c_p]_{1}\right]_{0} + \mathrm{h.c.} \right), \notag \\
&=
\frac{1}{\sqrt{6}}\Big(c^\dagger_{s} c^\dagger_{p,-1} c_{p,0} c_{p,-1} + c^\dagger_{s} c^\dagger_{p,0} c_{p,1} c_{p,-1} + c^\dagger_{s} c^\dagger_{p,1} c_{p,1} c_{p,0} \notag \\
&\qquad + c^\dagger_{p,0} c^\dagger_{p,-1} c_{s} c_{p,-1} + c^\dagger_{p,1} c^\dagger_{p,-1} c_{s} c_{p,0} + c^\dagger_{p,1} c^\dagger_{p,0} c_{s} c_{p,1}\Big), \notag  \\
&=
\bm{T}_1^{sp}  \cdot  \bm{M}_1^{p}.
\end{align}
Here $[A \times B]_{k,q}$ denotes the $q$ component of the rank-$k$ irreducible tensor operator obtained by coupling the tensor operators $A$ and $B$ using Clebsch--Gordan coefficients.
In particular, we use the notation
$[A \times B]_{0,0} = [A \times B]_{0} = \bm{A}\cdot\bm{B}$.
A general one-body multipole operator can then be written as
\begin{align}
O_{k,q}^{ab}
&= \frac{i^{\frac{1-\alpha}{2}}}{2}
\left(
[c^\dagger_a \times \tilde c_b]_{k,q}
+ \alpha\,(-1)^{l_a+l_b}\,[c^\dagger_b \times \tilde c_a]_{k,q}
\right),
\end{align}
\red{where $t=0$ and $t=1$ label $\mathcal{T}$-even and $\mathcal{T}$-odd multipoles, respectively, and $\alpha := (-1)^{k+t}$.
A concrete realization without spin--orbit coupling is provided in the Supplemental Material~\cite{SM}: a minimal spinless two-electron $s$-$p$ model exhibits $\langle G_0^{(2)}\rangle\neq0$ only when both an inversion-odd $s$-$p$ hybridization and a two-body interaction are present.}

Meanwhile, in the $pd$-hybrid systems, $G_0^{(2)}$ can be formed through three distinct channels:
(i) (hybrid MT dipole) $\times$ (single-orbital M dipole),
(ii) (hybrid ET quadrupole) $\times$ (single-orbital E quadrupole), and
(iii) (hybrid MT octupole) $\times$ (single-orbital M octupole).
As an example, $G_0^{(2)}$ in the $pd$-hybrid system is given by
\begin{align}
G_{0}^{(2)} &= 
\frac{1}{2}  \left(\left[[c^\dagger_p \times c^\dagger_d]_{1} \times[\tilde c_d \times \tilde c_d]_{1}\right]_{0} + {\rm h.c.} \right)\notag \\
&=
\frac{3}{10} {\bm T}_{1}^{pd}   \cdot   {\bm M}_{1}^{d}
 +  \frac{\sqrt{35}}{10} {\bm G}_{2}^{pd}   \cdot   {\bm Q}_{2}^{d}
 +  \frac{\sqrt{14}}{5}  {\bm T}_{3}^{pd}   \cdot   {\bm M}_{3}^{d}.
\end{align}
The coefficients in the one-body tensor-product decomposition are determined by Wigner-$9j$ symbols.
Thus, the two-body ET monopole operator is constructed as the direct product of a hybrid one-body multipole, MT or ET, and a single-orbital one-body multipole, M or E.
Although Eq.~(18) involves three channels, they are not linearly independent, and only two independent $G_0^{(2)}$ can be defined.
While Table~\ref{tab:two-body} lists three $G_0^{(2)}$ for the $pd$-hybrid system, one of them belongs to a different channel, namely $[[c_p^\dagger \times c_p^\dagger]_1 \times [c_p \times c_d]_1]_0$.
It is noted that the two-body ET monopole can be constructed for three-orbital hybrid systems ($spd$, $spf$, $spf$, $sdf$, and $pdf$), as presented in Supplemental Material~\cite{SM}.

In the $pf$-hybrid systems, the two-body MT monopole $T_0^{(2)}$ is also constructed by the scalar coupling between the hybrid multipole and single-orbital multipole:
\begin{align}
T_0^{(2)}
&=  
  \frac{i}{2} \left(\left[[c^\dagger_p \times c^\dagger_f]_{3} \times[\tilde c_f \times \tilde c_f]_{3}\right]_{0} -{\rm  h.c.} \right) \notag\\
&=
    -\frac{\sqrt{70}}{14}
    {\bm T}_{2}^{pf}  \cdot  {\bm Q}_{2}^{f}    
    -\frac{1}{2}
    {\bm G}_{3}^{pf}  \cdot  {\bm M}_{3}^{f}
    +\frac{\sqrt{77}}{14}
    {\bm T}_{4}^{pf}  \cdot  {\bm Q}_{4}^{f}.
\end{align}
More generally, in a spinless two-orbital system, $T_0^{(2)}$ arises from coupling between time-pair hybrid multipoles, MT and E or M and ET.
In contrast to $G_{0}^{(2)}$, $T_{0}^{(2)}$ is described by four-orbital hybrid systems ($spdf$) as well as three-orbital ones ($spd$, $sdf$, and $pdf$).

\red{
Although two-body operators can be generated from products of one-body operators, their genuine two-body components are determined only after normal ordering and projection onto antisymmetrized two-fermion sectors.
Apart from lower-body contractions, they can be recoupled into pair tensors
$
[
(c_{l_1}^\dagger \times c_{l_2}^\dagger)_{L}
\times
(c_{l_3} \times c_{l_4})_{L'}
]_{K}.
$
A two-body multipole channel therefore survives only when both pair tensors have nonvanishing antisymmetrized components. For spinless identical orbital tensors, this gives
$
(c_l^{(\dagger)}\times c_l^{(\dagger)})_{L}=0
$
for $(-1)^{2l-L}=+1$, and the tensor survives only for $(-1)^{2l-L}=-1$.
Thus, in the $sd$-hybrid system, the hybrid E monopole naively expected from $Q_{2,sd}^{(1)}\times Q_{2,d}^{(1)}$ is absent because it would require the forbidden rank-$2$ component of $c_d\times c_d$.
Consequently, after imposing antisymmetry and Hermiticity, the only monopole remaining in this sector is
$
n_s n_d =
c_s^\dagger c_s
\left(\sum_m c_{d,m}^\dagger c_{d,m}\right).
$
}

To summarize, we have developed a systematic framework for constructing and labeling many-body multipole operators based on exterior-power representations and a Fredholm-determinant character generating function.
This approach yields a nonredundant irreducible decomposition of operator sectors, covering both particle-number--conserving and nonconserving channels, and thereby extends the augmented-multipole framework beyond the one-body level.
As a concrete application, we have classified the particle-number--conserving monopoles in spinless multi-orbital hybrid systems.
We showed that the two-body sector activates monopoles including the toroidal monopoles, $T_0^{(2)}$ and $G_0^{(2)}$, which are absent in the spinless one-body hybrid orbital space (see Table~\ref{tab:two-body}).

Our framework provides symmetry-adapted many-body multipole operators that can be directly used to analyze interacting Hamiltonians and many-body order parameters~\cite{Hoshino_PhysRevLett.107.247202, Iwazaki_PhysRevB.108.L241108}, and it naturally interfaces with the one-body multipole description and its numerical evaluation in realistic materials.
In particular, the emergence of the spinless two-body ET monopole $G_{0}^{(2)}$ offers a microscopic route to interaction-induced electronic chirality without spin--orbit coupling.
Moreover, clarifying the physical implications of the spinless two-body MT monopole $T_{0}^{(2)}$ and identifying its signatures in correlated systems are important directions for future study.

We also note a related recent preprint that analyzes spinless electronic chirality in the atomic limit in terms of the hydrodynamic helicity, a two-body $\mathcal{T}$-even pseudoscalar quantity~\cite{Miki_arXiv_2026}.
While that work focuses on the microscopic origin of such chirality measures in a specific minimal atomic model, the present Letter provides a general symmetry-based construction and irreducible classification of many-body multipole operators.
In this sense, our results complement Ref.~\citen{Miki_arXiv_2026} by clarifying how the spinless two-body toroidal monopoles $G_{0}^{(2)}$ and $T_{0}^{(2)}$ emerge as interaction-enabled many-body degrees of freedom at the operator level.
\red{
%Although the hydrodynamic helicity discussed in Ref.~\citen{Miki_arXiv_2026} is not identical to the local spinless operator $G_{0}^{(2)}$ considered here, it\blue{s} %shares the 
%interaction-induced nature %of the corresponding
%\blue{as a} spinless two-body chirality measure is qualitatively consistent with our Supplemental Material~\cite{SM}.
% \cyan{
% Although the hydrodynamic helicity discussed in Ref. 43 is not identical to the local spinless operator $G_{0}^{(2)}$ considered here, both quantities represent interaction-induced spinless pseudoscalar measures of chirality.
% A minimal realization of such an interaction-induced spinless pseudoscalar is presented in Supplemental Material~\cite{SM}.
% }
Although the hydrodynamic helicity discussed in Ref.~\green{\citen{Miki_arXiv_2026}} is not identical to the local spinless operator $G_{0}^{(2)}$ considered here, both quantities represent interaction-induced spinless pseudoscalar measures of chirality.
A minimal realization of such an interaction-induced spinless pseudoscalar is presented in Supplemental Material~\cite{SM}.
}

\begin{acknowledgments}
The authors thank H. Kusunose, T. Ishitobi, and Y. Suzuki for fruitful discussions.
S.K. and S.H. thank 
T. Miki
% \final{T. Miki}
% for sharing unpublished theoretical results on hydrodynamic helicity and for insightful discussions{~\cite{Miki_arXiv_2026}}.
for sharing unpublished theoretical results on hydrodynamic helicity and for insightful discussions.
Clarifying the connection between hydrodynamic helicity as a two-body pseudoscalar and the present two-body multipole formulation is an interesting direction for future study.

This work was supported by JSPS KAKENHI Grants Numbers JP22H00101, JP22H01183, JP23H04869, JP23K03288, and by JST CREST (JPMJCR23O4) and JST FOREST (JPMJFR2366).
\end{acknowledgments}

% ---------------- References ----------------
\bibliography{Multipole_theory}
\bibliographystyle{jpsj}

\end{document}

\maketitle

\section{Characters and multipoles in the spinless $p$ system}
\subsection{Derivation of the character}

As discussed in the main text, the Fredholm determinant for the $l=1$ creation operators is given by
\begin{align}
 \det\!\left(I + z\,\rho_+(\theta,\phi)\right)
 = (1+z e^{-i\phi}e^{-i\theta})(1+z e^{-i\phi})(1+z e^{-i\phi}e^{+i\theta}) .
\end{align}
Similarly, for the annihilation operators one finds
\begin{align}
 \det\!\left(I + z\,\rho_-(\theta,\phi)\right)
 = (1+z e^{i\phi}e^{-i\theta})(1+z e^{i\phi})(1+z e^{i\phi}e^{+i\theta}) .
\end{align}
Therefore, the representation generated by products of creation and annihilation operators is obtained from the product of the two Fredholm determinants:
\begin{align}
 &\det\!\left(I + z\,\rho_+(\theta,\phi)\right)\,
  \det\!\left(I + z\,\rho_-(\theta,\phi)\right) \nonumber\\
 &=(1+z e^{-i\phi}e^{-i\theta})(1+z e^{-i\phi})(1+z e^{-i\phi}e^{+i\theta})
   (1+z e^{i\phi}e^{-i\theta})(1+z e^{i\phi})(1+z e^{i\phi}e^{+i\theta}) \nonumber\\
 &= \chi_{\Lambda^0}(\theta,\phi)
  + z\,\chi_{\Lambda^1}(\theta,\phi)
  + z^2\chi_{\Lambda^2}(\theta,\phi)
  + z^3\chi_{\Lambda^3}(\theta,\phi)
  + z^4\chi_{\Lambda^4}(\theta,\phi)
  + z^5\chi_{\Lambda^5}(\theta,\phi)
  + z^6\chi_{\Lambda^6}(\theta,\phi).
\end{align}
By comparing the coefficients of each power of $z$, we obtain the characters of the operator spaces that can be constructed from spinless $p$-orbital creation and annihilation operators.
\begin{align}
  \chi_{\Lambda^0}{(\theta {, \phi})} &= 1,\\ 
  \chi_{\Lambda^1} {(\theta {, \phi})} &= 
  (e^{-i\phi}+e^{i\phi})\, \chi_{1},\\
  \chi_{\Lambda^2} {(\theta {, \phi})} 
  &= 
  e^{-2i\phi}\chi_{1} + e^{2i\phi}\chi_{1} + \chi_{0} + \chi_{1}+ \chi_{2}\\
  \chi_{\Lambda^3} {(\theta {, \phi})} 
  &=   
  e^{-3i\phi} + e^{3i\phi} + e^{-i\phi}(\chi_{0} + \chi_{1}+ \chi_{2}) + e^{i\phi}(\chi_{0} + \chi_{1}+ \chi_{2})\\
  \chi_{\Lambda^4} {(\theta {, \phi})} 
  &=  
  \chi_{\Lambda^2} {(\theta {, \phi})} \\
  \chi_{\Lambda^5} {(\theta {, \phi})} 
  &=  \chi_{\Lambda^1} {(\theta {, \phi})} \\
  \chi_{\Lambda^6} {(\theta {, \phi})} 
  &=  \chi_{\Lambda^0} {(\theta {, \phi})} 
\end{align}

\subsection{Irreducible operators}
\subsubsection*{grade 0}
constants
\subsubsection*{grade 1}
% 
% \begin{align}
%   &\underline{e^{-i\phi}\, \chi_{1} ~:~ c^\dagger_{m}},\\
%   &\underline{e^{i\phi}\, \chi_{1} ~:~ \tilde c_m}
% \end{align}  
\begin{align}
  &\underline{e^{-i\phi}\, \chi_{1} ~:~ c^\dagger_{m}}\qquad\qquad \notag\\
  &\quad q=-1:{c^\dagger_{-1}}\\
  &\quad q=0:{c^\dagger_{0}}\\
  &\quad q=1:{c^\dagger_{1}}\\
  &\underline{e^{i\phi}\, \chi_{1} ~:~ \tilde c_m}\qquad\qquad \notag\\
  &\quad q=-1:{\tilde c_{-1}} = {c_{1}}\\
  &\quad q=0:{\tilde c_{0}} = {-c_{0}}\\
  &\quad q=1:{\tilde c_{1}} = {c_{-1}}
\end{align}          
 
\subsubsection*{grade 2}
\begin{align}
&\underline{e^{-2i\phi}\, \chi_{1}:  \sum C^{1,q}_{1,n;1,m} c^\dagger_n c^\dagger_m }  \qquad\qquad \notag\\
&\quad q=-1:\sum C^{1,-1}_{1,n;1,m} c^\dagger_n c^\dagger_m = \frac{\sqrt{2}}{2}{c^\dagger_{0}}{c^\dagger_{-1}} - \frac{\sqrt{2}}{2}{c^\dagger_{-1}}{c^\dagger_{0}}=\sqrt{2}{c^\dagger_{0}}{c^\dagger_{-1}}\\
&\quad q=0:\sum C^{1,0}_{1,n;1,m} c^\dagger_n c^\dagger_m = \frac{\sqrt{2}}{2}{c^\dagger_{1}}{c^\dagger_{-1}} - \frac{\sqrt{2}}{2}{c^\dagger_{-1}}{c^\dagger_{1}}=\sqrt{2}{c^\dagger_{1}}{c^\dagger_{-1}}\\
&\quad q=1:\sum C^{1,1}_{1,n;1,m} c^\dagger_n c^\dagger_m = \frac{\sqrt{2}}{2}{c^\dagger_{1}}{c^\dagger_{0}} - \frac{\sqrt{2}}{2}{c^\dagger_{0}}{c^\dagger_{1}}=\sqrt{2}{c^\dagger_{1}}{c^\dagger_{0}}\\
&\underline{e^{2i\phi}\, \chi_{1} : \sum C^{1,q}_{1,n;1,m} \tilde c_n \tilde c_m }  \qquad\qquad \notag\\
&\quad q=-1:\sum C^{1,-1}_{1,n;1,m} \tilde c_n \tilde c_m = \frac{\sqrt{2}}{2}{\tilde c_{0}}{\tilde c_{-1}} - \frac{\sqrt{2}}{2}{\tilde c_{-1}}{\tilde c_{0}}=\sqrt{2}{\tilde c_{0}}{\tilde c_{-1}}\\
&\quad q=  0:\sum C^{1,0}_{1,n;1,m} \tilde c_n \tilde c_m = \frac{\sqrt{2}}{2}{\tilde c_{1}}{\tilde c_{-1}} - \frac{\sqrt{2}}{2}{\tilde c_{-1}}{\tilde c_{1}}=\sqrt{2}{\tilde c_{1}}{\tilde c_{-1}}\\
&\quad q=  1:\sum C^{1,1}_{1,n;1,m} \tilde c_n \tilde c_m = \frac{\sqrt{2}}{2}{\tilde c_{1}}{\tilde c_{0}} - \frac{\sqrt{2}}{2}{\tilde c_{0}}{\tilde c_{1}}=\sqrt{2}{\tilde c_{1}}{\tilde c_{0}}\\
&\underline{\chi_{0}  : \sum C^{0,0}_{1,n;1,m} c_n^\dagger \tilde c_m  }  \qquad\qquad \notag\\
&\quad q=    0:\sum C^{0,0}_{1,n;1,m} c_n^\dagger \tilde c_m = \frac{\sqrt{3}}{3}{c^\dagger_{1}}{\tilde c_{-1}} - \frac{\sqrt{3}}{3}{c^\dagger_{0}}{\tilde c_{0}} + \frac{\sqrt{3}}{3}{c^\dagger_{-1}}{\tilde c_{1}} \propto n_p\\
&\underline{\chi_{1} : \sum C^{1,q}_{1,n;1,m} c_n^\dagger \tilde c_m  }  \qquad\qquad \notag\\
  &\quad q=-1:\sum C^{1,-1}_{1,n;1,m} c_n^\dagger \tilde c_m = \frac{\sqrt{2}}{2}{c^\dagger_{0}}{\tilde c_{-1}} - \frac{\sqrt{2}}{2}{c^\dagger_{-1}}{\tilde c_{0}}\\
&\quad q=  0:\sum C^{1,0}_{1,n;1,m} c_n^\dagger \tilde c_m = \frac{\sqrt{2}}{2}{c^\dagger_{1}}{\tilde c_{-1}} - \frac{\sqrt{2}}{2}{c^\dagger_{-1}}{\tilde c_{1}}\\
&\quad q=  1:\sum C^{1,1}_{1,n;1,m} c_n^\dagger \tilde c_m = \frac{\sqrt{2}}{2}{c^\dagger_{1}}{\tilde c_{0}} - \frac{\sqrt{2}}{2}{c^\dagger_{0}}{\tilde c_{1}}\\
&\underline{\chi_{2}: \sum C^{2,q}_{1,n;1,m} c_n^\dagger \tilde c_m }  \qquad\qquad \notag\\
&\quad q=-2:\sum C^{2,-2}_{1,n;1,m} c_n^\dagger \tilde c_m = {c^\dagger_{-1}}{\tilde c_{-1}}\\
&\quad q=-1:\sum C^{2,-1}_{1,n;1,m} c_n^\dagger \tilde c_m = \frac{\sqrt{2}}{2}{c^\dagger_{0}}{\tilde c_{-1}} + \frac{\sqrt{2}}{2}{c^\dagger_{-1}}{\tilde c_{0}}\\
&\quad q=      0:\sum C^{2,0}_{1,n;1,m} c_n^\dagger \tilde c_m = \frac{\sqrt{6}}{6}{c^\dagger_{1}}{\tilde c_{-1}} + \frac{\sqrt{6}}{3}{c^\dagger_{0}}{\tilde c_{0}} + \frac{\sqrt{6}}{6}{c^\dagger_{-1}}{\tilde c_{1}}\\
&\quad q=      1:\sum C^{2,1}_{1,n;1,m} c_n^\dagger \tilde c_m = \frac{\sqrt{2}}{2}{c^\dagger_{1}}{\tilde c_{0}} + \frac{\sqrt{2}}{2}{c^\dagger_{0}}{\tilde c_{1}}\\
&\quad q=      2:\sum C^{2,2}_{1,n;1,m} c_n^\dagger \tilde c_m = {c^\dagger_{1}}{\tilde c_{1}}
\end{align}

\subsubsection*{grade 3}
\begin{align}
&\underline{e^{-3i\phi}:\sum C^{0,q}_{1,n;1,m} {C_n^{2}}^\dagger c^\dagger_m}\qquad\left(  {C_q^{2}}^\dagger:= \sqrt{2} \sum C^{1,q}_{1,n;1,m} c^\dagger_n c^\dagger_m    \right)\notag \\
  &\quad q=0:\sum C^{0,0}_{1,n;1,m} {C_n^{2}}^\dagger c^\dagger_m   
  = \frac{\sqrt{3}}{3}{c^\dagger_{1}c^\dagger_{0}}{c^\dagger_{-1}} - \frac{\sqrt{3}}{3}{c^\dagger_{1}c^\dagger_{-1}}{c^\dagger_{0}} + \frac{\sqrt{3}}{3}{c^\dagger_{0}c^\dagger_{-1}}{c^\dagger_{1}}
  = \sqrt{3}c^\dagger_{1}c^\dagger_{0}c^\dagger_{-1}\\
&\underline{e^{3i\phi}:\sum C^{0,0}_{1,n;1,m} { {\widetilde C}_n^2} \tilde c_m} \qquad\left(  { {\widetilde C}_q^2} := \sqrt{2} \sum C^{1,q}_{1,n;1,m} \tilde c_n \tilde c_m    \right)\notag \\
  &\quad q=0:\sum C^{0,0}_{1,n;1,m} { {\widetilde C}_n^2} \tilde c_m = \frac{\sqrt{3}}{3}{\tilde c_{1}\tilde c_{0}}{\tilde c_{-1}} - \frac{\sqrt{3}}{3}{\tilde c_{1}\tilde c_{-1}}{\tilde c_{0}} + \frac{\sqrt{3}}{3}{\tilde c_{0}\tilde c_{-1}}{\tilde c_{1}}
  = \sqrt{3}\tilde c_{1}\tilde c_{0}\tilde c_{-1}\\
&\underline{ e^{-i\phi}\chi_{0} :\sum C^{0,q}_{1,n;1,m}{C_n^{2}}^\dagger \tilde c_m} \notag \\
  &\quad q=0:\sum C^{0,0}_{1,n;1,m}{C_n^{2}}^\dagger \tilde c_m = \frac{\sqrt{3}}{3}{c^\dagger_{1}c^\dagger_{0}}{\tilde c_{-1}} - \frac{\sqrt{3}}{3}{c^\dagger_{1}c^\dagger_{-1}}{\tilde c_{0}} + \frac{\sqrt{3}}{3}{c^\dagger_{0}c^\dagger_{-1}}{\tilde c_{1}}\\
&\underline{e^{-i\phi}\chi_{1} :\sum C^{1,q}_{1,n;1,m}{C_n^{2}}^\dagger \tilde c_m} \notag \\
    &\quad q=-1:\sum C^{1,-1}_{1,n;1,m}{C_n^{2}}^\dagger \tilde c_m = \frac{\sqrt{2}}{2}{c^\dagger_{1}c^\dagger_{-1}}{\tilde c_{-1}} - \frac{\sqrt{2}}{2}{c^\dagger_{0}c^\dagger_{-1}}{\tilde c_{0}}\\
  &\quad q=0:\sum C^{1,0}_{1,n;1,m}{C_n^{2}}^\dagger \tilde c_m = \frac{\sqrt{2}}{2}{c^\dagger_{1}c^\dagger_{0}}{\tilde c_{-1}} - \frac{\sqrt{2}}{2}{c^\dagger_{0}c^\dagger_{-1}}{\tilde c_{1}}\\
  &\quad q=1:\sum C^{1,1}_{1,n;1,m}{C_n^{2}}^\dagger \tilde c_m = \frac{\sqrt{2}}{2}{c^\dagger_{1}c^\dagger_{0}}{\tilde c_{0}} - \frac{\sqrt{2}}{2}{c^\dagger_{1}c^\dagger_{-1}}{\tilde c_{1}}\\
&\underline{e^{-i\phi}\chi_{2} :\sum C^{2,q}_{1,n;1,m}{C_n^{2}}^\dagger \tilde c_m} \notag \\
    &\quad q=-2:\sum C^{2,-2}_{1,n;1,m}{C_n^{2}}^\dagger \tilde c_m = {c^\dagger_{0}c^\dagger_{-1}}{\tilde c_{-1}}\\
    &\quad q=-1:\sum C^{2,-1}_{1,n;1,m}{C_n^{2}}^\dagger \tilde c_m = \frac{\sqrt{2}}{2}{c^\dagger_{1}c^\dagger_{-1}}{\tilde c_{-1}} + \frac{\sqrt{2}}{2}{c^\dagger_{0}c^\dagger_{-1}}{\tilde c_{0}}\\
    &\quad q=0:\sum C^{2,0}_{1,n;1,m}{C_n^{2}}^\dagger \tilde c_m = \frac{\sqrt{6}}{6}{c^\dagger_{1}c^\dagger_{0}}{\tilde c_{-1}} + \frac{\sqrt{6}}{3}{c^\dagger_{1}c^\dagger_{-1}}{\tilde c_{0}} + \frac{\sqrt{6}}{6}{c^\dagger_{0}c^\dagger_{-1}}{\tilde c_{1}}\\
    &\quad q=1:\sum C^{2,1}_{1,n;1,m}{C_n^{2}}^\dagger \tilde c_m = \frac{\sqrt{2}}{2}{c^\dagger_{1}c^\dagger_{0}}{\tilde c_{0}} + \frac{\sqrt{2}}{2}{c^\dagger_{1}c^\dagger_{-1}}{\tilde c_{1}}\\
    &\quad q=2:\sum C^{2,2}_{1,n;1,m}{C_n^{2}}^\dagger \tilde c_m = {c^\dagger_{1}c^\dagger_{0}}{\tilde c_{1}}\\
&\underline{  e^{i\phi}\chi_{0} : \sum C^{0,q}_{1,n;1,m} {c_n}^\dagger { {\widetilde C}_m^2}} \notag \\
 &\quad q=0:\sum C^{0,0}_{1,n;1,m} {c_n}^\dagger { {\widetilde C}_m^2} = \frac{\sqrt{3}}{3}{c^\dagger_{-1}} {\tilde  c_{1}\tilde  c_{0}} - \frac{\sqrt{3}}{3}{c^\dagger_{0}}{\tilde  c_{1}\tilde  c_{-1}} + \frac{\sqrt{3}}{3}{c^\dagger_{1}}{\tilde  c_{0}\tilde  c_{-1}}\\
&\underline{e^{i\phi}\chi_{1} :\sum C^{1,q}_{1,n;1,m}c^\dagger_m{\widetilde C_n^{2}} } \notag \\
 &\quad q=-1:\sum C^{1,-1}_{1,n;1,m}c^\dagger_m{\widetilde C_n^{2}} = \frac{\sqrt{2}}{2}{c^\dagger_{0}}{\tilde c_{0}\tilde c_{-1}} - \frac{\sqrt{2}}{2}{c^\dagger_{-1}}{\tilde c_{1}\tilde c_{-1}}\\
 &\quad q=0:\sum C^{1,0}_{1,n;1,m}c^\dagger_m{\widetilde C_n^{2}} = \frac{\sqrt{2}}{2}{c^\dagger_{1}}{\tilde c_{0}\tilde c_{-1}} - \frac{\sqrt{2}}{2}{c^\dagger_{-1}}{\tilde c_{1}\tilde c_{0}}\\
 &\quad q=1:\sum C^{1,1}_{1,n;1,m}c^\dagger_m{\widetilde C_n^{2}} = \frac{\sqrt{2}}{2}{c^\dagger_{1}}{\tilde c_{1}\tilde c_{-1}} - \frac{\sqrt{2}}{2}{c^\dagger_{0}}{\tilde c_{1}\tilde c_{0}}\\
&\underline{e^{i\phi}\chi_{2} : \sum C^{2,q}_{1,n;1,m}c^\dagger_m{\widetilde C_n^{2}}} \notag \\
 &\quad q=-2:\sum C^{2,-2}_{1,n;1,m}c^\dagger_m{\widetilde C_n^{2}} = {c^\dagger_{-1}}{\tilde c_{0}\tilde c_{-1}}\\
 &\quad q=-1:\sum C^{2,-1}_{1,n;1,m}c^\dagger_m{\widetilde C_n^{2}} = \frac{\sqrt{2}}{2}{c^\dagger_{0}}{\tilde c_{0}\tilde c_{-1}} + \frac{\sqrt{2}}{2}{c^\dagger_{-1}}{\tilde c_{1}\tilde c_{-1}}\\
 &\quad q=0:\sum C^{2,0}_{1,n;1,m}c^\dagger_m{\widetilde C_n^{2}} = \frac{\sqrt{6}}{6}{c^\dagger_{1}}{\tilde c_{0}\tilde c_{-1}} + \frac{\sqrt{6}}{3}{c^\dagger_{0}}{\tilde c_{1}\tilde c_{-1}} + \frac{\sqrt{6}}{6}{c^\dagger_{-1}}{\tilde c_{1}\tilde c_{0}}\\
 &\quad q=1:\sum C^{2,1}_{1,n;1,m}c^\dagger_m{\widetilde C_n^{2}} = \frac{\sqrt{2}}{2}{c^\dagger_{1}}{\tilde c_{1}\tilde c_{-1}} + \frac{\sqrt{2}}{2}{c^\dagger_{0}}{\tilde c_{1}\tilde c_{0}}\\
 &\quad q=2:\sum C^{2,2}_{1,n;1,m}c^\dagger_m{\widetilde C_n^{2}} = {c^\dagger_{1}}{\tilde c_{1}\tilde c_{0}}
\end{align}

\subsubsection*{grade 4}
\begin{align}
  &\underline{e^{-2i\phi}\chi_{1}:\sum C^{1,q}_{0,n;1,m}{C_n^{3}}^\dagger \tilde c_m} \qquad \left( {C_q^{3}}^\dagger := \sum C^{0,0}_{1,n;1,m} {C_n^{2}}^\dagger c^\dagger_m \right)\notag \\
  &\quad q=-1:\sum C^{1,-1}_{0,n;1,m}{C_n^{3}}^\dagger \tilde c_m = {c^\dagger_{1}c^\dagger_{0}c^\dagger_{-1}}{\tilde c_{-1}}\\
    &\quad q=0:\sum C^{1,0}_{0,n;1,m}{C_n^{3}}^\dagger \tilde c_m = {c^\dagger_{1}c^\dagger_{0}c^\dagger_{-1}}{\tilde c_{0}}\\
    &\quad q=1:\sum C^{1,1}_{0,n;1,m}{C_n^{3}}^\dagger \tilde c_m = {c^\dagger_{1}c^\dagger_{0}c^\dagger_{-1}}{\tilde c_{1}}\\
  &\underline{e^{2i\phi}\chi_{1}:\sum C^{1,q}_{1,n;0,m}c^\dagger_n{\widetilde C_m^{3}}} \qquad \left( { {\widetilde C}_q^3} := \sum C^{0,0}_{1,n;1,m} { {\widetilde C}_n^2} \tilde c_m  \right)\notag \\
  &\quad q=-1:\sum C^{1,-1}_{1,n;0,m}c^\dagger_n{\widetilde C_m^{3}} = {c^\dagger_{-1}}{\tilde c_{1}\tilde c_{0}\tilde c_{-1}}\\
    &\quad q=0:\sum C^{1,0}_{1,n;0,m}c^\dagger_n{\widetilde C_m^{3}} = {c^\dagger_{0}}{\tilde c_{1}\tilde c_{0}\tilde c_{-1}}\\
    &\quad q=1:\sum C^{1,1}_{1,n;0,m}c^\dagger_n{\widetilde C_m^{3}} = {c^\dagger_{1}}{\tilde c_{1}\tilde c_{0}\tilde c_{-1}}\\
   &\underline{\chi_{0}:\sum C^{0,q}_{1,n;1,m}{C_n^{2}}^\dagger {\widetilde C_m^{2}}} \notag \\
   &\quad q=0:\sum C^{0,0}_{1,n;1,m}{C_n^{2}}^\dagger {\widetilde C_m^{2}} = \frac{\sqrt{3}}{3}{c^\dagger_{1}c^\dagger_{0}}{\tilde c_{0}\tilde c_{-1}} - \frac{\sqrt{3}}{3}{c^\dagger_{1}c^\dagger_{-1}}{\tilde c_{1}\tilde c_{-1}} + \frac{\sqrt{3}}{3}{c^\dagger_{0}c^\dagger_{-1}}{\tilde c_{1}\tilde c_{0}}\\
   &\underline{\chi_{1}:\sum C^{1,q}_{1,n;1,m}{C_n^{2}}^\dagger {\widetilde C_m^{2}}} \notag \\
   &\quad q=-1:\sum C^{1,-1}_{1,n;1,m}{C_n^{2}}^\dagger {\widetilde C_m^{2}} = \frac{\sqrt{2}}{2}{c^\dagger_{1}c^\dagger_{-1}}{\tilde c_{0}\tilde c_{-1}} - \frac{\sqrt{2}}{2}{c^\dagger_{0}c^\dagger_{-1}}{\tilde c_{1}\tilde c_{-1}}\\
    &\quad q=0:\sum C^{1,0}_{1,n;1,m}{C_n^{2}}^\dagger {\widetilde C_m^{2}} = \frac{\sqrt{2}}{2}{c^\dagger_{1}c^\dagger_{0}}{\tilde c_{0}\tilde c_{-1}} - \frac{\sqrt{2}}{2}{c^\dagger_{0}c^\dagger_{-1}}{\tilde c_{1}\tilde c_{0}}\\
    &\quad q=1:\sum C^{1,1}_{1,n;1,m}{C_n^{2}}^\dagger {\widetilde C_m^{2}} = \frac{\sqrt{2}}{2}{c^\dagger_{1}c^\dagger_{0}}{\tilde c_{1}\tilde c_{-1}} - \frac{\sqrt{2}}{2}{c^\dagger_{1}c^\dagger_{-1}}{\tilde c_{1}\tilde c_{0}}\\
   &\underline{\chi_{2}:\sum C^{2,q}_{1,n;1,m}{C_n^{2}}^\dagger {\widetilde C_m^{2}}} \notag \\
    &\quad q=-2:\sum C^{2,-2}_{1,n;1,m}{C_n^{2}}^\dagger {\widetilde C_m^{2}} = {c^\dagger_{0}c^\dagger_{-1}}{\tilde c_{0}\tilde c_{-1}}\\
    &\quad q=-1:\sum C^{2,-1}_{1,n;1,m}{C_n^{2}}^\dagger {\widetilde C_m^{2}} = \frac{\sqrt{2}}{2}{c^\dagger_{1}c^\dagger_{-1}}{\tilde c_{0}\tilde c_{-1}} + \frac{\sqrt{2}}{2}{c^\dagger_{0}c^\dagger_{-1}}{\tilde c_{1}\tilde c_{-1}}\\
    &\quad q=0:\sum C^{2,0}_{1,n;1,m}{C_n^{2}}^\dagger {\widetilde C_m^{2}} = \frac{\sqrt{6}}{6}{c^\dagger_{1}c^\dagger_{0}}{\tilde c_{0}\tilde c_{-1}} + \frac{\sqrt{6}}{3}{c^\dagger_{1}c^\dagger_{-1}}{\tilde c_{1}\tilde c_{-1}} + \frac{\sqrt{6}}{6}{c^\dagger_{0}c^\dagger_{-1}}{\tilde c_{1}\tilde c_{0}}\\
    &\quad q=1:\sum C^{2,1}_{1,n;1,m}{C_n^{2}}^\dagger {\widetilde C_m^{2}} = \frac{\sqrt{2}}{2}{c^\dagger_{1}c^\dagger_{0}}{\tilde c_{1}\tilde c_{-1}} + \frac{\sqrt{2}}{2}{c^\dagger_{1}c^\dagger_{-1}}{\tilde c_{1}\tilde c_{0}}\\
    &\quad q=2:\sum C^{2,2}_{1,n;1,m}{C_n^{2}}^\dagger {\widetilde C_m^{2}} = {c^\dagger_{1}c^\dagger_{0}}{\tilde c_{1}\tilde c_{0}}
\end{align}

\subsubsection*{grade 5}
\begin{align}
  &\underline{e^{-i\phi}\chi_{1}:\sum C^{1,q}_{0,n;1,m}{C_n^{3}}^\dagger {\widetilde C_m^{2}}} \notag\\
  &\quad q=-1:\sum C^{1,-1}_{0,n;1,m}{C_n^{3}}^\dagger {\widetilde C_m^{2}} = {c^\dagger_{1}c^\dagger_{0}c^\dagger_{-1}}{\tilde c_{0}\tilde c_{-1}}\\
  &\quad q=0:\sum C^{1,0}_{0,n;1,m}{C_n^{3}}^\dagger {\widetilde C_m^{2}} = {c^\dagger_{1}c^\dagger_{0}c^\dagger_{-1}}{\tilde c_{1}\tilde c_{-1}}\\
  &\quad q=1:\sum C^{1,1}_{0,n;1,m}{C_n^{3}}^\dagger {\widetilde C_m^{2}} = {c^\dagger_{1}c^\dagger_{0}c^\dagger_{-1}}{\tilde c_{1}\tilde c_{0}}\\
  &\underline{e^{i\phi}\chi_{1}:\sum C^{1,q}_{1,n;0,m}{C_n^{2}}^\dagger {\widetilde C_m^{3}}} \notag\\
    &\quad q=-1:\sum C^{1,-1}_{1,n;0,m}{C_n^{2}}^\dagger {\widetilde C_m^{3}} = {c^\dagger_{0}c^\dagger_{-1}}{\tilde c_{1}\tilde c_{0}\tilde c_{-1}}\\
  &\quad q=0:\sum C^{1,0}_{1,n;0,m}{C_n^{2}}^\dagger {\widetilde C_m^{3}} = {c^\dagger_{1}c^\dagger_{-1}}{\tilde c_{1}\tilde c_{0}\tilde c_{-1}}\\
  &\quad q=1:\sum C^{1,1}_{1,n;0,m}{C_n^{2}}^\dagger {\widetilde C_m^{3}} = {c^\dagger_{1}c^\dagger_{0}}{\tilde c_{1}\tilde c_{0}\tilde c_{-1}}
\end{align}

\subsubsection*{grade 6}
\begin{align}
  &\underline{\chi_{0}:\sum C^{0,q}_{0,n;0,m}{C_n^{3}}^\dagger {\widetilde C_m^{3}}} \notag\\
  &\quad q=0:\sum C^{0,0}_{0,n;0,m}{C_n^{3}}^\dagger {\widetilde C_m^{3}} = {c^\dagger_{1}c^\dagger_{0}c^\dagger_{-1}}{\tilde c_{1}\tilde c_{0}\tilde c_{-1}}
\end{align}
\subsection{Real (tesseral) basis}
Given an irreducible spherical-tensor operator $O_{k,q}$ in the complex basis, which is in general non-Hermitian,
we construct a real basis by forming Hermitian operators with definite time-reversal parity.
A convenient procedure is as follows:
(i) first form Hermitian (and anti-Hermitian) combinations, and
(ii) then project them onto the $\mathcal T$-even and $\mathcal T$-odd sectors.
Explicitly,
\begin{align}
  &\mathcal T\text{-even}:\qquad
  (O_{k,q}+O_{k,q}^\dagger)
  +\mathcal T\,(O_{k,q}+O_{k,q}^\dagger)\,\mathcal T^{-1},
  \label{eq:tesseral_even1}
  \\
  &\mathcal T\text{-even}:\qquad
  i\,(O_{k,q}-O_{k,q}^\dagger)
  +\mathcal T\!\left[i\,(O_{k,q}-O_{k,q}^\dagger)\right]\mathcal T^{-1},
  \label{eq:tesseral_even2}
  \\
  &\mathcal T\text{-odd}:\qquad
  (O_{k,q}+O_{k,q}^\dagger)
  -\mathcal T\,(O_{k,q}+O_{k,q}^\dagger)\,\mathcal T^{-1},
  \label{eq:tesseral_odd1}
  \\
  &\mathcal T\text{-odd}:\qquad
  i\,(O_{k,q}-O_{k,q}^\dagger)
  -\mathcal T\!\left[i\,(O_{k,q}-O_{k,q}^\dagger)\right]\mathcal T^{-1}.
  \label{eq:tesseral_odd2}
\end{align}
If $O_{k,q}$ is already Hermitian and has a definite time-reversal parity, these projections are redundant, and no additional ``time-reversal partner'' needs to be introduced.

\medskip
\noindent
\textbf{Example 1: rank-1 one-body operators (grade 2)}\quad\\
As a first example, consider the rank-1 operator
$\sum_{n,m} C^{1,q}_{1,n;\,1,m}\,c_n^\dagger \tilde c_m$.
One finds
\begin{align}
 O_{1,-1 } &= {c^\dagger_{0}}{\tilde c_{-1}} -  {c^\dagger_{-1}}{\tilde c_{0}}
      = {c^\dagger_{0}}{ c_{1}} + {c^\dagger_{-1}}{ c_{0}},\\
 O_{1,  0} &=  {c^\dagger_{1}}{\tilde c_{-1}} -  {c^\dagger_{-1}}{\tilde c_{1}}
   =  {c^\dagger_{1}}{ c_{1}} -  {c^\dagger_{-1}}{ c_{-1}},\\
 O_{1,  1} &=  {c^\dagger_{1}}{\tilde c_{0}} -  {c^\dagger_{0}}{\tilde c_{1}}
    =  - {c^\dagger_{1}}{ c_{0}} -  {c^\dagger_{0}}{ c_{-1}}.
\end{align}
Here, the normalization is omitted.
Their Hermitian conjugates and time-reversal transforms satisfy
\begin{align}
 O_{1,-1 }^\dagger  = - O_{1,  1},\qquad
 O_{1,  0}^\dagger &=  O_{1,  0},\qquad
 O_{1,  1}^\dagger  = - O_{1, -1},\\
 \mathcal{T}\, O_{1,q}\, \mathcal{T}^{-1} &= - O_{1,-q}.
\end{align}
In this case, the independent Hermitian operators are all $\mathcal T$-odd, namely
\begin{align}
  M_x:O_{1,1}-O_{1,-1},\qquad
  M_y:i\,(O_{1,1}+O_{1,-1}),\qquad
  M_z:O_{1,0}.
\end{align}

\medskip
\noindent
\textbf{Example 2: rank-1 particle number non-conserving operators}\quad\\
Next, consider the rank-1 particle number non-conserving operators:
\begin{align}
  P_{1,-1} &= {c^\dagger_{0}}{c^\dagger_{-1}},\qquad
  &P_{1,0} &= {c^\dagger_{1}}{c^\dagger_{-1}},\qquad
  &P_{1,1} &= {c^\dagger_{1}}{c^\dagger_{0}},\\
  {\bar P}_{1,-1}&={\tilde c_{0}}{\tilde c_{-1}} = -c_0 c_{1},\qquad
  &{\bar P}_{1,0}&={\tilde c_{1}}{\tilde c_{-1}} = c_{-1}c_1,\qquad
  &{\bar P}_{1,1}&={\tilde c_{1}}{\tilde c_{0}} = -c_{-1}c_0.
\end{align}
For $q=\pm 1$, the real ($\mathcal T$-even/$\mathcal T$-odd) basis becomes
\begin{align*}
  &G_y:P_{1,1} + P_{1,1}^\dagger + \mathcal T\,(P_{1,1}+P_{1,1}^\dagger)\,\mathcal T^{-1}
  = P_{1,1} - {\bar P}_{1,-1} + P_{1,-1} - {\bar P}_{1,1},\\
  &G_x:i\,(P_{1,1} - P_{1,1}^\dagger) + \mathcal T\!\left[i\,(P_{1,1}-P_{1,1}^\dagger)\right]\mathcal T^{-1}
  = i\left(P_{1,1} + {\bar P}_{1,-1} - P_{1,-1} - {\bar P}_{1,1}\right),\\
  &M_x:P_{1,1} + P_{1,1}^\dagger - \mathcal T\,(P_{1,1}+P_{1,1}^\dagger)\,\mathcal T^{-1}
  = P_{1,1} - {\bar P}_{1,-1} - P_{1,-1} + {\bar P}_{1,1},\\
  &M_y:i\,(P_{1,1} - P_{1,1}^\dagger) - \mathcal T\!\left[i\,(P_{1,1}-P_{1,1}^\dagger)\right]\mathcal T^{-1}
  = i\left(P_{1,1} + {\bar P}_{1,-1} + P_{1,-1} + {\bar P}_{1,1}\right).
\end{align*}
For $q=0$, one obtains
\begin{align*}
  &G_z:i\,(P_{1,0} - P_{1,0}^\dagger) + \mathcal T\!\left[i\,(P_{1,0}-P_{1,0}^\dagger)\right]\mathcal T^{-1}
  = i\left(P_{1,0} - {\bar P}_{1,0} + P_{1,0} - {\bar P}_{1,0}\right),\\
  &M_z:P_{1,0} + P_{1,0}^\dagger - \mathcal T\,(P_{1,0}+P_{1,0}^\dagger)\,\mathcal T^{-1}
  = P_{1,0} + {\bar P}_{1,0} + P_{1,0} + {\bar P}_{1,0},\\
  &\qquad P_{1,0} + P_{1,0}^\dagger + \mathcal T\,(P_{1,0}+P_{1,0}^\dagger)\,\mathcal T^{-1}
  = 0,\\
  &\qquad i\,(P_{1,0} - P_{1,0}^\dagger) - \mathcal T\!\left[i\,(P_{1,0}-P_{1,0}^\dagger)\right]\mathcal T^{-1}
  = 0.
\end{align*}
In this way, irreducible operators in the complex basis can be systematically converted into a real (tesseral) basis.
% 
% 
% 
% 
% 
% 
% 
% 
% 
% 
% 
% %
\newpage 
\section{Two-body hybrid-monopoles}
A one-body multipole can be written as
\begin{align}
O_{k,q}^{ab}
&= \frac{i^{\frac{1-(-1)^{k+t}}{2}}}{2}
\left(
[c^\dagger_a \times \tilde c_b]_{k,q}
+ (-1)^{k+t} (-1)^{l_a+l_b}\,[c^\dagger_b \times \tilde c_a]_{k,q}
\right),
\end{align}
where $t=0$ for electric-type and $t=1$ for magnetic-type multipoles.
For even rank $k$, we obtain
\begin{align}
Q_{k,q}^{ab},\,G_{k,q}^{ab}
&= \frac{1}{2}
\left(
[c^\dagger_a \times \tilde c_b]_{k,q}
+ (-1)^{l_a+l_b}\,[c^\dagger_b \times \tilde c_a]_{k,q}
\right),\\
M_{k,q}^{ab},\,T_{k,q}^{ab}
&= \frac{i}{2}
\left(
[c^\dagger_a \times \tilde c_b]_{k,q}
- (-1)^{l_a+l_b}\,[c^\dagger_b \times \tilde c_a]_{k,q}
\right).
\end{align}
Whether the multipole is classified as $Q$ or $G$, and as $M$ or $T$, is determined by the orbital parities of $a$ and $b$.
For odd rank $k$, we instead have
\begin{align}
M_{k,q}^{ab},\,T_{k,q}^{ab}
&= \frac{1}{2}
\left(
[c^\dagger_a \times \tilde c_b]_{k,q}
+ (-1)^{l_a+l_b}\,[c^\dagger_b \times \tilde c_a]_{k,q}
\right),\\
Q_{k,q}^{ab},\,G_{k,q}^{ab}
&= \frac{i}{2}
\left(
[c^\dagger_a \times \tilde c_b]_{k,q}
- (-1)^{l_a+l_b}\,[c^\dagger_b \times \tilde c_a]_{k,q}
\right).
\end{align}

Two-body operators can be constructed by successive CG couplings.
Here we focus on two-body monopoles of the form
\begin{align}
  O^{(2)}_{0}
  := \bigl[[c^\dagger \times c^\dagger]\times[c \times c]\bigr]_{0}.
\end{align}
In general, $O^{(2)}_{0}$ is not Hermitian.
We therefore construct Hermitian operators by taking the real and imaginary parts,
\begin{align}
  \mathcal T\text{-even}:\quad
  O^{(2)}_{0,+} &:= \frac{1}{2}\Bigl(O^{(2)}_{0}+ \text{h.c.}\Bigr),\\
  \mathcal T\text{-odd}:\quad
  O^{(2)}_{0,-} &:= \frac{i}{2}\Bigl(O^{(2)}_{0}- \text{h.c.}\Bigr).
\end{align}
We use Wigner-$9j$ symbols to recouple the four-fermion tensor product and decompose $O^{(2)}_{0}$
into products of one-body irreducible tensor operators.
% , including the antisymmetry signs.
% 
Below we summarize the two-body hybrid monopoles and their one-body decompositions.

\subsection*{$sp$ system}
\begin{align*}
G_0^{(2)} &=  
\frac{1}{2} \left(\left[[c^\dagger_s \times c^\dagger_p]_{1} \times[\tilde c_p \times \tilde c_p]_{1}\right]_{0} + h.c. \right)\\
% &= \left[T^{sp}_{1} \times M_1^{p }\right]_{0}\notag \\ 
&= {\bm T}^{sp}_{1} \cdot {\bm M}_1^{p }\notag \\ 
&=
\frac{\sqrt{6}}{6}\Big(c^\dagger_{s,0}\,c^\dagger_{p,-1}\,c_{p,0}\,c_{p,-1} + c^\dagger_{s,0}\,c^\dagger_{p,0}\,c_{p,1}\,c_{p,-1} + c^\dagger_{s,0}\,c^\dagger_{p,1}\,c_{p,1}\,c_{p,0} \\
&\qquad + c^\dagger_{p,0}\,c^\dagger_{p,-1}\,c_{s,0}\,c_{p,-1} + c^\dagger_{p,1}\,c^\dagger_{p,-1}\,c_{s,0}\,c_{p,0} + c^\dagger_{p,1}\,c^\dagger_{p,0}\,c_{s,0}\,c_{p,1}\Big)\\
M_0^{(2)} &=  
\frac{i}{2} \left(\left[[c^\dagger_s \times c^\dagger_p]_{1} \times[\tilde c_p \times \tilde c_p]_{1}\right]_{0} - h.c. \right)\\
% &= \left[Q^{sp}_{1} \times M_1^{p }\right]_{0}\notag \\ 
&= {\bm Q}^{sp}_{1} \cdot {\bm M}_1^{p }\notag \\ 
&=
\frac{\sqrt{6}\, i}{6}
\Big(c^\dagger_{s,0}\,c^\dagger_{p,-1}\,c_{p,0}\,c_{p,-1} + c^\dagger_{s,0}\,c^\dagger_{p,0}\,c_{p,1}\,c_{p,-1} + c^\dagger_{s,0}\,c^\dagger_{p,1}\,c_{p,1}\,c_{p,0} \\
&\qquad - c^\dagger_{p,0}\,c^\dagger_{p,-1}\,c_{s,0}\,c_{p,-1} - c^\dagger_{p,1}\,c^\dagger_{p,-1}\,c_{s,0}\,c_{p,0} - c^\dagger_{p,1}\,c^\dagger_{p,0}\,c_{s,0}\,c_{p,1}\Big)
\end{align*}
\subsection*{$sf$ system}
\begin{align*}
G_0^{(2)} &=  
\frac{1}{2} \left(\left[[c^\dagger_s \times c^\dagger_f]_{3} \times[\tilde c_f \times \tilde c_f]_{3}\right]_{0} + h.c. \right)\\
% &= \left[T^{sf}_{3} \times M_3^{f}\right]_{0}\notag \\ 
&= {\bm T}_3^{sf} \cdot {\bm M}_3^{f}\notag \\ 
&= - \frac{\sqrt{21}}{21}\Big(- \frac{\sqrt{2}}{2}\,c^\dagger_{s,0}\,c^\dagger_{f,-3}\,c_{f,0}\,c_{f,-3} + c^\dagger_{s,0}\,c^\dagger_{f,-3}\,c_{f,-1}\,c_{f,-2} - c^\dagger_{s,0}\,c^\dagger_{f,-2}\,c_{f,1}\,c_{f,-3}
\\
&\qquad + \frac{\sqrt{2}}{2}\,c^\dagger_{s,0}\,c^\dagger_{f,-2}\,c_{f,0}\,c_{f,-2} - c^\dagger_{s,0}\,c^\dagger_{f,-1}\,c_{f,2}\,c_{f,-3} + \frac{\sqrt{2}}{2}\,c^\dagger_{s,0}\,c^\dagger_{f,-1}\,c_{f,0}\,c_{f,-1}
\\
&\qquad - \frac{\sqrt{2}}{2}\,c^\dagger_{s,0}\,c^\dagger_{f,0}\,c_{f,3}\,c_{f,-3} - \frac{\sqrt{2}}{2}\,c^\dagger_{s,0}\,c^\dagger_{f,0}\,c_{f,2}\,c_{f,-2} + \frac{\sqrt{2}}{2}\,c^\dagger_{s,0}\,c^\dagger_{f,0}\,c_{f,1}\,c_{f,-1}
\\
&\qquad - c^\dagger_{s,0}\,c^\dagger_{f,1}\,c_{f,3}\,c_{f,-2} + \frac{\sqrt{2}}{2}\,c^\dagger_{s,0}\,c^\dagger_{f,1}\,c_{f,1}\,c_{f,0} - c^\dagger_{s,0}\,c^\dagger_{f,2}\,c_{f,3}\,c_{f,-1}
\\
&\qquad + \frac{\sqrt{2}}{2}\,c^\dagger_{s,0}\,c^\dagger_{f,2}\,c_{f,2}\,c_{f,0} - \frac{\sqrt{2}}{2}\,c^\dagger_{s,0}\,c^\dagger_{f,3}\,c_{f,3}\,c_{f,0} + c^\dagger_{s,0}\,c^\dagger_{f,3}\,c_{f,2}\,c_{f,1}
\\
&\qquad - \frac{\sqrt{2}}{2}\,c^\dagger_{f,0}\,c^\dagger_{f,-3}\,c_{s,0}\,c_{f,-3} + c^\dagger_{f,-1}\,c^\dagger_{f,-2}\,c_{s,0}\,c_{f,-3} - c^\dagger_{f,1}\,c^\dagger_{f,-3}\,c_{s,0}\,c_{f,-2}
\\
&\qquad + \frac{\sqrt{2}}{2}\,c^\dagger_{f,0}\,c^\dagger_{f,-2}\,c_{s,0}\,c_{f,-2} - c^\dagger_{f,2}\,c^\dagger_{f,-3}\,c_{s,0}\,c_{f,-1} + \frac{\sqrt{2}}{2}\,c^\dagger_{f,0}\,c^\dagger_{f,-1}\,c_{s,0}\,c_{f,-1}
\\
&\qquad - \frac{\sqrt{2}}{2}\,c^\dagger_{f,3}\,c^\dagger_{f,-3}\,c_{s,0}\,c_{f,0} - \frac{\sqrt{2}}{2}\,c^\dagger_{f,2}\,c^\dagger_{f,-2}\,c_{s,0}\,c_{f,0} + \frac{\sqrt{2}}{2}\,c^\dagger_{f,1}\,c^\dagger_{f,-1}\,c_{s,0}\,c_{f,0}
\\
&\qquad - c^\dagger_{f,3}\,c^\dagger_{f,-2}\,c_{s,0}\,c_{f,1} + \frac{\sqrt{2}}{2}\,c^\dagger_{f,1}\,c^\dagger_{f,0}\,c_{s,0}\,c_{f,1} - c^\dagger_{f,3}\,c^\dagger_{f,-1}\,c_{s,0}\,c_{f,2}
\\
&\qquad + \frac{\sqrt{2}}{2}\,c^\dagger_{f,2}\,c^\dagger_{f,0}\,c_{s,0}\,c_{f,2} - \frac{\sqrt{2}}{2}\,c^\dagger_{f,3}\,c^\dagger_{f,0}\,c_{s,0}\,c_{f,3} + c^\dagger_{f,2}\,c^\dagger_{f,1}\,c_{s,0}\,c_{f,3}\Big)
\\
M_0^{(2)} &=  
\frac{i}{2} \left(\left[[c^\dagger_s \times c^\dagger_f]_{3} \times[\tilde c_f \times \tilde c_f]_{3}\right]_{0} - h.c. \right)\\
% &= \left[Q^{sf}_{3} \times M_3^{f}\right]_{0}\notag \\ 
&= {\bm Q}_3^{sf} \cdot {\bm M}_3^{f}\notag \\ 
&=
- \frac{\sqrt{21} i}{21}\Big(- \frac{\sqrt{2}}{2}\,c^\dagger_{s,0}\,c^\dagger_{f,-3}\,c_{f,0}\,c_{f,-3} + c^\dagger_{s,0}\,c^\dagger_{f,-3}\,c_{f,-1}\,c_{f,-2} - c^\dagger_{s,0}\,c^\dagger_{f,-2}\,c_{f,1}\,c_{f,-3}
\\
&\qquad + \frac{\sqrt{2}}{2}\,c^\dagger_{s,0}\,c^\dagger_{f,-2}\,c_{f,0}\,c_{f,-2} - c^\dagger_{s,0}\,c^\dagger_{f,-1}\,c_{f,2}\,c_{f,-3} + \frac{\sqrt{2}}{2}\,c^\dagger_{s,0}\,c^\dagger_{f,-1}\,c_{f,0}\,c_{f,-1}
\\
&\qquad - \frac{\sqrt{2}}{2}\,c^\dagger_{s,0}\,c^\dagger_{f,0}\,c_{f,3}\,c_{f,-3} - \frac{\sqrt{2}}{2}\,c^\dagger_{s,0}\,c^\dagger_{f,0}\,c_{f,2}\,c_{f,-2} + \frac{\sqrt{2}}{2}\,c^\dagger_{s,0}\,c^\dagger_{f,0}\,c_{f,1}\,c_{f,-1}
\\
&\qquad - c^\dagger_{s,0}\,c^\dagger_{f,1}\,c_{f,3}\,c_{f,-2} + \frac{\sqrt{2}}{2}\,c^\dagger_{s,0}\,c^\dagger_{f,1}\,c_{f,1}\,c_{f,0} - c^\dagger_{s,0}\,c^\dagger_{f,2}\,c_{f,3}\,c_{f,-1}
\\
&\qquad + \frac{\sqrt{2}}{2}\,c^\dagger_{s,0}\,c^\dagger_{f,2}\,c_{f,2}\,c_{f,0} - \frac{\sqrt{2}}{2}\,c^\dagger_{s,0}\,c^\dagger_{f,3}\,c_{f,3}\,c_{f,0} + c^\dagger_{s,0}\,c^\dagger_{f,3}\,c_{f,2}\,c_{f,1}
\\
&\qquad + \frac{\sqrt{2}}{2}\,c^\dagger_{f,0}\,c^\dagger_{f,-3}\,c_{s,0}\,c_{f,-3} - c^\dagger_{f,-1}\,c^\dagger_{f,-2}\,c_{s,0}\,c_{f,-3} + c^\dagger_{f,1}\,c^\dagger_{f,-3}\,c_{s,0}\,c_{f,-2}
\\
&\qquad - \frac{\sqrt{2}}{2}\,c^\dagger_{f,0}\,c^\dagger_{f,-2}\,c_{s,0}\,c_{f,-2} + c^\dagger_{f,2}\,c^\dagger_{f,-3}\,c_{s,0}\,c_{f,-1} - \frac{\sqrt{2}}{2}\,c^\dagger_{f,0}\,c^\dagger_{f,-1}\,c_{s,0}\,c_{f,-1}
\\
&\qquad + \frac{\sqrt{2}}{2}\,c^\dagger_{f,3}\,c^\dagger_{f,-3}\,c_{s,0}\,c_{f,0} + \frac{\sqrt{2}}{2}\,c^\dagger_{f,2}\,c^\dagger_{f,-2}\,c_{s,0}\,c_{f,0} - \frac{\sqrt{2}}{2}\,c^\dagger_{f,1}\,c^\dagger_{f,-1}\,c_{s,0}\,c_{f,0}
\\
&\qquad + c^\dagger_{f,3}\,c^\dagger_{f,-2}\,c_{s,0}\,c_{f,1} - \frac{\sqrt{2}}{2}\,c^\dagger_{f,1}\,c^\dagger_{f,0}\,c_{s,0}\,c_{f,1} + c^\dagger_{f,3}\,c^\dagger_{f,-1}\,c_{s,0}\,c_{f,2}
\\
&\qquad - \frac{\sqrt{2}}{2}\,c^\dagger_{f,2}\,c^\dagger_{f,0}\,c_{s,0}\,c_{f,2} + \frac{\sqrt{2}}{2}\,c^\dagger_{f,3}\,c^\dagger_{f,0}\,c_{s,0}\,c_{f,3} - c^\dagger_{f,2}\,c^\dagger_{f,1}\,c_{s,0}\,c_{f,3}\Big)
% \\
% 
% 
\end{align*}
\subsection*{$pd$ system}
\begin{align*}
Q_0^{(2)} 
&=  
  \frac{1}{2} \left(\left[[c^\dagger_p \times c^\dagger_p]_{1} \times[\tilde c_d \times \tilde c_d]_{1}\right]_{0} + h.c. \right)\\
&=
    -\frac{3\sqrt{5}}{10}
    \left( 
    {\bm T}^{pd}_1 \cdot {\bm T}_1^{pd}
    -
    {\bm Q}^{pd}_1 \cdot {\bm Q}_1^{pd}
    \right)    
    -\frac{\sqrt{3}}{6}
    \left( 
    {\bm G}^{pd}_2 \cdot {\bm G}_2^{pd}
    -
    {\bm M}^{pd}_2 \cdot {\bm M}_2^{pd}
    \right)
    \\&\qquad 
    +\frac{\sqrt{105}}{15}
    \left( 
    {\bm T}^{pd}_3 \cdot {\bm T}_3^{pd}
    -
    {\bm Q}^{pd}_3 \cdot {\bm Q}_3^{pd}
    \right)
  \\
&=
\frac{2 \sqrt{15}}{15}\Big(\frac{\sqrt{2}}{2}\,c^\dagger_{p,0}\,c^\dagger_{p,-1}\,c_{d,1}\,c_{d,-2} - \frac{\sqrt{3}}{2}\,c^\dagger_{p,0}\,c^\dagger_{p,-1}\,c_{d,0}\,c_{d,-1} + c^\dagger_{p,1}\,c^\dagger_{p,-1}\,c_{d,2}\,c_{d,-2}
\\
&\qquad - \frac{1}{2}\,c^\dagger_{p,1}\,c^\dagger_{p,-1}\,c_{d,1}\,c_{d,-1} + \frac{\sqrt{2}}{2}\,c^\dagger_{p,1}\,c^\dagger_{p,0}\,c_{d,2}\,c_{d,-1} - \frac{\sqrt{3}}{2}\,c^\dagger_{p,1}\,c^\dagger_{p,0}\,c_{d,1}\,c_{d,0}
\\
&\qquad + \frac{\sqrt{2}}{2}\,c^\dagger_{d,1}\,c^\dagger_{d,-2}\,c_{p,0}\,c_{p,-1} - \frac{\sqrt{3}}{2}\,c^\dagger_{d,0}\,c^\dagger_{d,-1}\,c_{p,0}\,c_{p,-1} + c^\dagger_{d,2}\,c^\dagger_{d,-2}\,c_{p,1}\,c_{p,-1}
\\
&\qquad - \frac{1}{2}\,c^\dagger_{d,1}\,c^\dagger_{d,-1}\,c_{p,1}\,c_{p,-1} + \frac{\sqrt{2}}{2}\,c^\dagger_{d,2}\,c^\dagger_{d,-1}\,c_{p,1}\,c_{p,0} - \frac{\sqrt{3}}{2}\,c^\dagger_{d,1}\,c^\dagger_{d,0}\,c_{p,1}\,c_{p,0}\Big)
\\
  T_0^{(2)} 
&=  
  \frac{i}{2} \left(\left[[c^\dagger_p \times c^\dagger_p]_{1} \times[\tilde c_d \times \tilde c_d]_{1}\right]_{0} - h.c. \right)\\
&=
    -\frac{3\sqrt{5}}{10}
    \left( 
    {\bm Q}^{pd}_1 \cdot {\bm T}_1^{pd}
    +
    {\bm T}^{pd}_1 \cdot {\bm Q}_1^{pd}
    \right)    
    -\frac{\sqrt{3}}{6}
    \left( 
    {\bm M}^{pd}_2 \cdot {\bm G}_2^{pd}
    +
    {\bm G}^{pd}_2 \cdot {\bm M}_2^{pd}
    \right)
    \\&\qquad 
    +\frac{\sqrt{105}}{15}
    \left( 
    {\bm Q}^{pd}_3 \cdot {\bm T}_3^{pd}
    +
    {\bm T}^{pd}_3 \cdot {\bm Q}_3^{pd}
    \right)
\\
% &=
%     -\frac{3\sqrt{5}}{5}
%     \left[Q^{pd}_1 \times T_1^{pd}\right]_{0} 
%     -\frac{\sqrt{3}}{3}
%     \left[M^{pd}_2 \times G_2^{pd}\right]_{0} 
%     +\frac{2\sqrt{105}}{15}
%     \left[Q^{pd}_3 \times T_3^{pd}\right]_{0} 
% \\
&=
\frac{2 \sqrt{15} i}{15}\Big(\frac{\sqrt{2}}{2}\,c^\dagger_{p,0}\,c^\dagger_{p,-1}\,c_{d,1}\,c_{d,-2} - \frac{\sqrt{3}}{2}\,c^\dagger_{p,0}\,c^\dagger_{p,-1}\,c_{d,0}\,c_{d,-1} + c^\dagger_{p,1}\,c^\dagger_{p,-1}\,c_{d,2}\,c_{d,-2}
\\
&\qquad - \frac{1}{2}\,c^\dagger_{p,1}\,c^\dagger_{p,-1}\,c_{d,1}\,c_{d,-1} + \frac{\sqrt{2}}{2}\,c^\dagger_{p,1}\,c^\dagger_{p,0}\,c_{d,2}\,c_{d,-1} - \frac{\sqrt{3}}{2}\,c^\dagger_{p,1}\,c^\dagger_{p,0}\,c_{d,1}\,c_{d,0}
\\
&\qquad - \frac{\sqrt{2}}{2}\,c^\dagger_{d,1}\,c^\dagger_{d,-2}\,c_{p,0}\,c_{p,-1} + \frac{\sqrt{3}}{2}\,c^\dagger_{d,0}\,c^\dagger_{d,-1}\,c_{p,0}\,c_{p,-1} - c^\dagger_{d,2}\,c^\dagger_{d,-2}\,c_{p,1}\,c_{p,-1}
\\
&\qquad + \frac{1}{2}\,c^\dagger_{d,1}\,c^\dagger_{d,-1}\,c_{p,1}\,c_{p,-1} - \frac{\sqrt{2}}{2}\,c^\dagger_{d,2}\,c^\dagger_{d,-1}\,c_{p,1}\,c_{p,0} + \frac{\sqrt{3}}{2}\,c^\dagger_{d,1}\,c^\dagger_{d,0}\,c_{p,1}\,c_{p,0}\Big)
\end{align*}

% ppdd : 1×1
% j13,j24,A wigner_9j： 1 , 1 , 3*sqrt(5)/10
% j13,j24,A wigner_9j： 2 , 2 , sqrt(3)/6
% j13,j24,A wigner_9j： 3 , 3 , -sqrt(105)/15

\begin{align*}
{G_0^{(2)}}_{[1]} &=  
\frac{1}{2} \left(\left[[c^\dagger_p \times c^\dagger_d]_{1} \times[\tilde c_d \times \tilde c_d]_{1}\right]_{0} + h.c. \right)\\
&= 
\frac{3}{10} {\bm T}^{pd}_1 \cdot {\bm M}_1^{d}
+ \frac{\sqrt{35}}{10} {\bm G}^{pd}_2 \cdot {\bm Q}_2^{d}
+ \frac{\sqrt{14}}{5}  {\bm T}^{pd}_3 \cdot {\bm M}_3^{d}\\ 
&=
- \frac{\sqrt{6}}{10}\Big(- \frac{1}{3}\,c^\dagger_{p,-1}\,c^\dagger_{d,0}\,c_{d,1}\,c_{d,-2} + \frac{\sqrt{6}}{6}\,c^\dagger_{p,-1}\,c^\dagger_{d,0}\,c_{d,0}\,c_{d,-1} + \frac{\sqrt{3}}{3}\,c^\dagger_{p,0}\,c^\dagger_{d,-1}\,c_{d,1}\,c_{d,-2}
\\
&\qquad - \frac{\sqrt{2}}{2}\,c^\dagger_{p,0}\,c^\dagger_{d,-1}\,c_{d,0}\,c_{d,-1} - \frac{\sqrt{6}}{3}\,c^\dagger_{p,1}\,c^\dagger_{d,-2}\,c_{d,1}\,c_{d,-2} + c^\dagger_{p,1}\,c^\dagger_{d,-2}\,c_{d,0}\,c_{d,-1}
\\
&\qquad - \frac{\sqrt{6}}{3}\,c^\dagger_{p,-1}\,c^\dagger_{d,1}\,c_{d,2}\,c_{d,-2} + \frac{\sqrt{6}}{6}\,c^\dagger_{p,-1}\,c^\dagger_{d,1}\,c_{d,1}\,c_{d,-1} + \frac{2 \sqrt{2}}{3}\,c^\dagger_{p,0}\,c^\dagger_{d,0}\,c_{d,2}\,c_{d,-2}
\\
&\qquad - \frac{\sqrt{2}}{3}\,c^\dagger_{p,0}\,c^\dagger_{d,0}\,c_{d,1}\,c_{d,-1} - \frac{\sqrt{6}}{3}\,c^\dagger_{p,1}\,c^\dagger_{d,-1}\,c_{d,2}\,c_{d,-2} + \frac{\sqrt{6}}{6}\,c^\dagger_{p,1}\,c^\dagger_{d,-1}\,c_{d,1}\,c_{d,-1}
\\
&\qquad - \frac{\sqrt{6}}{3}\,c^\dagger_{p,-1}\,c^\dagger_{d,2}\,c_{d,2}\,c_{d,-1} + c^\dagger_{p,-1}\,c^\dagger_{d,2}\,c_{d,1}\,c_{d,0} + \frac{\sqrt{3}}{3}\,c^\dagger_{p,0}\,c^\dagger_{d,1}\,c_{d,2}\,c_{d,-1}
\\
&\qquad - \frac{\sqrt{2}}{2}\,c^\dagger_{p,0}\,c^\dagger_{d,1}\,c_{d,1}\,c_{d,0} - \frac{1}{3}\,c^\dagger_{p,1}\,c^\dagger_{d,0}\,c_{d,2}\,c_{d,-1} + \frac{\sqrt{6}}{6}\,c^\dagger_{p,1}\,c^\dagger_{d,0}\,c_{d,1}\,c_{d,0}
\\
&\qquad - \frac{1}{3}\,c^\dagger_{d,1}\,c^\dagger_{d,-2}\,c_{p,-1}\,c_{d,0} + \frac{\sqrt{6}}{6}\,c^\dagger_{d,0}\,c^\dagger_{d,-1}\,c_{p,-1}\,c_{d,0} + \frac{\sqrt{3}}{3}\,c^\dagger_{d,1}\,c^\dagger_{d,-2}\,c_{p,0}\,c_{d,-1}
\\
&\qquad - \frac{\sqrt{2}}{2}\,c^\dagger_{d,0}\,c^\dagger_{d,-1}\,c_{p,0}\,c_{d,-1} - \frac{\sqrt{6}}{3}\,c^\dagger_{d,1}\,c^\dagger_{d,-2}\,c_{p,1}\,c_{d,-2} + c^\dagger_{d,0}\,c^\dagger_{d,-1}\,c_{p,1}\,c_{d,-2}
\\
&\qquad - \frac{\sqrt{6}}{3}\,c^\dagger_{d,2}\,c^\dagger_{d,-2}\,c_{p,-1}\,c_{d,1} + \frac{\sqrt{6}}{6}\,c^\dagger_{d,1}\,c^\dagger_{d,-1}\,c_{p,-1}\,c_{d,1} + \frac{2 \sqrt{2}}{3}\,c^\dagger_{d,2}\,c^\dagger_{d,-2}\,c_{p,0}\,c_{d,0}
\\
&\qquad - \frac{\sqrt{2}}{3}\,c^\dagger_{d,1}\,c^\dagger_{d,-1}\,c_{p,0}\,c_{d,0} - \frac{\sqrt{6}}{3}\,c^\dagger_{d,2}\,c^\dagger_{d,-2}\,c_{p,1}\,c_{d,-1} + \frac{\sqrt{6}}{6}\,c^\dagger_{d,1}\,c^\dagger_{d,-1}\,c_{p,1}\,c_{d,-1}
\\
&\qquad - \frac{\sqrt{6}}{3}\,c^\dagger_{d,2}\,c^\dagger_{d,-1}\,c_{p,-1}\,c_{d,2} + c^\dagger_{d,1}\,c^\dagger_{d,0}\,c_{p,-1}\,c_{d,2} + \frac{\sqrt{3}}{3}\,c^\dagger_{d,2}\,c^\dagger_{d,-1}\,c_{p,0}\,c_{d,1}
\\
&\qquad - \frac{\sqrt{2}}{2}\,c^\dagger_{d,1}\,c^\dagger_{d,0}\,c_{p,0}\,c_{d,1} - \frac{1}{3}\,c^\dagger_{d,2}\,c^\dagger_{d,-1}\,c_{p,1}\,c_{d,0} + \frac{\sqrt{6}}{6}\,c^\dagger_{d,1}\,c^\dagger_{d,0}\,c_{p,1}\,c_{d,0}\Big)
\\
{M_0^{(2)}}_{[1]} &=  
\frac{i}{2} \left(\left[[c^\dagger_p \times c^\dagger_d]_{3} \times[\tilde c_d \times \tilde c_d]_{3}\right]_{0} - h.c. \right)\\
&= 
\frac{3}{10} {\bm Q}^{pd}_1 \cdot {\bm M}_1^{d}
+ \frac{\sqrt{35}}{10} {\bm M}^{pd}_2 \cdot {\bm Q}_2^{d}
+ \frac{\sqrt{14}}{5}  {\bm Q}^{pd}_3 \cdot {\bm M}_3^{d}\\ 
&=
- \frac{\sqrt{6} i}{10}\Big(- \frac{1}{3}\,c^\dagger_{p,-1}\,c^\dagger_{d,0}\,c_{d,1}\,c_{d,-2} + \frac{\sqrt{6}}{6}\,c^\dagger_{p,-1}\,c^\dagger_{d,0}\,c_{d,0}\,c_{d,-1} + \frac{\sqrt{3}}{3}\,c^\dagger_{p,0}\,c^\dagger_{d,-1}\,c_{d,1}\,c_{d,-2}
\\
&\qquad - \frac{\sqrt{2}}{2}\,c^\dagger_{p,0}\,c^\dagger_{d,-1}\,c_{d,0}\,c_{d,-1} - \frac{\sqrt{6}}{3}\,c^\dagger_{p,1}\,c^\dagger_{d,-2}\,c_{d,1}\,c_{d,-2} + c^\dagger_{p,1}\,c^\dagger_{d,-2}\,c_{d,0}\,c_{d,-1}
\\
&\qquad - \frac{\sqrt{6}}{3}\,c^\dagger_{p,-1}\,c^\dagger_{d,1}\,c_{d,2}\,c_{d,-2} + \frac{\sqrt{6}}{6}\,c^\dagger_{p,-1}\,c^\dagger_{d,1}\,c_{d,1}\,c_{d,-1} + \frac{2 \sqrt{2}}{3}\,c^\dagger_{p,0}\,c^\dagger_{d,0}\,c_{d,2}\,c_{d,-2}
\\
&\qquad - \frac{\sqrt{2}}{3}\,c^\dagger_{p,0}\,c^\dagger_{d,0}\,c_{d,1}\,c_{d,-1} - \frac{\sqrt{6}}{3}\,c^\dagger_{p,1}\,c^\dagger_{d,-1}\,c_{d,2}\,c_{d,-2} + \frac{\sqrt{6}}{6}\,c^\dagger_{p,1}\,c^\dagger_{d,-1}\,c_{d,1}\,c_{d,-1}
\\
&\qquad - \frac{\sqrt{6}}{3}\,c^\dagger_{p,-1}\,c^\dagger_{d,2}\,c_{d,2}\,c_{d,-1} + c^\dagger_{p,-1}\,c^\dagger_{d,2}\,c_{d,1}\,c_{d,0} + \frac{\sqrt{3}}{3}\,c^\dagger_{p,0}\,c^\dagger_{d,1}\,c_{d,2}\,c_{d,-1}
\\
&\qquad - \frac{\sqrt{2}}{2}\,c^\dagger_{p,0}\,c^\dagger_{d,1}\,c_{d,1}\,c_{d,0} - \frac{1}{3}\,c^\dagger_{p,1}\,c^\dagger_{d,0}\,c_{d,2}\,c_{d,-1} + \frac{\sqrt{6}}{6}\,c^\dagger_{p,1}\,c^\dagger_{d,0}\,c_{d,1}\,c_{d,0}
\\
&\qquad + \frac{1}{3}\,c^\dagger_{d,1}\,c^\dagger_{d,-2}\,c_{p,-1}\,c_{d,0} - \frac{\sqrt{6}}{6}\,c^\dagger_{d,0}\,c^\dagger_{d,-1}\,c_{p,-1}\,c_{d,0} - \frac{\sqrt{3}}{3}\,c^\dagger_{d,1}\,c^\dagger_{d,-2}\,c_{p,0}\,c_{d,-1}
\\
&\qquad + \frac{\sqrt{2}}{2}\,c^\dagger_{d,0}\,c^\dagger_{d,-1}\,c_{p,0}\,c_{d,-1} + \frac{\sqrt{6}}{3}\,c^\dagger_{d,1}\,c^\dagger_{d,-2}\,c_{p,1}\,c_{d,-2} - c^\dagger_{d,0}\,c^\dagger_{d,-1}\,c_{p,1}\,c_{d,-2}
\\
&\qquad + \frac{\sqrt{6}}{3}\,c^\dagger_{d,2}\,c^\dagger_{d,-2}\,c_{p,-1}\,c_{d,1} - \frac{\sqrt{6}}{6}\,c^\dagger_{d,1}\,c^\dagger_{d,-1}\,c_{p,-1}\,c_{d,1} - \frac{2 \sqrt{2}}{3}\,c^\dagger_{d,2}\,c^\dagger_{d,-2}\,c_{p,0}\,c_{d,0}
\\
&\qquad + \frac{\sqrt{2}}{3}\,c^\dagger_{d,1}\,c^\dagger_{d,-1}\,c_{p,0}\,c_{d,0} + \frac{\sqrt{6}}{3}\,c^\dagger_{d,2}\,c^\dagger_{d,-2}\,c_{p,1}\,c_{d,-1} - \frac{\sqrt{6}}{6}\,c^\dagger_{d,1}\,c^\dagger_{d,-1}\,c_{p,1}\,c_{d,-1}
\\
&\qquad + \frac{\sqrt{6}}{3}\,c^\dagger_{d,2}\,c^\dagger_{d,-1}\,c_{p,-1}\,c_{d,2} - c^\dagger_{d,1}\,c^\dagger_{d,0}\,c_{p,-1}\,c_{d,2} - \frac{\sqrt{3}}{3}\,c^\dagger_{d,2}\,c^\dagger_{d,-1}\,c_{p,0}\,c_{d,1}
\\
&\qquad + \frac{\sqrt{2}}{2}\,c^\dagger_{d,1}\,c^\dagger_{d,0}\,c_{p,0}\,c_{d,1} + \frac{1}{3}\,c^\dagger_{d,2}\,c^\dagger_{d,-1}\,c_{p,1}\,c_{d,0} - \frac{\sqrt{6}}{6}\,c^\dagger_{d,1}\,c^\dagger_{d,0}\,c_{p,1}\,c_{d,0}\Big)
% \\
% 
% 
\end{align*}

\begin{align*}
{G_0^{(2)}}_{[2]} &=  
\frac{1}{2} \left(\left[[c^\dagger_p \times c^\dagger_d]_{3} \times[\tilde c_d \times \tilde c_d]_{3}\right]_{0} + h.c. \right)\\
&= 
\frac{\sqrt{14}}{5} {\bm T}^{pd}_1 \cdot {\bm M}_1^{d}
- \frac{\sqrt{10}}{5} {\bm G}^{pd}_2 \cdot {\bm Q}_2^{d}
+ \frac{1}{5}  {\bm T}^{pd}_3 \cdot {\bm M}_3^{d}\\ 
&= 
\frac{\sqrt{14}}{14}\Big(c^\dagger_{p,-1}\,c^\dagger_{d,-2}\,c_{d,-1}\,c_{d,-2} + \frac{\sqrt{6}}{3}\,c^\dagger_{p,-1}\,c^\dagger_{d,-1}\,c_{d,0}\,c_{d,-2} + \frac{\sqrt{3}}{3}\,c^\dagger_{p,0}\,c^\dagger_{d,-2}\,c_{d,0}\,c_{d,-2}
\\
&\qquad + \frac{\sqrt{6}}{5}\,c^\dagger_{p,-1}\,c^\dagger_{d,0}\,c_{d,1}\,c_{d,-2} + \frac{2}{5}\,c^\dagger_{p,-1}\,c^\dagger_{d,0}\,c_{d,0}\,c_{d,-1} + \frac{2 \sqrt{2}}{5}\,c^\dagger_{p,0}\,c^\dagger_{d,-1}\,c_{d,1}\,c_{d,-2}
\\
&\qquad + \frac{4 \sqrt{3}}{15}\,c^\dagger_{p,0}\,c^\dagger_{d,-1}\,c_{d,0}\,c_{d,-1} + \frac{1}{5}\,c^\dagger_{p,1}\,c^\dagger_{d,-2}\,c_{d,1}\,c_{d,-2} + \frac{\sqrt{6}}{15}\,c^\dagger_{p,1}\,c^\dagger_{d,-2}\,c_{d,0}\,c_{d,-1}
\\
&\qquad + \frac{1}{5}\,c^\dagger_{p,-1}\,c^\dagger_{d,1}\,c_{d,2}\,c_{d,-2} + \frac{2}{5}\,c^\dagger_{p,-1}\,c^\dagger_{d,1}\,c_{d,1}\,c_{d,-1} + \frac{\sqrt{3}}{5}\,c^\dagger_{p,0}\,c^\dagger_{d,0}\,c_{d,2}\,c_{d,-2}
\\
&\qquad + \frac{2 \sqrt{3}}{5}\,c^\dagger_{p,0}\,c^\dagger_{d,0}\,c_{d,1}\,c_{d,-1} + \frac{1}{5}\,c^\dagger_{p,1}\,c^\dagger_{d,-1}\,c_{d,2}\,c_{d,-2} + \frac{2}{5}\,c^\dagger_{p,1}\,c^\dagger_{d,-1}\,c_{d,1}\,c_{d,-1}
\\
&\qquad + \frac{1}{5}\,c^\dagger_{p,-1}\,c^\dagger_{d,2}\,c_{d,2}\,c_{d,-1} + \frac{\sqrt{6}}{15}\,c^\dagger_{p,-1}\,c^\dagger_{d,2}\,c_{d,1}\,c_{d,0} + \frac{2 \sqrt{2}}{5}\,c^\dagger_{p,0}\,c^\dagger_{d,1}\,c_{d,2}\,c_{d,-1}
\\
&\qquad + \frac{4 \sqrt{3}}{15}\,c^\dagger_{p,0}\,c^\dagger_{d,1}\,c_{d,1}\,c_{d,0} + \frac{\sqrt{6}}{5}\,c^\dagger_{p,1}\,c^\dagger_{d,0}\,c_{d,2}\,c_{d,-1} + \frac{2}{5}\,c^\dagger_{p,1}\,c^\dagger_{d,0}\,c_{d,1}\,c_{d,0}
\\
&\qquad + \frac{\sqrt{3}}{3}\,c^\dagger_{p,0}\,c^\dagger_{d,2}\,c_{d,2}\,c_{d,0} + \frac{\sqrt{6}}{3}\,c^\dagger_{p,1}\,c^\dagger_{d,1}\,c_{d,2}\,c_{d,0} + c^\dagger_{p,1}\,c^\dagger_{d,2}\,c_{d,2}\,c_{d,1}
\\
&\qquad + c^\dagger_{d,-1}\,c^\dagger_{d,-2}\,c_{p,-1}\,c_{d,-2} + \frac{\sqrt{6}}{3}\,c^\dagger_{d,0}\,c^\dagger_{d,-2}\,c_{p,-1}\,c_{d,-1} + \frac{\sqrt{3}}{3}\,c^\dagger_{d,0}\,c^\dagger_{d,-2}\,c_{p,0}\,c_{d,-2}
\\
&\qquad + \frac{\sqrt{6}}{5}\,c^\dagger_{d,1}\,c^\dagger_{d,-2}\,c_{p,-1}\,c_{d,0} + \frac{2}{5}\,c^\dagger_{d,0}\,c^\dagger_{d,-1}\,c_{p,-1}\,c_{d,0} + \frac{2 \sqrt{2}}{5}\,c^\dagger_{d,1}\,c^\dagger_{d,-2}\,c_{p,0}\,c_{d,-1}
\\
&\qquad + \frac{4 \sqrt{3}}{15}\,c^\dagger_{d,0}\,c^\dagger_{d,-1}\,c_{p,0}\,c_{d,-1} + \frac{1}{5}\,c^\dagger_{d,1}\,c^\dagger_{d,-2}\,c_{p,1}\,c_{d,-2} + \frac{\sqrt{6}}{15}\,c^\dagger_{d,0}\,c^\dagger_{d,-1}\,c_{p,1}\,c_{d,-2}
\\
&\qquad + \frac{1}{5}\,c^\dagger_{d,2}\,c^\dagger_{d,-2}\,c_{p,-1}\,c_{d,1} + \frac{2}{5}\,c^\dagger_{d,1}\,c^\dagger_{d,-1}\,c_{p,-1}\,c_{d,1} + \frac{\sqrt{3}}{5}\,c^\dagger_{d,2}\,c^\dagger_{d,-2}\,c_{p,0}\,c_{d,0}
\\
&\qquad + \frac{2 \sqrt{3}}{5}\,c^\dagger_{d,1}\,c^\dagger_{d,-1}\,c_{p,0}\,c_{d,0} + \frac{1}{5}\,c^\dagger_{d,2}\,c^\dagger_{d,-2}\,c_{p,1}\,c_{d,-1} + \frac{2}{5}\,c^\dagger_{d,1}\,c^\dagger_{d,-1}\,c_{p,1}\,c_{d,-1}
\\
&\qquad + \frac{1}{5}\,c^\dagger_{d,2}\,c^\dagger_{d,-1}\,c_{p,-1}\,c_{d,2} + \frac{\sqrt{6}}{15}\,c^\dagger_{d,1}\,c^\dagger_{d,0}\,c_{p,-1}\,c_{d,2} + \frac{2 \sqrt{2}}{5}\,c^\dagger_{d,2}\,c^\dagger_{d,-1}\,c_{p,0}\,c_{d,1}
\\
&\qquad + \frac{4 \sqrt{3}}{15}\,c^\dagger_{d,1}\,c^\dagger_{d,0}\,c_{p,0}\,c_{d,1} + \frac{\sqrt{6}}{5}\,c^\dagger_{d,2}\,c^\dagger_{d,-1}\,c_{p,1}\,c_{d,0} + \frac{2}{5}\,c^\dagger_{d,1}\,c^\dagger_{d,0}\,c_{p,1}\,c_{d,0}
\\
&\qquad + \frac{\sqrt{3}}{3}\,c^\dagger_{d,2}\,c^\dagger_{d,0}\,c_{p,0}\,c_{d,2} + \frac{\sqrt{6}}{3}\,c^\dagger_{d,2}\,c^\dagger_{d,0}\,c_{p,1}\,c_{d,1} + c^\dagger_{d,2}\,c^\dagger_{d,1}\,c_{p,1}\,c_{d,2}\Big)
\\
{M_0^{(2)}}_{[2]} &=  
\frac{i}{2} \left(\left[[c^\dagger_p \times c^\dagger_d]_{3} \times[\tilde c_d \times \tilde c_d]_{3}\right]_{0} - h.c. \right)\\
&= 
\frac{\sqrt{14}}{5} {\bm Q}^{pd}_1 \cdot {\bm M}_1^{d}
- \frac{\sqrt{10}}{5} {\bm M}^{pd}_2 \cdot {\bm Q}_2^{d}
+ \frac{1}{5}  {\bm Q}^{pd}_3 \cdot {\bm M}_3^{d}
\\ 
% 
% sqrt(14)*I/14  |max|= 0.267261241912424
% \begin{aligned}
&= \frac{\sqrt{14} i}{14}\Big(c^\dagger_{p,-1}\,c^\dagger_{d,-2}\,c_{d,-1}\,c_{d,-2} + \frac{\sqrt{6}}{3}\,c^\dagger_{p,-1}\,c^\dagger_{d,-1}\,c_{d,0}\,c_{d,-2} + \frac{\sqrt{3}}{3}\,c^\dagger_{p,0}\,c^\dagger_{d,-2}\,c_{d,0}\,c_{d,-2}
\\
&\qquad + \frac{\sqrt{6}}{5}\,c^\dagger_{p,-1}\,c^\dagger_{d,0}\,c_{d,1}\,c_{d,-2} + \frac{2}{5}\,c^\dagger_{p,-1}\,c^\dagger_{d,0}\,c_{d,0}\,c_{d,-1} + \frac{2 \sqrt{2}}{5}\,c^\dagger_{p,0}\,c^\dagger_{d,-1}\,c_{d,1}\,c_{d,-2}
\\
&\qquad + \frac{4 \sqrt{3}}{15}\,c^\dagger_{p,0}\,c^\dagger_{d,-1}\,c_{d,0}\,c_{d,-1} + \frac{1}{5}\,c^\dagger_{p,1}\,c^\dagger_{d,-2}\,c_{d,1}\,c_{d,-2} + \frac{\sqrt{6}}{15}\,c^\dagger_{p,1}\,c^\dagger_{d,-2}\,c_{d,0}\,c_{d,-1}
\\
&\qquad + \frac{1}{5}\,c^\dagger_{p,-1}\,c^\dagger_{d,1}\,c_{d,2}\,c_{d,-2} + \frac{2}{5}\,c^\dagger_{p,-1}\,c^\dagger_{d,1}\,c_{d,1}\,c_{d,-1} + \frac{\sqrt{3}}{5}\,c^\dagger_{p,0}\,c^\dagger_{d,0}\,c_{d,2}\,c_{d,-2}
\\
&\qquad + \frac{2 \sqrt{3}}{5}\,c^\dagger_{p,0}\,c^\dagger_{d,0}\,c_{d,1}\,c_{d,-1} + \frac{1}{5}\,c^\dagger_{p,1}\,c^\dagger_{d,-1}\,c_{d,2}\,c_{d,-2} + \frac{2}{5}\,c^\dagger_{p,1}\,c^\dagger_{d,-1}\,c_{d,1}\,c_{d,-1}
\\
&\qquad + \frac{1}{5}\,c^\dagger_{p,-1}\,c^\dagger_{d,2}\,c_{d,2}\,c_{d,-1} + \frac{\sqrt{6}}{15}\,c^\dagger_{p,-1}\,c^\dagger_{d,2}\,c_{d,1}\,c_{d,0} + \frac{2 \sqrt{2}}{5}\,c^\dagger_{p,0}\,c^\dagger_{d,1}\,c_{d,2}\,c_{d,-1}
\\
&\qquad + \frac{4 \sqrt{3}}{15}\,c^\dagger_{p,0}\,c^\dagger_{d,1}\,c_{d,1}\,c_{d,0} + \frac{\sqrt{6}}{5}\,c^\dagger_{p,1}\,c^\dagger_{d,0}\,c_{d,2}\,c_{d,-1} + \frac{2}{5}\,c^\dagger_{p,1}\,c^\dagger_{d,0}\,c_{d,1}\,c_{d,0}
\\
&\qquad + \frac{\sqrt{3}}{3}\,c^\dagger_{p,0}\,c^\dagger_{d,2}\,c_{d,2}\,c_{d,0} + \frac{\sqrt{6}}{3}\,c^\dagger_{p,1}\,c^\dagger_{d,1}\,c_{d,2}\,c_{d,0} + c^\dagger_{p,1}\,c^\dagger_{d,2}\,c_{d,2}\,c_{d,1}
\\
&\qquad - c^\dagger_{d,-1}\,c^\dagger_{d,-2}\,c_{p,-1}\,c_{d,-2} - \frac{\sqrt{6}}{3}\,c^\dagger_{d,0}\,c^\dagger_{d,-2}\,c_{p,-1}\,c_{d,-1} - \frac{\sqrt{3}}{3}\,c^\dagger_{d,0}\,c^\dagger_{d,-2}\,c_{p,0}\,c_{d,-2}
\\
&\qquad - \frac{\sqrt{6}}{5}\,c^\dagger_{d,1}\,c^\dagger_{d,-2}\,c_{p,-1}\,c_{d,0} - \frac{2}{5}\,c^\dagger_{d,0}\,c^\dagger_{d,-1}\,c_{p,-1}\,c_{d,0} - \frac{2 \sqrt{2}}{5}\,c^\dagger_{d,1}\,c^\dagger_{d,-2}\,c_{p,0}\,c_{d,-1}
\\
&\qquad - \frac{4 \sqrt{3}}{15}\,c^\dagger_{d,0}\,c^\dagger_{d,-1}\,c_{p,0}\,c_{d,-1} - \frac{1}{5}\,c^\dagger_{d,1}\,c^\dagger_{d,-2}\,c_{p,1}\,c_{d,-2} - \frac{\sqrt{6}}{15}\,c^\dagger_{d,0}\,c^\dagger_{d,-1}\,c_{p,1}\,c_{d,-2}
\\
&\qquad - \frac{1}{5}\,c^\dagger_{d,2}\,c^\dagger_{d,-2}\,c_{p,-1}\,c_{d,1} - \frac{2}{5}\,c^\dagger_{d,1}\,c^\dagger_{d,-1}\,c_{p,-1}\,c_{d,1} - \frac{\sqrt{3}}{5}\,c^\dagger_{d,2}\,c^\dagger_{d,-2}\,c_{p,0}\,c_{d,0}
\\
&\qquad - \frac{2 \sqrt{3}}{5}\,c^\dagger_{d,1}\,c^\dagger_{d,-1}\,c_{p,0}\,c_{d,0} - \frac{1}{5}\,c^\dagger_{d,2}\,c^\dagger_{d,-2}\,c_{p,1}\,c_{d,-1} - \frac{2}{5}\,c^\dagger_{d,1}\,c^\dagger_{d,-1}\,c_{p,1}\,c_{d,-1}
\\
&\qquad - \frac{1}{5}\,c^\dagger_{d,2}\,c^\dagger_{d,-1}\,c_{p,-1}\,c_{d,2} - \frac{\sqrt{6}}{15}\,c^\dagger_{d,1}\,c^\dagger_{d,0}\,c_{p,-1}\,c_{d,2} - \frac{2 \sqrt{2}}{5}\,c^\dagger_{d,2}\,c^\dagger_{d,-1}\,c_{p,0}\,c_{d,1}
\\
&\qquad - \frac{4 \sqrt{3}}{15}\,c^\dagger_{d,1}\,c^\dagger_{d,0}\,c_{p,0}\,c_{d,1} - \frac{\sqrt{6}}{5}\,c^\dagger_{d,2}\,c^\dagger_{d,-1}\,c_{p,1}\,c_{d,0} - \frac{2}{5}\,c^\dagger_{d,1}\,c^\dagger_{d,0}\,c_{p,1}\,c_{d,0}
\\
&\qquad - \frac{\sqrt{3}}{3}\,c^\dagger_{d,2}\,c^\dagger_{d,0}\,c_{p,0}\,c_{d,2} - \frac{\sqrt{6}}{3}\,c^\dagger_{d,2}\,c^\dagger_{d,0}\,c_{p,1}\,c_{d,1} - c^\dagger_{d,2}\,c^\dagger_{d,1}\,c_{p,1}\,c_{d,2}\Big)
\end{align*}

\begin{align*}
{G_0^{(2)}}_{[3]} &=  
\frac{1}{2} \left(\left[[c^\dagger_p \times c^\dagger_p]_{1} \times[\tilde c_d \times \tilde c_p]_{1}\right]_{0} + h.c. \right)\\
&= \frac{1}{2}   {\bm T}^{pd}_1 \cdot {\bm M}_1^{p}  
+  \frac{\sqrt{3}}{2}   {\bm G}^{pd}_2 \cdot {\bm Q}_2^{p}\notag \\ 
&=
- \frac{\sqrt{10}}{10}\Big(\frac{\sqrt{6}}{6}\,c^\dagger_{p,0}\,c^\dagger_{p,-1}\,c_{p,-1}\,c_{d,0} - \frac{\sqrt{2}}{2}\,c^\dagger_{p,0}\,c^\dagger_{p,-1}\,c_{p,0}\,c_{d,-1} + c^\dagger_{p,0}\,c^\dagger_{p,-1}\,c_{p,1}\,c_{d,-2}
\\
&\qquad + \frac{\sqrt{2}}{2}\,c^\dagger_{p,1}\,c^\dagger_{p,-1}\,c_{p,-1}\,c_{d,1} - \frac{\sqrt{6}}{3}\,c^\dagger_{p,1}\,c^\dagger_{p,-1}\,c_{p,0}\,c_{d,0} + \frac{\sqrt{2}}{2}\,c^\dagger_{p,1}\,c^\dagger_{p,-1}\,c_{p,1}\,c_{d,-1}
\\
&\qquad + c^\dagger_{p,1}\,c^\dagger_{p,0}\,c_{p,-1}\,c_{d,2} - \frac{\sqrt{2}}{2}\,c^\dagger_{p,1}\,c^\dagger_{p,0}\,c_{p,0}\,c_{d,1} + \frac{\sqrt{6}}{6}\,c^\dagger_{p,1}\,c^\dagger_{p,0}\,c_{p,1}\,c_{d,0}
\\
&\qquad + \frac{\sqrt{6}}{6}\,c^\dagger_{p,-1}\,c^\dagger_{d,0}\,c_{p,0}\,c_{p,-1} - \frac{\sqrt{2}}{2}\,c^\dagger_{p,0}\,c^\dagger_{d,-1}\,c_{p,0}\,c_{p,-1} + c^\dagger_{p,1}\,c^\dagger_{d,-2}\,c_{p,0}\,c_{p,-1}
\\
&\qquad + \frac{\sqrt{2}}{2}\,c^\dagger_{p,-1}\,c^\dagger_{d,1}\,c_{p,1}\,c_{p,-1} - \frac{\sqrt{6}}{3}\,c^\dagger_{p,0}\,c^\dagger_{d,0}\,c_{p,1}\,c_{p,-1} + \frac{\sqrt{2}}{2}\,c^\dagger_{p,1}\,c^\dagger_{d,-1}\,c_{p,1}\,c_{p,-1}
\\
&\qquad + c^\dagger_{p,-1}\,c^\dagger_{d,2}\,c_{p,1}\,c_{p,0} - \frac{\sqrt{2}}{2}\,c^\dagger_{p,0}\,c^\dagger_{d,1}\,c_{p,1}\,c_{p,0} + \frac{\sqrt{6}}{6}\,c^\dagger_{p,1}\,c^\dagger_{d,0}\,c_{p,1}\,c_{p,0}\Big)
\\
{M_0^{(2)}}_{[3]} &=  
\frac{i}{2} \left(\left[[c^\dagger_p \times c^\dagger_p]_{1} \times[\tilde c_d \times \tilde c_p]_{1}\right]_{0} - h.c. \right)\\
&= \frac{1}{2}   {\bm Q}^{pd}_1 \cdot {\bm M}_1^{p}  
+  \frac{\sqrt{3}}{2}   {\bm M}^{pd}_2 \cdot {\bm Q}_2^{p}\notag \\ 
&=
- \frac{\sqrt{10} i}{10}\Big(\frac{\sqrt{6}}{6}\,c^\dagger_{p,0}\,c^\dagger_{p,-1}\,c_{p,-1}\,c_{d,0} - \frac{\sqrt{2}}{2}\,c^\dagger_{p,0}\,c^\dagger_{p,-1}\,c_{p,0}\,c_{d,-1} + c^\dagger_{p,0}\,c^\dagger_{p,-1}\,c_{p,1}\,c_{d,-2}
\\
&\qquad + \frac{\sqrt{2}}{2}\,c^\dagger_{p,1}\,c^\dagger_{p,-1}\,c_{p,-1}\,c_{d,1} - \frac{\sqrt{6}}{3}\,c^\dagger_{p,1}\,c^\dagger_{p,-1}\,c_{p,0}\,c_{d,0} + \frac{\sqrt{2}}{2}\,c^\dagger_{p,1}\,c^\dagger_{p,-1}\,c_{p,1}\,c_{d,-1}
\\
&\qquad + c^\dagger_{p,1}\,c^\dagger_{p,0}\,c_{p,-1}\,c_{d,2} - \frac{\sqrt{2}}{2}\,c^\dagger_{p,1}\,c^\dagger_{p,0}\,c_{p,0}\,c_{d,1} + \frac{\sqrt{6}}{6}\,c^\dagger_{p,1}\,c^\dagger_{p,0}\,c_{p,1}\,c_{d,0}
\\
&\qquad - \frac{\sqrt{6}}{6}\,c^\dagger_{p,-1}\,c^\dagger_{d,0}\,c_{p,0}\,c_{p,-1} + \frac{\sqrt{2}}{2}\,c^\dagger_{p,0}\,c^\dagger_{d,-1}\,c_{p,0}\,c_{p,-1} - c^\dagger_{p,1}\,c^\dagger_{d,-2}\,c_{p,0}\,c_{p,-1}
\\
&\qquad - \frac{\sqrt{2}}{2}\,c^\dagger_{p,-1}\,c^\dagger_{d,1}\,c_{p,1}\,c_{p,-1} + \frac{\sqrt{6}}{3}\,c^\dagger_{p,0}\,c^\dagger_{d,0}\,c_{p,1}\,c_{p,-1} - \frac{\sqrt{2}}{2}\,c^\dagger_{p,1}\,c^\dagger_{d,-1}\,c_{p,1}\,c_{p,-1}
\\
&\qquad - c^\dagger_{p,-1}\,c^\dagger_{d,2}\,c_{p,1}\,c_{p,0} + \frac{\sqrt{2}}{2}\,c^\dagger_{p,0}\,c^\dagger_{d,1}\,c_{p,1}\,c_{p,0} - \frac{\sqrt{6}}{6}\,c^\dagger_{p,1}\,c^\dagger_{d,0}\,c_{p,1}\,c_{p,0}\Big)
% \\
\end{align*}

\subsection*{$pf$ system}
\begin{align*}
{Q_0^{(2)}}_{[1]}
&=  
  \frac{1}{2} \left(\left[[c^\dagger_p \times c^\dagger_p]_{1} \times[\tilde c_f \times \tilde c_f]_{1}\right]_{0} + h.c. \right)\\
&=
    -\frac{\sqrt{210}}{21}
    \left( 
    {\bm Q}^{pf}_2 \cdot {\bm Q}_2^{pf}
    -
    {\bm T}^{pf}_2 \cdot {\bm T}_2^{pf}
    \right)
    -\frac{\sqrt{6}}{12}
    \left( 
    {\bm M}^{pf}_3 \cdot {\bm M}_3^{pf}
    -
    {\bm G}^{pf}_3 \cdot {\bm G}_3^{pf}
    \right)
    \\&\qquad 
    +\frac{3\sqrt{42}}{28}
    \left( 
    {\bm Q}^{pf}_4 \cdot {\bm Q}_4^{pf}
    -
    {\bm T}^{pf}_4 \cdot {\bm T}_4^{pf}
    \right)
  \\
&=
\frac{\sqrt{42}}{14}\Big(\frac{\sqrt{3}}{3}\,c^\dagger_{p,0}\,c^\dagger_{p,-1}\,c_{f,2}\,c_{f,-3} - \frac{\sqrt{5}}{3}\,c^\dagger_{p,0}\,c^\dagger_{p,-1}\,c_{f,1}\,c_{f,-2} + \frac{\sqrt{6}}{3}\,c^\dagger_{p,0}\,c^\dagger_{p,-1}\,c_{f,0}\,c_{f,-1}
\\
&\qquad + c^\dagger_{p,1}\,c^\dagger_{p,-1}\,c_{f,3}\,c_{f,-3} - \frac{2}{3}\,c^\dagger_{p,1}\,c^\dagger_{p,-1}\,c_{f,2}\,c_{f,-2} + \frac{1}{3}\,c^\dagger_{p,1}\,c^\dagger_{p,-1}\,c_{f,1}\,c_{f,-1}
\\
&\qquad + \frac{\sqrt{3}}{3}\,c^\dagger_{p,1}\,c^\dagger_{p,0}\,c_{f,3}\,c_{f,-2} - \frac{\sqrt{5}}{3}\,c^\dagger_{p,1}\,c^\dagger_{p,0}\,c_{f,2}\,c_{f,-1} + \frac{\sqrt{6}}{3}\,c^\dagger_{p,1}\,c^\dagger_{p,0}\,c_{f,1}\,c_{f,0}
\\
&\qquad + \frac{\sqrt{3}}{3}\,c^\dagger_{f,2}\,c^\dagger_{f,-3}\,c_{p,0}\,c_{p,-1} - \frac{\sqrt{5}}{3}\,c^\dagger_{f,1}\,c^\dagger_{f,-2}\,c_{p,0}\,c_{p,-1} + \frac{\sqrt{6}}{3}\,c^\dagger_{f,0}\,c^\dagger_{f,-1}\,c_{p,0}\,c_{p,-1}
\\
&\qquad + c^\dagger_{f,3}\,c^\dagger_{f,-3}\,c_{p,1}\,c_{p,-1} - \frac{2}{3}\,c^\dagger_{f,2}\,c^\dagger_{f,-2}\,c_{p,1}\,c_{p,-1} + \frac{1}{3}\,c^\dagger_{f,1}\,c^\dagger_{f,-1}\,c_{p,1}\,c_{p,-1}
\\
&\qquad + \frac{\sqrt{3}}{3}\,c^\dagger_{f,3}\,c^\dagger_{f,-2}\,c_{p,1}\,c_{p,0} - \frac{\sqrt{5}}{3}\,c^\dagger_{f,2}\,c^\dagger_{f,-1}\,c_{p,1}\,c_{p,0} + \frac{\sqrt{6}}{3}\,c^\dagger_{f,1}\,c^\dagger_{f,0}\,c_{p,1}\,c_{p,0}\Big)
  \\
  {T_0^{(2)}}_{[1]}
&=  
  \frac{i}{2} \left(\left[[c^\dagger_p \times c^\dagger_p]_{1} \times[\tilde c_f \times \tilde c_f]_{1}\right]_{0} - h.c. \right)\\
&=
    -\frac{\sqrt{210}}{21}
    \left( 
    {\bm T}^{pf}_2 \cdot {\bm Q}_2^{pf}
    +
    {\bm Q}^{pf}_2 \cdot {\bm T}_2^{pf}
    \right)
    -\frac{\sqrt{6}}{12}
    \left( 
    {\bm G}^{pf}_3 \cdot {\bm M}_3^{pf}
    +
    {\bm M}^{pf}_3 \cdot {\bm G}_3^{pf}
    \right)
    \\&\qquad 
    +\frac{3\sqrt{42}}{28}
    \left( 
    {\bm T}^{pf}_4 \cdot {\bm Q}_4^{pf}
    +
    {\bm Q}^{pf}_4 \cdot {\bm T}_4^{pf}
    \right)
  \\
&=
\frac{\sqrt{42} i}{14}\Big(\frac{\sqrt{3}}{3}\,c^\dagger_{p,0}\,c^\dagger_{p,-1}\,c_{f,2}\,c_{f,-3} - \frac{\sqrt{5}}{3}\,c^\dagger_{p,0}\,c^\dagger_{p,-1}\,c_{f,1}\,c_{f,-2} + \frac{\sqrt{6}}{3}\,c^\dagger_{p,0}\,c^\dagger_{p,-1}\,c_{f,0}\,c_{f,-1}
\\
&\qquad + c^\dagger_{p,1}\,c^\dagger_{p,-1}\,c_{f,3}\,c_{f,-3} - \frac{2}{3}\,c^\dagger_{p,1}\,c^\dagger_{p,-1}\,c_{f,2}\,c_{f,-2} + \frac{1}{3}\,c^\dagger_{p,1}\,c^\dagger_{p,-1}\,c_{f,1}\,c_{f,-1}
\\
&\qquad + \frac{\sqrt{3}}{3}\,c^\dagger_{p,1}\,c^\dagger_{p,0}\,c_{f,3}\,c_{f,-2} - \frac{\sqrt{5}}{3}\,c^\dagger_{p,1}\,c^\dagger_{p,0}\,c_{f,2}\,c_{f,-1} + \frac{\sqrt{6}}{3}\,c^\dagger_{p,1}\,c^\dagger_{p,0}\,c_{f,1}\,c_{f,0}
\\
&\qquad - \frac{\sqrt{3}}{3}\,c^\dagger_{f,2}\,c^\dagger_{f,-3}\,c_{p,0}\,c_{p,-1} + \frac{\sqrt{5}}{3}\,c^\dagger_{f,1}\,c^\dagger_{f,-2}\,c_{p,0}\,c_{p,-1} - \frac{\sqrt{6}}{3}\,c^\dagger_{f,0}\,c^\dagger_{f,-1}\,c_{p,0}\,c_{p,-1}
\\
&\qquad - c^\dagger_{f,3}\,c^\dagger_{f,-3}\,c_{p,1}\,c_{p,-1} + \frac{2}{3}\,c^\dagger_{f,2}\,c^\dagger_{f,-2}\,c_{p,1}\,c_{p,-1} - \frac{1}{3}\,c^\dagger_{f,1}\,c^\dagger_{f,-1}\,c_{p,1}\,c_{p,-1}
\\
&\qquad - \frac{\sqrt{3}}{3}\,c^\dagger_{f,3}\,c^\dagger_{f,-2}\,c_{p,1}\,c_{p,0} + \frac{\sqrt{5}}{3}\,c^\dagger_{f,2}\,c^\dagger_{f,-1}\,c_{p,1}\,c_{p,0} - \frac{\sqrt{6}}{3}\,c^\dagger_{f,1}\,c^\dagger_{f,0}\,c_{p,1}\,c_{p,0}\Big)
\end{align*}
% ppff : 1×1
% j13,j24,A wigner_9j： 2 , 2 , sqrt(210)/21
% j13,j24,A wigner_9j： 3 , 3 , sqrt(6)/12
% j13,j24,A wigner_9j： 4 , 4 , -3*sqrt(42)/28

\begin{align*}
{Q_0^{(2)} }_{[2]}
&=  
  \frac{1}{2} \left(\left[[c^\dagger_p \times c^\dagger_f]_{3} \times[\tilde c_f \times \tilde c_f]_{3}\right]_{0} + h.c. \right)\\
&=
    -\frac{\sqrt{70}}{14}
    {\bm Q}^{pf}_2 \cdot {\bm Q}_2^{f}
    -\frac{1}{2}
    {\bm M}^{pf}_3 \cdot {\bm M}_3^{f}
    % \\&\qquad 
    +\frac{\sqrt{77}}{14}
    % \left( 
    {\bm Q}^{pf}_4 \cdot {\bm Q}_4^{f}
    % -
    % \left[T^{pf}_4 \times T_4^{pf}\right]_{0} 
    % \right)
  \\
&=
- \frac{\sqrt{7}}{14}\Big(\frac{\sqrt{6}}{6}\,c^\dagger_{p,-1}\,c^\dagger_{f,-2}\,c_{f,0}\,c_{f,-3} - \frac{\sqrt{3}}{3}\,c^\dagger_{p,-1}\,c^\dagger_{f,-2}\,c_{f,-1}\,c_{f,-2} - \frac{\sqrt{2}}{2}\,c^\dagger_{p,0}\,c^\dagger_{f,-3}\,c_{f,0}\,c_{f,-3}
\\
&\qquad + c^\dagger_{p,0}\,c^\dagger_{f,-3}\,c_{f,-1}\,c_{f,-2} + \frac{\sqrt{5}}{3}\,c^\dagger_{p,-1}\,c^\dagger_{f,-1}\,c_{f,1}\,c_{f,-3} - \frac{\sqrt{10}}{6}\,c^\dagger_{p,-1}\,c^\dagger_{f,-1}\,c_{f,0}\,c_{f,-2}
\\
&\qquad - \frac{2}{3}\,c^\dagger_{p,0}\,c^\dagger_{f,-2}\,c_{f,1}\,c_{f,-3} + \frac{\sqrt{2}}{3}\,c^\dagger_{p,0}\,c^\dagger_{f,-2}\,c_{f,0}\,c_{f,-2} - \frac{\sqrt{3}}{3}\,c^\dagger_{p,1}\,c^\dagger_{f,-3}\,c_{f,1}\,c_{f,-3}
\\
&\qquad + \frac{\sqrt{6}}{6}\,c^\dagger_{p,1}\,c^\dagger_{f,-3}\,c_{f,0}\,c_{f,-2} + \frac{\sqrt{6}}{3}\,c^\dagger_{p,-1}\,c^\dagger_{f,0}\,c_{f,2}\,c_{f,-3} - \frac{\sqrt{3}}{3}\,c^\dagger_{p,-1}\,c^\dagger_{f,0}\,c_{f,0}\,c_{f,-1}
\\
&\qquad - \frac{1}{3}\,c^\dagger_{p,0}\,c^\dagger_{f,-1}\,c_{f,2}\,c_{f,-3} + \frac{\sqrt{2}}{6}\,c^\dagger_{p,0}\,c^\dagger_{f,-1}\,c_{f,0}\,c_{f,-1} - \frac{\sqrt{5}}{3}\,c^\dagger_{p,1}\,c^\dagger_{f,-2}\,c_{f,2}\,c_{f,-3}
\\
&\qquad + \frac{\sqrt{10}}{6}\,c^\dagger_{p,1}\,c^\dagger_{f,-2}\,c_{f,0}\,c_{f,-1} + \frac{\sqrt{3}}{3}\,c^\dagger_{p,-1}\,c^\dagger_{f,1}\,c_{f,3}\,c_{f,-3} + \frac{\sqrt{3}}{3}\,c^\dagger_{p,-1}\,c^\dagger_{f,1}\,c_{f,2}\,c_{f,-2}
\\
&\qquad - \frac{\sqrt{3}}{3}\,c^\dagger_{p,-1}\,c^\dagger_{f,1}\,c_{f,1}\,c_{f,-1} - \frac{\sqrt{3}}{3}\,c^\dagger_{p,1}\,c^\dagger_{f,-1}\,c_{f,3}\,c_{f,-3} - \frac{\sqrt{3}}{3}\,c^\dagger_{p,1}\,c^\dagger_{f,-1}\,c_{f,2}\,c_{f,-2}
\\
&\qquad + \frac{\sqrt{3}}{3}\,c^\dagger_{p,1}\,c^\dagger_{f,-1}\,c_{f,1}\,c_{f,-1} + \frac{\sqrt{5}}{3}\,c^\dagger_{p,-1}\,c^\dagger_{f,2}\,c_{f,3}\,c_{f,-2} - \frac{\sqrt{10}}{6}\,c^\dagger_{p,-1}\,c^\dagger_{f,2}\,c_{f,1}\,c_{f,0}
\\
&\qquad + \frac{1}{3}\,c^\dagger_{p,0}\,c^\dagger_{f,1}\,c_{f,3}\,c_{f,-2} - \frac{\sqrt{2}}{6}\,c^\dagger_{p,0}\,c^\dagger_{f,1}\,c_{f,1}\,c_{f,0} - \frac{\sqrt{6}}{3}\,c^\dagger_{p,1}\,c^\dagger_{f,0}\,c_{f,3}\,c_{f,-2}
\\
&\qquad + \frac{\sqrt{3}}{3}\,c^\dagger_{p,1}\,c^\dagger_{f,0}\,c_{f,1}\,c_{f,0} + \frac{\sqrt{3}}{3}\,c^\dagger_{p,-1}\,c^\dagger_{f,3}\,c_{f,3}\,c_{f,-1} - \frac{\sqrt{6}}{6}\,c^\dagger_{p,-1}\,c^\dagger_{f,3}\,c_{f,2}\,c_{f,0}
\\
&\qquad + \frac{2}{3}\,c^\dagger_{p,0}\,c^\dagger_{f,2}\,c_{f,3}\,c_{f,-1} - \frac{\sqrt{2}}{3}\,c^\dagger_{p,0}\,c^\dagger_{f,2}\,c_{f,2}\,c_{f,0} - \frac{\sqrt{5}}{3}\,c^\dagger_{p,1}\,c^\dagger_{f,1}\,c_{f,3}\,c_{f,-1}
\\
&\qquad + \frac{\sqrt{10}}{6}\,c^\dagger_{p,1}\,c^\dagger_{f,1}\,c_{f,2}\,c_{f,0} + \frac{\sqrt{2}}{2}\,c^\dagger_{p,0}\,c^\dagger_{f,3}\,c_{f,3}\,c_{f,0} - c^\dagger_{p,0}\,c^\dagger_{f,3}\,c_{f,2}\,c_{f,1}
\\
&\qquad - \frac{\sqrt{6}}{6}\,c^\dagger_{p,1}\,c^\dagger_{f,2}\,c_{f,3}\,c_{f,0} + \frac{\sqrt{3}}{3}\,c^\dagger_{p,1}\,c^\dagger_{f,2}\,c_{f,2}\,c_{f,1} + \frac{\sqrt{6}}{6}\,c^\dagger_{f,0}\,c^\dagger_{f,-3}\,c_{p,-1}\,c_{f,-2}
\\
&\qquad - \frac{\sqrt{3}}{3}\,c^\dagger_{f,-1}\,c^\dagger_{f,-2}\,c_{p,-1}\,c_{f,-2} - \frac{\sqrt{2}}{2}\,c^\dagger_{f,0}\,c^\dagger_{f,-3}\,c_{p,0}\,c_{f,-3} + c^\dagger_{f,-1}\,c^\dagger_{f,-2}\,c_{p,0}\,c_{f,-3}
\\
&\qquad + \frac{\sqrt{5}}{3}\,c^\dagger_{f,1}\,c^\dagger_{f,-3}\,c_{p,-1}\,c_{f,-1} - \frac{\sqrt{10}}{6}\,c^\dagger_{f,0}\,c^\dagger_{f,-2}\,c_{p,-1}\,c_{f,-1} - \frac{2}{3}\,c^\dagger_{f,1}\,c^\dagger_{f,-3}\,c_{p,0}\,c_{f,-2}
\\
&\qquad + \frac{\sqrt{2}}{3}\,c^\dagger_{f,0}\,c^\dagger_{f,-2}\,c_{p,0}\,c_{f,-2} - \frac{\sqrt{3}}{3}\,c^\dagger_{f,1}\,c^\dagger_{f,-3}\,c_{p,1}\,c_{f,-3} + \frac{\sqrt{6}}{6}\,c^\dagger_{f,0}\,c^\dagger_{f,-2}\,c_{p,1}\,c_{f,-3}
\\
&\qquad + \frac{\sqrt{6}}{3}\,c^\dagger_{f,2}\,c^\dagger_{f,-3}\,c_{p,-1}\,c_{f,0} - \frac{\sqrt{3}}{3}\,c^\dagger_{f,0}\,c^\dagger_{f,-1}\,c_{p,-1}\,c_{f,0} - \frac{1}{3}\,c^\dagger_{f,2}\,c^\dagger_{f,-3}\,c_{p,0}\,c_{f,-1}
\\
&\qquad + \frac{\sqrt{2}}{6}\,c^\dagger_{f,0}\,c^\dagger_{f,-1}\,c_{p,0}\,c_{f,-1} - \frac{\sqrt{5}}{3}\,c^\dagger_{f,2}\,c^\dagger_{f,-3}\,c_{p,1}\,c_{f,-2} + \frac{\sqrt{10}}{6}\,c^\dagger_{f,0}\,c^\dagger_{f,-1}\,c_{p,1}\,c_{f,-2}
\\
&\qquad + \frac{\sqrt{3}}{3}\,c^\dagger_{f,3}\,c^\dagger_{f,-3}\,c_{p,-1}\,c_{f,1} + \frac{\sqrt{3}}{3}\,c^\dagger_{f,2}\,c^\dagger_{f,-2}\,c_{p,-1}\,c_{f,1} - \frac{\sqrt{3}}{3}\,c^\dagger_{f,1}\,c^\dagger_{f,-1}\,c_{p,-1}\,c_{f,1}
\\
&\qquad - \frac{\sqrt{3}}{3}\,c^\dagger_{f,3}\,c^\dagger_{f,-3}\,c_{p,1}\,c_{f,-1} - \frac{\sqrt{3}}{3}\,c^\dagger_{f,2}\,c^\dagger_{f,-2}\,c_{p,1}\,c_{f,-1} + \frac{\sqrt{3}}{3}\,c^\dagger_{f,1}\,c^\dagger_{f,-1}\,c_{p,1}\,c_{f,-1}
\\
&\qquad + \frac{\sqrt{5}}{3}\,c^\dagger_{f,3}\,c^\dagger_{f,-2}\,c_{p,-1}\,c_{f,2} - \frac{\sqrt{10}}{6}\,c^\dagger_{f,1}\,c^\dagger_{f,0}\,c_{p,-1}\,c_{f,2} + \frac{1}{3}\,c^\dagger_{f,3}\,c^\dagger_{f,-2}\,c_{p,0}\,c_{f,1}
\\
&\qquad - \frac{\sqrt{2}}{6}\,c^\dagger_{f,1}\,c^\dagger_{f,0}\,c_{p,0}\,c_{f,1} - \frac{\sqrt{6}}{3}\,c^\dagger_{f,3}\,c^\dagger_{f,-2}\,c_{p,1}\,c_{f,0} + \frac{\sqrt{3}}{3}\,c^\dagger_{f,1}\,c^\dagger_{f,0}\,c_{p,1}\,c_{f,0}
\\
&\qquad + \frac{\sqrt{3}}{3}\,c^\dagger_{f,3}\,c^\dagger_{f,-1}\,c_{p,-1}\,c_{f,3} - \frac{\sqrt{6}}{6}\,c^\dagger_{f,2}\,c^\dagger_{f,0}\,c_{p,-1}\,c_{f,3} + \frac{2}{3}\,c^\dagger_{f,3}\,c^\dagger_{f,-1}\,c_{p,0}\,c_{f,2}
\\
&\qquad - \frac{\sqrt{2}}{3}\,c^\dagger_{f,2}\,c^\dagger_{f,0}\,c_{p,0}\,c_{f,2} - \frac{\sqrt{5}}{3}\,c^\dagger_{f,3}\,c^\dagger_{f,-1}\,c_{p,1}\,c_{f,1} + \frac{\sqrt{10}}{6}\,c^\dagger_{f,2}\,c^\dagger_{f,0}\,c_{p,1}\,c_{f,1}
\\
&\qquad + \frac{\sqrt{2}}{2}\,c^\dagger_{f,3}\,c^\dagger_{f,0}\,c_{p,0}\,c_{f,3} - c^\dagger_{f,2}\,c^\dagger_{f,1}\,c_{p,0}\,c_{f,3} - \frac{\sqrt{6}}{6}\,c^\dagger_{f,3}\,c^\dagger_{f,0}\,c_{p,1}\,c_{f,2}
\\
&\qquad + \frac{\sqrt{3}}{3}\,c^\dagger_{f,2}\,c^\dagger_{f,1}\,c_{p,1}\,c_{f,2}\Big)
 \\
  {T_0^{(2)} }_{[2]}
&=  
  \frac{i}{2} \left(\left[[c^\dagger_p \times c^\dagger_f]_{3} \times[\tilde c_f \times \tilde c_f]_{3}\right]_{0} - h.c. \right)\\
&=
    -\frac{\sqrt{70}}{14}
    % \left( 
    {\bm T}^{pf}_2 \cdot {\bm Q}_2^{f}
    % +
    % \left[Q^{pf}_2 \times T_2^{f}\right]_{0} 
    % \right)
    -\frac{1}{2}
    % \left( 
    {\bm G}^{pf}_3 \cdot {\bm M}_3^{f}
    % +
    % \left[M^{pf}_3 \times G_3^{f}\right]_{0} 
    % \right)
    % \\&\qquad 
    +\frac{\sqrt{77}}{14}
    % \left( 
    {\bm T}^{pf}_4 \cdot {\bm Q}_4^{f}
    % +
    % \left[Q^{pf}_4 \times T_4^{f}\right]_{0} 
    % \right)
  \\
&=
- \frac{\sqrt{7} i}{14}\Big(\frac{\sqrt{6}}{6}\,c^\dagger_{p,-1}\,c^\dagger_{f,-2}\,c_{f,0}\,c_{f,-3} - \frac{\sqrt{3}}{3}\,c^\dagger_{p,-1}\,c^\dagger_{f,-2}\,c_{f,-1}\,c_{f,-2} - \frac{\sqrt{2}}{2}\,c^\dagger_{p,0}\,c^\dagger_{f,-3}\,c_{f,0}\,c_{f,-3}
\\
&\qquad + c^\dagger_{p,0}\,c^\dagger_{f,-3}\,c_{f,-1}\,c_{f,-2} + \frac{\sqrt{5}}{3}\,c^\dagger_{p,-1}\,c^\dagger_{f,-1}\,c_{f,1}\,c_{f,-3} - \frac{\sqrt{10}}{6}\,c^\dagger_{p,-1}\,c^\dagger_{f,-1}\,c_{f,0}\,c_{f,-2}
\\
&\qquad - \frac{2}{3}\,c^\dagger_{p,0}\,c^\dagger_{f,-2}\,c_{f,1}\,c_{f,-3} + \frac{\sqrt{2}}{3}\,c^\dagger_{p,0}\,c^\dagger_{f,-2}\,c_{f,0}\,c_{f,-2} - \frac{\sqrt{3}}{3}\,c^\dagger_{p,1}\,c^\dagger_{f,-3}\,c_{f,1}\,c_{f,-3}
\\
&\qquad + \frac{\sqrt{6}}{6}\,c^\dagger_{p,1}\,c^\dagger_{f,-3}\,c_{f,0}\,c_{f,-2} + \frac{\sqrt{6}}{3}\,c^\dagger_{p,-1}\,c^\dagger_{f,0}\,c_{f,2}\,c_{f,-3} - \frac{\sqrt{3}}{3}\,c^\dagger_{p,-1}\,c^\dagger_{f,0}\,c_{f,0}\,c_{f,-1}
\\
&\qquad - \frac{1}{3}\,c^\dagger_{p,0}\,c^\dagger_{f,-1}\,c_{f,2}\,c_{f,-3} + \frac{\sqrt{2}}{6}\,c^\dagger_{p,0}\,c^\dagger_{f,-1}\,c_{f,0}\,c_{f,-1} - \frac{\sqrt{5}}{3}\,c^\dagger_{p,1}\,c^\dagger_{f,-2}\,c_{f,2}\,c_{f,-3}
\\
&\qquad + \frac{\sqrt{10}}{6}\,c^\dagger_{p,1}\,c^\dagger_{f,-2}\,c_{f,0}\,c_{f,-1} + \frac{\sqrt{3}}{3}\,c^\dagger_{p,-1}\,c^\dagger_{f,1}\,c_{f,3}\,c_{f,-3} + \frac{\sqrt{3}}{3}\,c^\dagger_{p,-1}\,c^\dagger_{f,1}\,c_{f,2}\,c_{f,-2}
\\
&\qquad - \frac{\sqrt{3}}{3}\,c^\dagger_{p,-1}\,c^\dagger_{f,1}\,c_{f,1}\,c_{f,-1} - \frac{\sqrt{3}}{3}\,c^\dagger_{p,1}\,c^\dagger_{f,-1}\,c_{f,3}\,c_{f,-3} - \frac{\sqrt{3}}{3}\,c^\dagger_{p,1}\,c^\dagger_{f,-1}\,c_{f,2}\,c_{f,-2}
\\
&\qquad + \frac{\sqrt{3}}{3}\,c^\dagger_{p,1}\,c^\dagger_{f,-1}\,c_{f,1}\,c_{f,-1} + \frac{\sqrt{5}}{3}\,c^\dagger_{p,-1}\,c^\dagger_{f,2}\,c_{f,3}\,c_{f,-2} - \frac{\sqrt{10}}{6}\,c^\dagger_{p,-1}\,c^\dagger_{f,2}\,c_{f,1}\,c_{f,0}
\\
&\qquad + \frac{1}{3}\,c^\dagger_{p,0}\,c^\dagger_{f,1}\,c_{f,3}\,c_{f,-2} - \frac{\sqrt{2}}{6}\,c^\dagger_{p,0}\,c^\dagger_{f,1}\,c_{f,1}\,c_{f,0} - \frac{\sqrt{6}}{3}\,c^\dagger_{p,1}\,c^\dagger_{f,0}\,c_{f,3}\,c_{f,-2}
\\
&\qquad + \frac{\sqrt{3}}{3}\,c^\dagger_{p,1}\,c^\dagger_{f,0}\,c_{f,1}\,c_{f,0} + \frac{\sqrt{3}}{3}\,c^\dagger_{p,-1}\,c^\dagger_{f,3}\,c_{f,3}\,c_{f,-1} - \frac{\sqrt{6}}{6}\,c^\dagger_{p,-1}\,c^\dagger_{f,3}\,c_{f,2}\,c_{f,0}
\\
&\qquad + \frac{2}{3}\,c^\dagger_{p,0}\,c^\dagger_{f,2}\,c_{f,3}\,c_{f,-1} - \frac{\sqrt{2}}{3}\,c^\dagger_{p,0}\,c^\dagger_{f,2}\,c_{f,2}\,c_{f,0} - \frac{\sqrt{5}}{3}\,c^\dagger_{p,1}\,c^\dagger_{f,1}\,c_{f,3}\,c_{f,-1}
\\
&\qquad + \frac{\sqrt{10}}{6}\,c^\dagger_{p,1}\,c^\dagger_{f,1}\,c_{f,2}\,c_{f,0} + \frac{\sqrt{2}}{2}\,c^\dagger_{p,0}\,c^\dagger_{f,3}\,c_{f,3}\,c_{f,0} - c^\dagger_{p,0}\,c^\dagger_{f,3}\,c_{f,2}\,c_{f,1}
\\
&\qquad - \frac{\sqrt{6}}{6}\,c^\dagger_{p,1}\,c^\dagger_{f,2}\,c_{f,3}\,c_{f,0} + \frac{\sqrt{3}}{3}\,c^\dagger_{p,1}\,c^\dagger_{f,2}\,c_{f,2}\,c_{f,1} - \frac{\sqrt{6}}{6}\,c^\dagger_{f,0}\,c^\dagger_{f,-3}\,c_{p,-1}\,c_{f,-2}
\\
&\qquad + \frac{\sqrt{3}}{3}\,c^\dagger_{f,-1}\,c^\dagger_{f,-2}\,c_{p,-1}\,c_{f,-2} + \frac{\sqrt{2}}{2}\,c^\dagger_{f,0}\,c^\dagger_{f,-3}\,c_{p,0}\,c_{f,-3} - c^\dagger_{f,-1}\,c^\dagger_{f,-2}\,c_{p,0}\,c_{f,-3}
\\
&\qquad - \frac{\sqrt{5}}{3}\,c^\dagger_{f,1}\,c^\dagger_{f,-3}\,c_{p,-1}\,c_{f,-1} + \frac{\sqrt{10}}{6}\,c^\dagger_{f,0}\,c^\dagger_{f,-2}\,c_{p,-1}\,c_{f,-1} + \frac{2}{3}\,c^\dagger_{f,1}\,c^\dagger_{f,-3}\,c_{p,0}\,c_{f,-2}
\\
&\qquad - \frac{\sqrt{2}}{3}\,c^\dagger_{f,0}\,c^\dagger_{f,-2}\,c_{p,0}\,c_{f,-2} + \frac{\sqrt{3}}{3}\,c^\dagger_{f,1}\,c^\dagger_{f,-3}\,c_{p,1}\,c_{f,-3} - \frac{\sqrt{6}}{6}\,c^\dagger_{f,0}\,c^\dagger_{f,-2}\,c_{p,1}\,c_{f,-3}
\\
&\qquad - \frac{\sqrt{6}}{3}\,c^\dagger_{f,2}\,c^\dagger_{f,-3}\,c_{p,-1}\,c_{f,0} + \frac{\sqrt{3}}{3}\,c^\dagger_{f,0}\,c^\dagger_{f,-1}\,c_{p,-1}\,c_{f,0} + \frac{1}{3}\,c^\dagger_{f,2}\,c^\dagger_{f,-3}\,c_{p,0}\,c_{f,-1}
\\
&\qquad - \frac{\sqrt{2}}{6}\,c^\dagger_{f,0}\,c^\dagger_{f,-1}\,c_{p,0}\,c_{f,-1} + \frac{\sqrt{5}}{3}\,c^\dagger_{f,2}\,c^\dagger_{f,-3}\,c_{p,1}\,c_{f,-2} - \frac{\sqrt{10}}{6}\,c^\dagger_{f,0}\,c^\dagger_{f,-1}\,c_{p,1}\,c_{f,-2}
\\
&\qquad - \frac{\sqrt{3}}{3}\,c^\dagger_{f,3}\,c^\dagger_{f,-3}\,c_{p,-1}\,c_{f,1} - \frac{\sqrt{3}}{3}\,c^\dagger_{f,2}\,c^\dagger_{f,-2}\,c_{p,-1}\,c_{f,1} + \frac{\sqrt{3}}{3}\,c^\dagger_{f,1}\,c^\dagger_{f,-1}\,c_{p,-1}\,c_{f,1}
\\
&\qquad + \frac{\sqrt{3}}{3}\,c^\dagger_{f,3}\,c^\dagger_{f,-3}\,c_{p,1}\,c_{f,-1} + \frac{\sqrt{3}}{3}\,c^\dagger_{f,2}\,c^\dagger_{f,-2}\,c_{p,1}\,c_{f,-1} - \frac{\sqrt{3}}{3}\,c^\dagger_{f,1}\,c^\dagger_{f,-1}\,c_{p,1}\,c_{f,-1}
\\
&\qquad - \frac{\sqrt{5}}{3}\,c^\dagger_{f,3}\,c^\dagger_{f,-2}\,c_{p,-1}\,c_{f,2} + \frac{\sqrt{10}}{6}\,c^\dagger_{f,1}\,c^\dagger_{f,0}\,c_{p,-1}\,c_{f,2} - \frac{1}{3}\,c^\dagger_{f,3}\,c^\dagger_{f,-2}\,c_{p,0}\,c_{f,1}
\\
&\qquad + \frac{\sqrt{2}}{6}\,c^\dagger_{f,1}\,c^\dagger_{f,0}\,c_{p,0}\,c_{f,1} + \frac{\sqrt{6}}{3}\,c^\dagger_{f,3}\,c^\dagger_{f,-2}\,c_{p,1}\,c_{f,0} - \frac{\sqrt{3}}{3}\,c^\dagger_{f,1}\,c^\dagger_{f,0}\,c_{p,1}\,c_{f,0}
\\
&\qquad - \frac{\sqrt{3}}{3}\,c^\dagger_{f,3}\,c^\dagger_{f,-1}\,c_{p,-1}\,c_{f,3} + \frac{\sqrt{6}}{6}\,c^\dagger_{f,2}\,c^\dagger_{f,0}\,c_{p,-1}\,c_{f,3} - \frac{2}{3}\,c^\dagger_{f,3}\,c^\dagger_{f,-1}\,c_{p,0}\,c_{f,2}
\\
&\qquad + \frac{\sqrt{2}}{3}\,c^\dagger_{f,2}\,c^\dagger_{f,0}\,c_{p,0}\,c_{f,2} + \frac{\sqrt{5}}{3}\,c^\dagger_{f,3}\,c^\dagger_{f,-1}\,c_{p,1}\,c_{f,1} - \frac{\sqrt{10}}{6}\,c^\dagger_{f,2}\,c^\dagger_{f,0}\,c_{p,1}\,c_{f,1}
\\
&\qquad - \frac{\sqrt{2}}{2}\,c^\dagger_{f,3}\,c^\dagger_{f,0}\,c_{p,0}\,c_{f,3} + c^\dagger_{f,2}\,c^\dagger_{f,1}\,c_{p,0}\,c_{f,3} + \frac{\sqrt{6}}{6}\,c^\dagger_{f,3}\,c^\dagger_{f,0}\,c_{p,1}\,c_{f,2}
\\
&\qquad - \frac{\sqrt{3}}{3}\,c^\dagger_{f,2}\,c^\dagger_{f,1}\,c_{p,1}\,c_{f,2}\Big)
\end{align*}
% pfff : 3×3
% j13,j24,A wigner_9j： 2 , 2 , sqrt(70)/14
% j13,j24,A wigner_9j： 3 , 3 , 1/2
% j13,j24,A wigner_9j： 4 , 4 , -sqrt(77)/14

\subsection*{$df$ system}
% ddff : 1×1
% j13,j24,A wigner_9j： 1 , 1 , 2*sqrt(70)/35
% j13,j24,A wigner_9j： 2 , 2 , sqrt(42)/14
% j13,j24,A wigner_9j： 3 , 3 , sqrt(30)/20
% j13,j24,A wigner_9j： 4 , 4 , -sqrt(210)/140
% j13,j24,A wigner_9j： 5 , 5 , -sqrt(2310)/70
\begin{align*}
{Q_0^{(2)}}_{[1]} 
&=  
  \frac{1}{2} \left(\left[[c^\dagger_d \times c^\dagger_d]_{1} \times[\tilde c_f \times \tilde c_f]_{1}\right]_{0} + h.c. \right)\\
&=
-\frac{2\sqrt{70}}{35}
    \left( {\bm Q}^{df}_1 \cdot {\bm Q}^{df}_1 - {\bm T}^{df}_1 \cdot {\bm T}^{df}_1  \right)
-\frac{\sqrt{42}}{14}    
    \left( {\bm G}^{df}_2 \cdot {\bm G}^{df}_2 - {\bm M}^{df}_2 \cdot {\bm M}^{df}_2  \right)
    \\&\qquad
-\frac{\sqrt{30}}{20}    
    \left( {\bm Q}^{df}_3 \cdot {\bm Q}^{df}_3  -  {\bm T}^{df}_3 \cdot {\bm T}^{df}_3      \right)  
+\frac{\sqrt{210}}{140}    
    \left( {\bm G}^{df}_4 \cdot {\bm G}^{df}_4 - {\bm M}^{df}_4 \cdot {\bm M}^{df}_4  \right)
\\&\qquad    
+\frac{\sqrt{2310}}{70}    
    \left( {\bm Q}^{df}_5 \cdot {\bm Q}^{df}_5  -  {\bm T}^{df}_5 \cdot {\bm T}^{df}_5      \right)  
\\     
&=
\frac{\sqrt{210}}{35}\Big(\frac{\sqrt{6}}{6}\,c^\dagger_{d,1}\,c^\dagger_{d,-2}\,c_{f,2}\,c_{f,-3} - \frac{\sqrt{10}}{6}\,c^\dagger_{d,1}\,c^\dagger_{d,-2}\,c_{f,1}\,c_{f,-2} + \frac{\sqrt{3}}{3}\,c^\dagger_{d,1}\,c^\dagger_{d,-2}\,c_{f,0}\,c_{f,-1}
\\
&\qquad - \frac{1}{2}\,c^\dagger_{d,0}\,c^\dagger_{d,-1}\,c_{f,2}\,c_{f,-3} + \frac{\sqrt{15}}{6}\,c^\dagger_{d,0}\,c^\dagger_{d,-1}\,c_{f,1}\,c_{f,-2} - \frac{\sqrt{2}}{2}\,c^\dagger_{d,0}\,c^\dagger_{d,-1}\,c_{f,0}\,c_{f,-1}
\\
&\qquad + c^\dagger_{d,2}\,c^\dagger_{d,-2}\,c_{f,3}\,c_{f,-3} - \frac{2}{3}\,c^\dagger_{d,2}\,c^\dagger_{d,-2}\,c_{f,2}\,c_{f,-2} + \frac{1}{3}\,c^\dagger_{d,2}\,c^\dagger_{d,-2}\,c_{f,1}\,c_{f,-1}
\\
&\qquad - \frac{1}{2}\,c^\dagger_{d,1}\,c^\dagger_{d,-1}\,c_{f,3}\,c_{f,-3} + \frac{1}{3}\,c^\dagger_{d,1}\,c^\dagger_{d,-1}\,c_{f,2}\,c_{f,-2} - \frac{1}{6}\,c^\dagger_{d,1}\,c^\dagger_{d,-1}\,c_{f,1}\,c_{f,-1}
\\
&\qquad + \frac{\sqrt{6}}{6}\,c^\dagger_{d,2}\,c^\dagger_{d,-1}\,c_{f,3}\,c_{f,-2} - \frac{\sqrt{10}}{6}\,c^\dagger_{d,2}\,c^\dagger_{d,-1}\,c_{f,2}\,c_{f,-1} + \frac{\sqrt{3}}{3}\,c^\dagger_{d,2}\,c^\dagger_{d,-1}\,c_{f,1}\,c_{f,0}
\\
&\qquad - \frac{1}{2}\,c^\dagger_{d,1}\,c^\dagger_{d,0}\,c_{f,3}\,c_{f,-2} + \frac{\sqrt{15}}{6}\,c^\dagger_{d,1}\,c^\dagger_{d,0}\,c_{f,2}\,c_{f,-1} - \frac{\sqrt{2}}{2}\,c^\dagger_{d,1}\,c^\dagger_{d,0}\,c_{f,1}\,c_{f,0}
\\
&\qquad + \frac{\sqrt{6}}{6}\,c^\dagger_{f,2}\,c^\dagger_{f,-3}\,c_{d,1}\,c_{d,-2} - \frac{\sqrt{10}}{6}\,c^\dagger_{f,1}\,c^\dagger_{f,-2}\,c_{d,1}\,c_{d,-2} + \frac{\sqrt{3}}{3}\,c^\dagger_{f,0}\,c^\dagger_{f,-1}\,c_{d,1}\,c_{d,-2}
\\
&\qquad - \frac{1}{2}\,c^\dagger_{f,2}\,c^\dagger_{f,-3}\,c_{d,0}\,c_{d,-1} + \frac{\sqrt{15}}{6}\,c^\dagger_{f,1}\,c^\dagger_{f,-2}\,c_{d,0}\,c_{d,-1} - \frac{\sqrt{2}}{2}\,c^\dagger_{f,0}\,c^\dagger_{f,-1}\,c_{d,0}\,c_{d,-1}
\\
&\qquad + c^\dagger_{f,3}\,c^\dagger_{f,-3}\,c_{d,2}\,c_{d,-2} - \frac{2}{3}\,c^\dagger_{f,2}\,c^\dagger_{f,-2}\,c_{d,2}\,c_{d,-2} + \frac{1}{3}\,c^\dagger_{f,1}\,c^\dagger_{f,-1}\,c_{d,2}\,c_{d,-2}
\\
&\qquad - \frac{1}{2}\,c^\dagger_{f,3}\,c^\dagger_{f,-3}\,c_{d,1}\,c_{d,-1} + \frac{1}{3}\,c^\dagger_{f,2}\,c^\dagger_{f,-2}\,c_{d,1}\,c_{d,-1} - \frac{1}{6}\,c^\dagger_{f,1}\,c^\dagger_{f,-1}\,c_{d,1}\,c_{d,-1}
\\
&\qquad + \frac{\sqrt{6}}{6}\,c^\dagger_{f,3}\,c^\dagger_{f,-2}\,c_{d,2}\,c_{d,-1} - \frac{\sqrt{10}}{6}\,c^\dagger_{f,2}\,c^\dagger_{f,-1}\,c_{d,2}\,c_{d,-1} + \frac{\sqrt{3}}{3}\,c^\dagger_{f,1}\,c^\dagger_{f,0}\,c_{d,2}\,c_{d,-1}
\\
&\qquad - \frac{1}{2}\,c^\dagger_{f,3}\,c^\dagger_{f,-2}\,c_{d,1}\,c_{d,0} + \frac{\sqrt{15}}{6}\,c^\dagger_{f,2}\,c^\dagger_{f,-1}\,c_{d,1}\,c_{d,0} - \frac{\sqrt{2}}{2}\,c^\dagger_{f,1}\,c^\dagger_{f,0}\,c_{d,1}\,c_{d,0}\Big)
 \\
  {T_0^{(2)}}_{[1]} 
&=  
  \frac{i}{2} \left(\left[[c^\dagger_d \times c^\dagger_d]_{1} \times[\tilde c_f \times \tilde c_f]_{1}\right]_{0} - h.c. \right)\\
&=
-\frac{2\sqrt{70}}{35}
    \left( {\bm T}^{df}_1 \cdot {\bm Q}^{df}_1 + {\bm Q}^{df}_1 \cdot {\bm T}^{df}_1  \right)
-\frac{\sqrt{42}}{14}    
    \left( {\bm M}^{df}_2 \cdot {\bm G}^{df}_2 + {\bm G}^{df}_2 \cdot {\bm M}^{df}_2  \right)
    \\&\qquad
-\frac{\sqrt{30}}{20}    
    \left( {\bm T}^{df}_3 \cdot {\bm Q}^{df}_3  + {\bm Q}^{df}_3 \cdot {\bm T}^{df}_3      \right)  
+\frac{\sqrt{210}}{140}    
    \left( {\bm M}^{df}_4 \cdot {\bm G}^{df}_4 + {\bm G}^{df}_4 \cdot {\bm M}^{df}_4  \right)
\\&\qquad    
+\frac{\sqrt{2310}}{70}    
    \left( {\bm T}^{df}_5 \cdot {\bm Q}^{df}_5  + {\bm Q}^{df}_5 \cdot {\bm T}^{df}_5      \right)  
\\     
&=
\frac{\sqrt{210} i}{35}\Big(\frac{\sqrt{6}}{6}\,c^\dagger_{d,1}\,c^\dagger_{d,-2}\,c_{f,2}\,c_{f,-3} - \frac{\sqrt{10}}{6}\,c^\dagger_{d,1}\,c^\dagger_{d,-2}\,c_{f,1}\,c_{f,-2} + \frac{\sqrt{3}}{3}\,c^\dagger_{d,1}\,c^\dagger_{d,-2}\,c_{f,0}\,c_{f,-1}
\\
&\qquad - \frac{1}{2}\,c^\dagger_{d,0}\,c^\dagger_{d,-1}\,c_{f,2}\,c_{f,-3} + \frac{\sqrt{15}}{6}\,c^\dagger_{d,0}\,c^\dagger_{d,-1}\,c_{f,1}\,c_{f,-2} - \frac{\sqrt{2}}{2}\,c^\dagger_{d,0}\,c^\dagger_{d,-1}\,c_{f,0}\,c_{f,-1}
\\
&\qquad + c^\dagger_{d,2}\,c^\dagger_{d,-2}\,c_{f,3}\,c_{f,-3} - \frac{2}{3}\,c^\dagger_{d,2}\,c^\dagger_{d,-2}\,c_{f,2}\,c_{f,-2} + \frac{1}{3}\,c^\dagger_{d,2}\,c^\dagger_{d,-2}\,c_{f,1}\,c_{f,-1}
\\
&\qquad - \frac{1}{2}\,c^\dagger_{d,1}\,c^\dagger_{d,-1}\,c_{f,3}\,c_{f,-3} + \frac{1}{3}\,c^\dagger_{d,1}\,c^\dagger_{d,-1}\,c_{f,2}\,c_{f,-2} - \frac{1}{6}\,c^\dagger_{d,1}\,c^\dagger_{d,-1}\,c_{f,1}\,c_{f,-1}
\\
&\qquad + \frac{\sqrt{6}}{6}\,c^\dagger_{d,2}\,c^\dagger_{d,-1}\,c_{f,3}\,c_{f,-2} - \frac{\sqrt{10}}{6}\,c^\dagger_{d,2}\,c^\dagger_{d,-1}\,c_{f,2}\,c_{f,-1} + \frac{\sqrt{3}}{3}\,c^\dagger_{d,2}\,c^\dagger_{d,-1}\,c_{f,1}\,c_{f,0}
\\
&\qquad - \frac{1}{2}\,c^\dagger_{d,1}\,c^\dagger_{d,0}\,c_{f,3}\,c_{f,-2} + \frac{\sqrt{15}}{6}\,c^\dagger_{d,1}\,c^\dagger_{d,0}\,c_{f,2}\,c_{f,-1} - \frac{\sqrt{2}}{2}\,c^\dagger_{d,1}\,c^\dagger_{d,0}\,c_{f,1}\,c_{f,0}
\\
&\qquad - \frac{\sqrt{6}}{6}\,c^\dagger_{f,2}\,c^\dagger_{f,-3}\,c_{d,1}\,c_{d,-2} + \frac{\sqrt{10}}{6}\,c^\dagger_{f,1}\,c^\dagger_{f,-2}\,c_{d,1}\,c_{d,-2} - \frac{\sqrt{3}}{3}\,c^\dagger_{f,0}\,c^\dagger_{f,-1}\,c_{d,1}\,c_{d,-2}
\\
&\qquad + \frac{1}{2}\,c^\dagger_{f,2}\,c^\dagger_{f,-3}\,c_{d,0}\,c_{d,-1} - \frac{\sqrt{15}}{6}\,c^\dagger_{f,1}\,c^\dagger_{f,-2}\,c_{d,0}\,c_{d,-1} + \frac{\sqrt{2}}{2}\,c^\dagger_{f,0}\,c^\dagger_{f,-1}\,c_{d,0}\,c_{d,-1}
\\
&\qquad - c^\dagger_{f,3}\,c^\dagger_{f,-3}\,c_{d,2}\,c_{d,-2} + \frac{2}{3}\,c^\dagger_{f,2}\,c^\dagger_{f,-2}\,c_{d,2}\,c_{d,-2} - \frac{1}{3}\,c^\dagger_{f,1}\,c^\dagger_{f,-1}\,c_{d,2}\,c_{d,-2}
\\
&\qquad + \frac{1}{2}\,c^\dagger_{f,3}\,c^\dagger_{f,-3}\,c_{d,1}\,c_{d,-1} - \frac{1}{3}\,c^\dagger_{f,2}\,c^\dagger_{f,-2}\,c_{d,1}\,c_{d,-1} + \frac{1}{6}\,c^\dagger_{f,1}\,c^\dagger_{f,-1}\,c_{d,1}\,c_{d,-1}
\\
&\qquad - \frac{\sqrt{6}}{6}\,c^\dagger_{f,3}\,c^\dagger_{f,-2}\,c_{d,2}\,c_{d,-1} + \frac{\sqrt{10}}{6}\,c^\dagger_{f,2}\,c^\dagger_{f,-1}\,c_{d,2}\,c_{d,-1} - \frac{\sqrt{3}}{3}\,c^\dagger_{f,1}\,c^\dagger_{f,0}\,c_{d,2}\,c_{d,-1}
\\
&\qquad + \frac{1}{2}\,c^\dagger_{f,3}\,c^\dagger_{f,-2}\,c_{d,1}\,c_{d,0} - \frac{\sqrt{15}}{6}\,c^\dagger_{f,2}\,c^\dagger_{f,-1}\,c_{d,1}\,c_{d,0} + \frac{\sqrt{2}}{2}\,c^\dagger_{f,1}\,c^\dagger_{f,0}\,c_{d,1}\,c_{d,0}\Big)
\end{align*}
% ddff : 1×1
% j13,j24,A wigner_9j： 1 , 1 , 2*sqrt(70)/35
% j13,j24,A wigner_9j： 2 , 2 , sqrt(42)/14
% j13,j24,A wigner_9j： 3 , 3 , sqrt(30)/20
% j13,j24,A wigner_9j： 4 , 4 , -sqrt(210)/140
% j13,j24,A wigner_9j： 5 , 5 , -sqrt(2310)/70

% ddff : 3×3
% j13,j24,A wigner_9j： 1 , 1 , 3*sqrt(35)/35
% j13,j24,A wigner_9j： 2 , 2 , -sqrt(21)/14
% j13,j24,A wigner_9j： 3 , 3 , -2*sqrt(15)/15
% j13,j24,A wigner_9j： 4 , 4 , 2*sqrt(105)/35
% j13,j24,A wigner_9j： 5 , 5 , -sqrt(1155)/210

\begin{align*}
{Q_0^{(2)}}_{[2]} 
&=  
  \frac{1}{2} \left(\left[[c^\dagger_d \times c^\dagger_d]_{3} \times[\tilde c_f \times \tilde c_f]_{3}\right]_{0} + h.c. \right)\\
&=
-\frac{3\sqrt{35}}{35}
    \left( {\bm Q}^{df}_1 \cdot {\bm Q}^{df}_1 - {\bm T}^{df}_1 \cdot  {\bm T}^{df}_1 \right)
+\frac{\sqrt{21}}{14}    
    \left( {\bm G}^{df}_2 \cdot {\bm G}^{df}_2 - {\bm M}^{df}_2 \cdot  {\bm M}^{df}_2 \right)
    \\&\qquad
\frac{2\sqrt{15}}{15}
    \left( {\bm Q}^{df}_3 \cdot {\bm Q}^{df}_3  -  {\bm T}^{df}_3 \cdot  {\bm T}^{df}_3     \right)  
-\frac{2\sqrt{105}}{35}    
    \left( {\bm G}^{df}_4 \cdot {\bm G}^{df}_4 - {\bm M}^{df}_4 \cdot  {\bm M}^{df}_4 \right)
\\&\qquad    
+\frac{\sqrt{1155}}{210}    
    \left( {\bm Q}^{df}_5 \cdot {\bm Q}^{df}_5  -  {\bm T}^{df}_5 \cdot  {\bm T}^{df}_5     \right)  
\\
&=
- \frac{\sqrt{42}}{21}\Big(- \frac{\sqrt{2}}{2}\,c^\dagger_{d,-1}\,c^\dagger_{d,-2}\,c_{f,0}\,c_{f,-3} + c^\dagger_{d,-1}\,c^\dagger_{d,-2}\,c_{f,-1}\,c_{f,-2} - c^\dagger_{d,0}\,c^\dagger_{d,-2}\,c_{f,1}\,c_{f,-3}
\\
&\qquad + \frac{\sqrt{2}}{2}\,c^\dagger_{d,0}\,c^\dagger_{d,-2}\,c_{f,0}\,c_{f,-2} - \frac{\sqrt{15}}{5}\,c^\dagger_{d,1}\,c^\dagger_{d,-2}\,c_{f,2}\,c_{f,-3} + \frac{\sqrt{30}}{10}\,c^\dagger_{d,1}\,c^\dagger_{d,-2}\,c_{f,0}\,c_{f,-1}
\\
&\qquad - \frac{\sqrt{10}}{5}\,c^\dagger_{d,0}\,c^\dagger_{d,-1}\,c_{f,2}\,c_{f,-3} + \frac{\sqrt{5}}{5}\,c^\dagger_{d,0}\,c^\dagger_{d,-1}\,c_{f,0}\,c_{f,-1} - \frac{\sqrt{10}}{10}\,c^\dagger_{d,2}\,c^\dagger_{d,-2}\,c_{f,3}\,c_{f,-3}
\\
&\qquad - \frac{\sqrt{10}}{10}\,c^\dagger_{d,2}\,c^\dagger_{d,-2}\,c_{f,2}\,c_{f,-2} + \frac{\sqrt{10}}{10}\,c^\dagger_{d,2}\,c^\dagger_{d,-2}\,c_{f,1}\,c_{f,-1} - \frac{\sqrt{10}}{5}\,c^\dagger_{d,1}\,c^\dagger_{d,-1}\,c_{f,3}\,c_{f,-3}
\\
&\qquad - \frac{\sqrt{10}}{5}\,c^\dagger_{d,1}\,c^\dagger_{d,-1}\,c_{f,2}\,c_{f,-2} + \frac{\sqrt{10}}{5}\,c^\dagger_{d,1}\,c^\dagger_{d,-1}\,c_{f,1}\,c_{f,-1} - \frac{\sqrt{15}}{5}\,c^\dagger_{d,2}\,c^\dagger_{d,-1}\,c_{f,3}\,c_{f,-2}
\\
&\qquad + \frac{\sqrt{30}}{10}\,c^\dagger_{d,2}\,c^\dagger_{d,-1}\,c_{f,1}\,c_{f,0} - \frac{\sqrt{10}}{5}\,c^\dagger_{d,1}\,c^\dagger_{d,0}\,c_{f,3}\,c_{f,-2} + \frac{\sqrt{5}}{5}\,c^\dagger_{d,1}\,c^\dagger_{d,0}\,c_{f,1}\,c_{f,0}
\\
&\qquad - c^\dagger_{d,2}\,c^\dagger_{d,0}\,c_{f,3}\,c_{f,-1} + \frac{\sqrt{2}}{2}\,c^\dagger_{d,2}\,c^\dagger_{d,0}\,c_{f,2}\,c_{f,0} - \frac{\sqrt{2}}{2}\,c^\dagger_{d,2}\,c^\dagger_{d,1}\,c_{f,3}\,c_{f,0}
\\
&\qquad + c^\dagger_{d,2}\,c^\dagger_{d,1}\,c_{f,2}\,c_{f,1} - \frac{\sqrt{2}}{2}\,c^\dagger_{f,0}\,c^\dagger_{f,-3}\,c_{d,-1}\,c_{d,-2} + c^\dagger_{f,-1}\,c^\dagger_{f,-2}\,c_{d,-1}\,c_{d,-2}
\\
&\qquad - c^\dagger_{f,1}\,c^\dagger_{f,-3}\,c_{d,0}\,c_{d,-2} + \frac{\sqrt{2}}{2}\,c^\dagger_{f,0}\,c^\dagger_{f,-2}\,c_{d,0}\,c_{d,-2} - \frac{\sqrt{15}}{5}\,c^\dagger_{f,2}\,c^\dagger_{f,-3}\,c_{d,1}\,c_{d,-2}
\\
&\qquad + \frac{\sqrt{30}}{10}\,c^\dagger_{f,0}\,c^\dagger_{f,-1}\,c_{d,1}\,c_{d,-2} - \frac{\sqrt{10}}{5}\,c^\dagger_{f,2}\,c^\dagger_{f,-3}\,c_{d,0}\,c_{d,-1} + \frac{\sqrt{5}}{5}\,c^\dagger_{f,0}\,c^\dagger_{f,-1}\,c_{d,0}\,c_{d,-1}
\\
&\qquad - \frac{\sqrt{10}}{10}\,c^\dagger_{f,3}\,c^\dagger_{f,-3}\,c_{d,2}\,c_{d,-2} - \frac{\sqrt{10}}{10}\,c^\dagger_{f,2}\,c^\dagger_{f,-2}\,c_{d,2}\,c_{d,-2} + \frac{\sqrt{10}}{10}\,c^\dagger_{f,1}\,c^\dagger_{f,-1}\,c_{d,2}\,c_{d,-2}
\\
&\qquad - \frac{\sqrt{10}}{5}\,c^\dagger_{f,3}\,c^\dagger_{f,-3}\,c_{d,1}\,c_{d,-1} - \frac{\sqrt{10}}{5}\,c^\dagger_{f,2}\,c^\dagger_{f,-2}\,c_{d,1}\,c_{d,-1} + \frac{\sqrt{10}}{5}\,c^\dagger_{f,1}\,c^\dagger_{f,-1}\,c_{d,1}\,c_{d,-1}
\\
&\qquad - \frac{\sqrt{15}}{5}\,c^\dagger_{f,3}\,c^\dagger_{f,-2}\,c_{d,2}\,c_{d,-1} + \frac{\sqrt{30}}{10}\,c^\dagger_{f,1}\,c^\dagger_{f,0}\,c_{d,2}\,c_{d,-1} - \frac{\sqrt{10}}{5}\,c^\dagger_{f,3}\,c^\dagger_{f,-2}\,c_{d,1}\,c_{d,0}
\\
&\qquad + \frac{\sqrt{5}}{5}\,c^\dagger_{f,1}\,c^\dagger_{f,0}\,c_{d,1}\,c_{d,0} - c^\dagger_{f,3}\,c^\dagger_{f,-1}\,c_{d,2}\,c_{d,0} + \frac{\sqrt{2}}{2}\,c^\dagger_{f,2}\,c^\dagger_{f,0}\,c_{d,2}\,c_{d,0}
\\
&\qquad - \frac{\sqrt{2}}{2}\,c^\dagger_{f,3}\,c^\dagger_{f,0}\,c_{d,2}\,c_{d,1} + c^\dagger_{f,2}\,c^\dagger_{f,1}\,c_{d,2}\,c_{d,1}\Big)
 \\
  {T_0^{(2)}}_{[2]} 
&=  
  \frac{i}{2} \left(\left[[c^\dagger_d \times c^\dagger_d]_{3} \times[\tilde c_f \times \tilde c_f]_{3}\right]_{0} - h.c. \right)\\
&=
-\frac{3\sqrt{35}}{35}
    \left( {\bm T}^{df}_1 \cdot {\bm Q}^{df}_1 + {\bm Q}^{df}_1 \cdot  {\bm T}^{df}_1 \right)
+\frac{\sqrt{21}}{14}    
    \left( {\bm M}^{df}_2 \cdot {\bm G}^{df}_2 + {\bm G}^{df}_2 \cdot  {\bm M}^{df}_2 \right)
    \\&\qquad
\frac{2\sqrt{15}}{15}
    \left( {\bm T}^{df}_3 \cdot {\bm Q}^{df}_3  + {\bm Q}^{df}_3 \cdot  {\bm T}^{df}_3     \right)  
-\frac{2\sqrt{105}}{35}    
    \left( {\bm M}^{df}_4 \cdot {\bm G}^{df}_4 + {\bm G}^{df}_4 \cdot  {\bm M}^{df}_4 \right)
\\&\qquad    
+\frac{\sqrt{1155}}{210}    
    \left( {\bm T}^{df}_5 \cdot {\bm Q}^{df}_5  + {\bm Q}^{df}_5 \cdot  {\bm T}^{df}_5     \right)  
\\     
&=
- \frac{\sqrt{42} i}{21}\Big(- \frac{\sqrt{2}}{2}\,c^\dagger_{d,-1}\,c^\dagger_{d,-2}\,c_{f,0}\,c_{f,-3} + c^\dagger_{d,-1}\,c^\dagger_{d,-2}\,c_{f,-1}\,c_{f,-2} - c^\dagger_{d,0}\,c^\dagger_{d,-2}\,c_{f,1}\,c_{f,-3}
\\
&\qquad + \frac{\sqrt{2}}{2}\,c^\dagger_{d,0}\,c^\dagger_{d,-2}\,c_{f,0}\,c_{f,-2} - \frac{\sqrt{15}}{5}\,c^\dagger_{d,1}\,c^\dagger_{d,-2}\,c_{f,2}\,c_{f,-3} + \frac{\sqrt{30}}{10}\,c^\dagger_{d,1}\,c^\dagger_{d,-2}\,c_{f,0}\,c_{f,-1}
\\
&\qquad - \frac{\sqrt{10}}{5}\,c^\dagger_{d,0}\,c^\dagger_{d,-1}\,c_{f,2}\,c_{f,-3} + \frac{\sqrt{5}}{5}\,c^\dagger_{d,0}\,c^\dagger_{d,-1}\,c_{f,0}\,c_{f,-1} - \frac{\sqrt{10}}{10}\,c^\dagger_{d,2}\,c^\dagger_{d,-2}\,c_{f,3}\,c_{f,-3}
\\
&\qquad - \frac{\sqrt{10}}{10}\,c^\dagger_{d,2}\,c^\dagger_{d,-2}\,c_{f,2}\,c_{f,-2} + \frac{\sqrt{10}}{10}\,c^\dagger_{d,2}\,c^\dagger_{d,-2}\,c_{f,1}\,c_{f,-1} - \frac{\sqrt{10}}{5}\,c^\dagger_{d,1}\,c^\dagger_{d,-1}\,c_{f,3}\,c_{f,-3}
\\
&\qquad - \frac{\sqrt{10}}{5}\,c^\dagger_{d,1}\,c^\dagger_{d,-1}\,c_{f,2}\,c_{f,-2} + \frac{\sqrt{10}}{5}\,c^\dagger_{d,1}\,c^\dagger_{d,-1}\,c_{f,1}\,c_{f,-1} - \frac{\sqrt{15}}{5}\,c^\dagger_{d,2}\,c^\dagger_{d,-1}\,c_{f,3}\,c_{f,-2}
\\
&\qquad + \frac{\sqrt{30}}{10}\,c^\dagger_{d,2}\,c^\dagger_{d,-1}\,c_{f,1}\,c_{f,0} - \frac{\sqrt{10}}{5}\,c^\dagger_{d,1}\,c^\dagger_{d,0}\,c_{f,3}\,c_{f,-2} + \frac{\sqrt{5}}{5}\,c^\dagger_{d,1}\,c^\dagger_{d,0}\,c_{f,1}\,c_{f,0}
\\
&\qquad - c^\dagger_{d,2}\,c^\dagger_{d,0}\,c_{f,3}\,c_{f,-1} + \frac{\sqrt{2}}{2}\,c^\dagger_{d,2}\,c^\dagger_{d,0}\,c_{f,2}\,c_{f,0} - \frac{\sqrt{2}}{2}\,c^\dagger_{d,2}\,c^\dagger_{d,1}\,c_{f,3}\,c_{f,0}
\\
&\qquad + c^\dagger_{d,2}\,c^\dagger_{d,1}\,c_{f,2}\,c_{f,1} + \frac{\sqrt{2}}{2}\,c^\dagger_{f,0}\,c^\dagger_{f,-3}\,c_{d,-1}\,c_{d,-2} - c^\dagger_{f,-1}\,c^\dagger_{f,-2}\,c_{d,-1}\,c_{d,-2}
\\
&\qquad + c^\dagger_{f,1}\,c^\dagger_{f,-3}\,c_{d,0}\,c_{d,-2} - \frac{\sqrt{2}}{2}\,c^\dagger_{f,0}\,c^\dagger_{f,-2}\,c_{d,0}\,c_{d,-2} + \frac{\sqrt{15}}{5}\,c^\dagger_{f,2}\,c^\dagger_{f,-3}\,c_{d,1}\,c_{d,-2}
\\
&\qquad - \frac{\sqrt{30}}{10}\,c^\dagger_{f,0}\,c^\dagger_{f,-1}\,c_{d,1}\,c_{d,-2} + \frac{\sqrt{10}}{5}\,c^\dagger_{f,2}\,c^\dagger_{f,-3}\,c_{d,0}\,c_{d,-1} - \frac{\sqrt{5}}{5}\,c^\dagger_{f,0}\,c^\dagger_{f,-1}\,c_{d,0}\,c_{d,-1}
\\
&\qquad + \frac{\sqrt{10}}{10}\,c^\dagger_{f,3}\,c^\dagger_{f,-3}\,c_{d,2}\,c_{d,-2} + \frac{\sqrt{10}}{10}\,c^\dagger_{f,2}\,c^\dagger_{f,-2}\,c_{d,2}\,c_{d,-2} - \frac{\sqrt{10}}{10}\,c^\dagger_{f,1}\,c^\dagger_{f,-1}\,c_{d,2}\,c_{d,-2}
\\
&\qquad + \frac{\sqrt{10}}{5}\,c^\dagger_{f,3}\,c^\dagger_{f,-3}\,c_{d,1}\,c_{d,-1} + \frac{\sqrt{10}}{5}\,c^\dagger_{f,2}\,c^\dagger_{f,-2}\,c_{d,1}\,c_{d,-1} - \frac{\sqrt{10}}{5}\,c^\dagger_{f,1}\,c^\dagger_{f,-1}\,c_{d,1}\,c_{d,-1}
\\
&\qquad + \frac{\sqrt{15}}{5}\,c^\dagger_{f,3}\,c^\dagger_{f,-2}\,c_{d,2}\,c_{d,-1} - \frac{\sqrt{30}}{10}\,c^\dagger_{f,1}\,c^\dagger_{f,0}\,c_{d,2}\,c_{d,-1} + \frac{\sqrt{10}}{5}\,c^\dagger_{f,3}\,c^\dagger_{f,-2}\,c_{d,1}\,c_{d,0}
\\
&\qquad - \frac{\sqrt{5}}{5}\,c^\dagger_{f,1}\,c^\dagger_{f,0}\,c_{d,1}\,c_{d,0} + c^\dagger_{f,3}\,c^\dagger_{f,-1}\,c_{d,2}\,c_{d,0} - \frac{\sqrt{2}}{2}\,c^\dagger_{f,2}\,c^\dagger_{f,0}\,c_{d,2}\,c_{d,0}
\\
&\qquad + \frac{\sqrt{2}}{2}\,c^\dagger_{f,3}\,c^\dagger_{f,0}\,c_{d,2}\,c_{d,1} - c^\dagger_{f,2}\,c^\dagger_{f,1}\,c_{d,2}\,c_{d,1}\Big)
\end{align*}
% ddff : 3×3
% j13,j24,A wigner_9j： 1 , 1 , 3*sqrt(35)/35
% j13,j24,A wigner_9j： 2 , 2 , -sqrt(21)/14
% j13,j24,A wigner_9j： 3 , 3 , -2*sqrt(15)/15
% j13,j24,A wigner_9j： 4 , 4 , 2*sqrt(105)/35
% j13,j24,A wigner_9j： 5 , 5 , -sqrt(1155)/210

% df3 : 1×1
% j13,j24,A wigner_9j： 1 , 1 , -1/7
% j13,j24,A wigner_9j： 2 , 2 , -3*sqrt(2)/14
% j13,j24,A wigner_9j： 3 , 3 , -sqrt(42)/14
% j13,j24,A wigner_9j： 4 , 4 , -sqrt(66)/14
% j13,j24,A wigner_9j： 5 , 5 , -sqrt(66)/14
% df3 : 3×3
% j13,j24,A wigner_9j： 1 , 1 , -sqrt(42)/14
% j13,j24,A wigner_9j： 2 , 2 , -2*sqrt(21)/21
% j13,j24,A wigner_9j： 3 , 3 , 1/6
% j13,j24,A wigner_9j： 4 , 4 , sqrt(77)/14
% j13,j24,A wigner_9j： 5 , 5 , -sqrt(77)/21
% df3 : 5×5
% j13,j24,A wigner_9j： 1 , 1 , -sqrt(66)/14
% j13,j24,A wigner_9j： 2 , 2 , 5*sqrt(33)/42
% j13,j24,A wigner_9j： 3 , 3 , -sqrt(77)/21
% j13,j24,A wigner_9j： 4 , 4 , 1/7
% j13,j24,A wigner_9j： 5 , 5 , -1/42
\begin{align*}
{G_{0}}_{[1]} &=  
\frac{1}{2} \left(\left[[c^\dagger_d \times c^\dagger_f]_{1} \times[\tilde c_f \times \tilde c_f]_{1}\right]_{0} + h.c. \right)\\
&= 
  \frac{1}{7} {\bm T}^{df}_1 \cdot {\bm M}_1^{f}
+ \frac{3\sqrt{2}}{14}  {\bm G}^{df}_2 \cdot {\bm Q}_2^{f}
+ \frac{\sqrt{42}}{14}  {\bm T}^{df}_3 \cdot {\bm M}_3^{f}
+ \frac{\sqrt{66}}{14}  {\bm G}^{df}_4 \cdot {\bm Q}_4^{f}
+ \frac{\sqrt{66}}{14}  {\bm T}^{df}_5 \cdot {\bm M}_5^{f}
\\ 
&= 
\frac{\sqrt{6}}{14}\Big(\frac{\sqrt{30}}{30}\,c^\dagger_{d,-2}\,c^\dagger_{f,1}\,c_{f,2}\,c_{f,-3} - \frac{\sqrt{2}}{6}\,c^\dagger_{d,-2}\,c^\dagger_{f,1}\,c_{f,1}\,c_{f,-2} + \frac{\sqrt{15}}{15}\,c^\dagger_{d,-2}\,c^\dagger_{f,1}\,c_{f,0}\,c_{f,-1}
\\
&\qquad - \frac{\sqrt{10}}{10}\,c^\dagger_{d,-1}\,c^\dagger_{f,0}\,c_{f,2}\,c_{f,-3} + \frac{\sqrt{6}}{6}\,c^\dagger_{d,-1}\,c^\dagger_{f,0}\,c_{f,1}\,c_{f,-2} - \frac{\sqrt{5}}{5}\,c^\dagger_{d,-1}\,c^\dagger_{f,0}\,c_{f,0}\,c_{f,-1}
\\
&\qquad + \frac{\sqrt{5}}{5}\,c^\dagger_{d,0}\,c^\dagger_{f,-1}\,c_{f,2}\,c_{f,-3} - \frac{\sqrt{3}}{3}\,c^\dagger_{d,0}\,c^\dagger_{f,-1}\,c_{f,1}\,c_{f,-2} + \frac{\sqrt{10}}{5}\,c^\dagger_{d,0}\,c^\dagger_{f,-1}\,c_{f,0}\,c_{f,-1}
\\
&\qquad - \frac{\sqrt{3}}{3}\,c^\dagger_{d,1}\,c^\dagger_{f,-2}\,c_{f,2}\,c_{f,-3} + \frac{\sqrt{5}}{3}\,c^\dagger_{d,1}\,c^\dagger_{f,-2}\,c_{f,1}\,c_{f,-2} - \frac{\sqrt{6}}{3}\,c^\dagger_{d,1}\,c^\dagger_{f,-2}\,c_{f,0}\,c_{f,-1}
\\
&\qquad + \frac{\sqrt{2}}{2}\,c^\dagger_{d,2}\,c^\dagger_{f,-3}\,c_{f,2}\,c_{f,-3} - \frac{\sqrt{30}}{6}\,c^\dagger_{d,2}\,c^\dagger_{f,-3}\,c_{f,1}\,c_{f,-2} + c^\dagger_{d,2}\,c^\dagger_{f,-3}\,c_{f,0}\,c_{f,-1}
\\
&\qquad + \frac{\sqrt{2}}{2}\,c^\dagger_{d,-2}\,c^\dagger_{f,2}\,c_{f,3}\,c_{f,-3} - \frac{\sqrt{2}}{3}\,c^\dagger_{d,-2}\,c^\dagger_{f,2}\,c_{f,2}\,c_{f,-2} + \frac{\sqrt{2}}{6}\,c^\dagger_{d,-2}\,c^\dagger_{f,2}\,c_{f,1}\,c_{f,-1}
\\
&\qquad - \frac{2 \sqrt{5}}{5}\,c^\dagger_{d,-1}\,c^\dagger_{f,1}\,c_{f,3}\,c_{f,-3} + \frac{4 \sqrt{5}}{15}\,c^\dagger_{d,-1}\,c^\dagger_{f,1}\,c_{f,2}\,c_{f,-2} - \frac{2 \sqrt{5}}{15}\,c^\dagger_{d,-1}\,c^\dagger_{f,1}\,c_{f,1}\,c_{f,-1}
\\
&\qquad + \frac{3 \sqrt{10}}{10}\,c^\dagger_{d,0}\,c^\dagger_{f,0}\,c_{f,3}\,c_{f,-3} - \frac{\sqrt{10}}{5}\,c^\dagger_{d,0}\,c^\dagger_{f,0}\,c_{f,2}\,c_{f,-2} + \frac{\sqrt{10}}{10}\,c^\dagger_{d,0}\,c^\dagger_{f,0}\,c_{f,1}\,c_{f,-1}
\\
&\qquad - \frac{2 \sqrt{5}}{5}\,c^\dagger_{d,1}\,c^\dagger_{f,-1}\,c_{f,3}\,c_{f,-3} + \frac{4 \sqrt{5}}{15}\,c^\dagger_{d,1}\,c^\dagger_{f,-1}\,c_{f,2}\,c_{f,-2} - \frac{2 \sqrt{5}}{15}\,c^\dagger_{d,1}\,c^\dagger_{f,-1}\,c_{f,1}\,c_{f,-1}
\\
&\qquad + \frac{\sqrt{2}}{2}\,c^\dagger_{d,2}\,c^\dagger_{f,-2}\,c_{f,3}\,c_{f,-3} - \frac{\sqrt{2}}{3}\,c^\dagger_{d,2}\,c^\dagger_{f,-2}\,c_{f,2}\,c_{f,-2} + \frac{\sqrt{2}}{6}\,c^\dagger_{d,2}\,c^\dagger_{f,-2}\,c_{f,1}\,c_{f,-1}
\\
&\qquad + \frac{\sqrt{2}}{2}\,c^\dagger_{d,-2}\,c^\dagger_{f,3}\,c_{f,3}\,c_{f,-2} - \frac{\sqrt{30}}{6}\,c^\dagger_{d,-2}\,c^\dagger_{f,3}\,c_{f,2}\,c_{f,-1} + c^\dagger_{d,-2}\,c^\dagger_{f,3}\,c_{f,1}\,c_{f,0}
\\
&\qquad - \frac{\sqrt{3}}{3}\,c^\dagger_{d,-1}\,c^\dagger_{f,2}\,c_{f,3}\,c_{f,-2} + \frac{\sqrt{5}}{3}\,c^\dagger_{d,-1}\,c^\dagger_{f,2}\,c_{f,2}\,c_{f,-1} - \frac{\sqrt{6}}{3}\,c^\dagger_{d,-1}\,c^\dagger_{f,2}\,c_{f,1}\,c_{f,0}
\\
&\qquad + \frac{\sqrt{5}}{5}\,c^\dagger_{d,0}\,c^\dagger_{f,1}\,c_{f,3}\,c_{f,-2} - \frac{\sqrt{3}}{3}\,c^\dagger_{d,0}\,c^\dagger_{f,1}\,c_{f,2}\,c_{f,-1} + \frac{\sqrt{10}}{5}\,c^\dagger_{d,0}\,c^\dagger_{f,1}\,c_{f,1}\,c_{f,0}
\\
&\qquad - \frac{\sqrt{10}}{10}\,c^\dagger_{d,1}\,c^\dagger_{f,0}\,c_{f,3}\,c_{f,-2} + \frac{\sqrt{6}}{6}\,c^\dagger_{d,1}\,c^\dagger_{f,0}\,c_{f,2}\,c_{f,-1} - \frac{\sqrt{5}}{5}\,c^\dagger_{d,1}\,c^\dagger_{f,0}\,c_{f,1}\,c_{f,0}
\\
&\qquad + \frac{\sqrt{30}}{30}\,c^\dagger_{d,2}\,c^\dagger_{f,-1}\,c_{f,3}\,c_{f,-2} - \frac{\sqrt{2}}{6}\,c^\dagger_{d,2}\,c^\dagger_{f,-1}\,c_{f,2}\,c_{f,-1} + \frac{\sqrt{15}}{15}\,c^\dagger_{d,2}\,c^\dagger_{f,-1}\,c_{f,1}\,c_{f,0}
\\
&\qquad + \frac{\sqrt{30}}{30}\,c^\dagger_{f,2}\,c^\dagger_{f,-3}\,c_{d,-2}\,c_{f,1} - \frac{\sqrt{2}}{6}\,c^\dagger_{f,1}\,c^\dagger_{f,-2}\,c_{d,-2}\,c_{f,1} + \frac{\sqrt{15}}{15}\,c^\dagger_{f,0}\,c^\dagger_{f,-1}\,c_{d,-2}\,c_{f,1}
\\
&\qquad - \frac{\sqrt{10}}{10}\,c^\dagger_{f,2}\,c^\dagger_{f,-3}\,c_{d,-1}\,c_{f,0} + \frac{\sqrt{6}}{6}\,c^\dagger_{f,1}\,c^\dagger_{f,-2}\,c_{d,-1}\,c_{f,0} - \frac{\sqrt{5}}{5}\,c^\dagger_{f,0}\,c^\dagger_{f,-1}\,c_{d,-1}\,c_{f,0}
\\
&\qquad + \frac{\sqrt{5}}{5}\,c^\dagger_{f,2}\,c^\dagger_{f,-3}\,c_{d,0}\,c_{f,-1} - \frac{\sqrt{3}}{3}\,c^\dagger_{f,1}\,c^\dagger_{f,-2}\,c_{d,0}\,c_{f,-1} + \frac{\sqrt{10}}{5}\,c^\dagger_{f,0}\,c^\dagger_{f,-1}\,c_{d,0}\,c_{f,-1}
\\
&\qquad - \frac{\sqrt{3}}{3}\,c^\dagger_{f,2}\,c^\dagger_{f,-3}\,c_{d,1}\,c_{f,-2} + \frac{\sqrt{5}}{3}\,c^\dagger_{f,1}\,c^\dagger_{f,-2}\,c_{d,1}\,c_{f,-2} - \frac{\sqrt{6}}{3}\,c^\dagger_{f,0}\,c^\dagger_{f,-1}\,c_{d,1}\,c_{f,-2}
\\
&\qquad + \frac{\sqrt{2}}{2}\,c^\dagger_{f,2}\,c^\dagger_{f,-3}\,c_{d,2}\,c_{f,-3} - \frac{\sqrt{30}}{6}\,c^\dagger_{f,1}\,c^\dagger_{f,-2}\,c_{d,2}\,c_{f,-3} + c^\dagger_{f,0}\,c^\dagger_{f,-1}\,c_{d,2}\,c_{f,-3}
\\
&\qquad + \frac{\sqrt{2}}{2}\,c^\dagger_{f,3}\,c^\dagger_{f,-3}\,c_{d,-2}\,c_{f,2} - \frac{\sqrt{2}}{3}\,c^\dagger_{f,2}\,c^\dagger_{f,-2}\,c_{d,-2}\,c_{f,2} + \frac{\sqrt{2}}{6}\,c^\dagger_{f,1}\,c^\dagger_{f,-1}\,c_{d,-2}\,c_{f,2}
\\
&\qquad - \frac{2 \sqrt{5}}{5}\,c^\dagger_{f,3}\,c^\dagger_{f,-3}\,c_{d,-1}\,c_{f,1} + \frac{4 \sqrt{5}}{15}\,c^\dagger_{f,2}\,c^\dagger_{f,-2}\,c_{d,-1}\,c_{f,1} - \frac{2 \sqrt{5}}{15}\,c^\dagger_{f,1}\,c^\dagger_{f,-1}\,c_{d,-1}\,c_{f,1}
\\
&\qquad + \frac{3 \sqrt{10}}{10}\,c^\dagger_{f,3}\,c^\dagger_{f,-3}\,c_{d,0}\,c_{f,0} - \frac{\sqrt{10}}{5}\,c^\dagger_{f,2}\,c^\dagger_{f,-2}\,c_{d,0}\,c_{f,0} + \frac{\sqrt{10}}{10}\,c^\dagger_{f,1}\,c^\dagger_{f,-1}\,c_{d,0}\,c_{f,0}
\\
&\qquad - \frac{2 \sqrt{5}}{5}\,c^\dagger_{f,3}\,c^\dagger_{f,-3}\,c_{d,1}\,c_{f,-1} + \frac{4 \sqrt{5}}{15}\,c^\dagger_{f,2}\,c^\dagger_{f,-2}\,c_{d,1}\,c_{f,-1} - \frac{2 \sqrt{5}}{15}\,c^\dagger_{f,1}\,c^\dagger_{f,-1}\,c_{d,1}\,c_{f,-1}
\\
&\qquad + \frac{\sqrt{2}}{2}\,c^\dagger_{f,3}\,c^\dagger_{f,-3}\,c_{d,2}\,c_{f,-2} - \frac{\sqrt{2}}{3}\,c^\dagger_{f,2}\,c^\dagger_{f,-2}\,c_{d,2}\,c_{f,-2} + \frac{\sqrt{2}}{6}\,c^\dagger_{f,1}\,c^\dagger_{f,-1}\,c_{d,2}\,c_{f,-2}
\\
&\qquad + \frac{\sqrt{2}}{2}\,c^\dagger_{f,3}\,c^\dagger_{f,-2}\,c_{d,-2}\,c_{f,3} - \frac{\sqrt{30}}{6}\,c^\dagger_{f,2}\,c^\dagger_{f,-1}\,c_{d,-2}\,c_{f,3} + c^\dagger_{f,1}\,c^\dagger_{f,0}\,c_{d,-2}\,c_{f,3}
\\
&\qquad - \frac{\sqrt{3}}{3}\,c^\dagger_{f,3}\,c^\dagger_{f,-2}\,c_{d,-1}\,c_{f,2} + \frac{\sqrt{5}}{3}\,c^\dagger_{f,2}\,c^\dagger_{f,-1}\,c_{d,-1}\,c_{f,2} - \frac{\sqrt{6}}{3}\,c^\dagger_{f,1}\,c^\dagger_{f,0}\,c_{d,-1}\,c_{f,2}
\\
&\qquad + \frac{\sqrt{5}}{5}\,c^\dagger_{f,3}\,c^\dagger_{f,-2}\,c_{d,0}\,c_{f,1} - \frac{\sqrt{3}}{3}\,c^\dagger_{f,2}\,c^\dagger_{f,-1}\,c_{d,0}\,c_{f,1} + \frac{\sqrt{10}}{5}\,c^\dagger_{f,1}\,c^\dagger_{f,0}\,c_{d,0}\,c_{f,1}
\\
&\qquad - \frac{\sqrt{10}}{10}\,c^\dagger_{f,3}\,c^\dagger_{f,-2}\,c_{d,1}\,c_{f,0} + \frac{\sqrt{6}}{6}\,c^\dagger_{f,2}\,c^\dagger_{f,-1}\,c_{d,1}\,c_{f,0} - \frac{\sqrt{5}}{5}\,c^\dagger_{f,1}\,c^\dagger_{f,0}\,c_{d,1}\,c_{f,0}
\\
&\qquad + \frac{\sqrt{30}}{30}\,c^\dagger_{f,3}\,c^\dagger_{f,-2}\,c_{d,2}\,c_{f,-1} - \frac{\sqrt{2}}{6}\,c^\dagger_{f,2}\,c^\dagger_{f,-1}\,c_{d,2}\,c_{f,-1} + \frac{\sqrt{15}}{15}\,c^\dagger_{f,1}\,c^\dagger_{f,0}\,c_{d,2}\,c_{f,-1}\Big)
\\
{M_{0}}_{[1]} &=  
\frac{i}{2} \left(\left[[c^\dagger_d \times c^\dagger_f]_{1} \times[\tilde c_f \times \tilde c_f]_{1}\right]_{0} - h.c. \right)\\
&= 
  \frac{1}{7} {\bm Q}^{df}_1 \cdot {\bm M}_1^{f}
+ \frac{3\sqrt{2}}{14}  {\bm M}^{df}_2 \cdot {\bm Q}_2^{f}
+ \frac{\sqrt{42}}{14}  {\bm Q}^{df}_3 \cdot {\bm M}_3^{f}
+ \frac{\sqrt{66}}{14}  {\bm M}^{df}_4 \cdot {\bm Q}_4^{f}
+ \frac{\sqrt{66}}{14}  {\bm Q}^{df}_5 \cdot {\bm M}_5^{f}
\\ 
&= 
\frac{\sqrt{6} i}{14}\Big(\frac{\sqrt{30}}{30}\,c^\dagger_{d,-2}\,c^\dagger_{f,1}\,c_{f,2}\,c_{f,-3} - \frac{\sqrt{2}}{6}\,c^\dagger_{d,-2}\,c^\dagger_{f,1}\,c_{f,1}\,c_{f,-2} + \frac{\sqrt{15}}{15}\,c^\dagger_{d,-2}\,c^\dagger_{f,1}\,c_{f,0}\,c_{f,-1}
\\
&\qquad - \frac{\sqrt{10}}{10}\,c^\dagger_{d,-1}\,c^\dagger_{f,0}\,c_{f,2}\,c_{f,-3} + \frac{\sqrt{6}}{6}\,c^\dagger_{d,-1}\,c^\dagger_{f,0}\,c_{f,1}\,c_{f,-2} - \frac{\sqrt{5}}{5}\,c^\dagger_{d,-1}\,c^\dagger_{f,0}\,c_{f,0}\,c_{f,-1}
\\
&\qquad + \frac{\sqrt{5}}{5}\,c^\dagger_{d,0}\,c^\dagger_{f,-1}\,c_{f,2}\,c_{f,-3} - \frac{\sqrt{3}}{3}\,c^\dagger_{d,0}\,c^\dagger_{f,-1}\,c_{f,1}\,c_{f,-2} + \frac{\sqrt{10}}{5}\,c^\dagger_{d,0}\,c^\dagger_{f,-1}\,c_{f,0}\,c_{f,-1}
\\
&\qquad - \frac{\sqrt{3}}{3}\,c^\dagger_{d,1}\,c^\dagger_{f,-2}\,c_{f,2}\,c_{f,-3} + \frac{\sqrt{5}}{3}\,c^\dagger_{d,1}\,c^\dagger_{f,-2}\,c_{f,1}\,c_{f,-2} - \frac{\sqrt{6}}{3}\,c^\dagger_{d,1}\,c^\dagger_{f,-2}\,c_{f,0}\,c_{f,-1}
\\
&\qquad + \frac{\sqrt{2}}{2}\,c^\dagger_{d,2}\,c^\dagger_{f,-3}\,c_{f,2}\,c_{f,-3} - \frac{\sqrt{30}}{6}\,c^\dagger_{d,2}\,c^\dagger_{f,-3}\,c_{f,1}\,c_{f,-2} + c^\dagger_{d,2}\,c^\dagger_{f,-3}\,c_{f,0}\,c_{f,-1}
\\
&\qquad + \frac{\sqrt{2}}{2}\,c^\dagger_{d,-2}\,c^\dagger_{f,2}\,c_{f,3}\,c_{f,-3} - \frac{\sqrt{2}}{3}\,c^\dagger_{d,-2}\,c^\dagger_{f,2}\,c_{f,2}\,c_{f,-2} + \frac{\sqrt{2}}{6}\,c^\dagger_{d,-2}\,c^\dagger_{f,2}\,c_{f,1}\,c_{f,-1}
\\
&\qquad - \frac{2 \sqrt{5}}{5}\,c^\dagger_{d,-1}\,c^\dagger_{f,1}\,c_{f,3}\,c_{f,-3} + \frac{4 \sqrt{5}}{15}\,c^\dagger_{d,-1}\,c^\dagger_{f,1}\,c_{f,2}\,c_{f,-2} - \frac{2 \sqrt{5}}{15}\,c^\dagger_{d,-1}\,c^\dagger_{f,1}\,c_{f,1}\,c_{f,-1}
\\
&\qquad + \frac{3 \sqrt{10}}{10}\,c^\dagger_{d,0}\,c^\dagger_{f,0}\,c_{f,3}\,c_{f,-3} - \frac{\sqrt{10}}{5}\,c^\dagger_{d,0}\,c^\dagger_{f,0}\,c_{f,2}\,c_{f,-2} + \frac{\sqrt{10}}{10}\,c^\dagger_{d,0}\,c^\dagger_{f,0}\,c_{f,1}\,c_{f,-1}
\\
&\qquad - \frac{2 \sqrt{5}}{5}\,c^\dagger_{d,1}\,c^\dagger_{f,-1}\,c_{f,3}\,c_{f,-3} + \frac{4 \sqrt{5}}{15}\,c^\dagger_{d,1}\,c^\dagger_{f,-1}\,c_{f,2}\,c_{f,-2} - \frac{2 \sqrt{5}}{15}\,c^\dagger_{d,1}\,c^\dagger_{f,-1}\,c_{f,1}\,c_{f,-1}
\\
&\qquad + \frac{\sqrt{2}}{2}\,c^\dagger_{d,2}\,c^\dagger_{f,-2}\,c_{f,3}\,c_{f,-3} - \frac{\sqrt{2}}{3}\,c^\dagger_{d,2}\,c^\dagger_{f,-2}\,c_{f,2}\,c_{f,-2} + \frac{\sqrt{2}}{6}\,c^\dagger_{d,2}\,c^\dagger_{f,-2}\,c_{f,1}\,c_{f,-1}
\\
&\qquad + \frac{\sqrt{2}}{2}\,c^\dagger_{d,-2}\,c^\dagger_{f,3}\,c_{f,3}\,c_{f,-2} - \frac{\sqrt{30}}{6}\,c^\dagger_{d,-2}\,c^\dagger_{f,3}\,c_{f,2}\,c_{f,-1} + c^\dagger_{d,-2}\,c^\dagger_{f,3}\,c_{f,1}\,c_{f,0}
\\
&\qquad - \frac{\sqrt{3}}{3}\,c^\dagger_{d,-1}\,c^\dagger_{f,2}\,c_{f,3}\,c_{f,-2} + \frac{\sqrt{5}}{3}\,c^\dagger_{d,-1}\,c^\dagger_{f,2}\,c_{f,2}\,c_{f,-1} - \frac{\sqrt{6}}{3}\,c^\dagger_{d,-1}\,c^\dagger_{f,2}\,c_{f,1}\,c_{f,0}
\\
&\qquad + \frac{\sqrt{5}}{5}\,c^\dagger_{d,0}\,c^\dagger_{f,1}\,c_{f,3}\,c_{f,-2} - \frac{\sqrt{3}}{3}\,c^\dagger_{d,0}\,c^\dagger_{f,1}\,c_{f,2}\,c_{f,-1} + \frac{\sqrt{10}}{5}\,c^\dagger_{d,0}\,c^\dagger_{f,1}\,c_{f,1}\,c_{f,0}
\\
&\qquad - \frac{\sqrt{10}}{10}\,c^\dagger_{d,1}\,c^\dagger_{f,0}\,c_{f,3}\,c_{f,-2} + \frac{\sqrt{6}}{6}\,c^\dagger_{d,1}\,c^\dagger_{f,0}\,c_{f,2}\,c_{f,-1} - \frac{\sqrt{5}}{5}\,c^\dagger_{d,1}\,c^\dagger_{f,0}\,c_{f,1}\,c_{f,0}
\\
&\qquad + \frac{\sqrt{30}}{30}\,c^\dagger_{d,2}\,c^\dagger_{f,-1}\,c_{f,3}\,c_{f,-2} - \frac{\sqrt{2}}{6}\,c^\dagger_{d,2}\,c^\dagger_{f,-1}\,c_{f,2}\,c_{f,-1} + \frac{\sqrt{15}}{15}\,c^\dagger_{d,2}\,c^\dagger_{f,-1}\,c_{f,1}\,c_{f,0}
\\
&\qquad - \frac{\sqrt{30}}{30}\,c^\dagger_{f,2}\,c^\dagger_{f,-3}\,c_{d,-2}\,c_{f,1} + \frac{\sqrt{2}}{6}\,c^\dagger_{f,1}\,c^\dagger_{f,-2}\,c_{d,-2}\,c_{f,1} - \frac{\sqrt{15}}{15}\,c^\dagger_{f,0}\,c^\dagger_{f,-1}\,c_{d,-2}\,c_{f,1}
\\
&\qquad + \frac{\sqrt{10}}{10}\,c^\dagger_{f,2}\,c^\dagger_{f,-3}\,c_{d,-1}\,c_{f,0} - \frac{\sqrt{6}}{6}\,c^\dagger_{f,1}\,c^\dagger_{f,-2}\,c_{d,-1}\,c_{f,0} + \frac{\sqrt{5}}{5}\,c^\dagger_{f,0}\,c^\dagger_{f,-1}\,c_{d,-1}\,c_{f,0}
\\
&\qquad - \frac{\sqrt{5}}{5}\,c^\dagger_{f,2}\,c^\dagger_{f,-3}\,c_{d,0}\,c_{f,-1} + \frac{\sqrt{3}}{3}\,c^\dagger_{f,1}\,c^\dagger_{f,-2}\,c_{d,0}\,c_{f,-1} - \frac{\sqrt{10}}{5}\,c^\dagger_{f,0}\,c^\dagger_{f,-1}\,c_{d,0}\,c_{f,-1}
\\
&\qquad + \frac{\sqrt{3}}{3}\,c^\dagger_{f,2}\,c^\dagger_{f,-3}\,c_{d,1}\,c_{f,-2} - \frac{\sqrt{5}}{3}\,c^\dagger_{f,1}\,c^\dagger_{f,-2}\,c_{d,1}\,c_{f,-2} + \frac{\sqrt{6}}{3}\,c^\dagger_{f,0}\,c^\dagger_{f,-1}\,c_{d,1}\,c_{f,-2}
\\
&\qquad - \frac{\sqrt{2}}{2}\,c^\dagger_{f,2}\,c^\dagger_{f,-3}\,c_{d,2}\,c_{f,-3} + \frac{\sqrt{30}}{6}\,c^\dagger_{f,1}\,c^\dagger_{f,-2}\,c_{d,2}\,c_{f,-3} - c^\dagger_{f,0}\,c^\dagger_{f,-1}\,c_{d,2}\,c_{f,-3}
\\
&\qquad - \frac{\sqrt{2}}{2}\,c^\dagger_{f,3}\,c^\dagger_{f,-3}\,c_{d,-2}\,c_{f,2} + \frac{\sqrt{2}}{3}\,c^\dagger_{f,2}\,c^\dagger_{f,-2}\,c_{d,-2}\,c_{f,2} - \frac{\sqrt{2}}{6}\,c^\dagger_{f,1}\,c^\dagger_{f,-1}\,c_{d,-2}\,c_{f,2}
\\
&\qquad + \frac{2 \sqrt{5}}{5}\,c^\dagger_{f,3}\,c^\dagger_{f,-3}\,c_{d,-1}\,c_{f,1} - \frac{4 \sqrt{5}}{15}\,c^\dagger_{f,2}\,c^\dagger_{f,-2}\,c_{d,-1}\,c_{f,1} + \frac{2 \sqrt{5}}{15}\,c^\dagger_{f,1}\,c^\dagger_{f,-1}\,c_{d,-1}\,c_{f,1}
\\
&\qquad - \frac{3 \sqrt{10}}{10}\,c^\dagger_{f,3}\,c^\dagger_{f,-3}\,c_{d,0}\,c_{f,0} + \frac{\sqrt{10}}{5}\,c^\dagger_{f,2}\,c^\dagger_{f,-2}\,c_{d,0}\,c_{f,0} - \frac{\sqrt{10}}{10}\,c^\dagger_{f,1}\,c^\dagger_{f,-1}\,c_{d,0}\,c_{f,0}
\\
&\qquad + \frac{2 \sqrt{5}}{5}\,c^\dagger_{f,3}\,c^\dagger_{f,-3}\,c_{d,1}\,c_{f,-1} - \frac{4 \sqrt{5}}{15}\,c^\dagger_{f,2}\,c^\dagger_{f,-2}\,c_{d,1}\,c_{f,-1} + \frac{2 \sqrt{5}}{15}\,c^\dagger_{f,1}\,c^\dagger_{f,-1}\,c_{d,1}\,c_{f,-1}
\\
&\qquad - \frac{\sqrt{2}}{2}\,c^\dagger_{f,3}\,c^\dagger_{f,-3}\,c_{d,2}\,c_{f,-2} + \frac{\sqrt{2}}{3}\,c^\dagger_{f,2}\,c^\dagger_{f,-2}\,c_{d,2}\,c_{f,-2} - \frac{\sqrt{2}}{6}\,c^\dagger_{f,1}\,c^\dagger_{f,-1}\,c_{d,2}\,c_{f,-2}
\\
&\qquad - \frac{\sqrt{2}}{2}\,c^\dagger_{f,3}\,c^\dagger_{f,-2}\,c_{d,-2}\,c_{f,3} + \frac{\sqrt{30}}{6}\,c^\dagger_{f,2}\,c^\dagger_{f,-1}\,c_{d,-2}\,c_{f,3} - c^\dagger_{f,1}\,c^\dagger_{f,0}\,c_{d,-2}\,c_{f,3}
\\
&\qquad + \frac{\sqrt{3}}{3}\,c^\dagger_{f,3}\,c^\dagger_{f,-2}\,c_{d,-1}\,c_{f,2} - \frac{\sqrt{5}}{3}\,c^\dagger_{f,2}\,c^\dagger_{f,-1}\,c_{d,-1}\,c_{f,2} + \frac{\sqrt{6}}{3}\,c^\dagger_{f,1}\,c^\dagger_{f,0}\,c_{d,-1}\,c_{f,2}
\\
&\qquad - \frac{\sqrt{5}}{5}\,c^\dagger_{f,3}\,c^\dagger_{f,-2}\,c_{d,0}\,c_{f,1} + \frac{\sqrt{3}}{3}\,c^\dagger_{f,2}\,c^\dagger_{f,-1}\,c_{d,0}\,c_{f,1} - \frac{\sqrt{10}}{5}\,c^\dagger_{f,1}\,c^\dagger_{f,0}\,c_{d,0}\,c_{f,1}
\\
&\qquad + \frac{\sqrt{10}}{10}\,c^\dagger_{f,3}\,c^\dagger_{f,-2}\,c_{d,1}\,c_{f,0} - \frac{\sqrt{6}}{6}\,c^\dagger_{f,2}\,c^\dagger_{f,-1}\,c_{d,1}\,c_{f,0} + \frac{\sqrt{5}}{5}\,c^\dagger_{f,1}\,c^\dagger_{f,0}\,c_{d,1}\,c_{f,0}
\\
&\qquad - \frac{\sqrt{30}}{30}\,c^\dagger_{f,3}\,c^\dagger_{f,-2}\,c_{d,2}\,c_{f,-1} + \frac{\sqrt{2}}{6}\,c^\dagger_{f,2}\,c^\dagger_{f,-1}\,c_{d,2}\,c_{f,-1} - \frac{\sqrt{15}}{15}\,c^\dagger_{f,1}\,c^\dagger_{f,0}\,c_{d,2}\,c_{f,-1}\Big)
\end{align*}
% 
% 
% 
% % df3 : 3×3
% j13,j24,A wigner_9j： 1 , 1 , -sqrt(42)/14
% j13,j24,A wigner_9j： 2 , 2 , -2*sqrt(21)/21
% j13,j24,A wigner_9j： 3 , 3 , 1/6
% j13,j24,A wigner_9j： 4 , 4 , sqrt(77)/14
% j13,j24,A wigner_9j： 5 , 5 , -sqrt(77)/21
% 
\begin{align*}
{G_{0}}_{[2]} &=  
\frac{1}{2} \left(\left[[c^\dagger_d \times c^\dagger_f]_{3} \times[\tilde c_f \times \tilde c_f]_{3}\right]_{0} + h.c. \right)\\
&= 
  \frac{\sqrt{42}}{14} {\bm T}^{df}_1 \cdot {\bm M}_1^{f}
+ \frac{2\sqrt{21}}{21}  {\bm G}^{df}_2 \cdot {\bm Q}_2^{f}
- \frac{1}{6}  {\bm T}^{df}_3 \cdot {\bm M}_3^{f}
-\frac{\sqrt{77}}{14}  {\bm G}^{df}_4 \cdot {\bm Q}_4^{f}
+ \frac{\sqrt{77}}{21}  {\bm T}^{df}_5 \cdot {\bm M}_5^{f}
\\ 
&= 
\frac{\sqrt{35}}{42}\Big(\frac{\sqrt{5}}{5}\,c^\dagger_{d,-2}\,c^\dagger_{f,-1}\,c_{f,0}\,c_{f,-3} - \frac{\sqrt{10}}{5}\,c^\dagger_{d,-2}\,c^\dagger_{f,-1}\,c_{f,-1}\,c_{f,-2} - \frac{\sqrt{2}}{2}\,c^\dagger_{d,-1}\,c^\dagger_{f,-2}\,c_{f,0}\,c_{f,-3}
\\
&\qquad + c^\dagger_{d,-1}\,c^\dagger_{f,-2}\,c_{f,-1}\,c_{f,-2} + \frac{\sqrt{2}}{2}\,c^\dagger_{d,0}\,c^\dagger_{f,-3}\,c_{f,0}\,c_{f,-3} - c^\dagger_{d,0}\,c^\dagger_{f,-3}\,c_{f,-1}\,c_{f,-2}
\\
&\qquad + \frac{2 \sqrt{5}}{5}\,c^\dagger_{d,-2}\,c^\dagger_{f,0}\,c_{f,1}\,c_{f,-3} - \frac{\sqrt{10}}{5}\,c^\dagger_{d,-2}\,c^\dagger_{f,0}\,c_{f,0}\,c_{f,-2} - \frac{\sqrt{15}}{5}\,c^\dagger_{d,-1}\,c^\dagger_{f,-1}\,c_{f,1}\,c_{f,-3}
\\
&\qquad + \frac{\sqrt{30}}{10}\,c^\dagger_{d,-1}\,c^\dagger_{f,-1}\,c_{f,0}\,c_{f,-2} + c^\dagger_{d,1}\,c^\dagger_{f,-3}\,c_{f,1}\,c_{f,-3} - \frac{\sqrt{2}}{2}\,c^\dagger_{d,1}\,c^\dagger_{f,-3}\,c_{f,0}\,c_{f,-2}
\\
&\qquad + \frac{2 \sqrt{6}}{5}\,c^\dagger_{d,-2}\,c^\dagger_{f,1}\,c_{f,2}\,c_{f,-3} - \frac{2 \sqrt{3}}{5}\,c^\dagger_{d,-2}\,c^\dagger_{f,1}\,c_{f,0}\,c_{f,-1} - \frac{\sqrt{2}}{5}\,c^\dagger_{d,-1}\,c^\dagger_{f,0}\,c_{f,2}\,c_{f,-3}
\\
&\qquad + \frac{1}{5}\,c^\dagger_{d,-1}\,c^\dagger_{f,0}\,c_{f,0}\,c_{f,-1} - \frac{3}{5}\,c^\dagger_{d,0}\,c^\dagger_{f,-1}\,c_{f,2}\,c_{f,-3} + \frac{3 \sqrt{2}}{10}\,c^\dagger_{d,0}\,c^\dagger_{f,-1}\,c_{f,0}\,c_{f,-1}
\\
&\qquad + \frac{\sqrt{15}}{5}\,c^\dagger_{d,1}\,c^\dagger_{f,-2}\,c_{f,2}\,c_{f,-3} - \frac{\sqrt{30}}{10}\,c^\dagger_{d,1}\,c^\dagger_{f,-2}\,c_{f,0}\,c_{f,-1} + \frac{\sqrt{10}}{5}\,c^\dagger_{d,2}\,c^\dagger_{f,-3}\,c_{f,2}\,c_{f,-3}
\\
&\qquad - \frac{\sqrt{5}}{5}\,c^\dagger_{d,2}\,c^\dagger_{f,-3}\,c_{f,0}\,c_{f,-1} + \frac{\sqrt{10}}{5}\,c^\dagger_{d,-2}\,c^\dagger_{f,2}\,c_{f,3}\,c_{f,-3} + \frac{\sqrt{10}}{5}\,c^\dagger_{d,-2}\,c^\dagger_{f,2}\,c_{f,2}\,c_{f,-2}
\\
&\qquad - \frac{\sqrt{10}}{5}\,c^\dagger_{d,-2}\,c^\dagger_{f,2}\,c_{f,1}\,c_{f,-1} + \frac{1}{5}\,c^\dagger_{d,-1}\,c^\dagger_{f,1}\,c_{f,3}\,c_{f,-3} + \frac{1}{5}\,c^\dagger_{d,-1}\,c^\dagger_{f,1}\,c_{f,2}\,c_{f,-2}
\\
&\qquad - \frac{1}{5}\,c^\dagger_{d,-1}\,c^\dagger_{f,1}\,c_{f,1}\,c_{f,-1} - \frac{2 \sqrt{2}}{5}\,c^\dagger_{d,0}\,c^\dagger_{f,0}\,c_{f,3}\,c_{f,-3} - \frac{2 \sqrt{2}}{5}\,c^\dagger_{d,0}\,c^\dagger_{f,0}\,c_{f,2}\,c_{f,-2}
\\
&\qquad + \frac{2 \sqrt{2}}{5}\,c^\dagger_{d,0}\,c^\dagger_{f,0}\,c_{f,1}\,c_{f,-1} + \frac{1}{5}\,c^\dagger_{d,1}\,c^\dagger_{f,-1}\,c_{f,3}\,c_{f,-3} + \frac{1}{5}\,c^\dagger_{d,1}\,c^\dagger_{f,-1}\,c_{f,2}\,c_{f,-2}
\\
&\qquad - \frac{1}{5}\,c^\dagger_{d,1}\,c^\dagger_{f,-1}\,c_{f,1}\,c_{f,-1} + \frac{\sqrt{10}}{5}\,c^\dagger_{d,2}\,c^\dagger_{f,-2}\,c_{f,3}\,c_{f,-3} + \frac{\sqrt{10}}{5}\,c^\dagger_{d,2}\,c^\dagger_{f,-2}\,c_{f,2}\,c_{f,-2}
\\
&\qquad - \frac{\sqrt{10}}{5}\,c^\dagger_{d,2}\,c^\dagger_{f,-2}\,c_{f,1}\,c_{f,-1} + \frac{\sqrt{10}}{5}\,c^\dagger_{d,-2}\,c^\dagger_{f,3}\,c_{f,3}\,c_{f,-2} - \frac{\sqrt{5}}{5}\,c^\dagger_{d,-2}\,c^\dagger_{f,3}\,c_{f,1}\,c_{f,0}
\\
&\qquad + \frac{\sqrt{15}}{5}\,c^\dagger_{d,-1}\,c^\dagger_{f,2}\,c_{f,3}\,c_{f,-2} - \frac{\sqrt{30}}{10}\,c^\dagger_{d,-1}\,c^\dagger_{f,2}\,c_{f,1}\,c_{f,0} - \frac{3}{5}\,c^\dagger_{d,0}\,c^\dagger_{f,1}\,c_{f,3}\,c_{f,-2}
\\
&\qquad + \frac{3 \sqrt{2}}{10}\,c^\dagger_{d,0}\,c^\dagger_{f,1}\,c_{f,1}\,c_{f,0} - \frac{\sqrt{2}}{5}\,c^\dagger_{d,1}\,c^\dagger_{f,0}\,c_{f,3}\,c_{f,-2} + \frac{1}{5}\,c^\dagger_{d,1}\,c^\dagger_{f,0}\,c_{f,1}\,c_{f,0}
\\
&\qquad + \frac{2 \sqrt{6}}{5}\,c^\dagger_{d,2}\,c^\dagger_{f,-1}\,c_{f,3}\,c_{f,-2} - \frac{2 \sqrt{3}}{5}\,c^\dagger_{d,2}\,c^\dagger_{f,-1}\,c_{f,1}\,c_{f,0} + c^\dagger_{d,-1}\,c^\dagger_{f,3}\,c_{f,3}\,c_{f,-1}
\\
&\qquad - \frac{\sqrt{2}}{2}\,c^\dagger_{d,-1}\,c^\dagger_{f,3}\,c_{f,2}\,c_{f,0} - \frac{\sqrt{15}}{5}\,c^\dagger_{d,1}\,c^\dagger_{f,1}\,c_{f,3}\,c_{f,-1} + \frac{\sqrt{30}}{10}\,c^\dagger_{d,1}\,c^\dagger_{f,1}\,c_{f,2}\,c_{f,0}
\\
&\qquad + \frac{2 \sqrt{5}}{5}\,c^\dagger_{d,2}\,c^\dagger_{f,0}\,c_{f,3}\,c_{f,-1} - \frac{\sqrt{10}}{5}\,c^\dagger_{d,2}\,c^\dagger_{f,0}\,c_{f,2}\,c_{f,0} + \frac{\sqrt{2}}{2}\,c^\dagger_{d,0}\,c^\dagger_{f,3}\,c_{f,3}\,c_{f,0}
\\
&\qquad - c^\dagger_{d,0}\,c^\dagger_{f,3}\,c_{f,2}\,c_{f,1} - \frac{\sqrt{2}}{2}\,c^\dagger_{d,1}\,c^\dagger_{f,2}\,c_{f,3}\,c_{f,0} + c^\dagger_{d,1}\,c^\dagger_{f,2}\,c_{f,2}\,c_{f,1}
\\
&\qquad + \frac{\sqrt{5}}{5}\,c^\dagger_{d,2}\,c^\dagger_{f,1}\,c_{f,3}\,c_{f,0} - \frac{\sqrt{10}}{5}\,c^\dagger_{d,2}\,c^\dagger_{f,1}\,c_{f,2}\,c_{f,1} + \frac{\sqrt{5}}{5}\,c^\dagger_{f,0}\,c^\dagger_{f,-3}\,c_{d,-2}\,c_{f,-1}
\\
&\qquad - \frac{\sqrt{10}}{5}\,c^\dagger_{f,-1}\,c^\dagger_{f,-2}\,c_{d,-2}\,c_{f,-1} - \frac{\sqrt{2}}{2}\,c^\dagger_{f,0}\,c^\dagger_{f,-3}\,c_{d,-1}\,c_{f,-2} + c^\dagger_{f,-1}\,c^\dagger_{f,-2}\,c_{d,-1}\,c_{f,-2}
\\
&\qquad + \frac{\sqrt{2}}{2}\,c^\dagger_{f,0}\,c^\dagger_{f,-3}\,c_{d,0}\,c_{f,-3} - c^\dagger_{f,-1}\,c^\dagger_{f,-2}\,c_{d,0}\,c_{f,-3} + \frac{2 \sqrt{5}}{5}\,c^\dagger_{f,1}\,c^\dagger_{f,-3}\,c_{d,-2}\,c_{f,0}
\\
&\qquad - \frac{\sqrt{10}}{5}\,c^\dagger_{f,0}\,c^\dagger_{f,-2}\,c_{d,-2}\,c_{f,0} - \frac{\sqrt{15}}{5}\,c^\dagger_{f,1}\,c^\dagger_{f,-3}\,c_{d,-1}\,c_{f,-1} + \frac{\sqrt{30}}{10}\,c^\dagger_{f,0}\,c^\dagger_{f,-2}\,c_{d,-1}\,c_{f,-1}
\\
&\qquad + c^\dagger_{f,1}\,c^\dagger_{f,-3}\,c_{d,1}\,c_{f,-3} - \frac{\sqrt{2}}{2}\,c^\dagger_{f,0}\,c^\dagger_{f,-2}\,c_{d,1}\,c_{f,-3} + \frac{2 \sqrt{6}}{5}\,c^\dagger_{f,2}\,c^\dagger_{f,-3}\,c_{d,-2}\,c_{f,1}
\\
&\qquad - \frac{2 \sqrt{3}}{5}\,c^\dagger_{f,0}\,c^\dagger_{f,-1}\,c_{d,-2}\,c_{f,1} - \frac{\sqrt{2}}{5}\,c^\dagger_{f,2}\,c^\dagger_{f,-3}\,c_{d,-1}\,c_{f,0} + \frac{1}{5}\,c^\dagger_{f,0}\,c^\dagger_{f,-1}\,c_{d,-1}\,c_{f,0}
\\
&\qquad - \frac{3}{5}\,c^\dagger_{f,2}\,c^\dagger_{f,-3}\,c_{d,0}\,c_{f,-1} + \frac{3 \sqrt{2}}{10}\,c^\dagger_{f,0}\,c^\dagger_{f,-1}\,c_{d,0}\,c_{f,-1} + \frac{\sqrt{15}}{5}\,c^\dagger_{f,2}\,c^\dagger_{f,-3}\,c_{d,1}\,c_{f,-2}
\\
&\qquad - \frac{\sqrt{30}}{10}\,c^\dagger_{f,0}\,c^\dagger_{f,-1}\,c_{d,1}\,c_{f,-2} + \frac{\sqrt{10}}{5}\,c^\dagger_{f,2}\,c^\dagger_{f,-3}\,c_{d,2}\,c_{f,-3} - \frac{\sqrt{5}}{5}\,c^\dagger_{f,0}\,c^\dagger_{f,-1}\,c_{d,2}\,c_{f,-3}
\\
&\qquad + \frac{\sqrt{10}}{5}\,c^\dagger_{f,3}\,c^\dagger_{f,-3}\,c_{d,-2}\,c_{f,2} + \frac{\sqrt{10}}{5}\,c^\dagger_{f,2}\,c^\dagger_{f,-2}\,c_{d,-2}\,c_{f,2} - \frac{\sqrt{10}}{5}\,c^\dagger_{f,1}\,c^\dagger_{f,-1}\,c_{d,-2}\,c_{f,2}
\\
&\qquad + \frac{1}{5}\,c^\dagger_{f,3}\,c^\dagger_{f,-3}\,c_{d,-1}\,c_{f,1} + \frac{1}{5}\,c^\dagger_{f,2}\,c^\dagger_{f,-2}\,c_{d,-1}\,c_{f,1} - \frac{1}{5}\,c^\dagger_{f,1}\,c^\dagger_{f,-1}\,c_{d,-1}\,c_{f,1}
\\
&\qquad - \frac{2 \sqrt{2}}{5}\,c^\dagger_{f,3}\,c^\dagger_{f,-3}\,c_{d,0}\,c_{f,0} - \frac{2 \sqrt{2}}{5}\,c^\dagger_{f,2}\,c^\dagger_{f,-2}\,c_{d,0}\,c_{f,0} + \frac{2 \sqrt{2}}{5}\,c^\dagger_{f,1}\,c^\dagger_{f,-1}\,c_{d,0}\,c_{f,0}
\\
&\qquad + \frac{1}{5}\,c^\dagger_{f,3}\,c^\dagger_{f,-3}\,c_{d,1}\,c_{f,-1} + \frac{1}{5}\,c^\dagger_{f,2}\,c^\dagger_{f,-2}\,c_{d,1}\,c_{f,-1} - \frac{1}{5}\,c^\dagger_{f,1}\,c^\dagger_{f,-1}\,c_{d,1}\,c_{f,-1}
\\
&\qquad + \frac{\sqrt{10}}{5}\,c^\dagger_{f,3}\,c^\dagger_{f,-3}\,c_{d,2}\,c_{f,-2} + \frac{\sqrt{10}}{5}\,c^\dagger_{f,2}\,c^\dagger_{f,-2}\,c_{d,2}\,c_{f,-2} - \frac{\sqrt{10}}{5}\,c^\dagger_{f,1}\,c^\dagger_{f,-1}\,c_{d,2}\,c_{f,-2}
\\
&\qquad + \frac{\sqrt{10}}{5}\,c^\dagger_{f,3}\,c^\dagger_{f,-2}\,c_{d,-2}\,c_{f,3} - \frac{\sqrt{5}}{5}\,c^\dagger_{f,1}\,c^\dagger_{f,0}\,c_{d,-2}\,c_{f,3} + \frac{\sqrt{15}}{5}\,c^\dagger_{f,3}\,c^\dagger_{f,-2}\,c_{d,-1}\,c_{f,2}
\\
&\qquad - \frac{\sqrt{30}}{10}\,c^\dagger_{f,1}\,c^\dagger_{f,0}\,c_{d,-1}\,c_{f,2} - \frac{3}{5}\,c^\dagger_{f,3}\,c^\dagger_{f,-2}\,c_{d,0}\,c_{f,1} + \frac{3 \sqrt{2}}{10}\,c^\dagger_{f,1}\,c^\dagger_{f,0}\,c_{d,0}\,c_{f,1}
\\
&\qquad - \frac{\sqrt{2}}{5}\,c^\dagger_{f,3}\,c^\dagger_{f,-2}\,c_{d,1}\,c_{f,0} + \frac{1}{5}\,c^\dagger_{f,1}\,c^\dagger_{f,0}\,c_{d,1}\,c_{f,0} + \frac{2 \sqrt{6}}{5}\,c^\dagger_{f,3}\,c^\dagger_{f,-2}\,c_{d,2}\,c_{f,-1}
\\
&\qquad - \frac{2 \sqrt{3}}{5}\,c^\dagger_{f,1}\,c^\dagger_{f,0}\,c_{d,2}\,c_{f,-1} + c^\dagger_{f,3}\,c^\dagger_{f,-1}\,c_{d,-1}\,c_{f,3} - \frac{\sqrt{2}}{2}\,c^\dagger_{f,2}\,c^\dagger_{f,0}\,c_{d,-1}\,c_{f,3}
\\
&\qquad - \frac{\sqrt{15}}{5}\,c^\dagger_{f,3}\,c^\dagger_{f,-1}\,c_{d,1}\,c_{f,1} + \frac{\sqrt{30}}{10}\,c^\dagger_{f,2}\,c^\dagger_{f,0}\,c_{d,1}\,c_{f,1} + \frac{2 \sqrt{5}}{5}\,c^\dagger_{f,3}\,c^\dagger_{f,-1}\,c_{d,2}\,c_{f,0}
\\
&\qquad - \frac{\sqrt{10}}{5}\,c^\dagger_{f,2}\,c^\dagger_{f,0}\,c_{d,2}\,c_{f,0} + \frac{\sqrt{2}}{2}\,c^\dagger_{f,3}\,c^\dagger_{f,0}\,c_{d,0}\,c_{f,3} - c^\dagger_{f,2}\,c^\dagger_{f,1}\,c_{d,0}\,c_{f,3}
\\
&\qquad - \frac{\sqrt{2}}{2}\,c^\dagger_{f,3}\,c^\dagger_{f,0}\,c_{d,1}\,c_{f,2} + c^\dagger_{f,2}\,c^\dagger_{f,1}\,c_{d,1}\,c_{f,2} + \frac{\sqrt{5}}{5}\,c^\dagger_{f,3}\,c^\dagger_{f,0}\,c_{d,2}\,c_{f,1}
\\
&\qquad - \frac{\sqrt{10}}{5}\,c^\dagger_{f,2}\,c^\dagger_{f,1}\,c_{d,2}\,c_{f,1}\Big)
\\
{M_{0}}_{[2]} &=  
\frac{i}{2} \left(\left[[c^\dagger_d \times c^\dagger_f]_{3} \times[\tilde c_f \times \tilde c_f]_{3}\right]_{0} - h.c. \right)\\
&= 
  \frac{\sqrt{42}}{14} {\bm Q}^{df}_1 \cdot {\bm M}_1^{f}
+ \frac{2\sqrt{21}}{21}  {\bm M}^{df}_2 \cdot {\bm Q}_2^{f}
- \frac{1}{6}  {\bm Q}^{df}_3 \cdot {\bm M}_3^{f}
-\frac{\sqrt{77}}{14}  {\bm M}^{df}_4 \cdot {\bm Q}_4^{f}
+ \frac{\sqrt{77}}{21}  {\bm Q}^{df}_5 \cdot {\bm M}_5^{f}
\\ 
&= 
% sqrt(35)*I/42  |max|= 0.140859042454753
% \begin{aligned}
\frac{\sqrt{35} i}{42}\Big(\frac{\sqrt{5}}{5}\,c^\dagger_{d,-2}\,c^\dagger_{f,-1}\,c_{f,0}\,c_{f,-3} - \frac{\sqrt{10}}{5}\,c^\dagger_{d,-2}\,c^\dagger_{f,-1}\,c_{f,-1}\,c_{f,-2} - \frac{\sqrt{2}}{2}\,c^\dagger_{d,-1}\,c^\dagger_{f,-2}\,c_{f,0}\,c_{f,-3}
\\
&\qquad + c^\dagger_{d,-1}\,c^\dagger_{f,-2}\,c_{f,-1}\,c_{f,-2} + \frac{\sqrt{2}}{2}\,c^\dagger_{d,0}\,c^\dagger_{f,-3}\,c_{f,0}\,c_{f,-3} - c^\dagger_{d,0}\,c^\dagger_{f,-3}\,c_{f,-1}\,c_{f,-2}
\\
&\qquad + \frac{2 \sqrt{5}}{5}\,c^\dagger_{d,-2}\,c^\dagger_{f,0}\,c_{f,1}\,c_{f,-3} - \frac{\sqrt{10}}{5}\,c^\dagger_{d,-2}\,c^\dagger_{f,0}\,c_{f,0}\,c_{f,-2} - \frac{\sqrt{15}}{5}\,c^\dagger_{d,-1}\,c^\dagger_{f,-1}\,c_{f,1}\,c_{f,-3}
\\
&\qquad + \frac{\sqrt{30}}{10}\,c^\dagger_{d,-1}\,c^\dagger_{f,-1}\,c_{f,0}\,c_{f,-2} + c^\dagger_{d,1}\,c^\dagger_{f,-3}\,c_{f,1}\,c_{f,-3} - \frac{\sqrt{2}}{2}\,c^\dagger_{d,1}\,c^\dagger_{f,-3}\,c_{f,0}\,c_{f,-2}
\\
&\qquad + \frac{2 \sqrt{6}}{5}\,c^\dagger_{d,-2}\,c^\dagger_{f,1}\,c_{f,2}\,c_{f,-3} - \frac{2 \sqrt{3}}{5}\,c^\dagger_{d,-2}\,c^\dagger_{f,1}\,c_{f,0}\,c_{f,-1} - \frac{\sqrt{2}}{5}\,c^\dagger_{d,-1}\,c^\dagger_{f,0}\,c_{f,2}\,c_{f,-3}
\\
&\qquad + \frac{1}{5}\,c^\dagger_{d,-1}\,c^\dagger_{f,0}\,c_{f,0}\,c_{f,-1} - \frac{3}{5}\,c^\dagger_{d,0}\,c^\dagger_{f,-1}\,c_{f,2}\,c_{f,-3} + \frac{3 \sqrt{2}}{10}\,c^\dagger_{d,0}\,c^\dagger_{f,-1}\,c_{f,0}\,c_{f,-1}
\\
&\qquad + \frac{\sqrt{15}}{5}\,c^\dagger_{d,1}\,c^\dagger_{f,-2}\,c_{f,2}\,c_{f,-3} - \frac{\sqrt{30}}{10}\,c^\dagger_{d,1}\,c^\dagger_{f,-2}\,c_{f,0}\,c_{f,-1} + \frac{\sqrt{10}}{5}\,c^\dagger_{d,2}\,c^\dagger_{f,-3}\,c_{f,2}\,c_{f,-3}
\\
&\qquad - \frac{\sqrt{5}}{5}\,c^\dagger_{d,2}\,c^\dagger_{f,-3}\,c_{f,0}\,c_{f,-1} + \frac{\sqrt{10}}{5}\,c^\dagger_{d,-2}\,c^\dagger_{f,2}\,c_{f,3}\,c_{f,-3} + \frac{\sqrt{10}}{5}\,c^\dagger_{d,-2}\,c^\dagger_{f,2}\,c_{f,2}\,c_{f,-2}
\\
&\qquad - \frac{\sqrt{10}}{5}\,c^\dagger_{d,-2}\,c^\dagger_{f,2}\,c_{f,1}\,c_{f,-1} + \frac{1}{5}\,c^\dagger_{d,-1}\,c^\dagger_{f,1}\,c_{f,3}\,c_{f,-3} + \frac{1}{5}\,c^\dagger_{d,-1}\,c^\dagger_{f,1}\,c_{f,2}\,c_{f,-2}
\\
&\qquad - \frac{1}{5}\,c^\dagger_{d,-1}\,c^\dagger_{f,1}\,c_{f,1}\,c_{f,-1} - \frac{2 \sqrt{2}}{5}\,c^\dagger_{d,0}\,c^\dagger_{f,0}\,c_{f,3}\,c_{f,-3} - \frac{2 \sqrt{2}}{5}\,c^\dagger_{d,0}\,c^\dagger_{f,0}\,c_{f,2}\,c_{f,-2}
\\
&\qquad + \frac{2 \sqrt{2}}{5}\,c^\dagger_{d,0}\,c^\dagger_{f,0}\,c_{f,1}\,c_{f,-1} + \frac{1}{5}\,c^\dagger_{d,1}\,c^\dagger_{f,-1}\,c_{f,3}\,c_{f,-3} + \frac{1}{5}\,c^\dagger_{d,1}\,c^\dagger_{f,-1}\,c_{f,2}\,c_{f,-2}
\\
&\qquad - \frac{1}{5}\,c^\dagger_{d,1}\,c^\dagger_{f,-1}\,c_{f,1}\,c_{f,-1} + \frac{\sqrt{10}}{5}\,c^\dagger_{d,2}\,c^\dagger_{f,-2}\,c_{f,3}\,c_{f,-3} + \frac{\sqrt{10}}{5}\,c^\dagger_{d,2}\,c^\dagger_{f,-2}\,c_{f,2}\,c_{f,-2}
\\
&\qquad - \frac{\sqrt{10}}{5}\,c^\dagger_{d,2}\,c^\dagger_{f,-2}\,c_{f,1}\,c_{f,-1} + \frac{\sqrt{10}}{5}\,c^\dagger_{d,-2}\,c^\dagger_{f,3}\,c_{f,3}\,c_{f,-2} - \frac{\sqrt{5}}{5}\,c^\dagger_{d,-2}\,c^\dagger_{f,3}\,c_{f,1}\,c_{f,0}
\\
&\qquad + \frac{\sqrt{15}}{5}\,c^\dagger_{d,-1}\,c^\dagger_{f,2}\,c_{f,3}\,c_{f,-2} - \frac{\sqrt{30}}{10}\,c^\dagger_{d,-1}\,c^\dagger_{f,2}\,c_{f,1}\,c_{f,0} - \frac{3}{5}\,c^\dagger_{d,0}\,c^\dagger_{f,1}\,c_{f,3}\,c_{f,-2}
\\
&\qquad + \frac{3 \sqrt{2}}{10}\,c^\dagger_{d,0}\,c^\dagger_{f,1}\,c_{f,1}\,c_{f,0} - \frac{\sqrt{2}}{5}\,c^\dagger_{d,1}\,c^\dagger_{f,0}\,c_{f,3}\,c_{f,-2} + \frac{1}{5}\,c^\dagger_{d,1}\,c^\dagger_{f,0}\,c_{f,1}\,c_{f,0}
\\
&\qquad + \frac{2 \sqrt{6}}{5}\,c^\dagger_{d,2}\,c^\dagger_{f,-1}\,c_{f,3}\,c_{f,-2} - \frac{2 \sqrt{3}}{5}\,c^\dagger_{d,2}\,c^\dagger_{f,-1}\,c_{f,1}\,c_{f,0} + c^\dagger_{d,-1}\,c^\dagger_{f,3}\,c_{f,3}\,c_{f,-1}
\\
&\qquad - \frac{\sqrt{2}}{2}\,c^\dagger_{d,-1}\,c^\dagger_{f,3}\,c_{f,2}\,c_{f,0} - \frac{\sqrt{15}}{5}\,c^\dagger_{d,1}\,c^\dagger_{f,1}\,c_{f,3}\,c_{f,-1} + \frac{\sqrt{30}}{10}\,c^\dagger_{d,1}\,c^\dagger_{f,1}\,c_{f,2}\,c_{f,0}
\\
&\qquad + \frac{2 \sqrt{5}}{5}\,c^\dagger_{d,2}\,c^\dagger_{f,0}\,c_{f,3}\,c_{f,-1} - \frac{\sqrt{10}}{5}\,c^\dagger_{d,2}\,c^\dagger_{f,0}\,c_{f,2}\,c_{f,0} + \frac{\sqrt{2}}{2}\,c^\dagger_{d,0}\,c^\dagger_{f,3}\,c_{f,3}\,c_{f,0}
\\
&\qquad - c^\dagger_{d,0}\,c^\dagger_{f,3}\,c_{f,2}\,c_{f,1} - \frac{\sqrt{2}}{2}\,c^\dagger_{d,1}\,c^\dagger_{f,2}\,c_{f,3}\,c_{f,0} + c^\dagger_{d,1}\,c^\dagger_{f,2}\,c_{f,2}\,c_{f,1}
\\
&\qquad + \frac{\sqrt{5}}{5}\,c^\dagger_{d,2}\,c^\dagger_{f,1}\,c_{f,3}\,c_{f,0} - \frac{\sqrt{10}}{5}\,c^\dagger_{d,2}\,c^\dagger_{f,1}\,c_{f,2}\,c_{f,1} - \frac{\sqrt{5}}{5}\,c^\dagger_{f,0}\,c^\dagger_{f,-3}\,c_{d,-2}\,c_{f,-1}
\\
&\qquad + \frac{\sqrt{10}}{5}\,c^\dagger_{f,-1}\,c^\dagger_{f,-2}\,c_{d,-2}\,c_{f,-1} + \frac{\sqrt{2}}{2}\,c^\dagger_{f,0}\,c^\dagger_{f,-3}\,c_{d,-1}\,c_{f,-2} - c^\dagger_{f,-1}\,c^\dagger_{f,-2}\,c_{d,-1}\,c_{f,-2}
\\
&\qquad - \frac{\sqrt{2}}{2}\,c^\dagger_{f,0}\,c^\dagger_{f,-3}\,c_{d,0}\,c_{f,-3} + c^\dagger_{f,-1}\,c^\dagger_{f,-2}\,c_{d,0}\,c_{f,-3} - \frac{2 \sqrt{5}}{5}\,c^\dagger_{f,1}\,c^\dagger_{f,-3}\,c_{d,-2}\,c_{f,0}
\\
&\qquad + \frac{\sqrt{10}}{5}\,c^\dagger_{f,0}\,c^\dagger_{f,-2}\,c_{d,-2}\,c_{f,0} + \frac{\sqrt{15}}{5}\,c^\dagger_{f,1}\,c^\dagger_{f,-3}\,c_{d,-1}\,c_{f,-1} - \frac{\sqrt{30}}{10}\,c^\dagger_{f,0}\,c^\dagger_{f,-2}\,c_{d,-1}\,c_{f,-1}
\\
&\qquad - c^\dagger_{f,1}\,c^\dagger_{f,-3}\,c_{d,1}\,c_{f,-3} + \frac{\sqrt{2}}{2}\,c^\dagger_{f,0}\,c^\dagger_{f,-2}\,c_{d,1}\,c_{f,-3} - \frac{2 \sqrt{6}}{5}\,c^\dagger_{f,2}\,c^\dagger_{f,-3}\,c_{d,-2}\,c_{f,1}
\\
&\qquad + \frac{2 \sqrt{3}}{5}\,c^\dagger_{f,0}\,c^\dagger_{f,-1}\,c_{d,-2}\,c_{f,1} + \frac{\sqrt{2}}{5}\,c^\dagger_{f,2}\,c^\dagger_{f,-3}\,c_{d,-1}\,c_{f,0} - \frac{1}{5}\,c^\dagger_{f,0}\,c^\dagger_{f,-1}\,c_{d,-1}\,c_{f,0}
\\
&\qquad + \frac{3}{5}\,c^\dagger_{f,2}\,c^\dagger_{f,-3}\,c_{d,0}\,c_{f,-1} - \frac{3 \sqrt{2}}{10}\,c^\dagger_{f,0}\,c^\dagger_{f,-1}\,c_{d,0}\,c_{f,-1} - \frac{\sqrt{15}}{5}\,c^\dagger_{f,2}\,c^\dagger_{f,-3}\,c_{d,1}\,c_{f,-2}
\\
&\qquad + \frac{\sqrt{30}}{10}\,c^\dagger_{f,0}\,c^\dagger_{f,-1}\,c_{d,1}\,c_{f,-2} - \frac{\sqrt{10}}{5}\,c^\dagger_{f,2}\,c^\dagger_{f,-3}\,c_{d,2}\,c_{f,-3} + \frac{\sqrt{5}}{5}\,c^\dagger_{f,0}\,c^\dagger_{f,-1}\,c_{d,2}\,c_{f,-3}
\\
&\qquad - \frac{\sqrt{10}}{5}\,c^\dagger_{f,3}\,c^\dagger_{f,-3}\,c_{d,-2}\,c_{f,2} - \frac{\sqrt{10}}{5}\,c^\dagger_{f,2}\,c^\dagger_{f,-2}\,c_{d,-2}\,c_{f,2} + \frac{\sqrt{10}}{5}\,c^\dagger_{f,1}\,c^\dagger_{f,-1}\,c_{d,-2}\,c_{f,2}
\\
&\qquad - \frac{1}{5}\,c^\dagger_{f,3}\,c^\dagger_{f,-3}\,c_{d,-1}\,c_{f,1} - \frac{1}{5}\,c^\dagger_{f,2}\,c^\dagger_{f,-2}\,c_{d,-1}\,c_{f,1} + \frac{1}{5}\,c^\dagger_{f,1}\,c^\dagger_{f,-1}\,c_{d,-1}\,c_{f,1}
\\
&\qquad + \frac{2 \sqrt{2}}{5}\,c^\dagger_{f,3}\,c^\dagger_{f,-3}\,c_{d,0}\,c_{f,0} + \frac{2 \sqrt{2}}{5}\,c^\dagger_{f,2}\,c^\dagger_{f,-2}\,c_{d,0}\,c_{f,0} - \frac{2 \sqrt{2}}{5}\,c^\dagger_{f,1}\,c^\dagger_{f,-1}\,c_{d,0}\,c_{f,0}
\\
&\qquad - \frac{1}{5}\,c^\dagger_{f,3}\,c^\dagger_{f,-3}\,c_{d,1}\,c_{f,-1} - \frac{1}{5}\,c^\dagger_{f,2}\,c^\dagger_{f,-2}\,c_{d,1}\,c_{f,-1} + \frac{1}{5}\,c^\dagger_{f,1}\,c^\dagger_{f,-1}\,c_{d,1}\,c_{f,-1}
\\
&\qquad - \frac{\sqrt{10}}{5}\,c^\dagger_{f,3}\,c^\dagger_{f,-3}\,c_{d,2}\,c_{f,-2} - \frac{\sqrt{10}}{5}\,c^\dagger_{f,2}\,c^\dagger_{f,-2}\,c_{d,2}\,c_{f,-2} + \frac{\sqrt{10}}{5}\,c^\dagger_{f,1}\,c^\dagger_{f,-1}\,c_{d,2}\,c_{f,-2}
\\
&\qquad - \frac{\sqrt{10}}{5}\,c^\dagger_{f,3}\,c^\dagger_{f,-2}\,c_{d,-2}\,c_{f,3} + \frac{\sqrt{5}}{5}\,c^\dagger_{f,1}\,c^\dagger_{f,0}\,c_{d,-2}\,c_{f,3} - \frac{\sqrt{15}}{5}\,c^\dagger_{f,3}\,c^\dagger_{f,-2}\,c_{d,-1}\,c_{f,2}
\\
&\qquad + \frac{\sqrt{30}}{10}\,c^\dagger_{f,1}\,c^\dagger_{f,0}\,c_{d,-1}\,c_{f,2} + \frac{3}{5}\,c^\dagger_{f,3}\,c^\dagger_{f,-2}\,c_{d,0}\,c_{f,1} - \frac{3 \sqrt{2}}{10}\,c^\dagger_{f,1}\,c^\dagger_{f,0}\,c_{d,0}\,c_{f,1}
\\
&\qquad + \frac{\sqrt{2}}{5}\,c^\dagger_{f,3}\,c^\dagger_{f,-2}\,c_{d,1}\,c_{f,0} - \frac{1}{5}\,c^\dagger_{f,1}\,c^\dagger_{f,0}\,c_{d,1}\,c_{f,0} - \frac{2 \sqrt{6}}{5}\,c^\dagger_{f,3}\,c^\dagger_{f,-2}\,c_{d,2}\,c_{f,-1}
\\
&\qquad + \frac{2 \sqrt{3}}{5}\,c^\dagger_{f,1}\,c^\dagger_{f,0}\,c_{d,2}\,c_{f,-1} - c^\dagger_{f,3}\,c^\dagger_{f,-1}\,c_{d,-1}\,c_{f,3} + \frac{\sqrt{2}}{2}\,c^\dagger_{f,2}\,c^\dagger_{f,0}\,c_{d,-1}\,c_{f,3}
\\
&\qquad + \frac{\sqrt{15}}{5}\,c^\dagger_{f,3}\,c^\dagger_{f,-1}\,c_{d,1}\,c_{f,1} - \frac{\sqrt{30}}{10}\,c^\dagger_{f,2}\,c^\dagger_{f,0}\,c_{d,1}\,c_{f,1} - \frac{2 \sqrt{5}}{5}\,c^\dagger_{f,3}\,c^\dagger_{f,-1}\,c_{d,2}\,c_{f,0}
\\
&\qquad + \frac{\sqrt{10}}{5}\,c^\dagger_{f,2}\,c^\dagger_{f,0}\,c_{d,2}\,c_{f,0} - \frac{\sqrt{2}}{2}\,c^\dagger_{f,3}\,c^\dagger_{f,0}\,c_{d,0}\,c_{f,3} + c^\dagger_{f,2}\,c^\dagger_{f,1}\,c_{d,0}\,c_{f,3}
\\
&\qquad + \frac{\sqrt{2}}{2}\,c^\dagger_{f,3}\,c^\dagger_{f,0}\,c_{d,1}\,c_{f,2} - c^\dagger_{f,2}\,c^\dagger_{f,1}\,c_{d,1}\,c_{f,2} - \frac{\sqrt{5}}{5}\,c^\dagger_{f,3}\,c^\dagger_{f,0}\,c_{d,2}\,c_{f,1}
\\
&\qquad + \frac{\sqrt{10}}{5}\,c^\dagger_{f,2}\,c^\dagger_{f,1}\,c_{d,2}\,c_{f,1}\Big)
% \end{aligned}
% 
\end{align*}
% 
% 
% 
% df3 : 5×5
% j13,j24,A wigner_9j： 1 , 1 , -sqrt(66)/14
% j13,j24,A wigner_9j： 2 , 2 , 5*sqrt(33)/42
% j13,j24,A wigner_9j： 3 , 3 , -sqrt(77)/21
% j13,j24,A wigner_9j： 4 , 4 , 1/7
% j13,j24,A wigner_9j： 5 , 5 , -1/42
% 


% d3f : 1×1
% j13,j24,A wigner_9j： 1 , 1 , -1/5
% j13,j24,A wigner_9j： 2 , 2 , -sqrt(210)/35
% j13,j24,A wigner_9j： 3 , 3 , -3/5
% j13,j24,A wigner_9j： 4 , 4 , -sqrt(21)/7
\begin{align*}
{G_{0}}_{[4]} &=  
\frac{1}{2} \left(\left[[c^\dagger_d \times c^\dagger_d]_{1} \times[\tilde c_f \times \tilde c_d]_{1}\right]_{0} + h.c. \right)\\
&= 
\frac{1}{5} {\bm T}^{df}_1 \cdot {\bm M}_1^{d}
+ \frac{\sqrt{210}}{35} {\bm G}^{df}_2 \cdot {\bm Q}_2^{d}
+ \frac{3}{5}  {\bm T}^{df}_3 \cdot {\bm M}_3^{d}
+ \frac{\sqrt{21}}{7} {\bm G}^{df}_4 \cdot {\bm Q}_4^{d}
\\ 
&= 
% sqrt(210)/70  |max|= 0.207019667802706
% \begin{aligned}
% &
\frac{\sqrt{210}}{70}\Big(- \frac{\sqrt{10}}{15}\,c^\dagger_{d,1}\,c^\dagger_{d,-2}\,c_{d,-2}\,c_{f,1} + \frac{\sqrt{30}}{15}\,c^\dagger_{d,1}\,c^\dagger_{d,-2}\,c_{d,-1}\,c_{f,0} - \frac{2 \sqrt{15}}{15}\,c^\dagger_{d,1}\,c^\dagger_{d,-2}\,c_{d,0}\,c_{f,-1}
\\
&\qquad + \frac{2}{3}\,c^\dagger_{d,1}\,c^\dagger_{d,-2}\,c_{d,1}\,c_{f,-2} - \frac{\sqrt{6}}{3}\,c^\dagger_{d,1}\,c^\dagger_{d,-2}\,c_{d,2}\,c_{f,-3} + \frac{\sqrt{15}}{15}\,c^\dagger_{d,0}\,c^\dagger_{d,-1}\,c_{d,-2}\,c_{f,1}
\\
&\qquad - \frac{\sqrt{5}}{5}\,c^\dagger_{d,0}\,c^\dagger_{d,-1}\,c_{d,-1}\,c_{f,0} + \frac{\sqrt{10}}{5}\,c^\dagger_{d,0}\,c^\dagger_{d,-1}\,c_{d,0}\,c_{f,-1} - \frac{\sqrt{6}}{3}\,c^\dagger_{d,0}\,c^\dagger_{d,-1}\,c_{d,1}\,c_{f,-2}
\\
&\qquad + c^\dagger_{d,0}\,c^\dagger_{d,-1}\,c_{d,2}\,c_{f,-3} - \frac{2}{3}\,c^\dagger_{d,2}\,c^\dagger_{d,-2}\,c_{d,-2}\,c_{f,2} + \frac{4 \sqrt{10}}{15}\,c^\dagger_{d,2}\,c^\dagger_{d,-2}\,c_{d,-1}\,c_{f,1}
\\
&\qquad - \frac{2 \sqrt{5}}{5}\,c^\dagger_{d,2}\,c^\dagger_{d,-2}\,c_{d,0}\,c_{f,0} + \frac{4 \sqrt{10}}{15}\,c^\dagger_{d,2}\,c^\dagger_{d,-2}\,c_{d,1}\,c_{f,-1} - \frac{2}{3}\,c^\dagger_{d,2}\,c^\dagger_{d,-2}\,c_{d,2}\,c_{f,-2}
\\
&\qquad + \frac{1}{3}\,c^\dagger_{d,1}\,c^\dagger_{d,-1}\,c_{d,-2}\,c_{f,2} - \frac{2 \sqrt{10}}{15}\,c^\dagger_{d,1}\,c^\dagger_{d,-1}\,c_{d,-1}\,c_{f,1} + \frac{\sqrt{5}}{5}\,c^\dagger_{d,1}\,c^\dagger_{d,-1}\,c_{d,0}\,c_{f,0}
\\
&\qquad - \frac{2 \sqrt{10}}{15}\,c^\dagger_{d,1}\,c^\dagger_{d,-1}\,c_{d,1}\,c_{f,-1} + \frac{1}{3}\,c^\dagger_{d,1}\,c^\dagger_{d,-1}\,c_{d,2}\,c_{f,-2} - \frac{\sqrt{6}}{3}\,c^\dagger_{d,2}\,c^\dagger_{d,-1}\,c_{d,-2}\,c_{f,3}
\\
&\qquad + \frac{2}{3}\,c^\dagger_{d,2}\,c^\dagger_{d,-1}\,c_{d,-1}\,c_{f,2} - \frac{2 \sqrt{15}}{15}\,c^\dagger_{d,2}\,c^\dagger_{d,-1}\,c_{d,0}\,c_{f,1} + \frac{\sqrt{30}}{15}\,c^\dagger_{d,2}\,c^\dagger_{d,-1}\,c_{d,1}\,c_{f,0}
\\
&\qquad - \frac{\sqrt{10}}{15}\,c^\dagger_{d,2}\,c^\dagger_{d,-1}\,c_{d,2}\,c_{f,-1} + c^\dagger_{d,1}\,c^\dagger_{d,0}\,c_{d,-2}\,c_{f,3} - \frac{\sqrt{6}}{3}\,c^\dagger_{d,1}\,c^\dagger_{d,0}\,c_{d,-1}\,c_{f,2}
\\
&\qquad + \frac{\sqrt{10}}{5}\,c^\dagger_{d,1}\,c^\dagger_{d,0}\,c_{d,0}\,c_{f,1} - \frac{\sqrt{5}}{5}\,c^\dagger_{d,1}\,c^\dagger_{d,0}\,c_{d,1}\,c_{f,0} + \frac{\sqrt{15}}{15}\,c^\dagger_{d,1}\,c^\dagger_{d,0}\,c_{d,2}\,c_{f,-1}
\\
&\qquad - \frac{\sqrt{10}}{15}\,c^\dagger_{d,-2}\,c^\dagger_{f,1}\,c_{d,1}\,c_{d,-2} + \frac{\sqrt{30}}{15}\,c^\dagger_{d,-1}\,c^\dagger_{f,0}\,c_{d,1}\,c_{d,-2} - \frac{2 \sqrt{15}}{15}\,c^\dagger_{d,0}\,c^\dagger_{f,-1}\,c_{d,1}\,c_{d,-2}
\\
&\qquad + \frac{2}{3}\,c^\dagger_{d,1}\,c^\dagger_{f,-2}\,c_{d,1}\,c_{d,-2} - \frac{\sqrt{6}}{3}\,c^\dagger_{d,2}\,c^\dagger_{f,-3}\,c_{d,1}\,c_{d,-2} + \frac{\sqrt{15}}{15}\,c^\dagger_{d,-2}\,c^\dagger_{f,1}\,c_{d,0}\,c_{d,-1}
\\
&\qquad - \frac{\sqrt{5}}{5}\,c^\dagger_{d,-1}\,c^\dagger_{f,0}\,c_{d,0}\,c_{d,-1} + \frac{\sqrt{10}}{5}\,c^\dagger_{d,0}\,c^\dagger_{f,-1}\,c_{d,0}\,c_{d,-1} - \frac{\sqrt{6}}{3}\,c^\dagger_{d,1}\,c^\dagger_{f,-2}\,c_{d,0}\,c_{d,-1}
\\
&\qquad + c^\dagger_{d,2}\,c^\dagger_{f,-3}\,c_{d,0}\,c_{d,-1} - \frac{2}{3}\,c^\dagger_{d,-2}\,c^\dagger_{f,2}\,c_{d,2}\,c_{d,-2} + \frac{4 \sqrt{10}}{15}\,c^\dagger_{d,-1}\,c^\dagger_{f,1}\,c_{d,2}\,c_{d,-2}
\\
&\qquad - \frac{2 \sqrt{5}}{5}\,c^\dagger_{d,0}\,c^\dagger_{f,0}\,c_{d,2}\,c_{d,-2} + \frac{4 \sqrt{10}}{15}\,c^\dagger_{d,1}\,c^\dagger_{f,-1}\,c_{d,2}\,c_{d,-2} - \frac{2}{3}\,c^\dagger_{d,2}\,c^\dagger_{f,-2}\,c_{d,2}\,c_{d,-2}
\\
&\qquad + \frac{1}{3}\,c^\dagger_{d,-2}\,c^\dagger_{f,2}\,c_{d,1}\,c_{d,-1} - \frac{2 \sqrt{10}}{15}\,c^\dagger_{d,-1}\,c^\dagger_{f,1}\,c_{d,1}\,c_{d,-1} + \frac{\sqrt{5}}{5}\,c^\dagger_{d,0}\,c^\dagger_{f,0}\,c_{d,1}\,c_{d,-1}
\\
&\qquad - \frac{2 \sqrt{10}}{15}\,c^\dagger_{d,1}\,c^\dagger_{f,-1}\,c_{d,1}\,c_{d,-1} + \frac{1}{3}\,c^\dagger_{d,2}\,c^\dagger_{f,-2}\,c_{d,1}\,c_{d,-1} - \frac{\sqrt{6}}{3}\,c^\dagger_{d,-2}\,c^\dagger_{f,3}\,c_{d,2}\,c_{d,-1}
\\
&\qquad + \frac{2}{3}\,c^\dagger_{d,-1}\,c^\dagger_{f,2}\,c_{d,2}\,c_{d,-1} - \frac{2 \sqrt{15}}{15}\,c^\dagger_{d,0}\,c^\dagger_{f,1}\,c_{d,2}\,c_{d,-1} + \frac{\sqrt{30}}{15}\,c^\dagger_{d,1}\,c^\dagger_{f,0}\,c_{d,2}\,c_{d,-1}
\\
&\qquad - \frac{\sqrt{10}}{15}\,c^\dagger_{d,2}\,c^\dagger_{f,-1}\,c_{d,2}\,c_{d,-1} + c^\dagger_{d,-2}\,c^\dagger_{f,3}\,c_{d,1}\,c_{d,0} - \frac{\sqrt{6}}{3}\,c^\dagger_{d,-1}\,c^\dagger_{f,2}\,c_{d,1}\,c_{d,0}
\\
&\qquad + \frac{\sqrt{10}}{5}\,c^\dagger_{d,0}\,c^\dagger_{f,1}\,c_{d,1}\,c_{d,0} - \frac{\sqrt{5}}{5}\,c^\dagger_{d,1}\,c^\dagger_{f,0}\,c_{d,1}\,c_{d,0} + \frac{\sqrt{15}}{15}\,c^\dagger_{d,2}\,c^\dagger_{f,-1}\,c_{d,1}\,c_{d,0}\Big)
% \end{aligned}
\\
{M_{0}}_{[4]} &=  
\frac{1}{2} \left(\left[[c^\dagger_d \times c^\dagger_d]_{1} \times[\tilde c_f \times \tilde c_d]_{1}\right]_{0} + h.c. \right)\\
&= 
\frac{1}{5} {\bm Q}^{df}_1 \cdot {\bm M}_1^{d}
+ \frac{\sqrt{210}}{35} {\bm M}^{df}_2 \cdot {\bm Q}_2^{d}
+ \frac{3}{5}  {\bm Q}^{df}_3 \cdot {\bm M}_3^{d}
+ \frac{\sqrt{21}}{7} {\bm M}^{df}_4 \cdot {\bm Q}_4^{d}
\\ 
&= 
% sqrt(210)*I/70  |max|= 0.207019667802706
% \begin{aligned}
% &
\frac{\sqrt{210} i}{70}\Big(- \frac{\sqrt{10}}{15}\,c^\dagger_{d,1}\,c^\dagger_{d,-2}\,c_{d,-2}\,c_{f,1} + \frac{\sqrt{30}}{15}\,c^\dagger_{d,1}\,c^\dagger_{d,-2}\,c_{d,-1}\,c_{f,0} - \frac{2 \sqrt{15}}{15}\,c^\dagger_{d,1}\,c^\dagger_{d,-2}\,c_{d,0}\,c_{f,-1}
\\
&\qquad + \frac{2}{3}\,c^\dagger_{d,1}\,c^\dagger_{d,-2}\,c_{d,1}\,c_{f,-2} - \frac{\sqrt{6}}{3}\,c^\dagger_{d,1}\,c^\dagger_{d,-2}\,c_{d,2}\,c_{f,-3} + \frac{\sqrt{15}}{15}\,c^\dagger_{d,0}\,c^\dagger_{d,-1}\,c_{d,-2}\,c_{f,1}
\\
&\qquad - \frac{\sqrt{5}}{5}\,c^\dagger_{d,0}\,c^\dagger_{d,-1}\,c_{d,-1}\,c_{f,0} + \frac{\sqrt{10}}{5}\,c^\dagger_{d,0}\,c^\dagger_{d,-1}\,c_{d,0}\,c_{f,-1} - \frac{\sqrt{6}}{3}\,c^\dagger_{d,0}\,c^\dagger_{d,-1}\,c_{d,1}\,c_{f,-2}
\\
&\qquad + c^\dagger_{d,0}\,c^\dagger_{d,-1}\,c_{d,2}\,c_{f,-3} - \frac{2}{3}\,c^\dagger_{d,2}\,c^\dagger_{d,-2}\,c_{d,-2}\,c_{f,2} + \frac{4 \sqrt{10}}{15}\,c^\dagger_{d,2}\,c^\dagger_{d,-2}\,c_{d,-1}\,c_{f,1}
\\
&\qquad - \frac{2 \sqrt{5}}{5}\,c^\dagger_{d,2}\,c^\dagger_{d,-2}\,c_{d,0}\,c_{f,0} + \frac{4 \sqrt{10}}{15}\,c^\dagger_{d,2}\,c^\dagger_{d,-2}\,c_{d,1}\,c_{f,-1} - \frac{2}{3}\,c^\dagger_{d,2}\,c^\dagger_{d,-2}\,c_{d,2}\,c_{f,-2}
\\
&\qquad + \frac{1}{3}\,c^\dagger_{d,1}\,c^\dagger_{d,-1}\,c_{d,-2}\,c_{f,2} - \frac{2 \sqrt{10}}{15}\,c^\dagger_{d,1}\,c^\dagger_{d,-1}\,c_{d,-1}\,c_{f,1} + \frac{\sqrt{5}}{5}\,c^\dagger_{d,1}\,c^\dagger_{d,-1}\,c_{d,0}\,c_{f,0}
\\
&\qquad - \frac{2 \sqrt{10}}{15}\,c^\dagger_{d,1}\,c^\dagger_{d,-1}\,c_{d,1}\,c_{f,-1} + \frac{1}{3}\,c^\dagger_{d,1}\,c^\dagger_{d,-1}\,c_{d,2}\,c_{f,-2} - \frac{\sqrt{6}}{3}\,c^\dagger_{d,2}\,c^\dagger_{d,-1}\,c_{d,-2}\,c_{f,3}
\\
&\qquad + \frac{2}{3}\,c^\dagger_{d,2}\,c^\dagger_{d,-1}\,c_{d,-1}\,c_{f,2} - \frac{2 \sqrt{15}}{15}\,c^\dagger_{d,2}\,c^\dagger_{d,-1}\,c_{d,0}\,c_{f,1} + \frac{\sqrt{30}}{15}\,c^\dagger_{d,2}\,c^\dagger_{d,-1}\,c_{d,1}\,c_{f,0}
\\
&\qquad - \frac{\sqrt{10}}{15}\,c^\dagger_{d,2}\,c^\dagger_{d,-1}\,c_{d,2}\,c_{f,-1} + c^\dagger_{d,1}\,c^\dagger_{d,0}\,c_{d,-2}\,c_{f,3} - \frac{\sqrt{6}}{3}\,c^\dagger_{d,1}\,c^\dagger_{d,0}\,c_{d,-1}\,c_{f,2}
\\
&\qquad + \frac{\sqrt{10}}{5}\,c^\dagger_{d,1}\,c^\dagger_{d,0}\,c_{d,0}\,c_{f,1} - \frac{\sqrt{5}}{5}\,c^\dagger_{d,1}\,c^\dagger_{d,0}\,c_{d,1}\,c_{f,0} + \frac{\sqrt{15}}{15}\,c^\dagger_{d,1}\,c^\dagger_{d,0}\,c_{d,2}\,c_{f,-1}
\\
&\qquad + \frac{\sqrt{10}}{15}\,c^\dagger_{d,-2}\,c^\dagger_{f,1}\,c_{d,1}\,c_{d,-2} - \frac{\sqrt{30}}{15}\,c^\dagger_{d,-1}\,c^\dagger_{f,0}\,c_{d,1}\,c_{d,-2} + \frac{2 \sqrt{15}}{15}\,c^\dagger_{d,0}\,c^\dagger_{f,-1}\,c_{d,1}\,c_{d,-2}
\\
&\qquad - \frac{2}{3}\,c^\dagger_{d,1}\,c^\dagger_{f,-2}\,c_{d,1}\,c_{d,-2} + \frac{\sqrt{6}}{3}\,c^\dagger_{d,2}\,c^\dagger_{f,-3}\,c_{d,1}\,c_{d,-2} - \frac{\sqrt{15}}{15}\,c^\dagger_{d,-2}\,c^\dagger_{f,1}\,c_{d,0}\,c_{d,-1}
\\
&\qquad + \frac{\sqrt{5}}{5}\,c^\dagger_{d,-1}\,c^\dagger_{f,0}\,c_{d,0}\,c_{d,-1} - \frac{\sqrt{10}}{5}\,c^\dagger_{d,0}\,c^\dagger_{f,-1}\,c_{d,0}\,c_{d,-1} + \frac{\sqrt{6}}{3}\,c^\dagger_{d,1}\,c^\dagger_{f,-2}\,c_{d,0}\,c_{d,-1}
\\
&\qquad - c^\dagger_{d,2}\,c^\dagger_{f,-3}\,c_{d,0}\,c_{d,-1} + \frac{2}{3}\,c^\dagger_{d,-2}\,c^\dagger_{f,2}\,c_{d,2}\,c_{d,-2} - \frac{4 \sqrt{10}}{15}\,c^\dagger_{d,-1}\,c^\dagger_{f,1}\,c_{d,2}\,c_{d,-2}
\\
&\qquad + \frac{2 \sqrt{5}}{5}\,c^\dagger_{d,0}\,c^\dagger_{f,0}\,c_{d,2}\,c_{d,-2} - \frac{4 \sqrt{10}}{15}\,c^\dagger_{d,1}\,c^\dagger_{f,-1}\,c_{d,2}\,c_{d,-2} + \frac{2}{3}\,c^\dagger_{d,2}\,c^\dagger_{f,-2}\,c_{d,2}\,c_{d,-2}
\\
&\qquad - \frac{1}{3}\,c^\dagger_{d,-2}\,c^\dagger_{f,2}\,c_{d,1}\,c_{d,-1} + \frac{2 \sqrt{10}}{15}\,c^\dagger_{d,-1}\,c^\dagger_{f,1}\,c_{d,1}\,c_{d,-1} - \frac{\sqrt{5}}{5}\,c^\dagger_{d,0}\,c^\dagger_{f,0}\,c_{d,1}\,c_{d,-1}
\\
&\qquad + \frac{2 \sqrt{10}}{15}\,c^\dagger_{d,1}\,c^\dagger_{f,-1}\,c_{d,1}\,c_{d,-1} - \frac{1}{3}\,c^\dagger_{d,2}\,c^\dagger_{f,-2}\,c_{d,1}\,c_{d,-1} + \frac{\sqrt{6}}{3}\,c^\dagger_{d,-2}\,c^\dagger_{f,3}\,c_{d,2}\,c_{d,-1}
\\
&\qquad - \frac{2}{3}\,c^\dagger_{d,-1}\,c^\dagger_{f,2}\,c_{d,2}\,c_{d,-1} + \frac{2 \sqrt{15}}{15}\,c^\dagger_{d,0}\,c^\dagger_{f,1}\,c_{d,2}\,c_{d,-1} - \frac{\sqrt{30}}{15}\,c^\dagger_{d,1}\,c^\dagger_{f,0}\,c_{d,2}\,c_{d,-1}
\\
&\qquad + \frac{\sqrt{10}}{15}\,c^\dagger_{d,2}\,c^\dagger_{f,-1}\,c_{d,2}\,c_{d,-1} - c^\dagger_{d,-2}\,c^\dagger_{f,3}\,c_{d,1}\,c_{d,0} + \frac{\sqrt{6}}{3}\,c^\dagger_{d,-1}\,c^\dagger_{f,2}\,c_{d,1}\,c_{d,0}
\\
&\qquad - \frac{\sqrt{10}}{5}\,c^\dagger_{d,0}\,c^\dagger_{f,1}\,c_{d,1}\,c_{d,0} + \frac{\sqrt{5}}{5}\,c^\dagger_{d,1}\,c^\dagger_{f,0}\,c_{d,1}\,c_{d,0} - \frac{\sqrt{15}}{15}\,c^\dagger_{d,2}\,c^\dagger_{f,-1}\,c_{d,1}\,c_{d,0}\Big)
% \end{aligned}
\end{align*}

% d3f : 3×3
% j13,j24,A wigner_9j： 1 , 1 , -3/5
% j13,j24,A wigner_9j： 2 , 2 , -sqrt(210)/70
% j13,j24,A wigner_9j： 3 , 3 , 7/10
% j13,j24,A wigner_9j： 4 , 4 , -sqrt(21)/14
\begin{align*}
{G_{0}}_{[5]} &=  
\frac{1}{2} \left(\left[[c^\dagger_d \times c^\dagger_d]_{3} \times[\tilde c_f \times \tilde c_d]_{3}\right]_{0} + h.c. \right)\\
&= 
  \frac{3}{5} {\bm T}^{df}_1 \cdot {\bm M}_1^{d}
+ \frac{\sqrt{210}}{70} {\bm G}^{df}_2 \cdot {\bm Q}_2^{d}
- \frac{7}{10}  {\bm T}^{df}_3 \cdot {\bm M}_3^{d}
+ \frac{\sqrt{21}}{14} {\bm G}^{df}_4 \cdot {\bm Q}_4^{d}
\\ 
&=
% sqrt(210)/84  |max|= 0.172516389835589
% \begin{aligned}
\frac{\sqrt{210}}{84}\Big(- \frac{\sqrt{10}}{5}\,c^\dagger_{d,-1}\,c^\dagger_{d,-2}\,c_{d,-2}\,c_{f,-1} + c^\dagger_{d,-1}\,c^\dagger_{d,-2}\,c_{d,-1}\,c_{f,-2} - c^\dagger_{d,-1}\,c^\dagger_{d,-2}\,c_{d,0}\,c_{f,-3}
\\
&\qquad - \frac{2 \sqrt{5}}{5}\,c^\dagger_{d,0}\,c^\dagger_{d,-2}\,c_{d,-2}\,c_{f,0} + \frac{\sqrt{15}}{5}\,c^\dagger_{d,0}\,c^\dagger_{d,-2}\,c_{d,-1}\,c_{f,-1} - c^\dagger_{d,0}\,c^\dagger_{d,-2}\,c_{d,1}\,c_{f,-3}
\\
&\qquad - \frac{6 \sqrt{10}}{25}\,c^\dagger_{d,1}\,c^\dagger_{d,-2}\,c_{d,-2}\,c_{f,1} + \frac{\sqrt{30}}{25}\,c^\dagger_{d,1}\,c^\dagger_{d,-2}\,c_{d,-1}\,c_{f,0} + \frac{3 \sqrt{15}}{25}\,c^\dagger_{d,1}\,c^\dagger_{d,-2}\,c_{d,0}\,c_{f,-1}
\\
&\qquad - \frac{3}{5}\,c^\dagger_{d,1}\,c^\dagger_{d,-2}\,c_{d,1}\,c_{f,-2} - \frac{\sqrt{6}}{5}\,c^\dagger_{d,1}\,c^\dagger_{d,-2}\,c_{d,2}\,c_{f,-3} - \frac{4 \sqrt{15}}{25}\,c^\dagger_{d,0}\,c^\dagger_{d,-1}\,c_{d,-2}\,c_{f,1}
\\
&\qquad + \frac{2 \sqrt{5}}{25}\,c^\dagger_{d,0}\,c^\dagger_{d,-1}\,c_{d,-1}\,c_{f,0} + \frac{3 \sqrt{10}}{25}\,c^\dagger_{d,0}\,c^\dagger_{d,-1}\,c_{d,0}\,c_{f,-1} - \frac{\sqrt{6}}{5}\,c^\dagger_{d,0}\,c^\dagger_{d,-1}\,c_{d,1}\,c_{f,-2}
\\
&\qquad - \frac{2}{5}\,c^\dagger_{d,0}\,c^\dagger_{d,-1}\,c_{d,2}\,c_{f,-3} - \frac{2}{5}\,c^\dagger_{d,2}\,c^\dagger_{d,-2}\,c_{d,-2}\,c_{f,2} - \frac{\sqrt{10}}{25}\,c^\dagger_{d,2}\,c^\dagger_{d,-2}\,c_{d,-1}\,c_{f,1}
\\
&\qquad + \frac{4 \sqrt{5}}{25}\,c^\dagger_{d,2}\,c^\dagger_{d,-2}\,c_{d,0}\,c_{f,0} - \frac{\sqrt{10}}{25}\,c^\dagger_{d,2}\,c^\dagger_{d,-2}\,c_{d,1}\,c_{f,-1} - \frac{2}{5}\,c^\dagger_{d,2}\,c^\dagger_{d,-2}\,c_{d,2}\,c_{f,-2}
\\
&\qquad - \frac{4}{5}\,c^\dagger_{d,1}\,c^\dagger_{d,-1}\,c_{d,-2}\,c_{f,2} - \frac{2 \sqrt{10}}{25}\,c^\dagger_{d,1}\,c^\dagger_{d,-1}\,c_{d,-1}\,c_{f,1} + \frac{8 \sqrt{5}}{25}\,c^\dagger_{d,1}\,c^\dagger_{d,-1}\,c_{d,0}\,c_{f,0}
\\
&\qquad - \frac{2 \sqrt{10}}{25}\,c^\dagger_{d,1}\,c^\dagger_{d,-1}\,c_{d,1}\,c_{f,-1} - \frac{4}{5}\,c^\dagger_{d,1}\,c^\dagger_{d,-1}\,c_{d,2}\,c_{f,-2} - \frac{\sqrt{6}}{5}\,c^\dagger_{d,2}\,c^\dagger_{d,-1}\,c_{d,-2}\,c_{f,3}
\\
&\qquad - \frac{3}{5}\,c^\dagger_{d,2}\,c^\dagger_{d,-1}\,c_{d,-1}\,c_{f,2} + \frac{3 \sqrt{15}}{25}\,c^\dagger_{d,2}\,c^\dagger_{d,-1}\,c_{d,0}\,c_{f,1} + \frac{\sqrt{30}}{25}\,c^\dagger_{d,2}\,c^\dagger_{d,-1}\,c_{d,1}\,c_{f,0}
\\
&\qquad - \frac{6 \sqrt{10}}{25}\,c^\dagger_{d,2}\,c^\dagger_{d,-1}\,c_{d,2}\,c_{f,-1} - \frac{2}{5}\,c^\dagger_{d,1}\,c^\dagger_{d,0}\,c_{d,-2}\,c_{f,3} - \frac{\sqrt{6}}{5}\,c^\dagger_{d,1}\,c^\dagger_{d,0}\,c_{d,-1}\,c_{f,2}
\\
&\qquad + \frac{3 \sqrt{10}}{25}\,c^\dagger_{d,1}\,c^\dagger_{d,0}\,c_{d,0}\,c_{f,1} + \frac{2 \sqrt{5}}{25}\,c^\dagger_{d,1}\,c^\dagger_{d,0}\,c_{d,1}\,c_{f,0} - \frac{4 \sqrt{15}}{25}\,c^\dagger_{d,1}\,c^\dagger_{d,0}\,c_{d,2}\,c_{f,-1}
\\
&\qquad - c^\dagger_{d,2}\,c^\dagger_{d,0}\,c_{d,-1}\,c_{f,3} + \frac{\sqrt{15}}{5}\,c^\dagger_{d,2}\,c^\dagger_{d,0}\,c_{d,1}\,c_{f,1} - \frac{2 \sqrt{5}}{5}\,c^\dagger_{d,2}\,c^\dagger_{d,0}\,c_{d,2}\,c_{f,0}
\\
&\qquad - c^\dagger_{d,2}\,c^\dagger_{d,1}\,c_{d,0}\,c_{f,3} + c^\dagger_{d,2}\,c^\dagger_{d,1}\,c_{d,1}\,c_{f,2} - \frac{\sqrt{10}}{5}\,c^\dagger_{d,2}\,c^\dagger_{d,1}\,c_{d,2}\,c_{f,1}
\\
&\qquad - \frac{\sqrt{10}}{5}\,c^\dagger_{d,-2}\,c^\dagger_{f,-1}\,c_{d,-1}\,c_{d,-2} + c^\dagger_{d,-1}\,c^\dagger_{f,-2}\,c_{d,-1}\,c_{d,-2} - c^\dagger_{d,0}\,c^\dagger_{f,-3}\,c_{d,-1}\,c_{d,-2}
\\
&\qquad - \frac{2 \sqrt{5}}{5}\,c^\dagger_{d,-2}\,c^\dagger_{f,0}\,c_{d,0}\,c_{d,-2} + \frac{\sqrt{15}}{5}\,c^\dagger_{d,-1}\,c^\dagger_{f,-1}\,c_{d,0}\,c_{d,-2} - c^\dagger_{d,1}\,c^\dagger_{f,-3}\,c_{d,0}\,c_{d,-2}
\\
&\qquad - \frac{6 \sqrt{10}}{25}\,c^\dagger_{d,-2}\,c^\dagger_{f,1}\,c_{d,1}\,c_{d,-2} + \frac{\sqrt{30}}{25}\,c^\dagger_{d,-1}\,c^\dagger_{f,0}\,c_{d,1}\,c_{d,-2} + \frac{3 \sqrt{15}}{25}\,c^\dagger_{d,0}\,c^\dagger_{f,-1}\,c_{d,1}\,c_{d,-2}
\\
&\qquad - \frac{3}{5}\,c^\dagger_{d,1}\,c^\dagger_{f,-2}\,c_{d,1}\,c_{d,-2} - \frac{\sqrt{6}}{5}\,c^\dagger_{d,2}\,c^\dagger_{f,-3}\,c_{d,1}\,c_{d,-2} - \frac{4 \sqrt{15}}{25}\,c^\dagger_{d,-2}\,c^\dagger_{f,1}\,c_{d,0}\,c_{d,-1}
\\
&\qquad + \frac{2 \sqrt{5}}{25}\,c^\dagger_{d,-1}\,c^\dagger_{f,0}\,c_{d,0}\,c_{d,-1} + \frac{3 \sqrt{10}}{25}\,c^\dagger_{d,0}\,c^\dagger_{f,-1}\,c_{d,0}\,c_{d,-1} - \frac{\sqrt{6}}{5}\,c^\dagger_{d,1}\,c^\dagger_{f,-2}\,c_{d,0}\,c_{d,-1}
\\
&\qquad - \frac{2}{5}\,c^\dagger_{d,2}\,c^\dagger_{f,-3}\,c_{d,0}\,c_{d,-1} - \frac{2}{5}\,c^\dagger_{d,-2}\,c^\dagger_{f,2}\,c_{d,2}\,c_{d,-2} - \frac{\sqrt{10}}{25}\,c^\dagger_{d,-1}\,c^\dagger_{f,1}\,c_{d,2}\,c_{d,-2}
\\
&\qquad + \frac{4 \sqrt{5}}{25}\,c^\dagger_{d,0}\,c^\dagger_{f,0}\,c_{d,2}\,c_{d,-2} - \frac{\sqrt{10}}{25}\,c^\dagger_{d,1}\,c^\dagger_{f,-1}\,c_{d,2}\,c_{d,-2} - \frac{2}{5}\,c^\dagger_{d,2}\,c^\dagger_{f,-2}\,c_{d,2}\,c_{d,-2}
\\
&\qquad - \frac{4}{5}\,c^\dagger_{d,-2}\,c^\dagger_{f,2}\,c_{d,1}\,c_{d,-1} - \frac{2 \sqrt{10}}{25}\,c^\dagger_{d,-1}\,c^\dagger_{f,1}\,c_{d,1}\,c_{d,-1} + \frac{8 \sqrt{5}}{25}\,c^\dagger_{d,0}\,c^\dagger_{f,0}\,c_{d,1}\,c_{d,-1}
\\
&\qquad - \frac{2 \sqrt{10}}{25}\,c^\dagger_{d,1}\,c^\dagger_{f,-1}\,c_{d,1}\,c_{d,-1} - \frac{4}{5}\,c^\dagger_{d,2}\,c^\dagger_{f,-2}\,c_{d,1}\,c_{d,-1} - \frac{\sqrt{6}}{5}\,c^\dagger_{d,-2}\,c^\dagger_{f,3}\,c_{d,2}\,c_{d,-1}
\\
&\qquad - \frac{3}{5}\,c^\dagger_{d,-1}\,c^\dagger_{f,2}\,c_{d,2}\,c_{d,-1} + \frac{3 \sqrt{15}}{25}\,c^\dagger_{d,0}\,c^\dagger_{f,1}\,c_{d,2}\,c_{d,-1} + \frac{\sqrt{30}}{25}\,c^\dagger_{d,1}\,c^\dagger_{f,0}\,c_{d,2}\,c_{d,-1}
\\
&\qquad - \frac{6 \sqrt{10}}{25}\,c^\dagger_{d,2}\,c^\dagger_{f,-1}\,c_{d,2}\,c_{d,-1} - \frac{2}{5}\,c^\dagger_{d,-2}\,c^\dagger_{f,3}\,c_{d,1}\,c_{d,0} - \frac{\sqrt{6}}{5}\,c^\dagger_{d,-1}\,c^\dagger_{f,2}\,c_{d,1}\,c_{d,0}
\\
&\qquad + \frac{3 \sqrt{10}}{25}\,c^\dagger_{d,0}\,c^\dagger_{f,1}\,c_{d,1}\,c_{d,0} + \frac{2 \sqrt{5}}{25}\,c^\dagger_{d,1}\,c^\dagger_{f,0}\,c_{d,1}\,c_{d,0} - \frac{4 \sqrt{15}}{25}\,c^\dagger_{d,2}\,c^\dagger_{f,-1}\,c_{d,1}\,c_{d,0}
\\
&\qquad - c^\dagger_{d,-1}\,c^\dagger_{f,3}\,c_{d,2}\,c_{d,0} + \frac{\sqrt{15}}{5}\,c^\dagger_{d,1}\,c^\dagger_{f,1}\,c_{d,2}\,c_{d,0} - \frac{2 \sqrt{5}}{5}\,c^\dagger_{d,2}\,c^\dagger_{f,0}\,c_{d,2}\,c_{d,0}
\\
&\qquad - c^\dagger_{d,0}\,c^\dagger_{f,3}\,c_{d,2}\,c_{d,1} + c^\dagger_{d,1}\,c^\dagger_{f,2}\,c_{d,2}\,c_{d,1} - \frac{\sqrt{10}}{5}\,c^\dagger_{d,2}\,c^\dagger_{f,1}\,c_{d,2}\,c_{d,1}\Big)
% \end{aligned}
\\ 
{M_{0}}_{[5]} &=  
\frac{1}{2} \left(\left[[c^\dagger_d \times c^\dagger_d]_{3} \times[\tilde c_f \times \tilde c_d]_{3}\right]_{0} + h.c. \right)\\
&= 
  \frac{3}{5} {\bm Q}^{df}_1 \cdot {\bm M}_1^{d}
+ \frac{\sqrt{210}}{70} {\bm M}^{df}_2 \cdot {\bm Q}_2^{d}
- \frac{7}{10}  {\bm Q}^{df}_3 \cdot {\bm M}_3^{d}
+ \frac{\sqrt{21}}{14} {\bm M}^{df}_4 \cdot {\bm Q}_4^{d}
\\ 
&= 
% sqrt(210)*I/84  |max|= 0.172516389835589
% \begin{aligned}
% &
\frac{\sqrt{210} i}{84}\Big(- \frac{\sqrt{10}}{5}\,c^\dagger_{d,-1}\,c^\dagger_{d,-2}\,c_{d,-2}\,c_{f,-1} + c^\dagger_{d,-1}\,c^\dagger_{d,-2}\,c_{d,-1}\,c_{f,-2} - c^\dagger_{d,-1}\,c^\dagger_{d,-2}\,c_{d,0}\,c_{f,-3}
\\
&\qquad - \frac{2 \sqrt{5}}{5}\,c^\dagger_{d,0}\,c^\dagger_{d,-2}\,c_{d,-2}\,c_{f,0} + \frac{\sqrt{15}}{5}\,c^\dagger_{d,0}\,c^\dagger_{d,-2}\,c_{d,-1}\,c_{f,-1} - c^\dagger_{d,0}\,c^\dagger_{d,-2}\,c_{d,1}\,c_{f,-3}
\\
&\qquad - \frac{6 \sqrt{10}}{25}\,c^\dagger_{d,1}\,c^\dagger_{d,-2}\,c_{d,-2}\,c_{f,1} + \frac{\sqrt{30}}{25}\,c^\dagger_{d,1}\,c^\dagger_{d,-2}\,c_{d,-1}\,c_{f,0} + \frac{3 \sqrt{15}}{25}\,c^\dagger_{d,1}\,c^\dagger_{d,-2}\,c_{d,0}\,c_{f,-1}
\\
&\qquad - \frac{3}{5}\,c^\dagger_{d,1}\,c^\dagger_{d,-2}\,c_{d,1}\,c_{f,-2} - \frac{\sqrt{6}}{5}\,c^\dagger_{d,1}\,c^\dagger_{d,-2}\,c_{d,2}\,c_{f,-3} - \frac{4 \sqrt{15}}{25}\,c^\dagger_{d,0}\,c^\dagger_{d,-1}\,c_{d,-2}\,c_{f,1}
\\
&\qquad + \frac{2 \sqrt{5}}{25}\,c^\dagger_{d,0}\,c^\dagger_{d,-1}\,c_{d,-1}\,c_{f,0} + \frac{3 \sqrt{10}}{25}\,c^\dagger_{d,0}\,c^\dagger_{d,-1}\,c_{d,0}\,c_{f,-1} - \frac{\sqrt{6}}{5}\,c^\dagger_{d,0}\,c^\dagger_{d,-1}\,c_{d,1}\,c_{f,-2}
\\
&\qquad - \frac{2}{5}\,c^\dagger_{d,0}\,c^\dagger_{d,-1}\,c_{d,2}\,c_{f,-3} - \frac{2}{5}\,c^\dagger_{d,2}\,c^\dagger_{d,-2}\,c_{d,-2}\,c_{f,2} - \frac{\sqrt{10}}{25}\,c^\dagger_{d,2}\,c^\dagger_{d,-2}\,c_{d,-1}\,c_{f,1}
\\
&\qquad + \frac{4 \sqrt{5}}{25}\,c^\dagger_{d,2}\,c^\dagger_{d,-2}\,c_{d,0}\,c_{f,0} - \frac{\sqrt{10}}{25}\,c^\dagger_{d,2}\,c^\dagger_{d,-2}\,c_{d,1}\,c_{f,-1} - \frac{2}{5}\,c^\dagger_{d,2}\,c^\dagger_{d,-2}\,c_{d,2}\,c_{f,-2}
\\
&\qquad - \frac{4}{5}\,c^\dagger_{d,1}\,c^\dagger_{d,-1}\,c_{d,-2}\,c_{f,2} - \frac{2 \sqrt{10}}{25}\,c^\dagger_{d,1}\,c^\dagger_{d,-1}\,c_{d,-1}\,c_{f,1} + \frac{8 \sqrt{5}}{25}\,c^\dagger_{d,1}\,c^\dagger_{d,-1}\,c_{d,0}\,c_{f,0}
\\
&\qquad - \frac{2 \sqrt{10}}{25}\,c^\dagger_{d,1}\,c^\dagger_{d,-1}\,c_{d,1}\,c_{f,-1} - \frac{4}{5}\,c^\dagger_{d,1}\,c^\dagger_{d,-1}\,c_{d,2}\,c_{f,-2} - \frac{\sqrt{6}}{5}\,c^\dagger_{d,2}\,c^\dagger_{d,-1}\,c_{d,-2}\,c_{f,3}
\\
&\qquad - \frac{3}{5}\,c^\dagger_{d,2}\,c^\dagger_{d,-1}\,c_{d,-1}\,c_{f,2} + \frac{3 \sqrt{15}}{25}\,c^\dagger_{d,2}\,c^\dagger_{d,-1}\,c_{d,0}\,c_{f,1} + \frac{\sqrt{30}}{25}\,c^\dagger_{d,2}\,c^\dagger_{d,-1}\,c_{d,1}\,c_{f,0}
\\
&\qquad - \frac{6 \sqrt{10}}{25}\,c^\dagger_{d,2}\,c^\dagger_{d,-1}\,c_{d,2}\,c_{f,-1} - \frac{2}{5}\,c^\dagger_{d,1}\,c^\dagger_{d,0}\,c_{d,-2}\,c_{f,3} - \frac{\sqrt{6}}{5}\,c^\dagger_{d,1}\,c^\dagger_{d,0}\,c_{d,-1}\,c_{f,2}
\\
&\qquad + \frac{3 \sqrt{10}}{25}\,c^\dagger_{d,1}\,c^\dagger_{d,0}\,c_{d,0}\,c_{f,1} + \frac{2 \sqrt{5}}{25}\,c^\dagger_{d,1}\,c^\dagger_{d,0}\,c_{d,1}\,c_{f,0} - \frac{4 \sqrt{15}}{25}\,c^\dagger_{d,1}\,c^\dagger_{d,0}\,c_{d,2}\,c_{f,-1}
\\
&\qquad - c^\dagger_{d,2}\,c^\dagger_{d,0}\,c_{d,-1}\,c_{f,3} + \frac{\sqrt{15}}{5}\,c^\dagger_{d,2}\,c^\dagger_{d,0}\,c_{d,1}\,c_{f,1} - \frac{2 \sqrt{5}}{5}\,c^\dagger_{d,2}\,c^\dagger_{d,0}\,c_{d,2}\,c_{f,0}
\\
&\qquad - c^\dagger_{d,2}\,c^\dagger_{d,1}\,c_{d,0}\,c_{f,3} + c^\dagger_{d,2}\,c^\dagger_{d,1}\,c_{d,1}\,c_{f,2} - \frac{\sqrt{10}}{5}\,c^\dagger_{d,2}\,c^\dagger_{d,1}\,c_{d,2}\,c_{f,1}
\\
&\qquad + \frac{\sqrt{10}}{5}\,c^\dagger_{d,-2}\,c^\dagger_{f,-1}\,c_{d,-1}\,c_{d,-2} - c^\dagger_{d,-1}\,c^\dagger_{f,-2}\,c_{d,-1}\,c_{d,-2} + c^\dagger_{d,0}\,c^\dagger_{f,-3}\,c_{d,-1}\,c_{d,-2}
\\
&\qquad + \frac{2 \sqrt{5}}{5}\,c^\dagger_{d,-2}\,c^\dagger_{f,0}\,c_{d,0}\,c_{d,-2} - \frac{\sqrt{15}}{5}\,c^\dagger_{d,-1}\,c^\dagger_{f,-1}\,c_{d,0}\,c_{d,-2} + c^\dagger_{d,1}\,c^\dagger_{f,-3}\,c_{d,0}\,c_{d,-2}
\\
&\qquad + \frac{6 \sqrt{10}}{25}\,c^\dagger_{d,-2}\,c^\dagger_{f,1}\,c_{d,1}\,c_{d,-2} - \frac{\sqrt{30}}{25}\,c^\dagger_{d,-1}\,c^\dagger_{f,0}\,c_{d,1}\,c_{d,-2} - \frac{3 \sqrt{15}}{25}\,c^\dagger_{d,0}\,c^\dagger_{f,-1}\,c_{d,1}\,c_{d,-2}
\\
&\qquad + \frac{3}{5}\,c^\dagger_{d,1}\,c^\dagger_{f,-2}\,c_{d,1}\,c_{d,-2} + \frac{\sqrt{6}}{5}\,c^\dagger_{d,2}\,c^\dagger_{f,-3}\,c_{d,1}\,c_{d,-2} + \frac{4 \sqrt{15}}{25}\,c^\dagger_{d,-2}\,c^\dagger_{f,1}\,c_{d,0}\,c_{d,-1}
\\
&\qquad - \frac{2 \sqrt{5}}{25}\,c^\dagger_{d,-1}\,c^\dagger_{f,0}\,c_{d,0}\,c_{d,-1} - \frac{3 \sqrt{10}}{25}\,c^\dagger_{d,0}\,c^\dagger_{f,-1}\,c_{d,0}\,c_{d,-1} + \frac{\sqrt{6}}{5}\,c^\dagger_{d,1}\,c^\dagger_{f,-2}\,c_{d,0}\,c_{d,-1}
\\
&\qquad + \frac{2}{5}\,c^\dagger_{d,2}\,c^\dagger_{f,-3}\,c_{d,0}\,c_{d,-1} + \frac{2}{5}\,c^\dagger_{d,-2}\,c^\dagger_{f,2}\,c_{d,2}\,c_{d,-2} + \frac{\sqrt{10}}{25}\,c^\dagger_{d,-1}\,c^\dagger_{f,1}\,c_{d,2}\,c_{d,-2}
\\
&\qquad - \frac{4 \sqrt{5}}{25}\,c^\dagger_{d,0}\,c^\dagger_{f,0}\,c_{d,2}\,c_{d,-2} + \frac{\sqrt{10}}{25}\,c^\dagger_{d,1}\,c^\dagger_{f,-1}\,c_{d,2}\,c_{d,-2} + \frac{2}{5}\,c^\dagger_{d,2}\,c^\dagger_{f,-2}\,c_{d,2}\,c_{d,-2}
\\
&\qquad + \frac{4}{5}\,c^\dagger_{d,-2}\,c^\dagger_{f,2}\,c_{d,1}\,c_{d,-1} + \frac{2 \sqrt{10}}{25}\,c^\dagger_{d,-1}\,c^\dagger_{f,1}\,c_{d,1}\,c_{d,-1} - \frac{8 \sqrt{5}}{25}\,c^\dagger_{d,0}\,c^\dagger_{f,0}\,c_{d,1}\,c_{d,-1}
\\
&\qquad + \frac{2 \sqrt{10}}{25}\,c^\dagger_{d,1}\,c^\dagger_{f,-1}\,c_{d,1}\,c_{d,-1} + \frac{4}{5}\,c^\dagger_{d,2}\,c^\dagger_{f,-2}\,c_{d,1}\,c_{d,-1} + \frac{\sqrt{6}}{5}\,c^\dagger_{d,-2}\,c^\dagger_{f,3}\,c_{d,2}\,c_{d,-1}
\\
&\qquad + \frac{3}{5}\,c^\dagger_{d,-1}\,c^\dagger_{f,2}\,c_{d,2}\,c_{d,-1} - \frac{3 \sqrt{15}}{25}\,c^\dagger_{d,0}\,c^\dagger_{f,1}\,c_{d,2}\,c_{d,-1} - \frac{\sqrt{30}}{25}\,c^\dagger_{d,1}\,c^\dagger_{f,0}\,c_{d,2}\,c_{d,-1}
\\
&\qquad + \frac{6 \sqrt{10}}{25}\,c^\dagger_{d,2}\,c^\dagger_{f,-1}\,c_{d,2}\,c_{d,-1} + \frac{2}{5}\,c^\dagger_{d,-2}\,c^\dagger_{f,3}\,c_{d,1}\,c_{d,0} + \frac{\sqrt{6}}{5}\,c^\dagger_{d,-1}\,c^\dagger_{f,2}\,c_{d,1}\,c_{d,0}
\\
&\qquad - \frac{3 \sqrt{10}}{25}\,c^\dagger_{d,0}\,c^\dagger_{f,1}\,c_{d,1}\,c_{d,0} - \frac{2 \sqrt{5}}{25}\,c^\dagger_{d,1}\,c^\dagger_{f,0}\,c_{d,1}\,c_{d,0} + \frac{4 \sqrt{15}}{25}\,c^\dagger_{d,2}\,c^\dagger_{f,-1}\,c_{d,1}\,c_{d,0}
\\
&\qquad + c^\dagger_{d,-1}\,c^\dagger_{f,3}\,c_{d,2}\,c_{d,0} - \frac{\sqrt{15}}{5}\,c^\dagger_{d,1}\,c^\dagger_{f,1}\,c_{d,2}\,c_{d,0} + \frac{2 \sqrt{5}}{5}\,c^\dagger_{d,2}\,c^\dagger_{f,0}\,c_{d,2}\,c_{d,0}
\\
&\qquad + c^\dagger_{d,0}\,c^\dagger_{f,3}\,c_{d,2}\,c_{d,1} - c^\dagger_{d,1}\,c^\dagger_{f,2}\,c_{d,2}\,c_{d,1} + \frac{\sqrt{10}}{5}\,c^\dagger_{d,2}\,c^\dagger_{f,1}\,c_{d,2}\,c_{d,1}\Big)
% \end{aligned}
% 
\end{align*}

\subsection*{$spd$ system}
% 
% sppd : 1×1
% j13,j24,A wigner_9j： 1 , 1 , 1
\begin{align*}
Q_0^{(2)} 
&=  
  \frac{1}{2} \left(\left[[c^\dagger_s \times c^\dagger_p]_{1} \times[\tilde c_p \times \tilde c_d]_{1}\right]_{0} + h.c. \right)\\
&=
   - {\bm T}^{sp}_1 \cdot {\bm T}^{pd}_1 
   +
   {\bm Q}^{sp}_1 \cdot {\bm Q}^{pd}_1 
\\
&=
\frac{\sqrt{5}}{10}\Big(c^\dagger_{s,0}\,c^\dagger_{p,-1}\,c_{p,1}\,c_{d,-2} - \frac{\sqrt{2}}{2}\,c^\dagger_{s,0}\,c^\dagger_{p,-1}\,c_{p,0}\,c_{d,-1} + \frac{\sqrt{6}}{6}\,c^\dagger_{s,0}\,c^\dagger_{p,-1}\,c_{p,-1}\,c_{d,0}
\\
&\qquad + \frac{\sqrt{2}}{2}\,c^\dagger_{s,0}\,c^\dagger_{p,0}\,c_{p,1}\,c_{d,-1} - \frac{\sqrt{6}}{3}\,c^\dagger_{s,0}\,c^\dagger_{p,0}\,c_{p,0}\,c_{d,0} + \frac{\sqrt{2}}{2}\,c^\dagger_{s,0}\,c^\dagger_{p,0}\,c_{p,-1}\,c_{d,1}
\\
&\qquad + \frac{\sqrt{6}}{6}\,c^\dagger_{s,0}\,c^\dagger_{p,1}\,c_{p,1}\,c_{d,0} - \frac{\sqrt{2}}{2}\,c^\dagger_{s,0}\,c^\dagger_{p,1}\,c_{p,0}\,c_{d,1} + c^\dagger_{s,0}\,c^\dagger_{p,1}\,c_{p,-1}\,c_{d,2}
\\
&\qquad + c^\dagger_{p,1}\,c^\dagger_{d,-2}\,c_{s,0}\,c_{p,-1} - \frac{\sqrt{2}}{2}\,c^\dagger_{p,0}\,c^\dagger_{d,-1}\,c_{s,0}\,c_{p,-1} + \frac{\sqrt{6}}{6}\,c^\dagger_{p,-1}\,c^\dagger_{d,0}\,c_{s,0}\,c_{p,-1}
\\
&\qquad + \frac{\sqrt{2}}{2}\,c^\dagger_{p,1}\,c^\dagger_{d,-1}\,c_{s,0}\,c_{p,0} - \frac{\sqrt{6}}{3}\,c^\dagger_{p,0}\,c^\dagger_{d,0}\,c_{s,0}\,c_{p,0} + \frac{\sqrt{2}}{2}\,c^\dagger_{p,-1}\,c^\dagger_{d,1}\,c_{s,0}\,c_{p,0}
\\
&\qquad + \frac{\sqrt{6}}{6}\,c^\dagger_{p,1}\,c^\dagger_{d,0}\,c_{s,0}\,c_{p,1} - \frac{\sqrt{2}}{2}\,c^\dagger_{p,0}\,c^\dagger_{d,1}\,c_{s,0}\,c_{p,1} + c^\dagger_{p,-1}\,c^\dagger_{d,2}\,c_{s,0}\,c_{p,1}\Big)
\\
T_0^{(2)} 
&=  
  \frac{i}{2} \left(\left[[c^\dagger_s \times c^\dagger_p]_{1} \times[\tilde c_p \times \tilde c_d]_{1}\right]_{0} - h.c. \right)\\
&=
   - {\bm Q}^{sp}_1 \cdot {\bm T}^{pd}_1 
   -
   {\bm T}^{sp}_1 \cdot {\bm Q}^{pd}_1 
\\
&=
\frac{\sqrt{5} i}{10}\Big(c^\dagger_{s,0}\,c^\dagger_{p,-1}\,c_{p,1}\,c_{d,-2} - \frac{\sqrt{2}}{2}\,c^\dagger_{s,0}\,c^\dagger_{p,-1}\,c_{p,0}\,c_{d,-1} + \frac{\sqrt{6}}{6}\,c^\dagger_{s,0}\,c^\dagger_{p,-1}\,c_{p,-1}\,c_{d,0}
\\
&\qquad + \frac{\sqrt{2}}{2}\,c^\dagger_{s,0}\,c^\dagger_{p,0}\,c_{p,1}\,c_{d,-1} - \frac{\sqrt{6}}{3}\,c^\dagger_{s,0}\,c^\dagger_{p,0}\,c_{p,0}\,c_{d,0} + \frac{\sqrt{2}}{2}\,c^\dagger_{s,0}\,c^\dagger_{p,0}\,c_{p,-1}\,c_{d,1}
\\
&\qquad + \frac{\sqrt{6}}{6}\,c^\dagger_{s,0}\,c^\dagger_{p,1}\,c_{p,1}\,c_{d,0} - \frac{\sqrt{2}}{2}\,c^\dagger_{s,0}\,c^\dagger_{p,1}\,c_{p,0}\,c_{d,1} + c^\dagger_{s,0}\,c^\dagger_{p,1}\,c_{p,-1}\,c_{d,2}
\\
&\qquad - c^\dagger_{p,1}\,c^\dagger_{d,-2}\,c_{s,0}\,c_{p,-1} + \frac{\sqrt{2}}{2}\,c^\dagger_{p,0}\,c^\dagger_{d,-1}\,c_{s,0}\,c_{p,-1} - \frac{\sqrt{6}}{6}\,c^\dagger_{p,-1}\,c^\dagger_{d,0}\,c_{s,0}\,c_{p,-1}
\\
&\qquad - \frac{\sqrt{2}}{2}\,c^\dagger_{p,1}\,c^\dagger_{d,-1}\,c_{s,0}\,c_{p,0} + \frac{\sqrt{6}}{3}\,c^\dagger_{p,0}\,c^\dagger_{d,0}\,c_{s,0}\,c_{p,0} - \frac{\sqrt{2}}{2}\,c^\dagger_{p,-1}\,c^\dagger_{d,1}\,c_{s,0}\,c_{p,0}
\\
&\qquad - \frac{\sqrt{6}}{6}\,c^\dagger_{p,1}\,c^\dagger_{d,0}\,c_{s,0}\,c_{p,1} + \frac{\sqrt{2}}{2}\,c^\dagger_{p,0}\,c^\dagger_{d,1}\,c_{s,0}\,c_{p,1} - c^\dagger_{p,-1}\,c^\dagger_{d,2}\,c_{s,0}\,c_{p,1}\Big)
\end{align*}
\begin{align*}
{G_0^{(2)}}_{[1]} &=  
  \frac{1}{2} \left(\left[[c^\dagger_s \times c^\dagger_p]_{1} \times[\tilde c_d \times \tilde c_d]_{1}\right]_{0} + h.c. \right)\\
  &=  {\bm Q}^{sd}_2 \cdot {\bm G}_2^{pd} -  {\bm T}^{sd}_2 \cdot {\bm M}_2^{pd}\\
&=
\frac{\sqrt{30}}{15}\Big(\frac{\sqrt{2}}{2}\,c^\dagger_{s,0}\,c^\dagger_{p,-1}\,c_{d,1}\,c_{d,-2} - \frac{\sqrt{3}}{2}\,c^\dagger_{s,0}\,c^\dagger_{p,-1}\,c_{d,0}\,c_{d,-1} + c^\dagger_{s,0}\,c^\dagger_{p,0}\,c_{d,2}\,c_{d,-2}
\\
&\qquad - \frac{1}{2}\,c^\dagger_{s,0}\,c^\dagger_{p,0}\,c_{d,1}\,c_{d,-1} + \frac{\sqrt{2}}{2}\,c^\dagger_{s,0}\,c^\dagger_{p,1}\,c_{d,2}\,c_{d,-1} - \frac{\sqrt{3}}{2}\,c^\dagger_{s,0}\,c^\dagger_{p,1}\,c_{d,1}\,c_{d,0}
\\
&\qquad + \frac{\sqrt{2}}{2}\,c^\dagger_{d,1}\,c^\dagger_{d,-2}\,c_{s,0}\,c_{p,-1} - \frac{\sqrt{3}}{2}\,c^\dagger_{d,0}\,c^\dagger_{d,-1}\,c_{s,0}\,c_{p,-1} + c^\dagger_{d,2}\,c^\dagger_{d,-2}\,c_{s,0}\,c_{p,0}
\\
&\qquad - \frac{1}{2}\,c^\dagger_{d,1}\,c^\dagger_{d,-1}\,c_{s,0}\,c_{p,0} + \frac{\sqrt{2}}{2}\,c^\dagger_{d,2}\,c^\dagger_{d,-1}\,c_{s,0}\,c_{p,1} - \frac{\sqrt{3}}{2}\,c^\dagger_{d,1}\,c^\dagger_{d,0}\,c_{s,0}\,c_{p,1}\Big)
\\
{M_0^{(2)}}_{[1]} &=  
  \frac{i}{2} \left(\left[[c^\dagger_s \times c^\dagger_p]_{1} \times[\tilde c_d \times \tilde c_d]_{1}\right]_{0} - h.c. \right)\\
  &=  {\bm Q}^{sd}_2 \cdot {\bm M}_2^{pd} +  {\bm T}^{sd}_2 \cdot {\bm G}_2^{pd}\\
&=
\frac{\sqrt{30} i}{15}\Big(\frac{\sqrt{2}}{2}\,c^\dagger_{s,0}\,c^\dagger_{p,-1}\,c_{d,1}\,c_{d,-2} - \frac{\sqrt{3}}{2}\,c^\dagger_{s,0}\,c^\dagger_{p,-1}\,c_{d,0}\,c_{d,-1} + c^\dagger_{s,0}\,c^\dagger_{p,0}\,c_{d,2}\,c_{d,-2}
\\
&\qquad - \frac{1}{2}\,c^\dagger_{s,0}\,c^\dagger_{p,0}\,c_{d,1}\,c_{d,-1} + \frac{\sqrt{2}}{2}\,c^\dagger_{s,0}\,c^\dagger_{p,1}\,c_{d,2}\,c_{d,-1} - \frac{\sqrt{3}}{2}\,c^\dagger_{s,0}\,c^\dagger_{p,1}\,c_{d,1}\,c_{d,0}
\\
&\qquad - \frac{\sqrt{2}}{2}\,c^\dagger_{d,1}\,c^\dagger_{d,-2}\,c_{s,0}\,c_{p,-1} + \frac{\sqrt{3}}{2}\,c^\dagger_{d,0}\,c^\dagger_{d,-1}\,c_{s,0}\,c_{p,-1} - c^\dagger_{d,2}\,c^\dagger_{d,-2}\,c_{s,0}\,c_{p,0}
\\
&\qquad + \frac{1}{2}\,c^\dagger_{d,1}\,c^\dagger_{d,-1}\,c_{s,0}\,c_{p,0} - \frac{\sqrt{2}}{2}\,c^\dagger_{d,2}\,c^\dagger_{d,-1}\,c_{s,0}\,c_{p,1} + \frac{\sqrt{3}}{2}\,c^\dagger_{d,1}\,c^\dagger_{d,0}\,c_{s,0}\,c_{p,1}\Big)
\end{align*}

\begin{align*}
{G_0^{(2)}}_{[2]} 
  &=  
  \frac{1}{2} \left(\left[[c^\dagger_s \times c^\dagger_d]_{2} \times[\tilde c_p \times \tilde c_d]_{2}\right]_{0}
  + h.c. \right)\\
  &=  
  \frac{1}{4} \left(\left[[c^\dagger_s \times c^\dagger_d]_{2} \times[\tilde c_p \times \tilde c_d]_{2}\right]_{0}
  +
  \left[[c^\dagger_s \times c^\dagger_d]_{2} \times[\tilde c_d \times \tilde c_p]_{2}\right]_{0}
  + h.c. \right)\\
  &= 
  \frac{1}{2}
  % \left( 
    % \left[Q^{sp}_1 \times G_1^{d}\right]_{0} - 
    {\bm T}^{sp}_1 \cdot {\bm M}_1^{d}  
  % \right)
  -
  \frac{1}{2}
  \left( 
     {\bm Q}^{sd}_2 \cdot {\bm G}_2^{pd}  + {\bm T}^{sd}_2 \cdot {\bm M}_2^{pd}  
  \right)
  \\
&=
\frac{\sqrt{30}}{30}\Big(c^\dagger_{s,0}\,c^\dagger_{d,-2}\,c_{p,0}\,c_{d,-2} - \frac{\sqrt{2}}{2}\,c^\dagger_{s,0}\,c^\dagger_{d,-2}\,c_{p,-1}\,c_{d,-1} + \frac{\sqrt{2}}{2}\,c^\dagger_{s,0}\,c^\dagger_{d,-1}\,c_{p,1}\,c_{d,-2}
\\
&\qquad + \frac{1}{2}\,c^\dagger_{s,0}\,c^\dagger_{d,-1}\,c_{p,0}\,c_{d,-1} - \frac{\sqrt{3}}{2}\,c^\dagger_{s,0}\,c^\dagger_{d,-1}\,c_{p,-1}\,c_{d,0} + \frac{\sqrt{3}}{2}\,c^\dagger_{s,0}\,c^\dagger_{d,0}\,c_{p,1}\,c_{d,-1}
\\
&\qquad - \frac{\sqrt{3}}{2}\,c^\dagger_{s,0}\,c^\dagger_{d,0}\,c_{p,-1}\,c_{d,1} + \frac{\sqrt{3}}{2}\,c^\dagger_{s,0}\,c^\dagger_{d,1}\,c_{p,1}\,c_{d,0} - \frac{1}{2}\,c^\dagger_{s,0}\,c^\dagger_{d,1}\,c_{p,0}\,c_{d,1}
\\
&\qquad - \frac{\sqrt{2}}{2}\,c^\dagger_{s,0}\,c^\dagger_{d,1}\,c_{p,-1}\,c_{d,2} + \frac{\sqrt{2}}{2}\,c^\dagger_{s,0}\,c^\dagger_{d,2}\,c_{p,1}\,c_{d,1} - c^\dagger_{s,0}\,c^\dagger_{d,2}\,c_{p,0}\,c_{d,2}
\\
&\qquad + c^\dagger_{p,0}\,c^\dagger_{d,-2}\,c_{s,0}\,c_{d,-2} - \frac{\sqrt{2}}{2}\,c^\dagger_{p,-1}\,c^\dagger_{d,-1}\,c_{s,0}\,c_{d,-2} + \frac{\sqrt{2}}{2}\,c^\dagger_{p,1}\,c^\dagger_{d,-2}\,c_{s,0}\,c_{d,-1}
\\
&\qquad + \frac{1}{2}\,c^\dagger_{p,0}\,c^\dagger_{d,-1}\,c_{s,0}\,c_{d,-1} - \frac{\sqrt{3}}{2}\,c^\dagger_{p,-1}\,c^\dagger_{d,0}\,c_{s,0}\,c_{d,-1} + \frac{\sqrt{3}}{2}\,c^\dagger_{p,1}\,c^\dagger_{d,-1}\,c_{s,0}\,c_{d,0}
\\
&\qquad - \frac{\sqrt{3}}{2}\,c^\dagger_{p,-1}\,c^\dagger_{d,1}\,c_{s,0}\,c_{d,0} + \frac{\sqrt{3}}{2}\,c^\dagger_{p,1}\,c^\dagger_{d,0}\,c_{s,0}\,c_{d,1} - \frac{1}{2}\,c^\dagger_{p,0}\,c^\dagger_{d,1}\,c_{s,0}\,c_{d,1}
\\
&\qquad - \frac{\sqrt{2}}{2}\,c^\dagger_{p,-1}\,c^\dagger_{d,2}\,c_{s,0}\,c_{d,1} + \frac{\sqrt{2}}{2}\,c^\dagger_{p,1}\,c^\dagger_{d,1}\,c_{s,0}\,c_{d,2} - c^\dagger_{p,0}\,c^\dagger_{d,2}\,c_{s,0}\,c_{d,2}\Big)
\\
{M_0^{(2)}}_{[2]} 
  &=  
  \frac{i}{2} \left(\left[[c^\dagger_s \times c^\dagger_d]_{2} \times[\tilde c_p \times \tilde c_d]_{2}\right]_{0}
  - h.c. \right)\\
  &=  
  \frac{i}{4} \left(\left[[c^\dagger_s \times c^\dagger_d]_{1} \times[\tilde c_p \times \tilde c_d]_{1}\right]_{0}
  +
  \left[[c^\dagger_s \times c^\dagger_d]_{1} \times[\tilde c_d \times \tilde c_p]_{1}\right]_{0}
  - h.c. \right)\\ 
&=
  \frac{1}{2}
   {\bm Q}^{sp}_1 \cdot {\bm M}_1^{d}   
  -
  \frac{1}{2}
  \left( 
      {\bm T}^{sd}_2 \cdot {\bm G}_2^{pd} -  {\bm Q}^{sd}_2 \cdot {\bm M}_2^{pd} 
  \right)
  \\
  &=
  \frac{\sqrt{30} i}{30}\Big(c^\dagger_{s,0}\,c^\dagger_{d,-2}\,c_{p,0}\,c_{d,-2} - \frac{\sqrt{2}}{2}\,c^\dagger_{s,0}\,c^\dagger_{d,-2}\,c_{p,-1}\,c_{d,-1} + \frac{\sqrt{2}}{2}\,c^\dagger_{s,0}\,c^\dagger_{d,-1}\,c_{p,1}\,c_{d,-2}
\\
&\qquad + \frac{1}{2}\,c^\dagger_{s,0}\,c^\dagger_{d,-1}\,c_{p,0}\,c_{d,-1} - \frac{\sqrt{3}}{2}\,c^\dagger_{s,0}\,c^\dagger_{d,-1}\,c_{p,-1}\,c_{d,0} + \frac{\sqrt{3}}{2}\,c^\dagger_{s,0}\,c^\dagger_{d,0}\,c_{p,1}\,c_{d,-1}
\\
&\qquad - \frac{\sqrt{3}}{2}\,c^\dagger_{s,0}\,c^\dagger_{d,0}\,c_{p,-1}\,c_{d,1} + \frac{\sqrt{3}}{2}\,c^\dagger_{s,0}\,c^\dagger_{d,1}\,c_{p,1}\,c_{d,0} - \frac{1}{2}\,c^\dagger_{s,0}\,c^\dagger_{d,1}\,c_{p,0}\,c_{d,1}
\\
&\qquad - \frac{\sqrt{2}}{2}\,c^\dagger_{s,0}\,c^\dagger_{d,1}\,c_{p,-1}\,c_{d,2} + \frac{\sqrt{2}}{2}\,c^\dagger_{s,0}\,c^\dagger_{d,2}\,c_{p,1}\,c_{d,1} - c^\dagger_{s,0}\,c^\dagger_{d,2}\,c_{p,0}\,c_{d,2}
\\
&\qquad - c^\dagger_{p,0}\,c^\dagger_{d,-2}\,c_{s,0}\,c_{d,-2} + \frac{\sqrt{2}}{2}\,c^\dagger_{p,-1}\,c^\dagger_{d,-1}\,c_{s,0}\,c_{d,-2} - \frac{\sqrt{2}}{2}\,c^\dagger_{p,1}\,c^\dagger_{d,-2}\,c_{s,0}\,c_{d,-1}
\\
&\qquad - \frac{1}{2}\,c^\dagger_{p,0}\,c^\dagger_{d,-1}\,c_{s,0}\,c_{d,-1} + \frac{\sqrt{3}}{2}\,c^\dagger_{p,-1}\,c^\dagger_{d,0}\,c_{s,0}\,c_{d,-1} - \frac{\sqrt{3}}{2}\,c^\dagger_{p,1}\,c^\dagger_{d,-1}\,c_{s,0}\,c_{d,0}
\\
&\qquad + \frac{\sqrt{3}}{2}\,c^\dagger_{p,-1}\,c^\dagger_{d,1}\,c_{s,0}\,c_{d,0} - \frac{\sqrt{3}}{2}\,c^\dagger_{p,1}\,c^\dagger_{d,0}\,c_{s,0}\,c_{d,1} + \frac{1}{2}\,c^\dagger_{p,0}\,c^\dagger_{d,1}\,c_{s,0}\,c_{d,1}
\\
&\qquad + \frac{\sqrt{2}}{2}\,c^\dagger_{p,-1}\,c^\dagger_{d,2}\,c_{s,0}\,c_{d,1} - \frac{\sqrt{2}}{2}\,c^\dagger_{p,1}\,c^\dagger_{d,1}\,c_{s,0}\,c_{d,2} + c^\dagger_{p,0}\,c^\dagger_{d,2}\,c_{s,0}\,c_{d,2}\Big)
\end{align*}
% 
% 
% 
% \begin{align}
%   [\tilde c_a \times \tilde c_b]_{k}
%   &= \sum C^{k,q}_{a,n;b,m} \tilde c_a \tilde c_b
%   = - \sum C^{k,q}_{a,n;b,m} \tilde c_b \tilde c_a
%   = - (-1)^{a+b-k} \sum C^{k,q}_{b,m;a,n} \tilde c_b \tilde c_a\\
%   &= - (-1)^{a+b-k} [\tilde c_b \times \tilde c_a]_{k},\quad (a \ne b)
% \end{align}

\subsection*{$spf$ system}
\begin{align*}
{G_0^{(2)}}_{[1]} &=  
  \frac{1}{2} \left(\left[[c^\dagger_s \times c^\dagger_p]_{1} \times[\tilde c_f \times \tilde c_f]_{1}\right]_{0} + h.c. \right)\\
  &=  {\bm T}^{sf}_3 \cdot {\bm M}_3^{pf} - {\bm Q}^{sf}_3 \cdot {\bm G}_3^{pf}\\
&=
\frac{\sqrt{21}}{14}\Big(\frac{\sqrt{3}}{3}\,c^\dagger_{s,0}\,c^\dagger_{p,-1}\,c_{f,2}\,c_{f,-3} - \frac{\sqrt{5}}{3}\,c^\dagger_{s,0}\,c^\dagger_{p,-1}\,c_{f,1}\,c_{f,-2} + \frac{\sqrt{6}}{3}\,c^\dagger_{s,0}\,c^\dagger_{p,-1}\,c_{f,0}\,c_{f,-1}
\\
&\qquad + c^\dagger_{s,0}\,c^\dagger_{p,0}\,c_{f,3}\,c_{f,-3} - \frac{2}{3}\,c^\dagger_{s,0}\,c^\dagger_{p,0}\,c_{f,2}\,c_{f,-2} + \frac{1}{3}\,c^\dagger_{s,0}\,c^\dagger_{p,0}\,c_{f,1}\,c_{f,-1}
\\
&\qquad + \frac{\sqrt{3}}{3}\,c^\dagger_{s,0}\,c^\dagger_{p,1}\,c_{f,3}\,c_{f,-2} - \frac{\sqrt{5}}{3}\,c^\dagger_{s,0}\,c^\dagger_{p,1}\,c_{f,2}\,c_{f,-1} + \frac{\sqrt{6}}{3}\,c^\dagger_{s,0}\,c^\dagger_{p,1}\,c_{f,1}\,c_{f,0}
\\
&\qquad + \frac{\sqrt{3}}{3}\,c^\dagger_{f,2}\,c^\dagger_{f,-3}\,c_{s,0}\,c_{p,-1} - \frac{\sqrt{5}}{3}\,c^\dagger_{f,1}\,c^\dagger_{f,-2}\,c_{s,0}\,c_{p,-1} + \frac{\sqrt{6}}{3}\,c^\dagger_{f,0}\,c^\dagger_{f,-1}\,c_{s,0}\,c_{p,-1}
\\
&\qquad + c^\dagger_{f,3}\,c^\dagger_{f,-3}\,c_{s,0}\,c_{p,0} - \frac{2}{3}\,c^\dagger_{f,2}\,c^\dagger_{f,-2}\,c_{s,0}\,c_{p,0} + \frac{1}{3}\,c^\dagger_{f,1}\,c^\dagger_{f,-1}\,c_{s,0}\,c_{p,0}
\\
&\qquad + \frac{\sqrt{3}}{3}\,c^\dagger_{f,3}\,c^\dagger_{f,-2}\,c_{s,0}\,c_{p,1} - \frac{\sqrt{5}}{3}\,c^\dagger_{f,2}\,c^\dagger_{f,-1}\,c_{s,0}\,c_{p,1} + \frac{\sqrt{6}}{3}\,c^\dagger_{f,1}\,c^\dagger_{f,0}\,c_{s,0}\,c_{p,1}\Big)
\\
{M_0^{(2)}}_{[1]} &=  
  \frac{i}{2} \left(\left[[c^\dagger_s \times c^\dagger_p]_{1} \times[\tilde c_f \times \tilde c_f]_{1}\right]_{0} - h.c. \right)\\
  &=  {\bm T}^{sf}_3 \cdot {\bm G}_3^{pf} + {\bm Q}^{sf}_3 \cdot {\bm M}_3^{pf}\\
&=
\frac{\sqrt{21} i}{14}\Big(\frac{\sqrt{3}}{3}\,c^\dagger_{s,0}\,c^\dagger_{p,-1}\,c_{f,2}\,c_{f,-3} - \frac{\sqrt{5}}{3}\,c^\dagger_{s,0}\,c^\dagger_{p,-1}\,c_{f,1}\,c_{f,-2} + \frac{\sqrt{6}}{3}\,c^\dagger_{s,0}\,c^\dagger_{p,-1}\,c_{f,0}\,c_{f,-1}
\\
&\qquad + c^\dagger_{s,0}\,c^\dagger_{p,0}\,c_{f,3}\,c_{f,-3} - \frac{2}{3}\,c^\dagger_{s,0}\,c^\dagger_{p,0}\,c_{f,2}\,c_{f,-2} + \frac{1}{3}\,c^\dagger_{s,0}\,c^\dagger_{p,0}\,c_{f,1}\,c_{f,-1}
\\
&\qquad + \frac{\sqrt{3}}{3}\,c^\dagger_{s,0}\,c^\dagger_{p,1}\,c_{f,3}\,c_{f,-2} - \frac{\sqrt{5}}{3}\,c^\dagger_{s,0}\,c^\dagger_{p,1}\,c_{f,2}\,c_{f,-1} + \frac{\sqrt{6}}{3}\,c^\dagger_{s,0}\,c^\dagger_{p,1}\,c_{f,1}\,c_{f,0}
\\
&\qquad - \frac{\sqrt{3}}{3}\,c^\dagger_{f,2}\,c^\dagger_{f,-3}\,c_{s,0}\,c_{p,-1} + \frac{\sqrt{5}}{3}\,c^\dagger_{f,1}\,c^\dagger_{f,-2}\,c_{s,0}\,c_{p,-1} - \frac{\sqrt{6}}{3}\,c^\dagger_{f,0}\,c^\dagger_{f,-1}\,c_{s,0}\,c_{p,-1}
\\
&\qquad - c^\dagger_{f,3}\,c^\dagger_{f,-3}\,c_{s,0}\,c_{p,0} + \frac{2}{3}\,c^\dagger_{f,2}\,c^\dagger_{f,-2}\,c_{s,0}\,c_{p,0} - \frac{1}{3}\,c^\dagger_{f,1}\,c^\dagger_{f,-1}\,c_{s,0}\,c_{p,0}
\\
&\qquad - \frac{\sqrt{3}}{3}\,c^\dagger_{f,3}\,c^\dagger_{f,-2}\,c_{s,0}\,c_{p,1} + \frac{\sqrt{5}}{3}\,c^\dagger_{f,2}\,c^\dagger_{f,-1}\,c_{s,0}\,c_{p,1} - \frac{\sqrt{6}}{3}\,c^\dagger_{f,1}\,c^\dagger_{f,0}\,c_{s,0}\,c_{p,1}\Big)
\end{align*}

\begin{align*}
{G_0^{(2)}}_{[2]} &=  
  \frac{1}{2} \left(\left[[c^\dagger_s \times c^\dagger_f]_{3} \times[\tilde c_p \times \tilde c_f]_{3}\right]_{0} + h.c. \right)\\
  &= 
  \frac{1}{4} \left(\left[[c^\dagger_s \times c^\dagger_f]_{3} \times[\tilde c_p \times \tilde c_f]_{3}\right]_{0} + \left[[c^\dagger_s \times c^\dagger_f]_{3} \times[\tilde c_f \times \tilde c_p]_{3}\right]_{0} + h.c. \right)\\
  &= 
  \frac{1}{2}{\bm T}^{sp}_1 \cdot {\bm M}_1^{f}
  +
  \frac{1}{2} 
  \left( 
  {\bm T}^{sf}_3 \cdot {\bm M}_3^{pf}
  +
  {\bm Q}^{sf}_3 \cdot {\bm G}_3^{pf}   
  \right)
  \\
&=
\frac{\sqrt{21}}{28}\Big(c^\dagger_{s,0}\,c^\dagger_{f,-3}\,c_{p,0}\,c_{f,-3} - \frac{\sqrt{3}}{3}\,c^\dagger_{s,0}\,c^\dagger_{f,-3}\,c_{p,-1}\,c_{f,-2} + \frac{\sqrt{3}}{3}\,c^\dagger_{s,0}\,c^\dagger_{f,-2}\,c_{p,1}\,c_{f,-3}
\\
&\qquad + \frac{2}{3}\,c^\dagger_{s,0}\,c^\dagger_{f,-2}\,c_{p,0}\,c_{f,-2} - \frac{\sqrt{5}}{3}\,c^\dagger_{s,0}\,c^\dagger_{f,-2}\,c_{p,-1}\,c_{f,-1} + \frac{\sqrt{5}}{3}\,c^\dagger_{s,0}\,c^\dagger_{f,-1}\,c_{p,1}\,c_{f,-2}
\\
&\qquad + \frac{1}{3}\,c^\dagger_{s,0}\,c^\dagger_{f,-1}\,c_{p,0}\,c_{f,-1} - \frac{\sqrt{6}}{3}\,c^\dagger_{s,0}\,c^\dagger_{f,-1}\,c_{p,-1}\,c_{f,0} + \frac{\sqrt{6}}{3}\,c^\dagger_{s,0}\,c^\dagger_{f,0}\,c_{p,1}\,c_{f,-1}
\\
&\qquad - \frac{\sqrt{6}}{3}\,c^\dagger_{s,0}\,c^\dagger_{f,0}\,c_{p,-1}\,c_{f,1} + \frac{\sqrt{6}}{3}\,c^\dagger_{s,0}\,c^\dagger_{f,1}\,c_{p,1}\,c_{f,0} - \frac{1}{3}\,c^\dagger_{s,0}\,c^\dagger_{f,1}\,c_{p,0}\,c_{f,1}
\\
&\qquad - \frac{\sqrt{5}}{3}\,c^\dagger_{s,0}\,c^\dagger_{f,1}\,c_{p,-1}\,c_{f,2} + \frac{\sqrt{5}}{3}\,c^\dagger_{s,0}\,c^\dagger_{f,2}\,c_{p,1}\,c_{f,1} - \frac{2}{3}\,c^\dagger_{s,0}\,c^\dagger_{f,2}\,c_{p,0}\,c_{f,2}
\\
&\qquad - \frac{\sqrt{3}}{3}\,c^\dagger_{s,0}\,c^\dagger_{f,2}\,c_{p,-1}\,c_{f,3} + \frac{\sqrt{3}}{3}\,c^\dagger_{s,0}\,c^\dagger_{f,3}\,c_{p,1}\,c_{f,2} - c^\dagger_{s,0}\,c^\dagger_{f,3}\,c_{p,0}\,c_{f,3}
\\
&\qquad + c^\dagger_{p,0}\,c^\dagger_{f,-3}\,c_{s,0}\,c_{f,-3} - \frac{\sqrt{3}}{3}\,c^\dagger_{p,-1}\,c^\dagger_{f,-2}\,c_{s,0}\,c_{f,-3} + \frac{\sqrt{3}}{3}\,c^\dagger_{p,1}\,c^\dagger_{f,-3}\,c_{s,0}\,c_{f,-2}
\\
&\qquad + \frac{2}{3}\,c^\dagger_{p,0}\,c^\dagger_{f,-2}\,c_{s,0}\,c_{f,-2} - \frac{\sqrt{5}}{3}\,c^\dagger_{p,-1}\,c^\dagger_{f,-1}\,c_{s,0}\,c_{f,-2} + \frac{\sqrt{5}}{3}\,c^\dagger_{p,1}\,c^\dagger_{f,-2}\,c_{s,0}\,c_{f,-1}
\\
&\qquad + \frac{1}{3}\,c^\dagger_{p,0}\,c^\dagger_{f,-1}\,c_{s,0}\,c_{f,-1} - \frac{\sqrt{6}}{3}\,c^\dagger_{p,-1}\,c^\dagger_{f,0}\,c_{s,0}\,c_{f,-1} + \frac{\sqrt{6}}{3}\,c^\dagger_{p,1}\,c^\dagger_{f,-1}\,c_{s,0}\,c_{f,0}
\\
&\qquad - \frac{\sqrt{6}}{3}\,c^\dagger_{p,-1}\,c^\dagger_{f,1}\,c_{s,0}\,c_{f,0} + \frac{\sqrt{6}}{3}\,c^\dagger_{p,1}\,c^\dagger_{f,0}\,c_{s,0}\,c_{f,1} - \frac{1}{3}\,c^\dagger_{p,0}\,c^\dagger_{f,1}\,c_{s,0}\,c_{f,1}
\\
&\qquad - \frac{\sqrt{5}}{3}\,c^\dagger_{p,-1}\,c^\dagger_{f,2}\,c_{s,0}\,c_{f,1} + \frac{\sqrt{5}}{3}\,c^\dagger_{p,1}\,c^\dagger_{f,1}\,c_{s,0}\,c_{f,2} - \frac{2}{3}\,c^\dagger_{p,0}\,c^\dagger_{f,2}\,c_{s,0}\,c_{f,2}
\\
&\qquad - \frac{\sqrt{3}}{3}\,c^\dagger_{p,-1}\,c^\dagger_{f,3}\,c_{s,0}\,c_{f,2} + \frac{\sqrt{3}}{3}\,c^\dagger_{p,1}\,c^\dagger_{f,2}\,c_{s,0}\,c_{f,3} - c^\dagger_{p,0}\,c^\dagger_{f,3}\,c_{s,0}\,c_{f,3}\Big)
\\
{M_0^{(2)}}_{[2]} &=  
  \frac{i}{2} \left(\left[[c^\dagger_s \times c^\dagger_p]_{1} \times[\tilde c_f \times \tilde c_f]_{1}\right]_{0} - h.c. \right)\\
  % &={\bm Q}^{sp}_3 \cdot {\bm M}_3^{f}\
  &= 
  \frac{1}{2}{\bm Q}^{sp}_1 \cdot {\bm M}_1^{f}
  +
  \frac{1}{2} 
  \left( 
   {\bm Q}^{sf}_3 \cdot {\bm M}_3^{pf}
   -
   {\bm T}^{sf}_3 \cdot {\bm G}_3^{pf}
  \right)
  \\
&=
\frac{\sqrt{21} i}{28}\Big(c^\dagger_{s,0}\,c^\dagger_{f,-3}\,c_{p,0}\,c_{f,-3} - \frac{\sqrt{3}}{3}\,c^\dagger_{s,0}\,c^\dagger_{f,-3}\,c_{p,-1}\,c_{f,-2} + \frac{\sqrt{3}}{3}\,c^\dagger_{s,0}\,c^\dagger_{f,-2}\,c_{p,1}\,c_{f,-3}
\\
&\qquad + \frac{2}{3}\,c^\dagger_{s,0}\,c^\dagger_{f,-2}\,c_{p,0}\,c_{f,-2} - \frac{\sqrt{5}}{3}\,c^\dagger_{s,0}\,c^\dagger_{f,-2}\,c_{p,-1}\,c_{f,-1} + \frac{\sqrt{5}}{3}\,c^\dagger_{s,0}\,c^\dagger_{f,-1}\,c_{p,1}\,c_{f,-2}
\\
&\qquad + \frac{1}{3}\,c^\dagger_{s,0}\,c^\dagger_{f,-1}\,c_{p,0}\,c_{f,-1} - \frac{\sqrt{6}}{3}\,c^\dagger_{s,0}\,c^\dagger_{f,-1}\,c_{p,-1}\,c_{f,0} + \frac{\sqrt{6}}{3}\,c^\dagger_{s,0}\,c^\dagger_{f,0}\,c_{p,1}\,c_{f,-1}
\\
&\qquad - \frac{\sqrt{6}}{3}\,c^\dagger_{s,0}\,c^\dagger_{f,0}\,c_{p,-1}\,c_{f,1} + \frac{\sqrt{6}}{3}\,c^\dagger_{s,0}\,c^\dagger_{f,1}\,c_{p,1}\,c_{f,0} - \frac{1}{3}\,c^\dagger_{s,0}\,c^\dagger_{f,1}\,c_{p,0}\,c_{f,1}
\\
&\qquad - \frac{\sqrt{5}}{3}\,c^\dagger_{s,0}\,c^\dagger_{f,1}\,c_{p,-1}\,c_{f,2} + \frac{\sqrt{5}}{3}\,c^\dagger_{s,0}\,c^\dagger_{f,2}\,c_{p,1}\,c_{f,1} - \frac{2}{3}\,c^\dagger_{s,0}\,c^\dagger_{f,2}\,c_{p,0}\,c_{f,2}
\\
&\qquad - \frac{\sqrt{3}}{3}\,c^\dagger_{s,0}\,c^\dagger_{f,2}\,c_{p,-1}\,c_{f,3} + \frac{\sqrt{3}}{3}\,c^\dagger_{s,0}\,c^\dagger_{f,3}\,c_{p,1}\,c_{f,2} - c^\dagger_{s,0}\,c^\dagger_{f,3}\,c_{p,0}\,c_{f,3}
\\
&\qquad - c^\dagger_{p,0}\,c^\dagger_{f,-3}\,c_{s,0}\,c_{f,-3} + \frac{\sqrt{3}}{3}\,c^\dagger_{p,-1}\,c^\dagger_{f,-2}\,c_{s,0}\,c_{f,-3} - \frac{\sqrt{3}}{3}\,c^\dagger_{p,1}\,c^\dagger_{f,-3}\,c_{s,0}\,c_{f,-2}
\\
&\qquad - \frac{2}{3}\,c^\dagger_{p,0}\,c^\dagger_{f,-2}\,c_{s,0}\,c_{f,-2} + \frac{\sqrt{5}}{3}\,c^\dagger_{p,-1}\,c^\dagger_{f,-1}\,c_{s,0}\,c_{f,-2} - \frac{\sqrt{5}}{3}\,c^\dagger_{p,1}\,c^\dagger_{f,-2}\,c_{s,0}\,c_{f,-1}
\\
&\qquad - \frac{1}{3}\,c^\dagger_{p,0}\,c^\dagger_{f,-1}\,c_{s,0}\,c_{f,-1} + \frac{\sqrt{6}}{3}\,c^\dagger_{p,-1}\,c^\dagger_{f,0}\,c_{s,0}\,c_{f,-1} - \frac{\sqrt{6}}{3}\,c^\dagger_{p,1}\,c^\dagger_{f,-1}\,c_{s,0}\,c_{f,0}
\\
&\qquad + \frac{\sqrt{6}}{3}\,c^\dagger_{p,-1}\,c^\dagger_{f,1}\,c_{s,0}\,c_{f,0} - \frac{\sqrt{6}}{3}\,c^\dagger_{p,1}\,c^\dagger_{f,0}\,c_{s,0}\,c_{f,1} + \frac{1}{3}\,c^\dagger_{p,0}\,c^\dagger_{f,1}\,c_{s,0}\,c_{f,1}
\\
&\qquad + \frac{\sqrt{5}}{3}\,c^\dagger_{p,-1}\,c^\dagger_{f,2}\,c_{s,0}\,c_{f,1} - \frac{\sqrt{5}}{3}\,c^\dagger_{p,1}\,c^\dagger_{f,1}\,c_{s,0}\,c_{f,2} + \frac{2}{3}\,c^\dagger_{p,0}\,c^\dagger_{f,2}\,c_{s,0}\,c_{f,2}
\\
&\qquad + \frac{\sqrt{3}}{3}\,c^\dagger_{p,-1}\,c^\dagger_{f,3}\,c_{s,0}\,c_{f,2} - \frac{\sqrt{3}}{3}\,c^\dagger_{p,1}\,c^\dagger_{f,2}\,c_{s,0}\,c_{f,3} + c^\dagger_{p,0}\,c^\dagger_{f,3}\,c_{s,0}\,c_{f,3}\Big)
\end{align*}

\subsection*{$sdf$ system}

% sfdf : 3×3
% j13,j24,A wigner_9j： 2 , 2 , 1
% sffd : 3×3
% j13,j24,A wigner_9j： 3 , 3 , 1
\begin{align*}
Q_0^{(2)} 
&=  
  \frac{1}{2} \left(\left[[c^\dagger_s \times c^\dagger_f]_{3} \times[\tilde c_d \times \tilde c_f]_{3}\right]_{0} + h.c. \right)\\
&=  
  \frac{1}{4} \left(\left[[c^\dagger_s \times c^\dagger_f]_{3} \times[\tilde c_d \times \tilde c_f]_{3}\right]_{0}
  -
  \left[[c^\dagger_s \times c^\dagger_f]_{3} \times[\tilde c_f \times \tilde c_d]_{3}\right]_{0}
  + h.c. \right)\\
&=
-\frac{1}{2} \left( {\bm Q}^{sd}_2 \cdot {\bm Q}^{f}_2  \right)
   -
\frac{1}{2} \left( {\bm T}^{sf}_3 \cdot {\bm T}^{df}_3 +
 {\bm Q}^{sf}_3 \cdot {\bm Q}^{df}_3 \right)
\\
&=
\frac{\sqrt{105}}{84}\Big(c^\dagger_{s,0}\,c^\dagger_{f,-3}\,c_{d,0}\,c_{f,-3} - c^\dagger_{s,0}\,c^\dagger_{f,-3}\,c_{d,-1}\,c_{f,-2} + \frac{\sqrt{10}}{5}\,c^\dagger_{s,0}\,c^\dagger_{f,-3}\,c_{d,-2}\,c_{f,-1}
\\
&\qquad + c^\dagger_{s,0}\,c^\dagger_{f,-2}\,c_{d,1}\,c_{f,-3} - \frac{\sqrt{15}}{5}\,c^\dagger_{s,0}\,c^\dagger_{f,-2}\,c_{d,-1}\,c_{f,-1} + \frac{2 \sqrt{5}}{5}\,c^\dagger_{s,0}\,c^\dagger_{f,-2}\,c_{d,-2}\,c_{f,0}
\\
&\qquad + \frac{\sqrt{10}}{5}\,c^\dagger_{s,0}\,c^\dagger_{f,-1}\,c_{d,2}\,c_{f,-3} + \frac{\sqrt{15}}{5}\,c^\dagger_{s,0}\,c^\dagger_{f,-1}\,c_{d,1}\,c_{f,-2} - \frac{3}{5}\,c^\dagger_{s,0}\,c^\dagger_{f,-1}\,c_{d,0}\,c_{f,-1}
\\
&\qquad - \frac{\sqrt{2}}{5}\,c^\dagger_{s,0}\,c^\dagger_{f,-1}\,c_{d,-1}\,c_{f,0} + \frac{2 \sqrt{6}}{5}\,c^\dagger_{s,0}\,c^\dagger_{f,-1}\,c_{d,-2}\,c_{f,1} + \frac{2 \sqrt{5}}{5}\,c^\dagger_{s,0}\,c^\dagger_{f,0}\,c_{d,2}\,c_{f,-2}
\\
&\qquad + \frac{\sqrt{2}}{5}\,c^\dagger_{s,0}\,c^\dagger_{f,0}\,c_{d,1}\,c_{f,-1} - \frac{4}{5}\,c^\dagger_{s,0}\,c^\dagger_{f,0}\,c_{d,0}\,c_{f,0} + \frac{\sqrt{2}}{5}\,c^\dagger_{s,0}\,c^\dagger_{f,0}\,c_{d,-1}\,c_{f,1}
\\
&\qquad + \frac{2 \sqrt{5}}{5}\,c^\dagger_{s,0}\,c^\dagger_{f,0}\,c_{d,-2}\,c_{f,2} + \frac{2 \sqrt{6}}{5}\,c^\dagger_{s,0}\,c^\dagger_{f,1}\,c_{d,2}\,c_{f,-1} - \frac{\sqrt{2}}{5}\,c^\dagger_{s,0}\,c^\dagger_{f,1}\,c_{d,1}\,c_{f,0}
\\
&\qquad - \frac{3}{5}\,c^\dagger_{s,0}\,c^\dagger_{f,1}\,c_{d,0}\,c_{f,1} + \frac{\sqrt{15}}{5}\,c^\dagger_{s,0}\,c^\dagger_{f,1}\,c_{d,-1}\,c_{f,2} + \frac{\sqrt{10}}{5}\,c^\dagger_{s,0}\,c^\dagger_{f,1}\,c_{d,-2}\,c_{f,3}
\\
&\qquad + \frac{2 \sqrt{5}}{5}\,c^\dagger_{s,0}\,c^\dagger_{f,2}\,c_{d,2}\,c_{f,0} - \frac{\sqrt{15}}{5}\,c^\dagger_{s,0}\,c^\dagger_{f,2}\,c_{d,1}\,c_{f,1} + c^\dagger_{s,0}\,c^\dagger_{f,2}\,c_{d,-1}\,c_{f,3}
\\
&\qquad + \frac{\sqrt{10}}{5}\,c^\dagger_{s,0}\,c^\dagger_{f,3}\,c_{d,2}\,c_{f,1} - c^\dagger_{s,0}\,c^\dagger_{f,3}\,c_{d,1}\,c_{f,2} + c^\dagger_{s,0}\,c^\dagger_{f,3}\,c_{d,0}\,c_{f,3}
\\
&\qquad + c^\dagger_{d,0}\,c^\dagger_{f,-3}\,c_{s,0}\,c_{f,-3} - c^\dagger_{d,-1}\,c^\dagger_{f,-2}\,c_{s,0}\,c_{f,-3} + \frac{\sqrt{10}}{5}\,c^\dagger_{d,-2}\,c^\dagger_{f,-1}\,c_{s,0}\,c_{f,-3}
\\
&\qquad + c^\dagger_{d,1}\,c^\dagger_{f,-3}\,c_{s,0}\,c_{f,-2} - \frac{\sqrt{15}}{5}\,c^\dagger_{d,-1}\,c^\dagger_{f,-1}\,c_{s,0}\,c_{f,-2} + \frac{2 \sqrt{5}}{5}\,c^\dagger_{d,-2}\,c^\dagger_{f,0}\,c_{s,0}\,c_{f,-2}
\\
&\qquad + \frac{\sqrt{10}}{5}\,c^\dagger_{d,2}\,c^\dagger_{f,-3}\,c_{s,0}\,c_{f,-1} + \frac{\sqrt{15}}{5}\,c^\dagger_{d,1}\,c^\dagger_{f,-2}\,c_{s,0}\,c_{f,-1} - \frac{3}{5}\,c^\dagger_{d,0}\,c^\dagger_{f,-1}\,c_{s,0}\,c_{f,-1}
\\
&\qquad - \frac{\sqrt{2}}{5}\,c^\dagger_{d,-1}\,c^\dagger_{f,0}\,c_{s,0}\,c_{f,-1} + \frac{2 \sqrt{6}}{5}\,c^\dagger_{d,-2}\,c^\dagger_{f,1}\,c_{s,0}\,c_{f,-1} + \frac{2 \sqrt{5}}{5}\,c^\dagger_{d,2}\,c^\dagger_{f,-2}\,c_{s,0}\,c_{f,0}
\\
&\qquad + \frac{\sqrt{2}}{5}\,c^\dagger_{d,1}\,c^\dagger_{f,-1}\,c_{s,0}\,c_{f,0} - \frac{4}{5}\,c^\dagger_{d,0}\,c^\dagger_{f,0}\,c_{s,0}\,c_{f,0} + \frac{\sqrt{2}}{5}\,c^\dagger_{d,-1}\,c^\dagger_{f,1}\,c_{s,0}\,c_{f,0}
\\
&\qquad + \frac{2 \sqrt{5}}{5}\,c^\dagger_{d,-2}\,c^\dagger_{f,2}\,c_{s,0}\,c_{f,0} + \frac{2 \sqrt{6}}{5}\,c^\dagger_{d,2}\,c^\dagger_{f,-1}\,c_{s,0}\,c_{f,1} - \frac{\sqrt{2}}{5}\,c^\dagger_{d,1}\,c^\dagger_{f,0}\,c_{s,0}\,c_{f,1}
\\
&\qquad - \frac{3}{5}\,c^\dagger_{d,0}\,c^\dagger_{f,1}\,c_{s,0}\,c_{f,1} + \frac{\sqrt{15}}{5}\,c^\dagger_{d,-1}\,c^\dagger_{f,2}\,c_{s,0}\,c_{f,1} + \frac{\sqrt{10}}{5}\,c^\dagger_{d,-2}\,c^\dagger_{f,3}\,c_{s,0}\,c_{f,1}
\\
&\qquad + \frac{2 \sqrt{5}}{5}\,c^\dagger_{d,2}\,c^\dagger_{f,0}\,c_{s,0}\,c_{f,2} - \frac{\sqrt{15}}{5}\,c^\dagger_{d,1}\,c^\dagger_{f,1}\,c_{s,0}\,c_{f,2} + c^\dagger_{d,-1}\,c^\dagger_{f,3}\,c_{s,0}\,c_{f,2}
\\
&\qquad + \frac{\sqrt{10}}{5}\,c^\dagger_{d,2}\,c^\dagger_{f,1}\,c_{s,0}\,c_{f,3} - c^\dagger_{d,1}\,c^\dagger_{f,2}\,c_{s,0}\,c_{f,3} + c^\dagger_{d,0}\,c^\dagger_{f,3}\,c_{s,0}\,c_{f,3}\Big)
\\
T_0^{(2)} 
&=  
  \frac{i}{2} \left(\left[[c^\dagger_s \times c^\dagger_p]_{1} \times[\tilde c_p \times \tilde c_d]_{1}\right]_{0} - h.c. \right)\\
&=
-\frac{1}{2} \left( {\bm T}^{sd}_2 \cdot {\bm Q}^{f}_2  \right)
   -
\frac{1}{2} \left( {\bm Q}^{sf}_3 \cdot {\bm T}^{df}_3 
-
 {\bm T}^{sf}_3 \cdot {\bm Q}^{df}_3 \right)
\\
&=
\frac{\sqrt{105} i}{84}\Big(c^\dagger_{s,0}\,c^\dagger_{f,-3}\,c_{d,0}\,c_{f,-3} - c^\dagger_{s,0}\,c^\dagger_{f,-3}\,c_{d,-1}\,c_{f,-2} + \frac{\sqrt{10}}{5}\,c^\dagger_{s,0}\,c^\dagger_{f,-3}\,c_{d,-2}\,c_{f,-1}
\\
&\qquad + c^\dagger_{s,0}\,c^\dagger_{f,-2}\,c_{d,1}\,c_{f,-3} - \frac{\sqrt{15}}{5}\,c^\dagger_{s,0}\,c^\dagger_{f,-2}\,c_{d,-1}\,c_{f,-1} + \frac{2 \sqrt{5}}{5}\,c^\dagger_{s,0}\,c^\dagger_{f,-2}\,c_{d,-2}\,c_{f,0}
\\
&\qquad + \frac{\sqrt{10}}{5}\,c^\dagger_{s,0}\,c^\dagger_{f,-1}\,c_{d,2}\,c_{f,-3} + \frac{\sqrt{15}}{5}\,c^\dagger_{s,0}\,c^\dagger_{f,-1}\,c_{d,1}\,c_{f,-2} - \frac{3}{5}\,c^\dagger_{s,0}\,c^\dagger_{f,-1}\,c_{d,0}\,c_{f,-1}
\\
&\qquad - \frac{\sqrt{2}}{5}\,c^\dagger_{s,0}\,c^\dagger_{f,-1}\,c_{d,-1}\,c_{f,0} + \frac{2 \sqrt{6}}{5}\,c^\dagger_{s,0}\,c^\dagger_{f,-1}\,c_{d,-2}\,c_{f,1} + \frac{2 \sqrt{5}}{5}\,c^\dagger_{s,0}\,c^\dagger_{f,0}\,c_{d,2}\,c_{f,-2}
\\
&\qquad + \frac{\sqrt{2}}{5}\,c^\dagger_{s,0}\,c^\dagger_{f,0}\,c_{d,1}\,c_{f,-1} - \frac{4}{5}\,c^\dagger_{s,0}\,c^\dagger_{f,0}\,c_{d,0}\,c_{f,0} + \frac{\sqrt{2}}{5}\,c^\dagger_{s,0}\,c^\dagger_{f,0}\,c_{d,-1}\,c_{f,1}
\\
&\qquad + \frac{2 \sqrt{5}}{5}\,c^\dagger_{s,0}\,c^\dagger_{f,0}\,c_{d,-2}\,c_{f,2} + \frac{2 \sqrt{6}}{5}\,c^\dagger_{s,0}\,c^\dagger_{f,1}\,c_{d,2}\,c_{f,-1} - \frac{\sqrt{2}}{5}\,c^\dagger_{s,0}\,c^\dagger_{f,1}\,c_{d,1}\,c_{f,0}
\\
&\qquad - \frac{3}{5}\,c^\dagger_{s,0}\,c^\dagger_{f,1}\,c_{d,0}\,c_{f,1} + \frac{\sqrt{15}}{5}\,c^\dagger_{s,0}\,c^\dagger_{f,1}\,c_{d,-1}\,c_{f,2} + \frac{\sqrt{10}}{5}\,c^\dagger_{s,0}\,c^\dagger_{f,1}\,c_{d,-2}\,c_{f,3}
\\
&\qquad + \frac{2 \sqrt{5}}{5}\,c^\dagger_{s,0}\,c^\dagger_{f,2}\,c_{d,2}\,c_{f,0} - \frac{\sqrt{15}}{5}\,c^\dagger_{s,0}\,c^\dagger_{f,2}\,c_{d,1}\,c_{f,1} + c^\dagger_{s,0}\,c^\dagger_{f,2}\,c_{d,-1}\,c_{f,3}
\\
&\qquad + \frac{\sqrt{10}}{5}\,c^\dagger_{s,0}\,c^\dagger_{f,3}\,c_{d,2}\,c_{f,1} - c^\dagger_{s,0}\,c^\dagger_{f,3}\,c_{d,1}\,c_{f,2} + c^\dagger_{s,0}\,c^\dagger_{f,3}\,c_{d,0}\,c_{f,3}
\\
&\qquad - c^\dagger_{d,0}\,c^\dagger_{f,-3}\,c_{s,0}\,c_{f,-3} + c^\dagger_{d,-1}\,c^\dagger_{f,-2}\,c_{s,0}\,c_{f,-3} - \frac{\sqrt{10}}{5}\,c^\dagger_{d,-2}\,c^\dagger_{f,-1}\,c_{s,0}\,c_{f,-3}
\\
&\qquad - c^\dagger_{d,1}\,c^\dagger_{f,-3}\,c_{s,0}\,c_{f,-2} + \frac{\sqrt{15}}{5}\,c^\dagger_{d,-1}\,c^\dagger_{f,-1}\,c_{s,0}\,c_{f,-2} - \frac{2 \sqrt{5}}{5}\,c^\dagger_{d,-2}\,c^\dagger_{f,0}\,c_{s,0}\,c_{f,-2}
\\
&\qquad - \frac{\sqrt{10}}{5}\,c^\dagger_{d,2}\,c^\dagger_{f,-3}\,c_{s,0}\,c_{f,-1} - \frac{\sqrt{15}}{5}\,c^\dagger_{d,1}\,c^\dagger_{f,-2}\,c_{s,0}\,c_{f,-1} + \frac{3}{5}\,c^\dagger_{d,0}\,c^\dagger_{f,-1}\,c_{s,0}\,c_{f,-1}
\\
&\qquad + \frac{\sqrt{2}}{5}\,c^\dagger_{d,-1}\,c^\dagger_{f,0}\,c_{s,0}\,c_{f,-1} - \frac{2 \sqrt{6}}{5}\,c^\dagger_{d,-2}\,c^\dagger_{f,1}\,c_{s,0}\,c_{f,-1} - \frac{2 \sqrt{5}}{5}\,c^\dagger_{d,2}\,c^\dagger_{f,-2}\,c_{s,0}\,c_{f,0}
\\
&\qquad - \frac{\sqrt{2}}{5}\,c^\dagger_{d,1}\,c^\dagger_{f,-1}\,c_{s,0}\,c_{f,0} + \frac{4}{5}\,c^\dagger_{d,0}\,c^\dagger_{f,0}\,c_{s,0}\,c_{f,0} - \frac{\sqrt{2}}{5}\,c^\dagger_{d,-1}\,c^\dagger_{f,1}\,c_{s,0}\,c_{f,0}
\\
&\qquad - \frac{2 \sqrt{5}}{5}\,c^\dagger_{d,-2}\,c^\dagger_{f,2}\,c_{s,0}\,c_{f,0} - \frac{2 \sqrt{6}}{5}\,c^\dagger_{d,2}\,c^\dagger_{f,-1}\,c_{s,0}\,c_{f,1} + \frac{\sqrt{2}}{5}\,c^\dagger_{d,1}\,c^\dagger_{f,0}\,c_{s,0}\,c_{f,1}
\\
&\qquad + \frac{3}{5}\,c^\dagger_{d,0}\,c^\dagger_{f,1}\,c_{s,0}\,c_{f,1} - \frac{\sqrt{15}}{5}\,c^\dagger_{d,-1}\,c^\dagger_{f,2}\,c_{s,0}\,c_{f,1} - \frac{\sqrt{10}}{5}\,c^\dagger_{d,-2}\,c^\dagger_{f,3}\,c_{s,0}\,c_{f,1}
\\
&\qquad - \frac{2 \sqrt{5}}{5}\,c^\dagger_{d,2}\,c^\dagger_{f,0}\,c_{s,0}\,c_{f,2} + \frac{\sqrt{15}}{5}\,c^\dagger_{d,1}\,c^\dagger_{f,1}\,c_{s,0}\,c_{f,2} - c^\dagger_{d,-1}\,c^\dagger_{f,3}\,c_{s,0}\,c_{f,2}
\\
&\qquad - \frac{\sqrt{10}}{5}\,c^\dagger_{d,2}\,c^\dagger_{f,1}\,c_{s,0}\,c_{f,3} + c^\dagger_{d,1}\,c^\dagger_{f,2}\,c_{s,0}\,c_{f,3} - c^\dagger_{d,0}\,c^\dagger_{f,3}\,c_{s,0}\,c_{f,3}\Big)
\end{align*}

\begin{align*}
{G_0^{(2)}}_{[1]} 
&=  
  \frac{1}{2} \left(\left[[c^\dagger_s \times c^\dagger_f]_{3} \times[\tilde c_d \times \tilde c_d]_{3}\right]_{0} + h.c. \right)\\
&= 
  - {\bm Q}^{sd}_2 \cdot {\bm G}_2^{df}  - {\bm T}^{sd}_2 \cdot {\bm M}_2^{df} \\
&=
\frac{\sqrt{14}}{14}\Big(c^\dagger_{s,0}\,c^\dagger_{f,-3}\,c_{d,-1}\,c_{d,-2} + c^\dagger_{s,0}\,c^\dagger_{f,-2}\,c_{d,0}\,c_{d,-2} + \frac{\sqrt{15}}{5}\,c^\dagger_{s,0}\,c^\dagger_{f,-1}\,c_{d,1}\,c_{d,-2}
\\
&\qquad + \frac{\sqrt{10}}{5}\,c^\dagger_{s,0}\,c^\dagger_{f,-1}\,c_{d,0}\,c_{d,-1} + \frac{\sqrt{5}}{5}\,c^\dagger_{s,0}\,c^\dagger_{f,0}\,c_{d,2}\,c_{d,-2} + \frac{2 \sqrt{5}}{5}\,c^\dagger_{s,0}\,c^\dagger_{f,0}\,c_{d,1}\,c_{d,-1}
\\
&\qquad + \frac{\sqrt{15}}{5}\,c^\dagger_{s,0}\,c^\dagger_{f,1}\,c_{d,2}\,c_{d,-1} + \frac{\sqrt{10}}{5}\,c^\dagger_{s,0}\,c^\dagger_{f,1}\,c_{d,1}\,c_{d,0} + c^\dagger_{s,0}\,c^\dagger_{f,2}\,c_{d,2}\,c_{d,0}
\\
&\qquad + c^\dagger_{s,0}\,c^\dagger_{f,3}\,c_{d,2}\,c_{d,1} + c^\dagger_{d,-1}\,c^\dagger_{d,-2}\,c_{s,0}\,c_{f,-3} + c^\dagger_{d,0}\,c^\dagger_{d,-2}\,c_{s,0}\,c_{f,-2}
\\
&\qquad + \frac{\sqrt{15}}{5}\,c^\dagger_{d,1}\,c^\dagger_{d,-2}\,c_{s,0}\,c_{f,-1} + \frac{\sqrt{10}}{5}\,c^\dagger_{d,0}\,c^\dagger_{d,-1}\,c_{s,0}\,c_{f,-1} + \frac{\sqrt{5}}{5}\,c^\dagger_{d,2}\,c^\dagger_{d,-2}\,c_{s,0}\,c_{f,0}
\\
&\qquad + \frac{2 \sqrt{5}}{5}\,c^\dagger_{d,1}\,c^\dagger_{d,-1}\,c_{s,0}\,c_{f,0} + \frac{\sqrt{15}}{5}\,c^\dagger_{d,2}\,c^\dagger_{d,-1}\,c_{s,0}\,c_{f,1} + \frac{\sqrt{10}}{5}\,c^\dagger_{d,1}\,c^\dagger_{d,0}\,c_{s,0}\,c_{f,1}
\\
&\qquad + c^\dagger_{d,2}\,c^\dagger_{d,0}\,c_{s,0}\,c_{f,2} + c^\dagger_{d,2}\,c^\dagger_{d,1}\,c_{s,0}\,c_{f,3}\Big)
\\
{M_0^{(2)}}_{[1]} 
&=  
  \frac{1}{2} \left(\left[[c^\dagger_s \times c^\dagger_f]_{3} \times[\tilde c_d \times \tilde c_d]_{3}\right]_{0} + h.c. \right)\\
&= 
  - {\bm T}^{sd}_2 \cdot {\bm G}_2^{df}  + {\bm Q}^{sd}_2 \cdot {\bm M}_2^{df} \\
&=
\frac{\sqrt{14} i}{14}\Big(c^\dagger_{s,0}\,c^\dagger_{f,-3}\,c_{d,-1}\,c_{d,-2} + c^\dagger_{s,0}\,c^\dagger_{f,-2}\,c_{d,0}\,c_{d,-2} + \frac{\sqrt{15}}{5}\,c^\dagger_{s,0}\,c^\dagger_{f,-1}\,c_{d,1}\,c_{d,-2}
\\
&\qquad + \frac{\sqrt{10}}{5}\,c^\dagger_{s,0}\,c^\dagger_{f,-1}\,c_{d,0}\,c_{d,-1} + \frac{\sqrt{5}}{5}\,c^\dagger_{s,0}\,c^\dagger_{f,0}\,c_{d,2}\,c_{d,-2} + \frac{2 \sqrt{5}}{5}\,c^\dagger_{s,0}\,c^\dagger_{f,0}\,c_{d,1}\,c_{d,-1}
\\
&\qquad + \frac{\sqrt{15}}{5}\,c^\dagger_{s,0}\,c^\dagger_{f,1}\,c_{d,2}\,c_{d,-1} + \frac{\sqrt{10}}{5}\,c^\dagger_{s,0}\,c^\dagger_{f,1}\,c_{d,1}\,c_{d,0} + c^\dagger_{s,0}\,c^\dagger_{f,2}\,c_{d,2}\,c_{d,0}
\\
&\qquad + c^\dagger_{s,0}\,c^\dagger_{f,3}\,c_{d,2}\,c_{d,1} - c^\dagger_{d,-1}\,c^\dagger_{d,-2}\,c_{s,0}\,c_{f,-3} - c^\dagger_{d,0}\,c^\dagger_{d,-2}\,c_{s,0}\,c_{f,-2}
\\
&\qquad - \frac{\sqrt{15}}{5}\,c^\dagger_{d,1}\,c^\dagger_{d,-2}\,c_{s,0}\,c_{f,-1} - \frac{\sqrt{10}}{5}\,c^\dagger_{d,0}\,c^\dagger_{d,-1}\,c_{s,0}\,c_{f,-1} - \frac{\sqrt{5}}{5}\,c^\dagger_{d,2}\,c^\dagger_{d,-2}\,c_{s,0}\,c_{f,0}
\\
&\qquad - \frac{2 \sqrt{5}}{5}\,c^\dagger_{d,1}\,c^\dagger_{d,-1}\,c_{s,0}\,c_{f,0} - \frac{\sqrt{15}}{5}\,c^\dagger_{d,2}\,c^\dagger_{d,-1}\,c_{s,0}\,c_{f,1} - \frac{\sqrt{10}}{5}\,c^\dagger_{d,1}\,c^\dagger_{d,0}\,c_{s,0}\,c_{f,1}
\\
&\qquad - c^\dagger_{d,2}\,c^\dagger_{d,0}\,c_{s,0}\,c_{f,2} - c^\dagger_{d,2}\,c^\dagger_{d,1}\,c_{s,0}\,c_{f,3}\Big)
\end{align*}

\begin{align*}
{G_0^{(2)}}_{[2]} 
&=  
  \frac{1}{2} \left(\left[[c^\dagger_s \times c^\dagger_d]_{2} \times[\tilde c_d \times \tilde c_f]_{2}\right]_{0} + h.c. \right)\\
&= 
  \frac{1}{4} \left(\left[[c^\dagger_s \times c^\dagger_d]_{2} \times[\tilde c_d \times \tilde c_f]_{2}\right]_{0}
  +
\left[[c^\dagger_s \times c^\dagger_d]_{2} \times[\tilde c_f \times \tilde c_d]_{2}\right]_{0}
  + h.c. \right)\\
&= 
  \frac{1}{2} \left( {\bm Q}^{sd}_2 \cdot {\bm G}_2^{df} - {\bm T}^{sd}_2 \cdot {\bm M}_2^{df} \right)
  +
  \frac{1}{2} {\bm T}^{sf}_3 \cdot {\bm M}_3^{d}
  \\
&=
\frac{\sqrt{14}}{28}\Big(c^\dagger_{s,0}\,c^\dagger_{d,-2}\,c_{d,1}\,c_{f,-3} - c^\dagger_{s,0}\,c^\dagger_{d,-2}\,c_{d,0}\,c_{f,-2} + \frac{\sqrt{15}}{5}\,c^\dagger_{s,0}\,c^\dagger_{d,-2}\,c_{d,-1}\,c_{f,-1}
\\
&\qquad - \frac{\sqrt{5}}{5}\,c^\dagger_{s,0}\,c^\dagger_{d,-2}\,c_{d,-2}\,c_{f,0} + c^\dagger_{s,0}\,c^\dagger_{d,-1}\,c_{d,2}\,c_{f,-3} - \frac{\sqrt{10}}{5}\,c^\dagger_{s,0}\,c^\dagger_{d,-1}\,c_{d,0}\,c_{f,-1}
\\
&\qquad + \frac{2 \sqrt{5}}{5}\,c^\dagger_{s,0}\,c^\dagger_{d,-1}\,c_{d,-1}\,c_{f,0} - \frac{\sqrt{15}}{5}\,c^\dagger_{s,0}\,c^\dagger_{d,-1}\,c_{d,-2}\,c_{f,1} + c^\dagger_{s,0}\,c^\dagger_{d,0}\,c_{d,2}\,c_{f,-2}
\\
&\qquad - \frac{\sqrt{10}}{5}\,c^\dagger_{s,0}\,c^\dagger_{d,0}\,c_{d,1}\,c_{f,-1} + \frac{\sqrt{10}}{5}\,c^\dagger_{s,0}\,c^\dagger_{d,0}\,c_{d,-1}\,c_{f,1} - c^\dagger_{s,0}\,c^\dagger_{d,0}\,c_{d,-2}\,c_{f,2}
\\
&\qquad + \frac{\sqrt{15}}{5}\,c^\dagger_{s,0}\,c^\dagger_{d,1}\,c_{d,2}\,c_{f,-1} - \frac{2 \sqrt{5}}{5}\,c^\dagger_{s,0}\,c^\dagger_{d,1}\,c_{d,1}\,c_{f,0} + \frac{\sqrt{10}}{5}\,c^\dagger_{s,0}\,c^\dagger_{d,1}\,c_{d,0}\,c_{f,1}
\\
&\qquad - c^\dagger_{s,0}\,c^\dagger_{d,1}\,c_{d,-2}\,c_{f,3} + \frac{\sqrt{5}}{5}\,c^\dagger_{s,0}\,c^\dagger_{d,2}\,c_{d,2}\,c_{f,0} - \frac{\sqrt{15}}{5}\,c^\dagger_{s,0}\,c^\dagger_{d,2}\,c_{d,1}\,c_{f,1}
\\
&\qquad + c^\dagger_{s,0}\,c^\dagger_{d,2}\,c_{d,0}\,c_{f,2} - c^\dagger_{s,0}\,c^\dagger_{d,2}\,c_{d,-1}\,c_{f,3} + c^\dagger_{d,1}\,c^\dagger_{f,-3}\,c_{s,0}\,c_{d,-2}
\\
&\qquad - c^\dagger_{d,0}\,c^\dagger_{f,-2}\,c_{s,0}\,c_{d,-2} + \frac{\sqrt{15}}{5}\,c^\dagger_{d,-1}\,c^\dagger_{f,-1}\,c_{s,0}\,c_{d,-2} - \frac{\sqrt{5}}{5}\,c^\dagger_{d,-2}\,c^\dagger_{f,0}\,c_{s,0}\,c_{d,-2}
\\
&\qquad + c^\dagger_{d,2}\,c^\dagger_{f,-3}\,c_{s,0}\,c_{d,-1} - \frac{\sqrt{10}}{5}\,c^\dagger_{d,0}\,c^\dagger_{f,-1}\,c_{s,0}\,c_{d,-1} + \frac{2 \sqrt{5}}{5}\,c^\dagger_{d,-1}\,c^\dagger_{f,0}\,c_{s,0}\,c_{d,-1}
\\
&\qquad - \frac{\sqrt{15}}{5}\,c^\dagger_{d,-2}\,c^\dagger_{f,1}\,c_{s,0}\,c_{d,-1} + c^\dagger_{d,2}\,c^\dagger_{f,-2}\,c_{s,0}\,c_{d,0} - \frac{\sqrt{10}}{5}\,c^\dagger_{d,1}\,c^\dagger_{f,-1}\,c_{s,0}\,c_{d,0}
\\
&\qquad + \frac{\sqrt{10}}{5}\,c^\dagger_{d,-1}\,c^\dagger_{f,1}\,c_{s,0}\,c_{d,0} - c^\dagger_{d,-2}\,c^\dagger_{f,2}\,c_{s,0}\,c_{d,0} + \frac{\sqrt{15}}{5}\,c^\dagger_{d,2}\,c^\dagger_{f,-1}\,c_{s,0}\,c_{d,1}
\\
&\qquad - \frac{2 \sqrt{5}}{5}\,c^\dagger_{d,1}\,c^\dagger_{f,0}\,c_{s,0}\,c_{d,1} + \frac{\sqrt{10}}{5}\,c^\dagger_{d,0}\,c^\dagger_{f,1}\,c_{s,0}\,c_{d,1} - c^\dagger_{d,-2}\,c^\dagger_{f,3}\,c_{s,0}\,c_{d,1}
\\
&\qquad + \frac{\sqrt{5}}{5}\,c^\dagger_{d,2}\,c^\dagger_{f,0}\,c_{s,0}\,c_{d,2} - \frac{\sqrt{15}}{5}\,c^\dagger_{d,1}\,c^\dagger_{f,1}\,c_{s,0}\,c_{d,2} + c^\dagger_{d,0}\,c^\dagger_{f,2}\,c_{s,0}\,c_{d,2}
\\
&\qquad - c^\dagger_{d,-1}\,c^\dagger_{f,3}\,c_{s,0}\,c_{d,2}\Big)
\\
{M_0^{(2)}}_{[2]} 
&=  
  \frac{i}{2} \left(\left[[c^\dagger_s \times c^\dagger_d]_{2} \times[\tilde c_d \times \tilde c_f]_{2}\right]_{0} - h.c. \right)\\
&= 
  \frac{i}{4} \left(\left[[c^\dagger_s \times c^\dagger_d]_{2} \times[\tilde c_d \times \tilde c_f]_{2}\right]_{0}
  +
\left[[c^\dagger_s \times c^\dagger_d]_{2} \times[\tilde c_f \times \tilde c_d]_{2}\right]_{0}
  - h.c. \right)\\
&= 
  \frac{1}{2} \left( {\bm T}^{sd}_2 \cdot {\bm G}_2^{df} + {\bm Q}^{sd}_2 \cdot {\bm M}_2^{df} \right)
  +
  \frac{1}{2} {\bm Q}^{sf}_3 \cdot {\bm M}_3^{d}
  \\
&=
\frac{\sqrt{14} i}{28}\Big(c^\dagger_{s,0}\,c^\dagger_{d,-2}\,c_{d,1}\,c_{f,-3} - c^\dagger_{s,0}\,c^\dagger_{d,-2}\,c_{d,0}\,c_{f,-2} + \frac{\sqrt{15}}{5}\,c^\dagger_{s,0}\,c^\dagger_{d,-2}\,c_{d,-1}\,c_{f,-1}
\\
&\qquad - \frac{\sqrt{5}}{5}\,c^\dagger_{s,0}\,c^\dagger_{d,-2}\,c_{d,-2}\,c_{f,0} + c^\dagger_{s,0}\,c^\dagger_{d,-1}\,c_{d,2}\,c_{f,-3} - \frac{\sqrt{10}}{5}\,c^\dagger_{s,0}\,c^\dagger_{d,-1}\,c_{d,0}\,c_{f,-1}
\\
&\qquad + \frac{2 \sqrt{5}}{5}\,c^\dagger_{s,0}\,c^\dagger_{d,-1}\,c_{d,-1}\,c_{f,0} - \frac{\sqrt{15}}{5}\,c^\dagger_{s,0}\,c^\dagger_{d,-1}\,c_{d,-2}\,c_{f,1} + c^\dagger_{s,0}\,c^\dagger_{d,0}\,c_{d,2}\,c_{f,-2}
\\
&\qquad - \frac{\sqrt{10}}{5}\,c^\dagger_{s,0}\,c^\dagger_{d,0}\,c_{d,1}\,c_{f,-1} + \frac{\sqrt{10}}{5}\,c^\dagger_{s,0}\,c^\dagger_{d,0}\,c_{d,-1}\,c_{f,1} - c^\dagger_{s,0}\,c^\dagger_{d,0}\,c_{d,-2}\,c_{f,2}
\\
&\qquad + \frac{\sqrt{15}}{5}\,c^\dagger_{s,0}\,c^\dagger_{d,1}\,c_{d,2}\,c_{f,-1} - \frac{2 \sqrt{5}}{5}\,c^\dagger_{s,0}\,c^\dagger_{d,1}\,c_{d,1}\,c_{f,0} + \frac{\sqrt{10}}{5}\,c^\dagger_{s,0}\,c^\dagger_{d,1}\,c_{d,0}\,c_{f,1}
\\
&\qquad - c^\dagger_{s,0}\,c^\dagger_{d,1}\,c_{d,-2}\,c_{f,3} + \frac{\sqrt{5}}{5}\,c^\dagger_{s,0}\,c^\dagger_{d,2}\,c_{d,2}\,c_{f,0} - \frac{\sqrt{15}}{5}\,c^\dagger_{s,0}\,c^\dagger_{d,2}\,c_{d,1}\,c_{f,1}
\\
&\qquad + c^\dagger_{s,0}\,c^\dagger_{d,2}\,c_{d,0}\,c_{f,2} - c^\dagger_{s,0}\,c^\dagger_{d,2}\,c_{d,-1}\,c_{f,3} - c^\dagger_{d,1}\,c^\dagger_{f,-3}\,c_{s,0}\,c_{d,-2}
\\
&\qquad + c^\dagger_{d,0}\,c^\dagger_{f,-2}\,c_{s,0}\,c_{d,-2} - \frac{\sqrt{15}}{5}\,c^\dagger_{d,-1}\,c^\dagger_{f,-1}\,c_{s,0}\,c_{d,-2} + \frac{\sqrt{5}}{5}\,c^\dagger_{d,-2}\,c^\dagger_{f,0}\,c_{s,0}\,c_{d,-2}
\\
&\qquad - c^\dagger_{d,2}\,c^\dagger_{f,-3}\,c_{s,0}\,c_{d,-1} + \frac{\sqrt{10}}{5}\,c^\dagger_{d,0}\,c^\dagger_{f,-1}\,c_{s,0}\,c_{d,-1} - \frac{2 \sqrt{5}}{5}\,c^\dagger_{d,-1}\,c^\dagger_{f,0}\,c_{s,0}\,c_{d,-1}
\\
&\qquad + \frac{\sqrt{15}}{5}\,c^\dagger_{d,-2}\,c^\dagger_{f,1}\,c_{s,0}\,c_{d,-1} - c^\dagger_{d,2}\,c^\dagger_{f,-2}\,c_{s,0}\,c_{d,0} + \frac{\sqrt{10}}{5}\,c^\dagger_{d,1}\,c^\dagger_{f,-1}\,c_{s,0}\,c_{d,0}
\\
&\qquad - \frac{\sqrt{10}}{5}\,c^\dagger_{d,-1}\,c^\dagger_{f,1}\,c_{s,0}\,c_{d,0} + c^\dagger_{d,-2}\,c^\dagger_{f,2}\,c_{s,0}\,c_{d,0} - \frac{\sqrt{15}}{5}\,c^\dagger_{d,2}\,c^\dagger_{f,-1}\,c_{s,0}\,c_{d,1}
\\
&\qquad + \frac{2 \sqrt{5}}{5}\,c^\dagger_{d,1}\,c^\dagger_{f,0}\,c_{s,0}\,c_{d,1} - \frac{\sqrt{10}}{5}\,c^\dagger_{d,0}\,c^\dagger_{f,1}\,c_{s,0}\,c_{d,1} + c^\dagger_{d,-2}\,c^\dagger_{f,3}\,c_{s,0}\,c_{d,1}
\\
&\qquad - \frac{\sqrt{5}}{5}\,c^\dagger_{d,2}\,c^\dagger_{f,0}\,c_{s,0}\,c_{d,2} + \frac{\sqrt{15}}{5}\,c^\dagger_{d,1}\,c^\dagger_{f,1}\,c_{s,0}\,c_{d,2} - c^\dagger_{d,0}\,c^\dagger_{f,2}\,c_{s,0}\,c_{d,2}
\\
&\qquad + c^\dagger_{d,-1}\,c^\dagger_{f,3}\,c_{s,0}\,c_{d,2}\Big)
\end{align*}

\subsection*{$pdf$ system}

\begin{align*}
{Q_0^{(2)}}_{[1]}
&=  
  \frac{1}{2} \left(
    \left[[c^\dagger_p \times c^\dagger_d]_{1} \times[\tilde c_d \times \tilde c_f]_{1}\right]_{0} 
    + h.c. 
    \right)\\
&=  
  \frac{1}{4} \left(
    \left[[c^\dagger_p \times c^\dagger_d]_{1} \times[\tilde c_d \times \tilde c_f]_{1}\right]_{0} 
    -
    \left[[c^\dagger_p \times c^\dagger_d]_{1} \times[\tilde c_f \times \tilde c_d]_{1}\right]_{0}         
  + h.c. \right)\\
&=
\frac{1}{2} \Big\{
  -\frac{3}{5}\left(  {\bm T}^{pd}_1 \cdot {\bm T}^{df}_1   
- {\bm Q}^{pd}_1 \cdot {\bm Q}^{df}_1   \right)
  -\frac{\sqrt{10}}{5}\left(  {\bm G}^{pd}_2 \cdot {\bm G}^{df}_2   
- {\bm M}^{pd}_2 \cdot {\bm M}^{df}_2   \right)
\\
  &\qquad 
  -\frac{\sqrt{6}}{5}\left(  {\bm T}^{pd}_3 \cdot {\bm T}^{df}_3   
- {\bm Q}^{pd}_3 \cdot {\bm Q}^{df}_3   \right) \Big\}
\\
&\quad 
- \frac{1}{2} \Big\{
  -\frac{\sqrt{35}}{35} {\bm Q}^{pf}_2 \cdot {\bm Q}^{d}_2 
  -\frac{\sqrt{5}}{5} {\bm M}^{pf}_3 \cdot {\bm M}^{d}_3 
  -\frac{3\sqrt{105}}{35} {\bm Q}^{pf}_4 \cdot {\bm Q}^{d}_4 
  \Big\}
\\
&=
\frac{\sqrt{105}}{70}\Big(\frac{\sqrt{6}}{6}\,c^\dagger_{p,-1}\,c^\dagger_{d,0}\,c_{d,2}\,c_{f,-3} - \frac{1}{3}\,c^\dagger_{p,-1}\,c^\dagger_{d,0}\,c_{d,1}\,c_{f,-2} + \frac{\sqrt{15}}{15}\,c^\dagger_{p,-1}\,c^\dagger_{d,0}\,c_{d,0}\,c_{f,-1}
\\
&\qquad - \frac{\sqrt{30}}{30}\,c^\dagger_{p,-1}\,c^\dagger_{d,0}\,c_{d,-1}\,c_{f,0} + \frac{\sqrt{10}}{30}\,c^\dagger_{p,-1}\,c^\dagger_{d,0}\,c_{d,-2}\,c_{f,1} - \frac{\sqrt{2}}{2}\,c^\dagger_{p,0}\,c^\dagger_{d,-1}\,c_{d,2}\,c_{f,-3}
\\
&\qquad + \frac{\sqrt{3}}{3}\,c^\dagger_{p,0}\,c^\dagger_{d,-1}\,c_{d,1}\,c_{f,-2} - \frac{\sqrt{5}}{5}\,c^\dagger_{p,0}\,c^\dagger_{d,-1}\,c_{d,0}\,c_{f,-1} + \frac{\sqrt{10}}{10}\,c^\dagger_{p,0}\,c^\dagger_{d,-1}\,c_{d,-1}\,c_{f,0}
\\
&\qquad - \frac{\sqrt{30}}{30}\,c^\dagger_{p,0}\,c^\dagger_{d,-1}\,c_{d,-2}\,c_{f,1} + c^\dagger_{p,1}\,c^\dagger_{d,-2}\,c_{d,2}\,c_{f,-3} - \frac{\sqrt{6}}{3}\,c^\dagger_{p,1}\,c^\dagger_{d,-2}\,c_{d,1}\,c_{f,-2}
\\
&\qquad + \frac{\sqrt{10}}{5}\,c^\dagger_{p,1}\,c^\dagger_{d,-2}\,c_{d,0}\,c_{f,-1} - \frac{\sqrt{5}}{5}\,c^\dagger_{p,1}\,c^\dagger_{d,-2}\,c_{d,-1}\,c_{f,0} + \frac{\sqrt{15}}{15}\,c^\dagger_{p,1}\,c^\dagger_{d,-2}\,c_{d,-2}\,c_{f,1}
\\
&\qquad + \frac{\sqrt{6}}{6}\,c^\dagger_{p,-1}\,c^\dagger_{d,1}\,c_{d,2}\,c_{f,-2} - \frac{2 \sqrt{15}}{15}\,c^\dagger_{p,-1}\,c^\dagger_{d,1}\,c_{d,1}\,c_{f,-1} + \frac{\sqrt{30}}{10}\,c^\dagger_{p,-1}\,c^\dagger_{d,1}\,c_{d,0}\,c_{f,0}
\\
&\qquad - \frac{2 \sqrt{15}}{15}\,c^\dagger_{p,-1}\,c^\dagger_{d,1}\,c_{d,-1}\,c_{f,1} + \frac{\sqrt{6}}{6}\,c^\dagger_{p,-1}\,c^\dagger_{d,1}\,c_{d,-2}\,c_{f,2} - \frac{\sqrt{2}}{3}\,c^\dagger_{p,0}\,c^\dagger_{d,0}\,c_{d,2}\,c_{f,-2}
\\
&\qquad + \frac{4 \sqrt{5}}{15}\,c^\dagger_{p,0}\,c^\dagger_{d,0}\,c_{d,1}\,c_{f,-1} - \frac{\sqrt{10}}{5}\,c^\dagger_{p,0}\,c^\dagger_{d,0}\,c_{d,0}\,c_{f,0} + \frac{4 \sqrt{5}}{15}\,c^\dagger_{p,0}\,c^\dagger_{d,0}\,c_{d,-1}\,c_{f,1}
\\
&\qquad - \frac{\sqrt{2}}{3}\,c^\dagger_{p,0}\,c^\dagger_{d,0}\,c_{d,-2}\,c_{f,2} + \frac{\sqrt{6}}{6}\,c^\dagger_{p,1}\,c^\dagger_{d,-1}\,c_{d,2}\,c_{f,-2} - \frac{2 \sqrt{15}}{15}\,c^\dagger_{p,1}\,c^\dagger_{d,-1}\,c_{d,1}\,c_{f,-1}
\\
&\qquad + \frac{\sqrt{30}}{10}\,c^\dagger_{p,1}\,c^\dagger_{d,-1}\,c_{d,0}\,c_{f,0} - \frac{2 \sqrt{15}}{15}\,c^\dagger_{p,1}\,c^\dagger_{d,-1}\,c_{d,-1}\,c_{f,1} + \frac{\sqrt{6}}{6}\,c^\dagger_{p,1}\,c^\dagger_{d,-1}\,c_{d,-2}\,c_{f,2}
\\
&\qquad + \frac{\sqrt{15}}{15}\,c^\dagger_{p,-1}\,c^\dagger_{d,2}\,c_{d,2}\,c_{f,-1} - \frac{\sqrt{5}}{5}\,c^\dagger_{p,-1}\,c^\dagger_{d,2}\,c_{d,1}\,c_{f,0} + \frac{\sqrt{10}}{5}\,c^\dagger_{p,-1}\,c^\dagger_{d,2}\,c_{d,0}\,c_{f,1}
\\
&\qquad - \frac{\sqrt{6}}{3}\,c^\dagger_{p,-1}\,c^\dagger_{d,2}\,c_{d,-1}\,c_{f,2} + c^\dagger_{p,-1}\,c^\dagger_{d,2}\,c_{d,-2}\,c_{f,3} - \frac{\sqrt{30}}{30}\,c^\dagger_{p,0}\,c^\dagger_{d,1}\,c_{d,2}\,c_{f,-1}
\\
&\qquad + \frac{\sqrt{10}}{10}\,c^\dagger_{p,0}\,c^\dagger_{d,1}\,c_{d,1}\,c_{f,0} - \frac{\sqrt{5}}{5}\,c^\dagger_{p,0}\,c^\dagger_{d,1}\,c_{d,0}\,c_{f,1} + \frac{\sqrt{3}}{3}\,c^\dagger_{p,0}\,c^\dagger_{d,1}\,c_{d,-1}\,c_{f,2}
\\
&\qquad - \frac{\sqrt{2}}{2}\,c^\dagger_{p,0}\,c^\dagger_{d,1}\,c_{d,-2}\,c_{f,3} + \frac{\sqrt{10}}{30}\,c^\dagger_{p,1}\,c^\dagger_{d,0}\,c_{d,2}\,c_{f,-1} - \frac{\sqrt{30}}{30}\,c^\dagger_{p,1}\,c^\dagger_{d,0}\,c_{d,1}\,c_{f,0}
\\
&\qquad + \frac{\sqrt{15}}{15}\,c^\dagger_{p,1}\,c^\dagger_{d,0}\,c_{d,0}\,c_{f,1} - \frac{1}{3}\,c^\dagger_{p,1}\,c^\dagger_{d,0}\,c_{d,-1}\,c_{f,2} + \frac{\sqrt{6}}{6}\,c^\dagger_{p,1}\,c^\dagger_{d,0}\,c_{d,-2}\,c_{f,3}
\\
&\qquad + \frac{\sqrt{6}}{6}\,c^\dagger_{d,2}\,c^\dagger_{f,-3}\,c_{p,-1}\,c_{d,0} - \frac{1}{3}\,c^\dagger_{d,1}\,c^\dagger_{f,-2}\,c_{p,-1}\,c_{d,0} + \frac{\sqrt{15}}{15}\,c^\dagger_{d,0}\,c^\dagger_{f,-1}\,c_{p,-1}\,c_{d,0}
\\
&\qquad - \frac{\sqrt{30}}{30}\,c^\dagger_{d,-1}\,c^\dagger_{f,0}\,c_{p,-1}\,c_{d,0} + \frac{\sqrt{10}}{30}\,c^\dagger_{d,-2}\,c^\dagger_{f,1}\,c_{p,-1}\,c_{d,0} - \frac{\sqrt{2}}{2}\,c^\dagger_{d,2}\,c^\dagger_{f,-3}\,c_{p,0}\,c_{d,-1}
\\
&\qquad + \frac{\sqrt{3}}{3}\,c^\dagger_{d,1}\,c^\dagger_{f,-2}\,c_{p,0}\,c_{d,-1} - \frac{\sqrt{5}}{5}\,c^\dagger_{d,0}\,c^\dagger_{f,-1}\,c_{p,0}\,c_{d,-1} + \frac{\sqrt{10}}{10}\,c^\dagger_{d,-1}\,c^\dagger_{f,0}\,c_{p,0}\,c_{d,-1}
\\
&\qquad - \frac{\sqrt{30}}{30}\,c^\dagger_{d,-2}\,c^\dagger_{f,1}\,c_{p,0}\,c_{d,-1} + c^\dagger_{d,2}\,c^\dagger_{f,-3}\,c_{p,1}\,c_{d,-2} - \frac{\sqrt{6}}{3}\,c^\dagger_{d,1}\,c^\dagger_{f,-2}\,c_{p,1}\,c_{d,-2}
\\
&\qquad + \frac{\sqrt{10}}{5}\,c^\dagger_{d,0}\,c^\dagger_{f,-1}\,c_{p,1}\,c_{d,-2} - \frac{\sqrt{5}}{5}\,c^\dagger_{d,-1}\,c^\dagger_{f,0}\,c_{p,1}\,c_{d,-2} + \frac{\sqrt{15}}{15}\,c^\dagger_{d,-2}\,c^\dagger_{f,1}\,c_{p,1}\,c_{d,-2}
\\
&\qquad + \frac{\sqrt{6}}{6}\,c^\dagger_{d,2}\,c^\dagger_{f,-2}\,c_{p,-1}\,c_{d,1} - \frac{2 \sqrt{15}}{15}\,c^\dagger_{d,1}\,c^\dagger_{f,-1}\,c_{p,-1}\,c_{d,1} + \frac{\sqrt{30}}{10}\,c^\dagger_{d,0}\,c^\dagger_{f,0}\,c_{p,-1}\,c_{d,1}
\\
&\qquad - \frac{2 \sqrt{15}}{15}\,c^\dagger_{d,-1}\,c^\dagger_{f,1}\,c_{p,-1}\,c_{d,1} + \frac{\sqrt{6}}{6}\,c^\dagger_{d,-2}\,c^\dagger_{f,2}\,c_{p,-1}\,c_{d,1} - \frac{\sqrt{2}}{3}\,c^\dagger_{d,2}\,c^\dagger_{f,-2}\,c_{p,0}\,c_{d,0}
\\
&\qquad + \frac{4 \sqrt{5}}{15}\,c^\dagger_{d,1}\,c^\dagger_{f,-1}\,c_{p,0}\,c_{d,0} - \frac{\sqrt{10}}{5}\,c^\dagger_{d,0}\,c^\dagger_{f,0}\,c_{p,0}\,c_{d,0} + \frac{4 \sqrt{5}}{15}\,c^\dagger_{d,-1}\,c^\dagger_{f,1}\,c_{p,0}\,c_{d,0}
\\
&\qquad - \frac{\sqrt{2}}{3}\,c^\dagger_{d,-2}\,c^\dagger_{f,2}\,c_{p,0}\,c_{d,0} + \frac{\sqrt{6}}{6}\,c^\dagger_{d,2}\,c^\dagger_{f,-2}\,c_{p,1}\,c_{d,-1} - \frac{2 \sqrt{15}}{15}\,c^\dagger_{d,1}\,c^\dagger_{f,-1}\,c_{p,1}\,c_{d,-1}
\\
&\qquad + \frac{\sqrt{30}}{10}\,c^\dagger_{d,0}\,c^\dagger_{f,0}\,c_{p,1}\,c_{d,-1} - \frac{2 \sqrt{15}}{15}\,c^\dagger_{d,-1}\,c^\dagger_{f,1}\,c_{p,1}\,c_{d,-1} + \frac{\sqrt{6}}{6}\,c^\dagger_{d,-2}\,c^\dagger_{f,2}\,c_{p,1}\,c_{d,-1}
\\
&\qquad + \frac{\sqrt{15}}{15}\,c^\dagger_{d,2}\,c^\dagger_{f,-1}\,c_{p,-1}\,c_{d,2} - \frac{\sqrt{5}}{5}\,c^\dagger_{d,1}\,c^\dagger_{f,0}\,c_{p,-1}\,c_{d,2} + \frac{\sqrt{10}}{5}\,c^\dagger_{d,0}\,c^\dagger_{f,1}\,c_{p,-1}\,c_{d,2}
\\
&\qquad - \frac{\sqrt{6}}{3}\,c^\dagger_{d,-1}\,c^\dagger_{f,2}\,c_{p,-1}\,c_{d,2} + c^\dagger_{d,-2}\,c^\dagger_{f,3}\,c_{p,-1}\,c_{d,2} - \frac{\sqrt{30}}{30}\,c^\dagger_{d,2}\,c^\dagger_{f,-1}\,c_{p,0}\,c_{d,1}
\\
&\qquad + \frac{\sqrt{10}}{10}\,c^\dagger_{d,1}\,c^\dagger_{f,0}\,c_{p,0}\,c_{d,1} - \frac{\sqrt{5}}{5}\,c^\dagger_{d,0}\,c^\dagger_{f,1}\,c_{p,0}\,c_{d,1} + \frac{\sqrt{3}}{3}\,c^\dagger_{d,-1}\,c^\dagger_{f,2}\,c_{p,0}\,c_{d,1}
\\
&\qquad - \frac{\sqrt{2}}{2}\,c^\dagger_{d,-2}\,c^\dagger_{f,3}\,c_{p,0}\,c_{d,1} + \frac{\sqrt{10}}{30}\,c^\dagger_{d,2}\,c^\dagger_{f,-1}\,c_{p,1}\,c_{d,0} - \frac{\sqrt{30}}{30}\,c^\dagger_{d,1}\,c^\dagger_{f,0}\,c_{p,1}\,c_{d,0}
\\
&\qquad + \frac{\sqrt{15}}{15}\,c^\dagger_{d,0}\,c^\dagger_{f,1}\,c_{p,1}\,c_{d,0} - \frac{1}{3}\,c^\dagger_{d,-1}\,c^\dagger_{f,2}\,c_{p,1}\,c_{d,0} + \frac{\sqrt{6}}{6}\,c^\dagger_{d,-2}\,c^\dagger_{f,3}\,c_{p,1}\,c_{d,0}\Big)
\\
{T_0^{(2)}}_{[1]}
&=  
  \frac{i}{2} \left(
    \left[[c^\dagger_p \times c^\dagger_d]_{1} \times[\tilde c_d \times \tilde c_f]_{1}\right]_{0} 
    - h.c. 
    \right)\\
&=  
  \frac{i}{4} \left(
    \left[[c^\dagger_p \times c^\dagger_d]_{1} \times[\tilde c_d \times \tilde c_f]_{1}\right]_{0} 
    -
    \left[[c^\dagger_p \times c^\dagger_d]_{1} \times[\tilde c_f \times \tilde c_d]_{1}\right]_{0}         
  - h.c. \right)\\
&=
\frac{1}{2} \Big\{
  -\frac{3}{5}\left(  {\bm Q}^{pd}_1 \cdot {\bm T}^{df}_1   
+ {\bm T}^{pd}_1 \cdot {\bm Q}^{df}_1   \right)
  -\frac{\sqrt{10}}{5}\left(  {\bm M}^{pd}_2 \cdot {\bm G}^{df}_2   
+ {\bm G}^{pd}_2 \cdot {\bm M}^{df}_2   \right)
\\
  &\qquad 
  -\frac{\sqrt{6}}{5}\left(  {\bm Q}^{pd}_3 \cdot {\bm T}^{df}_3   
+ {\bm T}^{pd}_3 \cdot {\bm Q}^{df}_3   \right) \Big\}
\\
&\quad 
- \frac{1}{2} \Big\{
  -\frac{\sqrt{35}}{35} {\bm T}^{pf}_2 \cdot {\bm Q}^{d}_2\  
  -\frac{\sqrt{5}}{5} {\bm G}^{pf}_3 \cdot {\bm M}^{d}_3\  
  -\frac{3\sqrt{105}}{35} {\bm T}^{pf}_4 \cdot {\bm Q}^{d}_4\  
  \Big\}
\\
&=
\frac{\sqrt{105} i}{70}\Big(\frac{\sqrt{6}}{6}\,c^\dagger_{p,-1}\,c^\dagger_{d,0}\,c_{d,2}\,c_{f,-3} - \frac{1}{3}\,c^\dagger_{p,-1}\,c^\dagger_{d,0}\,c_{d,1}\,c_{f,-2} + \frac{\sqrt{15}}{15}\,c^\dagger_{p,-1}\,c^\dagger_{d,0}\,c_{d,0}\,c_{f,-1}
\\
&\qquad - \frac{\sqrt{30}}{30}\,c^\dagger_{p,-1}\,c^\dagger_{d,0}\,c_{d,-1}\,c_{f,0} + \frac{\sqrt{10}}{30}\,c^\dagger_{p,-1}\,c^\dagger_{d,0}\,c_{d,-2}\,c_{f,1} - \frac{\sqrt{2}}{2}\,c^\dagger_{p,0}\,c^\dagger_{d,-1}\,c_{d,2}\,c_{f,-3}
\\
&\qquad + \frac{\sqrt{3}}{3}\,c^\dagger_{p,0}\,c^\dagger_{d,-1}\,c_{d,1}\,c_{f,-2} - \frac{\sqrt{5}}{5}\,c^\dagger_{p,0}\,c^\dagger_{d,-1}\,c_{d,0}\,c_{f,-1} + \frac{\sqrt{10}}{10}\,c^\dagger_{p,0}\,c^\dagger_{d,-1}\,c_{d,-1}\,c_{f,0}
\\
&\qquad - \frac{\sqrt{30}}{30}\,c^\dagger_{p,0}\,c^\dagger_{d,-1}\,c_{d,-2}\,c_{f,1} + c^\dagger_{p,1}\,c^\dagger_{d,-2}\,c_{d,2}\,c_{f,-3} - \frac{\sqrt{6}}{3}\,c^\dagger_{p,1}\,c^\dagger_{d,-2}\,c_{d,1}\,c_{f,-2}
\\
&\qquad + \frac{\sqrt{10}}{5}\,c^\dagger_{p,1}\,c^\dagger_{d,-2}\,c_{d,0}\,c_{f,-1} - \frac{\sqrt{5}}{5}\,c^\dagger_{p,1}\,c^\dagger_{d,-2}\,c_{d,-1}\,c_{f,0} + \frac{\sqrt{15}}{15}\,c^\dagger_{p,1}\,c^\dagger_{d,-2}\,c_{d,-2}\,c_{f,1}
\\
&\qquad + \frac{\sqrt{6}}{6}\,c^\dagger_{p,-1}\,c^\dagger_{d,1}\,c_{d,2}\,c_{f,-2} - \frac{2 \sqrt{15}}{15}\,c^\dagger_{p,-1}\,c^\dagger_{d,1}\,c_{d,1}\,c_{f,-1} + \frac{\sqrt{30}}{10}\,c^\dagger_{p,-1}\,c^\dagger_{d,1}\,c_{d,0}\,c_{f,0}
\\
&\qquad - \frac{2 \sqrt{15}}{15}\,c^\dagger_{p,-1}\,c^\dagger_{d,1}\,c_{d,-1}\,c_{f,1} + \frac{\sqrt{6}}{6}\,c^\dagger_{p,-1}\,c^\dagger_{d,1}\,c_{d,-2}\,c_{f,2} - \frac{\sqrt{2}}{3}\,c^\dagger_{p,0}\,c^\dagger_{d,0}\,c_{d,2}\,c_{f,-2}
\\
&\qquad + \frac{4 \sqrt{5}}{15}\,c^\dagger_{p,0}\,c^\dagger_{d,0}\,c_{d,1}\,c_{f,-1} - \frac{\sqrt{10}}{5}\,c^\dagger_{p,0}\,c^\dagger_{d,0}\,c_{d,0}\,c_{f,0} + \frac{4 \sqrt{5}}{15}\,c^\dagger_{p,0}\,c^\dagger_{d,0}\,c_{d,-1}\,c_{f,1}
\\
&\qquad - \frac{\sqrt{2}}{3}\,c^\dagger_{p,0}\,c^\dagger_{d,0}\,c_{d,-2}\,c_{f,2} + \frac{\sqrt{6}}{6}\,c^\dagger_{p,1}\,c^\dagger_{d,-1}\,c_{d,2}\,c_{f,-2} - \frac{2 \sqrt{15}}{15}\,c^\dagger_{p,1}\,c^\dagger_{d,-1}\,c_{d,1}\,c_{f,-1}
\\
&\qquad + \frac{\sqrt{30}}{10}\,c^\dagger_{p,1}\,c^\dagger_{d,-1}\,c_{d,0}\,c_{f,0} - \frac{2 \sqrt{15}}{15}\,c^\dagger_{p,1}\,c^\dagger_{d,-1}\,c_{d,-1}\,c_{f,1} + \frac{\sqrt{6}}{6}\,c^\dagger_{p,1}\,c^\dagger_{d,-1}\,c_{d,-2}\,c_{f,2}
\\
&\qquad + \frac{\sqrt{15}}{15}\,c^\dagger_{p,-1}\,c^\dagger_{d,2}\,c_{d,2}\,c_{f,-1} - \frac{\sqrt{5}}{5}\,c^\dagger_{p,-1}\,c^\dagger_{d,2}\,c_{d,1}\,c_{f,0} + \frac{\sqrt{10}}{5}\,c^\dagger_{p,-1}\,c^\dagger_{d,2}\,c_{d,0}\,c_{f,1}
\\
&\qquad - \frac{\sqrt{6}}{3}\,c^\dagger_{p,-1}\,c^\dagger_{d,2}\,c_{d,-1}\,c_{f,2} + c^\dagger_{p,-1}\,c^\dagger_{d,2}\,c_{d,-2}\,c_{f,3} - \frac{\sqrt{30}}{30}\,c^\dagger_{p,0}\,c^\dagger_{d,1}\,c_{d,2}\,c_{f,-1}
\\
&\qquad + \frac{\sqrt{10}}{10}\,c^\dagger_{p,0}\,c^\dagger_{d,1}\,c_{d,1}\,c_{f,0} - \frac{\sqrt{5}}{5}\,c^\dagger_{p,0}\,c^\dagger_{d,1}\,c_{d,0}\,c_{f,1} + \frac{\sqrt{3}}{3}\,c^\dagger_{p,0}\,c^\dagger_{d,1}\,c_{d,-1}\,c_{f,2}
\\
&\qquad - \frac{\sqrt{2}}{2}\,c^\dagger_{p,0}\,c^\dagger_{d,1}\,c_{d,-2}\,c_{f,3} + \frac{\sqrt{10}}{30}\,c^\dagger_{p,1}\,c^\dagger_{d,0}\,c_{d,2}\,c_{f,-1} - \frac{\sqrt{30}}{30}\,c^\dagger_{p,1}\,c^\dagger_{d,0}\,c_{d,1}\,c_{f,0}
\\
&\qquad + \frac{\sqrt{15}}{15}\,c^\dagger_{p,1}\,c^\dagger_{d,0}\,c_{d,0}\,c_{f,1} - \frac{1}{3}\,c^\dagger_{p,1}\,c^\dagger_{d,0}\,c_{d,-1}\,c_{f,2} + \frac{\sqrt{6}}{6}\,c^\dagger_{p,1}\,c^\dagger_{d,0}\,c_{d,-2}\,c_{f,3}
\\
&\qquad - \frac{\sqrt{6}}{6}\,c^\dagger_{d,2}\,c^\dagger_{f,-3}\,c_{p,-1}\,c_{d,0} + \frac{1}{3}\,c^\dagger_{d,1}\,c^\dagger_{f,-2}\,c_{p,-1}\,c_{d,0} - \frac{\sqrt{15}}{15}\,c^\dagger_{d,0}\,c^\dagger_{f,-1}\,c_{p,-1}\,c_{d,0}
\\
&\qquad + \frac{\sqrt{30}}{30}\,c^\dagger_{d,-1}\,c^\dagger_{f,0}\,c_{p,-1}\,c_{d,0} - \frac{\sqrt{10}}{30}\,c^\dagger_{d,-2}\,c^\dagger_{f,1}\,c_{p,-1}\,c_{d,0} + \frac{\sqrt{2}}{2}\,c^\dagger_{d,2}\,c^\dagger_{f,-3}\,c_{p,0}\,c_{d,-1}
\\
&\qquad - \frac{\sqrt{3}}{3}\,c^\dagger_{d,1}\,c^\dagger_{f,-2}\,c_{p,0}\,c_{d,-1} + \frac{\sqrt{5}}{5}\,c^\dagger_{d,0}\,c^\dagger_{f,-1}\,c_{p,0}\,c_{d,-1} - \frac{\sqrt{10}}{10}\,c^\dagger_{d,-1}\,c^\dagger_{f,0}\,c_{p,0}\,c_{d,-1}
\\
&\qquad + \frac{\sqrt{30}}{30}\,c^\dagger_{d,-2}\,c^\dagger_{f,1}\,c_{p,0}\,c_{d,-1} - c^\dagger_{d,2}\,c^\dagger_{f,-3}\,c_{p,1}\,c_{d,-2} + \frac{\sqrt{6}}{3}\,c^\dagger_{d,1}\,c^\dagger_{f,-2}\,c_{p,1}\,c_{d,-2}
\\
&\qquad - \frac{\sqrt{10}}{5}\,c^\dagger_{d,0}\,c^\dagger_{f,-1}\,c_{p,1}\,c_{d,-2} + \frac{\sqrt{5}}{5}\,c^\dagger_{d,-1}\,c^\dagger_{f,0}\,c_{p,1}\,c_{d,-2} - \frac{\sqrt{15}}{15}\,c^\dagger_{d,-2}\,c^\dagger_{f,1}\,c_{p,1}\,c_{d,-2}
\\
&\qquad - \frac{\sqrt{6}}{6}\,c^\dagger_{d,2}\,c^\dagger_{f,-2}\,c_{p,-1}\,c_{d,1} + \frac{2 \sqrt{15}}{15}\,c^\dagger_{d,1}\,c^\dagger_{f,-1}\,c_{p,-1}\,c_{d,1} - \frac{\sqrt{30}}{10}\,c^\dagger_{d,0}\,c^\dagger_{f,0}\,c_{p,-1}\,c_{d,1}
\\
&\qquad + \frac{2 \sqrt{15}}{15}\,c^\dagger_{d,-1}\,c^\dagger_{f,1}\,c_{p,-1}\,c_{d,1} - \frac{\sqrt{6}}{6}\,c^\dagger_{d,-2}\,c^\dagger_{f,2}\,c_{p,-1}\,c_{d,1} + \frac{\sqrt{2}}{3}\,c^\dagger_{d,2}\,c^\dagger_{f,-2}\,c_{p,0}\,c_{d,0}
\\
&\qquad - \frac{4 \sqrt{5}}{15}\,c^\dagger_{d,1}\,c^\dagger_{f,-1}\,c_{p,0}\,c_{d,0} + \frac{\sqrt{10}}{5}\,c^\dagger_{d,0}\,c^\dagger_{f,0}\,c_{p,0}\,c_{d,0} - \frac{4 \sqrt{5}}{15}\,c^\dagger_{d,-1}\,c^\dagger_{f,1}\,c_{p,0}\,c_{d,0}
\\
&\qquad + \frac{\sqrt{2}}{3}\,c^\dagger_{d,-2}\,c^\dagger_{f,2}\,c_{p,0}\,c_{d,0} - \frac{\sqrt{6}}{6}\,c^\dagger_{d,2}\,c^\dagger_{f,-2}\,c_{p,1}\,c_{d,-1} + \frac{2 \sqrt{15}}{15}\,c^\dagger_{d,1}\,c^\dagger_{f,-1}\,c_{p,1}\,c_{d,-1}
\\
&\qquad - \frac{\sqrt{30}}{10}\,c^\dagger_{d,0}\,c^\dagger_{f,0}\,c_{p,1}\,c_{d,-1} + \frac{2 \sqrt{15}}{15}\,c^\dagger_{d,-1}\,c^\dagger_{f,1}\,c_{p,1}\,c_{d,-1} - \frac{\sqrt{6}}{6}\,c^\dagger_{d,-2}\,c^\dagger_{f,2}\,c_{p,1}\,c_{d,-1}
\\
&\qquad - \frac{\sqrt{15}}{15}\,c^\dagger_{d,2}\,c^\dagger_{f,-1}\,c_{p,-1}\,c_{d,2} + \frac{\sqrt{5}}{5}\,c^\dagger_{d,1}\,c^\dagger_{f,0}\,c_{p,-1}\,c_{d,2} - \frac{\sqrt{10}}{5}\,c^\dagger_{d,0}\,c^\dagger_{f,1}\,c_{p,-1}\,c_{d,2}
\\
&\qquad + \frac{\sqrt{6}}{3}\,c^\dagger_{d,-1}\,c^\dagger_{f,2}\,c_{p,-1}\,c_{d,2} - c^\dagger_{d,-2}\,c^\dagger_{f,3}\,c_{p,-1}\,c_{d,2} + \frac{\sqrt{30}}{30}\,c^\dagger_{d,2}\,c^\dagger_{f,-1}\,c_{p,0}\,c_{d,1}
\\
&\qquad - \frac{\sqrt{10}}{10}\,c^\dagger_{d,1}\,c^\dagger_{f,0}\,c_{p,0}\,c_{d,1} + \frac{\sqrt{5}}{5}\,c^\dagger_{d,0}\,c^\dagger_{f,1}\,c_{p,0}\,c_{d,1} - \frac{\sqrt{3}}{3}\,c^\dagger_{d,-1}\,c^\dagger_{f,2}\,c_{p,0}\,c_{d,1}
\\
&\qquad + \frac{\sqrt{2}}{2}\,c^\dagger_{d,-2}\,c^\dagger_{f,3}\,c_{p,0}\,c_{d,1} - \frac{\sqrt{10}}{30}\,c^\dagger_{d,2}\,c^\dagger_{f,-1}\,c_{p,1}\,c_{d,0} + \frac{\sqrt{30}}{30}\,c^\dagger_{d,1}\,c^\dagger_{f,0}\,c_{p,1}\,c_{d,0}
\\
&\qquad - \frac{\sqrt{15}}{15}\,c^\dagger_{d,0}\,c^\dagger_{f,1}\,c_{p,1}\,c_{d,0} + \frac{1}{3}\,c^\dagger_{d,-1}\,c^\dagger_{f,2}\,c_{p,1}\,c_{d,0} - \frac{\sqrt{6}}{6}\,c^\dagger_{d,-2}\,c^\dagger_{f,3}\,c_{p,1}\,c_{d,0}\Big)
\end{align*}

% pddf : 2×2
% j13,j24,A wigner_9j： 1 , 1 , sqrt(10)/5
% j13,j24,A wigner_9j： 3 , 3 , -sqrt(15)/5

% pdfd : 2×2
% j13,j24,A wigner_9j： 2 , 2 , sqrt(14)/7
% j13,j24,A wigner_9j： 3 , 3 , sqrt(2)/2
% j13,j24,A wigner_9j： 4 , 4 , -sqrt(42)/14

\begin{align*}
{Q_0^{(2)}}_{[2]}
&=  
  \frac{1}{2} \left(
    \left[[c^\dagger_p \times c^\dagger_d]_{2} \times[\tilde c_d \times \tilde c_f]_{2}\right]_{0} 
    + h.c. 
    \right)\\
&=  
  \frac{1}{4} \left(
    \left[[c^\dagger_p \times c^\dagger_d]_{2} \times[\tilde c_d \times \tilde c_f]_{2}\right]_{0} 
    +
    \left[[c^\dagger_p \times c^\dagger_d]_{2} \times[\tilde c_f \times \tilde c_d]_{2}\right]_{0}         
  + h.c. \right)\\
&=
\frac{1}{2} \Big\{
  -\frac{\sqrt{10}}{5}\left(  {\bm T}^{pd}_1 \cdot  {\bm T}^{df}_1  
- {\bm Q}^{pd}_1 \cdot  {\bm Q}^{df}_1  \right)
  +\frac{\sqrt{15}}{5}\left(  {\bm T}^{pd}_3 \cdot  {\bm T}^{df}_3  
- {\bm Q}^{pd}_3 \cdot  {\bm Q}^{df}_3  \right) \Big\}
\\
&\quad 
+ \frac{1}{2} \Big\{
  -\frac{\sqrt{14}}{7} {\bm Q}^{pf}_2 \cdot  {\bm Q}^{d}_2
  -\frac{\sqrt{2}}{2} {\bm M}^{pf}_3 \cdot  {\bm M}^{d}_3
  +\frac{\sqrt{42}}{14} {\bm Q}^{pf}_4 \cdot  {\bm Q}^{d}_4
  \Big\}
\\
&=
\frac{\sqrt{21}}{42}\Big(- \frac{\sqrt{2}}{2}\,c^\dagger_{p,-1}\,c^\dagger_{d,-1}\,c_{d,1}\,c_{f,-3} + \frac{\sqrt{2}}{2}\,c^\dagger_{p,-1}\,c^\dagger_{d,-1}\,c_{d,0}\,c_{f,-2} - \frac{\sqrt{30}}{10}\,c^\dagger_{p,-1}\,c^\dagger_{d,-1}\,c_{d,-1}\,c_{f,-1}
\\
&\qquad + \frac{\sqrt{10}}{10}\,c^\dagger_{p,-1}\,c^\dagger_{d,-1}\,c_{d,-2}\,c_{f,0} + c^\dagger_{p,0}\,c^\dagger_{d,-2}\,c_{d,1}\,c_{f,-3} - c^\dagger_{p,0}\,c^\dagger_{d,-2}\,c_{d,0}\,c_{f,-2}
\\
&\qquad + \frac{\sqrt{15}}{5}\,c^\dagger_{p,0}\,c^\dagger_{d,-2}\,c_{d,-1}\,c_{f,-1} - \frac{\sqrt{5}}{5}\,c^\dagger_{p,0}\,c^\dagger_{d,-2}\,c_{d,-2}\,c_{f,0} - \frac{\sqrt{3}}{2}\,c^\dagger_{p,-1}\,c^\dagger_{d,0}\,c_{d,2}\,c_{f,-3}
\\
&\qquad + \frac{\sqrt{30}}{10}\,c^\dagger_{p,-1}\,c^\dagger_{d,0}\,c_{d,0}\,c_{f,-1} - \frac{\sqrt{15}}{5}\,c^\dagger_{p,-1}\,c^\dagger_{d,0}\,c_{d,-1}\,c_{f,0} + \frac{3 \sqrt{5}}{10}\,c^\dagger_{p,-1}\,c^\dagger_{d,0}\,c_{d,-2}\,c_{f,1}
\\
&\qquad + \frac{1}{2}\,c^\dagger_{p,0}\,c^\dagger_{d,-1}\,c_{d,2}\,c_{f,-3} - \frac{\sqrt{10}}{10}\,c^\dagger_{p,0}\,c^\dagger_{d,-1}\,c_{d,0}\,c_{f,-1} + \frac{\sqrt{5}}{5}\,c^\dagger_{p,0}\,c^\dagger_{d,-1}\,c_{d,-1}\,c_{f,0}
\\
&\qquad - \frac{\sqrt{15}}{10}\,c^\dagger_{p,0}\,c^\dagger_{d,-1}\,c_{d,-2}\,c_{f,1} + \frac{\sqrt{2}}{2}\,c^\dagger_{p,1}\,c^\dagger_{d,-2}\,c_{d,2}\,c_{f,-3} - \frac{\sqrt{5}}{5}\,c^\dagger_{p,1}\,c^\dagger_{d,-2}\,c_{d,0}\,c_{f,-1}
\\
&\qquad + \frac{\sqrt{10}}{5}\,c^\dagger_{p,1}\,c^\dagger_{d,-2}\,c_{d,-1}\,c_{f,0} - \frac{\sqrt{30}}{10}\,c^\dagger_{p,1}\,c^\dagger_{d,-2}\,c_{d,-2}\,c_{f,1} - \frac{\sqrt{3}}{2}\,c^\dagger_{p,-1}\,c^\dagger_{d,1}\,c_{d,2}\,c_{f,-2}
\\
&\qquad + \frac{\sqrt{30}}{10}\,c^\dagger_{p,-1}\,c^\dagger_{d,1}\,c_{d,1}\,c_{f,-1} - \frac{\sqrt{30}}{10}\,c^\dagger_{p,-1}\,c^\dagger_{d,1}\,c_{d,-1}\,c_{f,1} + \frac{\sqrt{3}}{2}\,c^\dagger_{p,-1}\,c^\dagger_{d,1}\,c_{d,-2}\,c_{f,2}
\\
&\qquad + \frac{\sqrt{3}}{2}\,c^\dagger_{p,1}\,c^\dagger_{d,-1}\,c_{d,2}\,c_{f,-2} - \frac{\sqrt{30}}{10}\,c^\dagger_{p,1}\,c^\dagger_{d,-1}\,c_{d,1}\,c_{f,-1} + \frac{\sqrt{30}}{10}\,c^\dagger_{p,1}\,c^\dagger_{d,-1}\,c_{d,-1}\,c_{f,1}
\\
&\qquad - \frac{\sqrt{3}}{2}\,c^\dagger_{p,1}\,c^\dagger_{d,-1}\,c_{d,-2}\,c_{f,2} - \frac{\sqrt{30}}{10}\,c^\dagger_{p,-1}\,c^\dagger_{d,2}\,c_{d,2}\,c_{f,-1} + \frac{\sqrt{10}}{5}\,c^\dagger_{p,-1}\,c^\dagger_{d,2}\,c_{d,1}\,c_{f,0}
\\
&\qquad - \frac{\sqrt{5}}{5}\,c^\dagger_{p,-1}\,c^\dagger_{d,2}\,c_{d,0}\,c_{f,1} + \frac{\sqrt{2}}{2}\,c^\dagger_{p,-1}\,c^\dagger_{d,2}\,c_{d,-2}\,c_{f,3} - \frac{\sqrt{15}}{10}\,c^\dagger_{p,0}\,c^\dagger_{d,1}\,c_{d,2}\,c_{f,-1}
\\
&\qquad + \frac{\sqrt{5}}{5}\,c^\dagger_{p,0}\,c^\dagger_{d,1}\,c_{d,1}\,c_{f,0} - \frac{\sqrt{10}}{10}\,c^\dagger_{p,0}\,c^\dagger_{d,1}\,c_{d,0}\,c_{f,1} + \frac{1}{2}\,c^\dagger_{p,0}\,c^\dagger_{d,1}\,c_{d,-2}\,c_{f,3}
\\
&\qquad + \frac{3 \sqrt{5}}{10}\,c^\dagger_{p,1}\,c^\dagger_{d,0}\,c_{d,2}\,c_{f,-1} - \frac{\sqrt{15}}{5}\,c^\dagger_{p,1}\,c^\dagger_{d,0}\,c_{d,1}\,c_{f,0} + \frac{\sqrt{30}}{10}\,c^\dagger_{p,1}\,c^\dagger_{d,0}\,c_{d,0}\,c_{f,1}
\\
&\qquad - \frac{\sqrt{3}}{2}\,c^\dagger_{p,1}\,c^\dagger_{d,0}\,c_{d,-2}\,c_{f,3} - \frac{\sqrt{5}}{5}\,c^\dagger_{p,0}\,c^\dagger_{d,2}\,c_{d,2}\,c_{f,0} + \frac{\sqrt{15}}{5}\,c^\dagger_{p,0}\,c^\dagger_{d,2}\,c_{d,1}\,c_{f,1}
\\
&\qquad - c^\dagger_{p,0}\,c^\dagger_{d,2}\,c_{d,0}\,c_{f,2} + c^\dagger_{p,0}\,c^\dagger_{d,2}\,c_{d,-1}\,c_{f,3} + \frac{\sqrt{10}}{10}\,c^\dagger_{p,1}\,c^\dagger_{d,1}\,c_{d,2}\,c_{f,0}
\\
&\qquad - \frac{\sqrt{30}}{10}\,c^\dagger_{p,1}\,c^\dagger_{d,1}\,c_{d,1}\,c_{f,1} + \frac{\sqrt{2}}{2}\,c^\dagger_{p,1}\,c^\dagger_{d,1}\,c_{d,0}\,c_{f,2} - \frac{\sqrt{2}}{2}\,c^\dagger_{p,1}\,c^\dagger_{d,1}\,c_{d,-1}\,c_{f,3}
\\
&\qquad - \frac{\sqrt{2}}{2}\,c^\dagger_{d,1}\,c^\dagger_{f,-3}\,c_{p,-1}\,c_{d,-1} + \frac{\sqrt{2}}{2}\,c^\dagger_{d,0}\,c^\dagger_{f,-2}\,c_{p,-1}\,c_{d,-1} - \frac{\sqrt{30}}{10}\,c^\dagger_{d,-1}\,c^\dagger_{f,-1}\,c_{p,-1}\,c_{d,-1}
\\
&\qquad + \frac{\sqrt{10}}{10}\,c^\dagger_{d,-2}\,c^\dagger_{f,0}\,c_{p,-1}\,c_{d,-1} + c^\dagger_{d,1}\,c^\dagger_{f,-3}\,c_{p,0}\,c_{d,-2} - c^\dagger_{d,0}\,c^\dagger_{f,-2}\,c_{p,0}\,c_{d,-2}
\\
&\qquad + \frac{\sqrt{15}}{5}\,c^\dagger_{d,-1}\,c^\dagger_{f,-1}\,c_{p,0}\,c_{d,-2} - \frac{\sqrt{5}}{5}\,c^\dagger_{d,-2}\,c^\dagger_{f,0}\,c_{p,0}\,c_{d,-2} - \frac{\sqrt{3}}{2}\,c^\dagger_{d,2}\,c^\dagger_{f,-3}\,c_{p,-1}\,c_{d,0}
\\
&\qquad + \frac{\sqrt{30}}{10}\,c^\dagger_{d,0}\,c^\dagger_{f,-1}\,c_{p,-1}\,c_{d,0} - \frac{\sqrt{15}}{5}\,c^\dagger_{d,-1}\,c^\dagger_{f,0}\,c_{p,-1}\,c_{d,0} + \frac{3 \sqrt{5}}{10}\,c^\dagger_{d,-2}\,c^\dagger_{f,1}\,c_{p,-1}\,c_{d,0}
\\
&\qquad + \frac{1}{2}\,c^\dagger_{d,2}\,c^\dagger_{f,-3}\,c_{p,0}\,c_{d,-1} - \frac{\sqrt{10}}{10}\,c^\dagger_{d,0}\,c^\dagger_{f,-1}\,c_{p,0}\,c_{d,-1} + \frac{\sqrt{5}}{5}\,c^\dagger_{d,-1}\,c^\dagger_{f,0}\,c_{p,0}\,c_{d,-1}
\\
&\qquad - \frac{\sqrt{15}}{10}\,c^\dagger_{d,-2}\,c^\dagger_{f,1}\,c_{p,0}\,c_{d,-1} + \frac{\sqrt{2}}{2}\,c^\dagger_{d,2}\,c^\dagger_{f,-3}\,c_{p,1}\,c_{d,-2} - \frac{\sqrt{5}}{5}\,c^\dagger_{d,0}\,c^\dagger_{f,-1}\,c_{p,1}\,c_{d,-2}
\\
&\qquad + \frac{\sqrt{10}}{5}\,c^\dagger_{d,-1}\,c^\dagger_{f,0}\,c_{p,1}\,c_{d,-2} - \frac{\sqrt{30}}{10}\,c^\dagger_{d,-2}\,c^\dagger_{f,1}\,c_{p,1}\,c_{d,-2} - \frac{\sqrt{3}}{2}\,c^\dagger_{d,2}\,c^\dagger_{f,-2}\,c_{p,-1}\,c_{d,1}
\\
&\qquad + \frac{\sqrt{30}}{10}\,c^\dagger_{d,1}\,c^\dagger_{f,-1}\,c_{p,-1}\,c_{d,1} - \frac{\sqrt{30}}{10}\,c^\dagger_{d,-1}\,c^\dagger_{f,1}\,c_{p,-1}\,c_{d,1} + \frac{\sqrt{3}}{2}\,c^\dagger_{d,-2}\,c^\dagger_{f,2}\,c_{p,-1}\,c_{d,1}
\\
&\qquad + \frac{\sqrt{3}}{2}\,c^\dagger_{d,2}\,c^\dagger_{f,-2}\,c_{p,1}\,c_{d,-1} - \frac{\sqrt{30}}{10}\,c^\dagger_{d,1}\,c^\dagger_{f,-1}\,c_{p,1}\,c_{d,-1} + \frac{\sqrt{30}}{10}\,c^\dagger_{d,-1}\,c^\dagger_{f,1}\,c_{p,1}\,c_{d,-1}
\\
&\qquad - \frac{\sqrt{3}}{2}\,c^\dagger_{d,-2}\,c^\dagger_{f,2}\,c_{p,1}\,c_{d,-1} - \frac{\sqrt{30}}{10}\,c^\dagger_{d,2}\,c^\dagger_{f,-1}\,c_{p,-1}\,c_{d,2} + \frac{\sqrt{10}}{5}\,c^\dagger_{d,1}\,c^\dagger_{f,0}\,c_{p,-1}\,c_{d,2}
\\
&\qquad - \frac{\sqrt{5}}{5}\,c^\dagger_{d,0}\,c^\dagger_{f,1}\,c_{p,-1}\,c_{d,2} + \frac{\sqrt{2}}{2}\,c^\dagger_{d,-2}\,c^\dagger_{f,3}\,c_{p,-1}\,c_{d,2} - \frac{\sqrt{15}}{10}\,c^\dagger_{d,2}\,c^\dagger_{f,-1}\,c_{p,0}\,c_{d,1}
\\
&\qquad + \frac{\sqrt{5}}{5}\,c^\dagger_{d,1}\,c^\dagger_{f,0}\,c_{p,0}\,c_{d,1} - \frac{\sqrt{10}}{10}\,c^\dagger_{d,0}\,c^\dagger_{f,1}\,c_{p,0}\,c_{d,1} + \frac{1}{2}\,c^\dagger_{d,-2}\,c^\dagger_{f,3}\,c_{p,0}\,c_{d,1}
\\
&\qquad + \frac{3 \sqrt{5}}{10}\,c^\dagger_{d,2}\,c^\dagger_{f,-1}\,c_{p,1}\,c_{d,0} - \frac{\sqrt{15}}{5}\,c^\dagger_{d,1}\,c^\dagger_{f,0}\,c_{p,1}\,c_{d,0} + \frac{\sqrt{30}}{10}\,c^\dagger_{d,0}\,c^\dagger_{f,1}\,c_{p,1}\,c_{d,0}
\\
&\qquad - \frac{\sqrt{3}}{2}\,c^\dagger_{d,-2}\,c^\dagger_{f,3}\,c_{p,1}\,c_{d,0} - \frac{\sqrt{5}}{5}\,c^\dagger_{d,2}\,c^\dagger_{f,0}\,c_{p,0}\,c_{d,2} + \frac{\sqrt{15}}{5}\,c^\dagger_{d,1}\,c^\dagger_{f,1}\,c_{p,0}\,c_{d,2}
\\
&\qquad - c^\dagger_{d,0}\,c^\dagger_{f,2}\,c_{p,0}\,c_{d,2} + c^\dagger_{d,-1}\,c^\dagger_{f,3}\,c_{p,0}\,c_{d,2} + \frac{\sqrt{10}}{10}\,c^\dagger_{d,2}\,c^\dagger_{f,0}\,c_{p,1}\,c_{d,1}
\\
&\qquad - \frac{\sqrt{30}}{10}\,c^\dagger_{d,1}\,c^\dagger_{f,1}\,c_{p,1}\,c_{d,1} + \frac{\sqrt{2}}{2}\,c^\dagger_{d,0}\,c^\dagger_{f,2}\,c_{p,1}\,c_{d,1} - \frac{\sqrt{2}}{2}\,c^\dagger_{d,-1}\,c^\dagger_{f,3}\,c_{p,1}\,c_{d,1}\Big)
\\
{T_0^{(2)}}_{[2]}
&=  
  \frac{i}{2} \left(
    \left[[c^\dagger_p \times c^\dagger_d]_{2} \times[\tilde c_d \times \tilde c_f]_{2}\right]_{0} 
    - h.c. 
    \right)\\
&=  
  \frac{i}{4} \left(
    \left[[c^\dagger_p \times c^\dagger_d]_{2} \times[\tilde c_d \times \tilde c_f]_{2}\right]_{0} 
    +
    \left[[c^\dagger_p \times c^\dagger_d]_{2} \times[\tilde c_f \times \tilde c_d]_{2}\right]_{0}         
  - h.c. \right)\\
&=
\frac{1}{2} \Big\{
  -\frac{\sqrt{10}}{5}\left(   {\bm Q}^{pd}_1 \cdot  {\bm T}^{df}_1  
+  {\bm T}^{pd}_1 \cdot  {\bm Q}^{df}_1  \right)
  +\frac{\sqrt{15}}{5}\left(   {\bm Q}^{pd}_3 \cdot  {\bm T}^{df}_3  
+  {\bm T}^{pd}_3 \cdot  {\bm Q}^{df}_3  \right) \Big\}
\\
&\quad 
+ \frac{1}{2} \Big\{
  -\frac{\sqrt{14}}{7}  {\bm T}^{pf}_2 \cdot  {\bm Q}^{d}_2
  -\frac{\sqrt{2}}{2}  {\bm G}^{pf}_3 \cdot  {\bm M}^{d}_3 
  +\frac{\sqrt{42}}{14}  {\bm T}^{pf}_4 \cdot  {\bm Q}^{d}_4
  \Big\}
\\
&=
\frac{\sqrt{21} i}{42}\Big(- \frac{\sqrt{2}}{2}\,c^\dagger_{p,-1}\,c^\dagger_{d,-1}\,c_{d,1}\,c_{f,-3} + \frac{\sqrt{2}}{2}\,c^\dagger_{p,-1}\,c^\dagger_{d,-1}\,c_{d,0}\,c_{f,-2} - \frac{\sqrt{30}}{10}\,c^\dagger_{p,-1}\,c^\dagger_{d,-1}\,c_{d,-1}\,c_{f,-1}
\\
&\qquad + \frac{\sqrt{10}}{10}\,c^\dagger_{p,-1}\,c^\dagger_{d,-1}\,c_{d,-2}\,c_{f,0} + c^\dagger_{p,0}\,c^\dagger_{d,-2}\,c_{d,1}\,c_{f,-3} - c^\dagger_{p,0}\,c^\dagger_{d,-2}\,c_{d,0}\,c_{f,-2}
\\
&\qquad + \frac{\sqrt{15}}{5}\,c^\dagger_{p,0}\,c^\dagger_{d,-2}\,c_{d,-1}\,c_{f,-1} - \frac{\sqrt{5}}{5}\,c^\dagger_{p,0}\,c^\dagger_{d,-2}\,c_{d,-2}\,c_{f,0} - \frac{\sqrt{3}}{2}\,c^\dagger_{p,-1}\,c^\dagger_{d,0}\,c_{d,2}\,c_{f,-3}
\\
&\qquad + \frac{\sqrt{30}}{10}\,c^\dagger_{p,-1}\,c^\dagger_{d,0}\,c_{d,0}\,c_{f,-1} - \frac{\sqrt{15}}{5}\,c^\dagger_{p,-1}\,c^\dagger_{d,0}\,c_{d,-1}\,c_{f,0} + \frac{3 \sqrt{5}}{10}\,c^\dagger_{p,-1}\,c^\dagger_{d,0}\,c_{d,-2}\,c_{f,1}
\\
&\qquad + \frac{1}{2}\,c^\dagger_{p,0}\,c^\dagger_{d,-1}\,c_{d,2}\,c_{f,-3} - \frac{\sqrt{10}}{10}\,c^\dagger_{p,0}\,c^\dagger_{d,-1}\,c_{d,0}\,c_{f,-1} + \frac{\sqrt{5}}{5}\,c^\dagger_{p,0}\,c^\dagger_{d,-1}\,c_{d,-1}\,c_{f,0}
\\
&\qquad - \frac{\sqrt{15}}{10}\,c^\dagger_{p,0}\,c^\dagger_{d,-1}\,c_{d,-2}\,c_{f,1} + \frac{\sqrt{2}}{2}\,c^\dagger_{p,1}\,c^\dagger_{d,-2}\,c_{d,2}\,c_{f,-3} - \frac{\sqrt{5}}{5}\,c^\dagger_{p,1}\,c^\dagger_{d,-2}\,c_{d,0}\,c_{f,-1}
\\
&\qquad + \frac{\sqrt{10}}{5}\,c^\dagger_{p,1}\,c^\dagger_{d,-2}\,c_{d,-1}\,c_{f,0} - \frac{\sqrt{30}}{10}\,c^\dagger_{p,1}\,c^\dagger_{d,-2}\,c_{d,-2}\,c_{f,1} - \frac{\sqrt{3}}{2}\,c^\dagger_{p,-1}\,c^\dagger_{d,1}\,c_{d,2}\,c_{f,-2}
\\
&\qquad + \frac{\sqrt{30}}{10}\,c^\dagger_{p,-1}\,c^\dagger_{d,1}\,c_{d,1}\,c_{f,-1} - \frac{\sqrt{30}}{10}\,c^\dagger_{p,-1}\,c^\dagger_{d,1}\,c_{d,-1}\,c_{f,1} + \frac{\sqrt{3}}{2}\,c^\dagger_{p,-1}\,c^\dagger_{d,1}\,c_{d,-2}\,c_{f,2}
\\
&\qquad + \frac{\sqrt{3}}{2}\,c^\dagger_{p,1}\,c^\dagger_{d,-1}\,c_{d,2}\,c_{f,-2} - \frac{\sqrt{30}}{10}\,c^\dagger_{p,1}\,c^\dagger_{d,-1}\,c_{d,1}\,c_{f,-1} + \frac{\sqrt{30}}{10}\,c^\dagger_{p,1}\,c^\dagger_{d,-1}\,c_{d,-1}\,c_{f,1}
\\
&\qquad - \frac{\sqrt{3}}{2}\,c^\dagger_{p,1}\,c^\dagger_{d,-1}\,c_{d,-2}\,c_{f,2} - \frac{\sqrt{30}}{10}\,c^\dagger_{p,-1}\,c^\dagger_{d,2}\,c_{d,2}\,c_{f,-1} + \frac{\sqrt{10}}{5}\,c^\dagger_{p,-1}\,c^\dagger_{d,2}\,c_{d,1}\,c_{f,0}
\\
&\qquad - \frac{\sqrt{5}}{5}\,c^\dagger_{p,-1}\,c^\dagger_{d,2}\,c_{d,0}\,c_{f,1} + \frac{\sqrt{2}}{2}\,c^\dagger_{p,-1}\,c^\dagger_{d,2}\,c_{d,-2}\,c_{f,3} - \frac{\sqrt{15}}{10}\,c^\dagger_{p,0}\,c^\dagger_{d,1}\,c_{d,2}\,c_{f,-1}
\\
&\qquad + \frac{\sqrt{5}}{5}\,c^\dagger_{p,0}\,c^\dagger_{d,1}\,c_{d,1}\,c_{f,0} - \frac{\sqrt{10}}{10}\,c^\dagger_{p,0}\,c^\dagger_{d,1}\,c_{d,0}\,c_{f,1} + \frac{1}{2}\,c^\dagger_{p,0}\,c^\dagger_{d,1}\,c_{d,-2}\,c_{f,3}
\\
&\qquad + \frac{3 \sqrt{5}}{10}\,c^\dagger_{p,1}\,c^\dagger_{d,0}\,c_{d,2}\,c_{f,-1} - \frac{\sqrt{15}}{5}\,c^\dagger_{p,1}\,c^\dagger_{d,0}\,c_{d,1}\,c_{f,0} + \frac{\sqrt{30}}{10}\,c^\dagger_{p,1}\,c^\dagger_{d,0}\,c_{d,0}\,c_{f,1}
\\
&\qquad - \frac{\sqrt{3}}{2}\,c^\dagger_{p,1}\,c^\dagger_{d,0}\,c_{d,-2}\,c_{f,3} - \frac{\sqrt{5}}{5}\,c^\dagger_{p,0}\,c^\dagger_{d,2}\,c_{d,2}\,c_{f,0} + \frac{\sqrt{15}}{5}\,c^\dagger_{p,0}\,c^\dagger_{d,2}\,c_{d,1}\,c_{f,1}
\\
&\qquad - c^\dagger_{p,0}\,c^\dagger_{d,2}\,c_{d,0}\,c_{f,2} + c^\dagger_{p,0}\,c^\dagger_{d,2}\,c_{d,-1}\,c_{f,3} + \frac{\sqrt{10}}{10}\,c^\dagger_{p,1}\,c^\dagger_{d,1}\,c_{d,2}\,c_{f,0}
\\
&\qquad - \frac{\sqrt{30}}{10}\,c^\dagger_{p,1}\,c^\dagger_{d,1}\,c_{d,1}\,c_{f,1} + \frac{\sqrt{2}}{2}\,c^\dagger_{p,1}\,c^\dagger_{d,1}\,c_{d,0}\,c_{f,2} - \frac{\sqrt{2}}{2}\,c^\dagger_{p,1}\,c^\dagger_{d,1}\,c_{d,-1}\,c_{f,3}
\\
&\qquad + \frac{\sqrt{2}}{2}\,c^\dagger_{d,1}\,c^\dagger_{f,-3}\,c_{p,-1}\,c_{d,-1} - \frac{\sqrt{2}}{2}\,c^\dagger_{d,0}\,c^\dagger_{f,-2}\,c_{p,-1}\,c_{d,-1} + \frac{\sqrt{30}}{10}\,c^\dagger_{d,-1}\,c^\dagger_{f,-1}\,c_{p,-1}\,c_{d,-1}
\\
&\qquad - \frac{\sqrt{10}}{10}\,c^\dagger_{d,-2}\,c^\dagger_{f,0}\,c_{p,-1}\,c_{d,-1} - c^\dagger_{d,1}\,c^\dagger_{f,-3}\,c_{p,0}\,c_{d,-2} + c^\dagger_{d,0}\,c^\dagger_{f,-2}\,c_{p,0}\,c_{d,-2}
\\
&\qquad - \frac{\sqrt{15}}{5}\,c^\dagger_{d,-1}\,c^\dagger_{f,-1}\,c_{p,0}\,c_{d,-2} + \frac{\sqrt{5}}{5}\,c^\dagger_{d,-2}\,c^\dagger_{f,0}\,c_{p,0}\,c_{d,-2} + \frac{\sqrt{3}}{2}\,c^\dagger_{d,2}\,c^\dagger_{f,-3}\,c_{p,-1}\,c_{d,0}
\\
&\qquad - \frac{\sqrt{30}}{10}\,c^\dagger_{d,0}\,c^\dagger_{f,-1}\,c_{p,-1}\,c_{d,0} + \frac{\sqrt{15}}{5}\,c^\dagger_{d,-1}\,c^\dagger_{f,0}\,c_{p,-1}\,c_{d,0} - \frac{3 \sqrt{5}}{10}\,c^\dagger_{d,-2}\,c^\dagger_{f,1}\,c_{p,-1}\,c_{d,0}
\\
&\qquad - \frac{1}{2}\,c^\dagger_{d,2}\,c^\dagger_{f,-3}\,c_{p,0}\,c_{d,-1} + \frac{\sqrt{10}}{10}\,c^\dagger_{d,0}\,c^\dagger_{f,-1}\,c_{p,0}\,c_{d,-1} - \frac{\sqrt{5}}{5}\,c^\dagger_{d,-1}\,c^\dagger_{f,0}\,c_{p,0}\,c_{d,-1}
\\
&\qquad + \frac{\sqrt{15}}{10}\,c^\dagger_{d,-2}\,c^\dagger_{f,1}\,c_{p,0}\,c_{d,-1} - \frac{\sqrt{2}}{2}\,c^\dagger_{d,2}\,c^\dagger_{f,-3}\,c_{p,1}\,c_{d,-2} + \frac{\sqrt{5}}{5}\,c^\dagger_{d,0}\,c^\dagger_{f,-1}\,c_{p,1}\,c_{d,-2}
\\
&\qquad - \frac{\sqrt{10}}{5}\,c^\dagger_{d,-1}\,c^\dagger_{f,0}\,c_{p,1}\,c_{d,-2} + \frac{\sqrt{30}}{10}\,c^\dagger_{d,-2}\,c^\dagger_{f,1}\,c_{p,1}\,c_{d,-2} + \frac{\sqrt{3}}{2}\,c^\dagger_{d,2}\,c^\dagger_{f,-2}\,c_{p,-1}\,c_{d,1}
\\
&\qquad - \frac{\sqrt{30}}{10}\,c^\dagger_{d,1}\,c^\dagger_{f,-1}\,c_{p,-1}\,c_{d,1} + \frac{\sqrt{30}}{10}\,c^\dagger_{d,-1}\,c^\dagger_{f,1}\,c_{p,-1}\,c_{d,1} - \frac{\sqrt{3}}{2}\,c^\dagger_{d,-2}\,c^\dagger_{f,2}\,c_{p,-1}\,c_{d,1}
\\
&\qquad - \frac{\sqrt{3}}{2}\,c^\dagger_{d,2}\,c^\dagger_{f,-2}\,c_{p,1}\,c_{d,-1} + \frac{\sqrt{30}}{10}\,c^\dagger_{d,1}\,c^\dagger_{f,-1}\,c_{p,1}\,c_{d,-1} - \frac{\sqrt{30}}{10}\,c^\dagger_{d,-1}\,c^\dagger_{f,1}\,c_{p,1}\,c_{d,-1}
\\
&\qquad + \frac{\sqrt{3}}{2}\,c^\dagger_{d,-2}\,c^\dagger_{f,2}\,c_{p,1}\,c_{d,-1} + \frac{\sqrt{30}}{10}\,c^\dagger_{d,2}\,c^\dagger_{f,-1}\,c_{p,-1}\,c_{d,2} - \frac{\sqrt{10}}{5}\,c^\dagger_{d,1}\,c^\dagger_{f,0}\,c_{p,-1}\,c_{d,2}
\\
&\qquad + \frac{\sqrt{5}}{5}\,c^\dagger_{d,0}\,c^\dagger_{f,1}\,c_{p,-1}\,c_{d,2} - \frac{\sqrt{2}}{2}\,c^\dagger_{d,-2}\,c^\dagger_{f,3}\,c_{p,-1}\,c_{d,2} + \frac{\sqrt{15}}{10}\,c^\dagger_{d,2}\,c^\dagger_{f,-1}\,c_{p,0}\,c_{d,1}
\\
&\qquad - \frac{\sqrt{5}}{5}\,c^\dagger_{d,1}\,c^\dagger_{f,0}\,c_{p,0}\,c_{d,1} + \frac{\sqrt{10}}{10}\,c^\dagger_{d,0}\,c^\dagger_{f,1}\,c_{p,0}\,c_{d,1} - \frac{1}{2}\,c^\dagger_{d,-2}\,c^\dagger_{f,3}\,c_{p,0}\,c_{d,1}
\\
&\qquad - \frac{3 \sqrt{5}}{10}\,c^\dagger_{d,2}\,c^\dagger_{f,-1}\,c_{p,1}\,c_{d,0} + \frac{\sqrt{15}}{5}\,c^\dagger_{d,1}\,c^\dagger_{f,0}\,c_{p,1}\,c_{d,0} - \frac{\sqrt{30}}{10}\,c^\dagger_{d,0}\,c^\dagger_{f,1}\,c_{p,1}\,c_{d,0}
\\
&\qquad + \frac{\sqrt{3}}{2}\,c^\dagger_{d,-2}\,c^\dagger_{f,3}\,c_{p,1}\,c_{d,0} + \frac{\sqrt{5}}{5}\,c^\dagger_{d,2}\,c^\dagger_{f,0}\,c_{p,0}\,c_{d,2} - \frac{\sqrt{15}}{5}\,c^\dagger_{d,1}\,c^\dagger_{f,1}\,c_{p,0}\,c_{d,2}
\\
&\qquad + c^\dagger_{d,0}\,c^\dagger_{f,2}\,c_{p,0}\,c_{d,2} - c^\dagger_{d,-1}\,c^\dagger_{f,3}\,c_{p,0}\,c_{d,2} - \frac{\sqrt{10}}{10}\,c^\dagger_{d,2}\,c^\dagger_{f,0}\,c_{p,1}\,c_{d,1}
\\
&\qquad + \frac{\sqrt{30}}{10}\,c^\dagger_{d,1}\,c^\dagger_{f,1}\,c_{p,1}\,c_{d,1} - \frac{\sqrt{2}}{2}\,c^\dagger_{d,0}\,c^\dagger_{f,2}\,c_{p,1}\,c_{d,1} + \frac{\sqrt{2}}{2}\,c^\dagger_{d,-1}\,c^\dagger_{f,3}\,c_{p,1}\,c_{d,1}\Big)
\end{align*}

% dpdf : 3×3
% j13,j24,A wigner_9j： 2 , 2 , 2*sqrt(210)/35
% j13,j24,A wigner_9j： 3 , 3 , -sqrt(30)/10
% j13,j24,A wigner_9j： 4 , 4 , sqrt(70)/70
% dpfd : 3×3
% j13,j24,A wigner_9j： 1 , 1 , sqrt(6)/5
% j13,j24,A wigner_9j： 2 , 2 , -sqrt(15)/5
% j13,j24,A wigner_9j： 3 , 3 , 2/5


% pfdd : 3×3
% j13,j24,A wigner_9j： 1 , 1 , sqrt(5)/5
% j13,j24,A wigner_9j： 2 , 2 , sqrt(2)/2
% j13,j24,A wigner_9j： 3 , 3 , -sqrt(30)/10
\begin{align*}
{Q_0^{(2)}}_{[4]}
&=  
  \frac{1}{2} \left(
    \left[[c^\dagger_p \times c^\dagger_f]_{3} \times[\tilde c_d \times \tilde c_d]_{3}\right]_{0} 
    + h.c. 
    \right)\\
&=  
  \frac{\sqrt{5}}{5}\left(  {\bm T}^{pd}_1 \cdot {\bm T}^{df}_1   
+ {\bm Q}^{pd}_1 \cdot {\bm Q}^{df}_1   \right)
+ \frac{\sqrt{2}}{2}\left(  {\bm G}^{pd}_2 \cdot {\bm G}^{df}_2   
+ {\bm M}^{pd}_2 \cdot {\bm M}^{df}_2   \right)
\\
  &\qquad 
  - \frac{\sqrt{30}}{10}\left(  {\bm T}^{pd}_3 \cdot {\bm T}^{df}_3   
+ {\bm Q}^{pd}_3 \cdot {\bm Q}^{df}_3   \right) 
\\
&=
\frac{\sqrt{42}}{28}\Big(- \frac{\sqrt{3}}{3}\,c^\dagger_{p,-1}\,c^\dagger_{f,-2}\,c_{d,-1}\,c_{d,-2} + c^\dagger_{p,0}\,c^\dagger_{f,-3}\,c_{d,-1}\,c_{d,-2} - \frac{\sqrt{5}}{3}\,c^\dagger_{p,-1}\,c^\dagger_{f,-1}\,c_{d,0}\,c_{d,-2}
\\
&\qquad + \frac{2}{3}\,c^\dagger_{p,0}\,c^\dagger_{f,-2}\,c_{d,0}\,c_{d,-2} + \frac{\sqrt{3}}{3}\,c^\dagger_{p,1}\,c^\dagger_{f,-3}\,c_{d,0}\,c_{d,-2} - \frac{\sqrt{10}}{5}\,c^\dagger_{p,-1}\,c^\dagger_{f,0}\,c_{d,1}\,c_{d,-2}
\\
&\qquad - \frac{2 \sqrt{15}}{15}\,c^\dagger_{p,-1}\,c^\dagger_{f,0}\,c_{d,0}\,c_{d,-1} + \frac{\sqrt{15}}{15}\,c^\dagger_{p,0}\,c^\dagger_{f,-1}\,c_{d,1}\,c_{d,-2} + \frac{\sqrt{10}}{15}\,c^\dagger_{p,0}\,c^\dagger_{f,-1}\,c_{d,0}\,c_{d,-1}
\\
&\qquad + \frac{\sqrt{3}}{3}\,c^\dagger_{p,1}\,c^\dagger_{f,-2}\,c_{d,1}\,c_{d,-2} + \frac{\sqrt{2}}{3}\,c^\dagger_{p,1}\,c^\dagger_{f,-2}\,c_{d,0}\,c_{d,-1} - \frac{\sqrt{30}}{15}\,c^\dagger_{p,-1}\,c^\dagger_{f,1}\,c_{d,2}\,c_{d,-2}
\\
&\qquad - \frac{2 \sqrt{30}}{15}\,c^\dagger_{p,-1}\,c^\dagger_{f,1}\,c_{d,1}\,c_{d,-1} + \frac{\sqrt{30}}{15}\,c^\dagger_{p,1}\,c^\dagger_{f,-1}\,c_{d,2}\,c_{d,-2} + \frac{2 \sqrt{30}}{15}\,c^\dagger_{p,1}\,c^\dagger_{f,-1}\,c_{d,1}\,c_{d,-1}
\\
&\qquad - \frac{\sqrt{3}}{3}\,c^\dagger_{p,-1}\,c^\dagger_{f,2}\,c_{d,2}\,c_{d,-1} - \frac{\sqrt{2}}{3}\,c^\dagger_{p,-1}\,c^\dagger_{f,2}\,c_{d,1}\,c_{d,0} - \frac{\sqrt{15}}{15}\,c^\dagger_{p,0}\,c^\dagger_{f,1}\,c_{d,2}\,c_{d,-1}
\\
&\qquad - \frac{\sqrt{10}}{15}\,c^\dagger_{p,0}\,c^\dagger_{f,1}\,c_{d,1}\,c_{d,0} + \frac{\sqrt{10}}{5}\,c^\dagger_{p,1}\,c^\dagger_{f,0}\,c_{d,2}\,c_{d,-1} + \frac{2 \sqrt{15}}{15}\,c^\dagger_{p,1}\,c^\dagger_{f,0}\,c_{d,1}\,c_{d,0}
\\
&\qquad - \frac{\sqrt{3}}{3}\,c^\dagger_{p,-1}\,c^\dagger_{f,3}\,c_{d,2}\,c_{d,0} - \frac{2}{3}\,c^\dagger_{p,0}\,c^\dagger_{f,2}\,c_{d,2}\,c_{d,0} + \frac{\sqrt{5}}{3}\,c^\dagger_{p,1}\,c^\dagger_{f,1}\,c_{d,2}\,c_{d,0}
\\
&\qquad - c^\dagger_{p,0}\,c^\dagger_{f,3}\,c_{d,2}\,c_{d,1} + \frac{\sqrt{3}}{3}\,c^\dagger_{p,1}\,c^\dagger_{f,2}\,c_{d,2}\,c_{d,1} - \frac{\sqrt{3}}{3}\,c^\dagger_{d,-1}\,c^\dagger_{d,-2}\,c_{p,-1}\,c_{f,-2}
\\
&\qquad + c^\dagger_{d,-1}\,c^\dagger_{d,-2}\,c_{p,0}\,c_{f,-3} - \frac{\sqrt{5}}{3}\,c^\dagger_{d,0}\,c^\dagger_{d,-2}\,c_{p,-1}\,c_{f,-1} + \frac{2}{3}\,c^\dagger_{d,0}\,c^\dagger_{d,-2}\,c_{p,0}\,c_{f,-2}
\\
&\qquad + \frac{\sqrt{3}}{3}\,c^\dagger_{d,0}\,c^\dagger_{d,-2}\,c_{p,1}\,c_{f,-3} - \frac{\sqrt{10}}{5}\,c^\dagger_{d,1}\,c^\dagger_{d,-2}\,c_{p,-1}\,c_{f,0} - \frac{2 \sqrt{15}}{15}\,c^\dagger_{d,0}\,c^\dagger_{d,-1}\,c_{p,-1}\,c_{f,0}
\\
&\qquad + \frac{\sqrt{15}}{15}\,c^\dagger_{d,1}\,c^\dagger_{d,-2}\,c_{p,0}\,c_{f,-1} + \frac{\sqrt{10}}{15}\,c^\dagger_{d,0}\,c^\dagger_{d,-1}\,c_{p,0}\,c_{f,-1} + \frac{\sqrt{3}}{3}\,c^\dagger_{d,1}\,c^\dagger_{d,-2}\,c_{p,1}\,c_{f,-2}
\\
&\qquad + \frac{\sqrt{2}}{3}\,c^\dagger_{d,0}\,c^\dagger_{d,-1}\,c_{p,1}\,c_{f,-2} - \frac{\sqrt{30}}{15}\,c^\dagger_{d,2}\,c^\dagger_{d,-2}\,c_{p,-1}\,c_{f,1} - \frac{2 \sqrt{30}}{15}\,c^\dagger_{d,1}\,c^\dagger_{d,-1}\,c_{p,-1}\,c_{f,1}
\\
&\qquad + \frac{\sqrt{30}}{15}\,c^\dagger_{d,2}\,c^\dagger_{d,-2}\,c_{p,1}\,c_{f,-1} + \frac{2 \sqrt{30}}{15}\,c^\dagger_{d,1}\,c^\dagger_{d,-1}\,c_{p,1}\,c_{f,-1} - \frac{\sqrt{3}}{3}\,c^\dagger_{d,2}\,c^\dagger_{d,-1}\,c_{p,-1}\,c_{f,2}
\\
&\qquad - \frac{\sqrt{2}}{3}\,c^\dagger_{d,1}\,c^\dagger_{d,0}\,c_{p,-1}\,c_{f,2} - \frac{\sqrt{15}}{15}\,c^\dagger_{d,2}\,c^\dagger_{d,-1}\,c_{p,0}\,c_{f,1} - \frac{\sqrt{10}}{15}\,c^\dagger_{d,1}\,c^\dagger_{d,0}\,c_{p,0}\,c_{f,1}
\\
&\qquad + \frac{\sqrt{10}}{5}\,c^\dagger_{d,2}\,c^\dagger_{d,-1}\,c_{p,1}\,c_{f,0} + \frac{2 \sqrt{15}}{15}\,c^\dagger_{d,1}\,c^\dagger_{d,0}\,c_{p,1}\,c_{f,0} - \frac{\sqrt{3}}{3}\,c^\dagger_{d,2}\,c^\dagger_{d,0}\,c_{p,-1}\,c_{f,3}
\\
&\qquad - \frac{2}{3}\,c^\dagger_{d,2}\,c^\dagger_{d,0}\,c_{p,0}\,c_{f,2} + \frac{\sqrt{5}}{3}\,c^\dagger_{d,2}\,c^\dagger_{d,0}\,c_{p,1}\,c_{f,1} - c^\dagger_{d,2}\,c^\dagger_{d,1}\,c_{p,0}\,c_{f,3}
\\
&\qquad + \frac{\sqrt{3}}{3}\,c^\dagger_{d,2}\,c^\dagger_{d,1}\,c_{p,1}\,c_{f,2}\Big)
\\
{T_0^{(2)}}_{[4]}
&=  
  \frac{i}{2} \left(
    \left[[c^\dagger_p \times c^\dagger_f]_{3} \times[\tilde c_d \times \tilde c_d]_{3}\right]_{0} 
    - h.c. 
    \right)\\
&=  
  \frac{\sqrt{5}}{5}\left(  {\bm Q}^{pd}_1 \cdot {\bm T}^{df}_1   
- {\bm T}^{pd}_1 \cdot {\bm Q}^{df}_1   \right)
+ \frac{\sqrt{2}}{2}\left(  {\bm M}^{pd}_2 \cdot {\bm G}^{df}_2   
- {\bm G}^{pd}_2 \cdot {\bm M}^{df}_2   \right)
\\
  &\qquad 
  - \frac{\sqrt{30}}{10}\left(  {\bm Q}^{pd}_3 \cdot {\bm T}^{df}_3   
- {\bm T}^{pd}_3 \cdot {\bm Q}^{df}_3   \right) 
\\
&=
\frac{\sqrt{42} i}{28}\Big(- \frac{\sqrt{3}}{3}\,c^\dagger_{p,-1}\,c^\dagger_{f,-2}\,c_{d,-1}\,c_{d,-2} + c^\dagger_{p,0}\,c^\dagger_{f,-3}\,c_{d,-1}\,c_{d,-2} - \frac{\sqrt{5}}{3}\,c^\dagger_{p,-1}\,c^\dagger_{f,-1}\,c_{d,0}\,c_{d,-2}
\\
&\qquad + \frac{2}{3}\,c^\dagger_{p,0}\,c^\dagger_{f,-2}\,c_{d,0}\,c_{d,-2} + \frac{\sqrt{3}}{3}\,c^\dagger_{p,1}\,c^\dagger_{f,-3}\,c_{d,0}\,c_{d,-2} - \frac{\sqrt{10}}{5}\,c^\dagger_{p,-1}\,c^\dagger_{f,0}\,c_{d,1}\,c_{d,-2}
\\
&\qquad - \frac{2 \sqrt{15}}{15}\,c^\dagger_{p,-1}\,c^\dagger_{f,0}\,c_{d,0}\,c_{d,-1} + \frac{\sqrt{15}}{15}\,c^\dagger_{p,0}\,c^\dagger_{f,-1}\,c_{d,1}\,c_{d,-2} + \frac{\sqrt{10}}{15}\,c^\dagger_{p,0}\,c^\dagger_{f,-1}\,c_{d,0}\,c_{d,-1}
\\
&\qquad + \frac{\sqrt{3}}{3}\,c^\dagger_{p,1}\,c^\dagger_{f,-2}\,c_{d,1}\,c_{d,-2} + \frac{\sqrt{2}}{3}\,c^\dagger_{p,1}\,c^\dagger_{f,-2}\,c_{d,0}\,c_{d,-1} - \frac{\sqrt{30}}{15}\,c^\dagger_{p,-1}\,c^\dagger_{f,1}\,c_{d,2}\,c_{d,-2}
\\
&\qquad - \frac{2 \sqrt{30}}{15}\,c^\dagger_{p,-1}\,c^\dagger_{f,1}\,c_{d,1}\,c_{d,-1} + \frac{\sqrt{30}}{15}\,c^\dagger_{p,1}\,c^\dagger_{f,-1}\,c_{d,2}\,c_{d,-2} + \frac{2 \sqrt{30}}{15}\,c^\dagger_{p,1}\,c^\dagger_{f,-1}\,c_{d,1}\,c_{d,-1}
\\
&\qquad - \frac{\sqrt{3}}{3}\,c^\dagger_{p,-1}\,c^\dagger_{f,2}\,c_{d,2}\,c_{d,-1} - \frac{\sqrt{2}}{3}\,c^\dagger_{p,-1}\,c^\dagger_{f,2}\,c_{d,1}\,c_{d,0} - \frac{\sqrt{15}}{15}\,c^\dagger_{p,0}\,c^\dagger_{f,1}\,c_{d,2}\,c_{d,-1}
\\
&\qquad - \frac{\sqrt{10}}{15}\,c^\dagger_{p,0}\,c^\dagger_{f,1}\,c_{d,1}\,c_{d,0} + \frac{\sqrt{10}}{5}\,c^\dagger_{p,1}\,c^\dagger_{f,0}\,c_{d,2}\,c_{d,-1} + \frac{2 \sqrt{15}}{15}\,c^\dagger_{p,1}\,c^\dagger_{f,0}\,c_{d,1}\,c_{d,0}
\\
&\qquad - \frac{\sqrt{3}}{3}\,c^\dagger_{p,-1}\,c^\dagger_{f,3}\,c_{d,2}\,c_{d,0} - \frac{2}{3}\,c^\dagger_{p,0}\,c^\dagger_{f,2}\,c_{d,2}\,c_{d,0} + \frac{\sqrt{5}}{3}\,c^\dagger_{p,1}\,c^\dagger_{f,1}\,c_{d,2}\,c_{d,0}
\\
&\qquad - c^\dagger_{p,0}\,c^\dagger_{f,3}\,c_{d,2}\,c_{d,1} + \frac{\sqrt{3}}{3}\,c^\dagger_{p,1}\,c^\dagger_{f,2}\,c_{d,2}\,c_{d,1} + \frac{\sqrt{3}}{3}\,c^\dagger_{d,-1}\,c^\dagger_{d,-2}\,c_{p,-1}\,c_{f,-2}
\\
&\qquad - c^\dagger_{d,-1}\,c^\dagger_{d,-2}\,c_{p,0}\,c_{f,-3} + \frac{\sqrt{5}}{3}\,c^\dagger_{d,0}\,c^\dagger_{d,-2}\,c_{p,-1}\,c_{f,-1} - \frac{2}{3}\,c^\dagger_{d,0}\,c^\dagger_{d,-2}\,c_{p,0}\,c_{f,-2}
\\
&\qquad - \frac{\sqrt{3}}{3}\,c^\dagger_{d,0}\,c^\dagger_{d,-2}\,c_{p,1}\,c_{f,-3} + \frac{\sqrt{10}}{5}\,c^\dagger_{d,1}\,c^\dagger_{d,-2}\,c_{p,-1}\,c_{f,0} + \frac{2 \sqrt{15}}{15}\,c^\dagger_{d,0}\,c^\dagger_{d,-1}\,c_{p,-1}\,c_{f,0}
\\
&\qquad - \frac{\sqrt{15}}{15}\,c^\dagger_{d,1}\,c^\dagger_{d,-2}\,c_{p,0}\,c_{f,-1} - \frac{\sqrt{10}}{15}\,c^\dagger_{d,0}\,c^\dagger_{d,-1}\,c_{p,0}\,c_{f,-1} - \frac{\sqrt{3}}{3}\,c^\dagger_{d,1}\,c^\dagger_{d,-2}\,c_{p,1}\,c_{f,-2}
\\
&\qquad - \frac{\sqrt{2}}{3}\,c^\dagger_{d,0}\,c^\dagger_{d,-1}\,c_{p,1}\,c_{f,-2} + \frac{\sqrt{30}}{15}\,c^\dagger_{d,2}\,c^\dagger_{d,-2}\,c_{p,-1}\,c_{f,1} + \frac{2 \sqrt{30}}{15}\,c^\dagger_{d,1}\,c^\dagger_{d,-1}\,c_{p,-1}\,c_{f,1}
\\
&\qquad - \frac{\sqrt{30}}{15}\,c^\dagger_{d,2}\,c^\dagger_{d,-2}\,c_{p,1}\,c_{f,-1} - \frac{2 \sqrt{30}}{15}\,c^\dagger_{d,1}\,c^\dagger_{d,-1}\,c_{p,1}\,c_{f,-1} + \frac{\sqrt{3}}{3}\,c^\dagger_{d,2}\,c^\dagger_{d,-1}\,c_{p,-1}\,c_{f,2}
\\
&\qquad + \frac{\sqrt{2}}{3}\,c^\dagger_{d,1}\,c^\dagger_{d,0}\,c_{p,-1}\,c_{f,2} + \frac{\sqrt{15}}{15}\,c^\dagger_{d,2}\,c^\dagger_{d,-1}\,c_{p,0}\,c_{f,1} + \frac{\sqrt{10}}{15}\,c^\dagger_{d,1}\,c^\dagger_{d,0}\,c_{p,0}\,c_{f,1}
\\
&\qquad - \frac{\sqrt{10}}{5}\,c^\dagger_{d,2}\,c^\dagger_{d,-1}\,c_{p,1}\,c_{f,0} - \frac{2 \sqrt{15}}{15}\,c^\dagger_{d,1}\,c^\dagger_{d,0}\,c_{p,1}\,c_{f,0} + \frac{\sqrt{3}}{3}\,c^\dagger_{d,2}\,c^\dagger_{d,0}\,c_{p,-1}\,c_{f,3}
\\
&\qquad + \frac{2}{3}\,c^\dagger_{d,2}\,c^\dagger_{d,0}\,c_{p,0}\,c_{f,2} - \frac{\sqrt{5}}{3}\,c^\dagger_{d,2}\,c^\dagger_{d,0}\,c_{p,1}\,c_{f,1} + c^\dagger_{d,2}\,c^\dagger_{d,1}\,c_{p,0}\,c_{f,3}
\\
&\qquad - \frac{\sqrt{3}}{3}\,c^\dagger_{d,2}\,c^\dagger_{d,1}\,c_{p,1}\,c_{f,2}\Big)
\end{align*}

\begin{align*}
{G_0^{(2)}}_{[1]} 
&=  
  \frac{1}{2} \left(\left[[c^\dagger_p \times c^\dagger_d]_{2} \times[\tilde c_f \times \tilde c_p]_{2}\right]_{0} + h.c. \right)\\
&=
\frac{1}{4} \left(\left[[c^\dagger_p \times c^\dagger_d]_{2} \times[\tilde c_f \times \tilde c_p]_{2}\right]_{0} 
-
\left[[c^\dagger_p \times c^\dagger_d]_{2} \times[\tilde c_p \times \tilde c_f]_{2}\right]_{0} 
+ h.c. \right)\\
&=
  \frac{1}{2} \left\{  
    -\frac{1}{3} \left( 
    {\bm Q}^{pf}_2 \cdot {\bm G}_2^{pd} + {\bm T}^{pf}_2 \cdot {\bm M}_2^{pd} 
    \right)
    % \\
    % &\qquad 
    -\frac{2\sqrt{2}}{3} \left( 
    {\bm M}^{pf}_3 \cdot {\bm T}_3^{pd} + {\bm G}^{pf}_3 \cdot {\bm Q}_3^{pd} 
    \right)
    % \\
    % &\qquad 
    % -\frac{\sqrt{42}}{14} \left( 
    % {\bm Q}^{pf}_4 \cdot {\bm G}_4^{pd} + {\bm T}^{pf}_4 \cdot {\bm M}_4^{pd} 
    % \right)
    \right\}\\
    &\qquad 
    +\frac{1}{2} \left\{  
    \frac{\sqrt{3}}{3} 
    {\bm M}^{p }_1 \cdot {\bm T}_1^{df} 
    +\frac{\sqrt{6}}{3}
     {\bm Q}^{p }_2 \cdot {\bm G}_2^{df} 
     \right\}
  \\
  &=
- \frac{\sqrt{42}}{42}\Big(- \frac{\sqrt{30}}{30}\,c^\dagger_{p,-1}\,c^\dagger_{d,-1}\,c_{p,-1}\,c_{f,-1} + \frac{\sqrt{6}}{6}\,c^\dagger_{p,-1}\,c^\dagger_{d,-1}\,c_{p,0}\,c_{f,-2} - \frac{\sqrt{2}}{2}\,c^\dagger_{p,-1}\,c^\dagger_{d,-1}\,c_{p,1}\,c_{f,-3}
\\
&\qquad + \frac{\sqrt{15}}{15}\,c^\dagger_{p,0}\,c^\dagger_{d,-2}\,c_{p,-1}\,c_{f,-1} - \frac{\sqrt{3}}{3}\,c^\dagger_{p,0}\,c^\dagger_{d,-2}\,c_{p,0}\,c_{f,-2} + c^\dagger_{p,0}\,c^\dagger_{d,-2}\,c_{p,1}\,c_{f,-3}
\\
&\qquad - \frac{\sqrt{15}}{10}\,c^\dagger_{p,-1}\,c^\dagger_{d,0}\,c_{p,-1}\,c_{f,0} + \frac{\sqrt{10}}{5}\,c^\dagger_{p,-1}\,c^\dagger_{d,0}\,c_{p,0}\,c_{f,-1} - \frac{\sqrt{2}}{2}\,c^\dagger_{p,-1}\,c^\dagger_{d,0}\,c_{p,1}\,c_{f,-2}
\\
&\qquad + \frac{\sqrt{5}}{10}\,c^\dagger_{p,0}\,c^\dagger_{d,-1}\,c_{p,-1}\,c_{f,0} - \frac{\sqrt{30}}{15}\,c^\dagger_{p,0}\,c^\dagger_{d,-1}\,c_{p,0}\,c_{f,-1} + \frac{\sqrt{6}}{6}\,c^\dagger_{p,0}\,c^\dagger_{d,-1}\,c_{p,1}\,c_{f,-2}
\\
&\qquad + \frac{\sqrt{10}}{10}\,c^\dagger_{p,1}\,c^\dagger_{d,-2}\,c_{p,-1}\,c_{f,0} - \frac{2 \sqrt{15}}{15}\,c^\dagger_{p,1}\,c^\dagger_{d,-2}\,c_{p,0}\,c_{f,-1} + \frac{\sqrt{3}}{3}\,c^\dagger_{p,1}\,c^\dagger_{d,-2}\,c_{p,1}\,c_{f,-2}
\\
&\qquad - \frac{\sqrt{30}}{10}\,c^\dagger_{p,-1}\,c^\dagger_{d,1}\,c_{p,-1}\,c_{f,1} + \frac{3 \sqrt{5}}{10}\,c^\dagger_{p,-1}\,c^\dagger_{d,1}\,c_{p,0}\,c_{f,0} - \frac{\sqrt{30}}{10}\,c^\dagger_{p,-1}\,c^\dagger_{d,1}\,c_{p,1}\,c_{f,-1}
\\
&\qquad + \frac{\sqrt{30}}{10}\,c^\dagger_{p,1}\,c^\dagger_{d,-1}\,c_{p,-1}\,c_{f,1} - \frac{3 \sqrt{5}}{10}\,c^\dagger_{p,1}\,c^\dagger_{d,-1}\,c_{p,0}\,c_{f,0} + \frac{\sqrt{30}}{10}\,c^\dagger_{p,1}\,c^\dagger_{d,-1}\,c_{p,1}\,c_{f,-1}
\\
&\qquad - \frac{\sqrt{3}}{3}\,c^\dagger_{p,-1}\,c^\dagger_{d,2}\,c_{p,-1}\,c_{f,2} + \frac{2 \sqrt{15}}{15}\,c^\dagger_{p,-1}\,c^\dagger_{d,2}\,c_{p,0}\,c_{f,1} - \frac{\sqrt{10}}{10}\,c^\dagger_{p,-1}\,c^\dagger_{d,2}\,c_{p,1}\,c_{f,0}
\\
&\qquad - \frac{\sqrt{6}}{6}\,c^\dagger_{p,0}\,c^\dagger_{d,1}\,c_{p,-1}\,c_{f,2} + \frac{\sqrt{30}}{15}\,c^\dagger_{p,0}\,c^\dagger_{d,1}\,c_{p,0}\,c_{f,1} - \frac{\sqrt{5}}{10}\,c^\dagger_{p,0}\,c^\dagger_{d,1}\,c_{p,1}\,c_{f,0}
\\
&\qquad + \frac{\sqrt{2}}{2}\,c^\dagger_{p,1}\,c^\dagger_{d,0}\,c_{p,-1}\,c_{f,2} - \frac{\sqrt{10}}{5}\,c^\dagger_{p,1}\,c^\dagger_{d,0}\,c_{p,0}\,c_{f,1} + \frac{\sqrt{15}}{10}\,c^\dagger_{p,1}\,c^\dagger_{d,0}\,c_{p,1}\,c_{f,0}
\\
&\qquad - c^\dagger_{p,0}\,c^\dagger_{d,2}\,c_{p,-1}\,c_{f,3} + \frac{\sqrt{3}}{3}\,c^\dagger_{p,0}\,c^\dagger_{d,2}\,c_{p,0}\,c_{f,2} - \frac{\sqrt{15}}{15}\,c^\dagger_{p,0}\,c^\dagger_{d,2}\,c_{p,1}\,c_{f,1}
\\
&\qquad + \frac{\sqrt{2}}{2}\,c^\dagger_{p,1}\,c^\dagger_{d,1}\,c_{p,-1}\,c_{f,3} - \frac{\sqrt{6}}{6}\,c^\dagger_{p,1}\,c^\dagger_{d,1}\,c_{p,0}\,c_{f,2} + \frac{\sqrt{30}}{30}\,c^\dagger_{p,1}\,c^\dagger_{d,1}\,c_{p,1}\,c_{f,1}
\\
&\qquad - \frac{\sqrt{30}}{30}\,c^\dagger_{p,-1}\,c^\dagger_{f,-1}\,c_{p,-1}\,c_{d,-1} + \frac{\sqrt{6}}{6}\,c^\dagger_{p,0}\,c^\dagger_{f,-2}\,c_{p,-1}\,c_{d,-1} - \frac{\sqrt{2}}{2}\,c^\dagger_{p,1}\,c^\dagger_{f,-3}\,c_{p,-1}\,c_{d,-1}
\\
&\qquad + \frac{\sqrt{15}}{15}\,c^\dagger_{p,-1}\,c^\dagger_{f,-1}\,c_{p,0}\,c_{d,-2} - \frac{\sqrt{3}}{3}\,c^\dagger_{p,0}\,c^\dagger_{f,-2}\,c_{p,0}\,c_{d,-2} + c^\dagger_{p,1}\,c^\dagger_{f,-3}\,c_{p,0}\,c_{d,-2}
\\
&\qquad - \frac{\sqrt{15}}{10}\,c^\dagger_{p,-1}\,c^\dagger_{f,0}\,c_{p,-1}\,c_{d,0} + \frac{\sqrt{10}}{5}\,c^\dagger_{p,0}\,c^\dagger_{f,-1}\,c_{p,-1}\,c_{d,0} - \frac{\sqrt{2}}{2}\,c^\dagger_{p,1}\,c^\dagger_{f,-2}\,c_{p,-1}\,c_{d,0}
\\
&\qquad + \frac{\sqrt{5}}{10}\,c^\dagger_{p,-1}\,c^\dagger_{f,0}\,c_{p,0}\,c_{d,-1} - \frac{\sqrt{30}}{15}\,c^\dagger_{p,0}\,c^\dagger_{f,-1}\,c_{p,0}\,c_{d,-1} + \frac{\sqrt{6}}{6}\,c^\dagger_{p,1}\,c^\dagger_{f,-2}\,c_{p,0}\,c_{d,-1}
\\
&\qquad + \frac{\sqrt{10}}{10}\,c^\dagger_{p,-1}\,c^\dagger_{f,0}\,c_{p,1}\,c_{d,-2} - \frac{2 \sqrt{15}}{15}\,c^\dagger_{p,0}\,c^\dagger_{f,-1}\,c_{p,1}\,c_{d,-2} + \frac{\sqrt{3}}{3}\,c^\dagger_{p,1}\,c^\dagger_{f,-2}\,c_{p,1}\,c_{d,-2}
\\
&\qquad - \frac{\sqrt{30}}{10}\,c^\dagger_{p,-1}\,c^\dagger_{f,1}\,c_{p,-1}\,c_{d,1} + \frac{3 \sqrt{5}}{10}\,c^\dagger_{p,0}\,c^\dagger_{f,0}\,c_{p,-1}\,c_{d,1} - \frac{\sqrt{30}}{10}\,c^\dagger_{p,1}\,c^\dagger_{f,-1}\,c_{p,-1}\,c_{d,1}
\\
&\qquad + \frac{\sqrt{30}}{10}\,c^\dagger_{p,-1}\,c^\dagger_{f,1}\,c_{p,1}\,c_{d,-1} - \frac{3 \sqrt{5}}{10}\,c^\dagger_{p,0}\,c^\dagger_{f,0}\,c_{p,1}\,c_{d,-1} + \frac{\sqrt{30}}{10}\,c^\dagger_{p,1}\,c^\dagger_{f,-1}\,c_{p,1}\,c_{d,-1}
\\
&\qquad - \frac{\sqrt{3}}{3}\,c^\dagger_{p,-1}\,c^\dagger_{f,2}\,c_{p,-1}\,c_{d,2} + \frac{2 \sqrt{15}}{15}\,c^\dagger_{p,0}\,c^\dagger_{f,1}\,c_{p,-1}\,c_{d,2} - \frac{\sqrt{10}}{10}\,c^\dagger_{p,1}\,c^\dagger_{f,0}\,c_{p,-1}\,c_{d,2}
\\
&\qquad - \frac{\sqrt{6}}{6}\,c^\dagger_{p,-1}\,c^\dagger_{f,2}\,c_{p,0}\,c_{d,1} + \frac{\sqrt{30}}{15}\,c^\dagger_{p,0}\,c^\dagger_{f,1}\,c_{p,0}\,c_{d,1} - \frac{\sqrt{5}}{10}\,c^\dagger_{p,1}\,c^\dagger_{f,0}\,c_{p,0}\,c_{d,1}
\\
&\qquad + \frac{\sqrt{2}}{2}\,c^\dagger_{p,-1}\,c^\dagger_{f,2}\,c_{p,1}\,c_{d,0} - \frac{\sqrt{10}}{5}\,c^\dagger_{p,0}\,c^\dagger_{f,1}\,c_{p,1}\,c_{d,0} + \frac{\sqrt{15}}{10}\,c^\dagger_{p,1}\,c^\dagger_{f,0}\,c_{p,1}\,c_{d,0}
\\
&\qquad - c^\dagger_{p,-1}\,c^\dagger_{f,3}\,c_{p,0}\,c_{d,2} + \frac{\sqrt{3}}{3}\,c^\dagger_{p,0}\,c^\dagger_{f,2}\,c_{p,0}\,c_{d,2} - \frac{\sqrt{15}}{15}\,c^\dagger_{p,1}\,c^\dagger_{f,1}\,c_{p,0}\,c_{d,2}
\\
&\qquad + \frac{\sqrt{2}}{2}\,c^\dagger_{p,-1}\,c^\dagger_{f,3}\,c_{p,1}\,c_{d,1} - \frac{\sqrt{6}}{6}\,c^\dagger_{p,0}\,c^\dagger_{f,2}\,c_{p,1}\,c_{d,1} + \frac{\sqrt{30}}{30}\,c^\dagger_{p,1}\,c^\dagger_{f,1}\,c_{p,1}\,c_{d,1}\Big)
\\
{M_0^{(2)}}_{[1]} 
&=  
  \frac{i}{2} \left(\left[[c^\dagger_p \times c^\dagger_d]_{2} \times[\tilde c_f \times \tilde c_p]_{2}\right]_{0} - h.c. \right)\\
&=
\frac{i}{4} \left(\left[[c^\dagger_p \times c^\dagger_d]_{2} \times[\tilde c_f \times \tilde c_p]_{2}\right]_{0} 
-
\left[[c^\dagger_p \times c^\dagger_d]_{2} \times[\tilde c_p \times \tilde c_f]_{2}\right]_{0} 
- h.c. \right)\\
&=
  \frac{1}{2} \left\{  
    -\frac{1}{3} \left( 
    {\bm T}^{pf}_2 \cdot {\bm G}_2^{pd} - {\bm Q}^{pf}_2 \cdot {\bm M}_2^{pd} 
    \right)
    -\frac{2\sqrt{2}}{3} \left( 
    {\bm G}^{pf}_3 \cdot {\bm T}_3^{pd} - {\bm M}^{pf}_3 \cdot {\bm Q}_3^{pd} 
    \right)
    \right\}\\
    &\qquad 
    +\frac{1}{2} \left\{  
    \frac{\sqrt{3}}{3} 
    {\bm M}^{p }_1 \cdot {\bm Q}_1^{df} 
    +\frac{\sqrt{6}}{3}
     {\bm Q}^{p }_2 \cdot {\bm M}_2^{df} 
     \right\}
  \\
  &=
- \frac{\sqrt{42} i}{42}\Big(- \frac{\sqrt{30}}{30}\,c^\dagger_{p,-1}\,c^\dagger_{d,-1}\,c_{p,-1}\,c_{f,-1} + \frac{\sqrt{6}}{6}\,c^\dagger_{p,-1}\,c^\dagger_{d,-1}\,c_{p,0}\,c_{f,-2} - \frac{\sqrt{2}}{2}\,c^\dagger_{p,-1}\,c^\dagger_{d,-1}\,c_{p,1}\,c_{f,-3}
\\
&\qquad + \frac{\sqrt{15}}{15}\,c^\dagger_{p,0}\,c^\dagger_{d,-2}\,c_{p,-1}\,c_{f,-1} - \frac{\sqrt{3}}{3}\,c^\dagger_{p,0}\,c^\dagger_{d,-2}\,c_{p,0}\,c_{f,-2} + c^\dagger_{p,0}\,c^\dagger_{d,-2}\,c_{p,1}\,c_{f,-3}
\\
&\qquad - \frac{\sqrt{15}}{10}\,c^\dagger_{p,-1}\,c^\dagger_{d,0}\,c_{p,-1}\,c_{f,0} + \frac{\sqrt{10}}{5}\,c^\dagger_{p,-1}\,c^\dagger_{d,0}\,c_{p,0}\,c_{f,-1} - \frac{\sqrt{2}}{2}\,c^\dagger_{p,-1}\,c^\dagger_{d,0}\,c_{p,1}\,c_{f,-2}
\\
&\qquad + \frac{\sqrt{5}}{10}\,c^\dagger_{p,0}\,c^\dagger_{d,-1}\,c_{p,-1}\,c_{f,0} - \frac{\sqrt{30}}{15}\,c^\dagger_{p,0}\,c^\dagger_{d,-1}\,c_{p,0}\,c_{f,-1} + \frac{\sqrt{6}}{6}\,c^\dagger_{p,0}\,c^\dagger_{d,-1}\,c_{p,1}\,c_{f,-2}
\\
&\qquad + \frac{\sqrt{10}}{10}\,c^\dagger_{p,1}\,c^\dagger_{d,-2}\,c_{p,-1}\,c_{f,0} - \frac{2 \sqrt{15}}{15}\,c^\dagger_{p,1}\,c^\dagger_{d,-2}\,c_{p,0}\,c_{f,-1} + \frac{\sqrt{3}}{3}\,c^\dagger_{p,1}\,c^\dagger_{d,-2}\,c_{p,1}\,c_{f,-2}
\\
&\qquad - \frac{\sqrt{30}}{10}\,c^\dagger_{p,-1}\,c^\dagger_{d,1}\,c_{p,-1}\,c_{f,1} + \frac{3 \sqrt{5}}{10}\,c^\dagger_{p,-1}\,c^\dagger_{d,1}\,c_{p,0}\,c_{f,0} - \frac{\sqrt{30}}{10}\,c^\dagger_{p,-1}\,c^\dagger_{d,1}\,c_{p,1}\,c_{f,-1}
\\
&\qquad + \frac{\sqrt{30}}{10}\,c^\dagger_{p,1}\,c^\dagger_{d,-1}\,c_{p,-1}\,c_{f,1} - \frac{3 \sqrt{5}}{10}\,c^\dagger_{p,1}\,c^\dagger_{d,-1}\,c_{p,0}\,c_{f,0} + \frac{\sqrt{30}}{10}\,c^\dagger_{p,1}\,c^\dagger_{d,-1}\,c_{p,1}\,c_{f,-1}
\\
&\qquad - \frac{\sqrt{3}}{3}\,c^\dagger_{p,-1}\,c^\dagger_{d,2}\,c_{p,-1}\,c_{f,2} + \frac{2 \sqrt{15}}{15}\,c^\dagger_{p,-1}\,c^\dagger_{d,2}\,c_{p,0}\,c_{f,1} - \frac{\sqrt{10}}{10}\,c^\dagger_{p,-1}\,c^\dagger_{d,2}\,c_{p,1}\,c_{f,0}
\\
&\qquad - \frac{\sqrt{6}}{6}\,c^\dagger_{p,0}\,c^\dagger_{d,1}\,c_{p,-1}\,c_{f,2} + \frac{\sqrt{30}}{15}\,c^\dagger_{p,0}\,c^\dagger_{d,1}\,c_{p,0}\,c_{f,1} - \frac{\sqrt{5}}{10}\,c^\dagger_{p,0}\,c^\dagger_{d,1}\,c_{p,1}\,c_{f,0}
\\
&\qquad + \frac{\sqrt{2}}{2}\,c^\dagger_{p,1}\,c^\dagger_{d,0}\,c_{p,-1}\,c_{f,2} - \frac{\sqrt{10}}{5}\,c^\dagger_{p,1}\,c^\dagger_{d,0}\,c_{p,0}\,c_{f,1} + \frac{\sqrt{15}}{10}\,c^\dagger_{p,1}\,c^\dagger_{d,0}\,c_{p,1}\,c_{f,0}
\\
&\qquad - c^\dagger_{p,0}\,c^\dagger_{d,2}\,c_{p,-1}\,c_{f,3} + \frac{\sqrt{3}}{3}\,c^\dagger_{p,0}\,c^\dagger_{d,2}\,c_{p,0}\,c_{f,2} - \frac{\sqrt{15}}{15}\,c^\dagger_{p,0}\,c^\dagger_{d,2}\,c_{p,1}\,c_{f,1}
\\
&\qquad + \frac{\sqrt{2}}{2}\,c^\dagger_{p,1}\,c^\dagger_{d,1}\,c_{p,-1}\,c_{f,3} - \frac{\sqrt{6}}{6}\,c^\dagger_{p,1}\,c^\dagger_{d,1}\,c_{p,0}\,c_{f,2} + \frac{\sqrt{30}}{30}\,c^\dagger_{p,1}\,c^\dagger_{d,1}\,c_{p,1}\,c_{f,1}
\\
&\qquad + \frac{\sqrt{30}}{30}\,c^\dagger_{p,-1}\,c^\dagger_{f,-1}\,c_{p,-1}\,c_{d,-1} - \frac{\sqrt{6}}{6}\,c^\dagger_{p,0}\,c^\dagger_{f,-2}\,c_{p,-1}\,c_{d,-1} + \frac{\sqrt{2}}{2}\,c^\dagger_{p,1}\,c^\dagger_{f,-3}\,c_{p,-1}\,c_{d,-1}
\\
&\qquad - \frac{\sqrt{15}}{15}\,c^\dagger_{p,-1}\,c^\dagger_{f,-1}\,c_{p,0}\,c_{d,-2} + \frac{\sqrt{3}}{3}\,c^\dagger_{p,0}\,c^\dagger_{f,-2}\,c_{p,0}\,c_{d,-2} - c^\dagger_{p,1}\,c^\dagger_{f,-3}\,c_{p,0}\,c_{d,-2}
\\
&\qquad + \frac{\sqrt{15}}{10}\,c^\dagger_{p,-1}\,c^\dagger_{f,0}\,c_{p,-1}\,c_{d,0} - \frac{\sqrt{10}}{5}\,c^\dagger_{p,0}\,c^\dagger_{f,-1}\,c_{p,-1}\,c_{d,0} + \frac{\sqrt{2}}{2}\,c^\dagger_{p,1}\,c^\dagger_{f,-2}\,c_{p,-1}\,c_{d,0}
\\
&\qquad - \frac{\sqrt{5}}{10}\,c^\dagger_{p,-1}\,c^\dagger_{f,0}\,c_{p,0}\,c_{d,-1} + \frac{\sqrt{30}}{15}\,c^\dagger_{p,0}\,c^\dagger_{f,-1}\,c_{p,0}\,c_{d,-1} - \frac{\sqrt{6}}{6}\,c^\dagger_{p,1}\,c^\dagger_{f,-2}\,c_{p,0}\,c_{d,-1}
\\
&\qquad - \frac{\sqrt{10}}{10}\,c^\dagger_{p,-1}\,c^\dagger_{f,0}\,c_{p,1}\,c_{d,-2} + \frac{2 \sqrt{15}}{15}\,c^\dagger_{p,0}\,c^\dagger_{f,-1}\,c_{p,1}\,c_{d,-2} - \frac{\sqrt{3}}{3}\,c^\dagger_{p,1}\,c^\dagger_{f,-2}\,c_{p,1}\,c_{d,-2}
\\
&\qquad + \frac{\sqrt{30}}{10}\,c^\dagger_{p,-1}\,c^\dagger_{f,1}\,c_{p,-1}\,c_{d,1} - \frac{3 \sqrt{5}}{10}\,c^\dagger_{p,0}\,c^\dagger_{f,0}\,c_{p,-1}\,c_{d,1} + \frac{\sqrt{30}}{10}\,c^\dagger_{p,1}\,c^\dagger_{f,-1}\,c_{p,-1}\,c_{d,1}
\\
&\qquad - \frac{\sqrt{30}}{10}\,c^\dagger_{p,-1}\,c^\dagger_{f,1}\,c_{p,1}\,c_{d,-1} + \frac{3 \sqrt{5}}{10}\,c^\dagger_{p,0}\,c^\dagger_{f,0}\,c_{p,1}\,c_{d,-1} - \frac{\sqrt{30}}{10}\,c^\dagger_{p,1}\,c^\dagger_{f,-1}\,c_{p,1}\,c_{d,-1}
\\
&\qquad + \frac{\sqrt{3}}{3}\,c^\dagger_{p,-1}\,c^\dagger_{f,2}\,c_{p,-1}\,c_{d,2} - \frac{2 \sqrt{15}}{15}\,c^\dagger_{p,0}\,c^\dagger_{f,1}\,c_{p,-1}\,c_{d,2} + \frac{\sqrt{10}}{10}\,c^\dagger_{p,1}\,c^\dagger_{f,0}\,c_{p,-1}\,c_{d,2}
\\
&\qquad + \frac{\sqrt{6}}{6}\,c^\dagger_{p,-1}\,c^\dagger_{f,2}\,c_{p,0}\,c_{d,1} - \frac{\sqrt{30}}{15}\,c^\dagger_{p,0}\,c^\dagger_{f,1}\,c_{p,0}\,c_{d,1} + \frac{\sqrt{5}}{10}\,c^\dagger_{p,1}\,c^\dagger_{f,0}\,c_{p,0}\,c_{d,1}
\\
&\qquad - \frac{\sqrt{2}}{2}\,c^\dagger_{p,-1}\,c^\dagger_{f,2}\,c_{p,1}\,c_{d,0} + \frac{\sqrt{10}}{5}\,c^\dagger_{p,0}\,c^\dagger_{f,1}\,c_{p,1}\,c_{d,0} - \frac{\sqrt{15}}{10}\,c^\dagger_{p,1}\,c^\dagger_{f,0}\,c_{p,1}\,c_{d,0}
\\
&\qquad + c^\dagger_{p,-1}\,c^\dagger_{f,3}\,c_{p,0}\,c_{d,2} - \frac{\sqrt{3}}{3}\,c^\dagger_{p,0}\,c^\dagger_{f,2}\,c_{p,0}\,c_{d,2} + \frac{\sqrt{15}}{15}\,c^\dagger_{p,1}\,c^\dagger_{f,1}\,c_{p,0}\,c_{d,2}
\\
&\qquad - \frac{\sqrt{2}}{2}\,c^\dagger_{p,-1}\,c^\dagger_{f,3}\,c_{p,1}\,c_{d,1} + \frac{\sqrt{6}}{6}\,c^\dagger_{p,0}\,c^\dagger_{f,2}\,c_{p,1}\,c_{d,1} - \frac{\sqrt{30}}{30}\,c^\dagger_{p,1}\,c^\dagger_{f,1}\,c_{p,1}\,c_{d,1}\Big)
\end{align*}

% pdfp : 2×2
% j13,j24,A wigner_9j： 2 , 2 , -1/3
% j13,j24,A wigner_9j： 3 , 3 , -2*sqrt(2)/3
% pdpf : 2×2
% j13,j24,A wigner_9j： 1 , 1 , sqrt(3)/3
% j13,j24,A wigner_9j： 2 , 2 , sqrt(6)/3

\begin{align*}
{G_0^{(2)}}_{[2]} 
&=  
  \frac{1}{2} \left(\left[[c^\dagger_p \times c^\dagger_d]_{3} \times[\tilde c_f \times \tilde c_p]_{3}\right]_{0} + h.c. \right)\\
&=
\frac{1}{4} \left(\left[[c^\dagger_p \times c^\dagger_d]_{3} \times[\tilde c_f \times \tilde c_p]_{3}\right]_{0} 
+
\left[[c^\dagger_p \times c^\dagger_d]_{3} \times[\tilde c_p \times \tilde c_f]_{3}\right]_{0} 
+ h.c. \right)\\
&=
  \frac{1}{2} \left\{  
    -\frac{2\sqrt{2}}{3} \left( 
    {\bm Q}^{pf}_2 \cdot {\bm G}_2^{pd} + {\bm T}^{pf}_2 \cdot {\bm M}_2^{pd} 
    \right)
    +\frac{1}{3} \left( 
    {\bm M}^{pf}_3 \cdot {\bm T}_3^{pd} + {\bm G}^{pf}_3 \cdot {\bm Q}_3^{pd} 
    \right)
    \right\}\\
    &\qquad 
    +\frac{1}{2} \left\{  
    -\frac{\sqrt{6}}{3} 
    {\bm M}^{p }_1 \cdot {\bm T}_1^{df} 
    +\frac{\sqrt{3}}{3}
     {\bm Q}^{p }_2 \cdot {\bm G}_2^{df} 
     \right\}
  \\
&=
\frac{\sqrt{21}}{28}\Big(- \frac{\sqrt{3}}{3}\,c^\dagger_{p,-1}\,c^\dagger_{d,-2}\,c_{p,-1}\,c_{f,-2} + c^\dagger_{p,-1}\,c^\dagger_{d,-2}\,c_{p,0}\,c_{f,-3} - \frac{\sqrt{30}}{9}\,c^\dagger_{p,-1}\,c^\dagger_{d,-1}\,c_{p,-1}\,c_{f,-1}
\\
&\qquad + \frac{2 \sqrt{6}}{9}\,c^\dagger_{p,-1}\,c^\dagger_{d,-1}\,c_{p,0}\,c_{f,-2} + \frac{\sqrt{2}}{3}\,c^\dagger_{p,-1}\,c^\dagger_{d,-1}\,c_{p,1}\,c_{f,-3} - \frac{\sqrt{15}}{9}\,c^\dagger_{p,0}\,c^\dagger_{d,-2}\,c_{p,-1}\,c_{f,-1}
\\
&\qquad + \frac{2 \sqrt{3}}{9}\,c^\dagger_{p,0}\,c^\dagger_{d,-2}\,c_{p,0}\,c_{f,-2} + \frac{1}{3}\,c^\dagger_{p,0}\,c^\dagger_{d,-2}\,c_{p,1}\,c_{f,-3} - \frac{2 \sqrt{15}}{15}\,c^\dagger_{p,-1}\,c^\dagger_{d,0}\,c_{p,-1}\,c_{f,0}
\\
&\qquad + \frac{\sqrt{10}}{15}\,c^\dagger_{p,-1}\,c^\dagger_{d,0}\,c_{p,0}\,c_{f,-1} + \frac{\sqrt{2}}{3}\,c^\dagger_{p,-1}\,c^\dagger_{d,0}\,c_{p,1}\,c_{f,-2} - \frac{4 \sqrt{5}}{15}\,c^\dagger_{p,0}\,c^\dagger_{d,-1}\,c_{p,-1}\,c_{f,0}
\\
&\qquad + \frac{2 \sqrt{30}}{45}\,c^\dagger_{p,0}\,c^\dagger_{d,-1}\,c_{p,0}\,c_{f,-1} + \frac{2 \sqrt{6}}{9}\,c^\dagger_{p,0}\,c^\dagger_{d,-1}\,c_{p,1}\,c_{f,-2} - \frac{\sqrt{10}}{15}\,c^\dagger_{p,1}\,c^\dagger_{d,-2}\,c_{p,-1}\,c_{f,0}
\\
&\qquad + \frac{\sqrt{15}}{45}\,c^\dagger_{p,1}\,c^\dagger_{d,-2}\,c_{p,0}\,c_{f,-1} + \frac{\sqrt{3}}{9}\,c^\dagger_{p,1}\,c^\dagger_{d,-2}\,c_{p,1}\,c_{f,-2} - \frac{\sqrt{30}}{15}\,c^\dagger_{p,-1}\,c^\dagger_{d,1}\,c_{p,-1}\,c_{f,1}
\\
&\qquad + \frac{\sqrt{30}}{15}\,c^\dagger_{p,-1}\,c^\dagger_{d,1}\,c_{p,1}\,c_{f,-1} - \frac{\sqrt{10}}{5}\,c^\dagger_{p,0}\,c^\dagger_{d,0}\,c_{p,-1}\,c_{f,1} + \frac{\sqrt{10}}{5}\,c^\dagger_{p,0}\,c^\dagger_{d,0}\,c_{p,1}\,c_{f,-1}
\\
&\qquad - \frac{\sqrt{30}}{15}\,c^\dagger_{p,1}\,c^\dagger_{d,-1}\,c_{p,-1}\,c_{f,1} + \frac{\sqrt{30}}{15}\,c^\dagger_{p,1}\,c^\dagger_{d,-1}\,c_{p,1}\,c_{f,-1} - \frac{\sqrt{3}}{9}\,c^\dagger_{p,-1}\,c^\dagger_{d,2}\,c_{p,-1}\,c_{f,2}
\\
&\qquad - \frac{\sqrt{15}}{45}\,c^\dagger_{p,-1}\,c^\dagger_{d,2}\,c_{p,0}\,c_{f,1} + \frac{\sqrt{10}}{15}\,c^\dagger_{p,-1}\,c^\dagger_{d,2}\,c_{p,1}\,c_{f,0} - \frac{2 \sqrt{6}}{9}\,c^\dagger_{p,0}\,c^\dagger_{d,1}\,c_{p,-1}\,c_{f,2}
\\
&\qquad - \frac{2 \sqrt{30}}{45}\,c^\dagger_{p,0}\,c^\dagger_{d,1}\,c_{p,0}\,c_{f,1} + \frac{4 \sqrt{5}}{15}\,c^\dagger_{p,0}\,c^\dagger_{d,1}\,c_{p,1}\,c_{f,0} - \frac{\sqrt{2}}{3}\,c^\dagger_{p,1}\,c^\dagger_{d,0}\,c_{p,-1}\,c_{f,2}
\\
&\qquad - \frac{\sqrt{10}}{15}\,c^\dagger_{p,1}\,c^\dagger_{d,0}\,c_{p,0}\,c_{f,1} + \frac{2 \sqrt{15}}{15}\,c^\dagger_{p,1}\,c^\dagger_{d,0}\,c_{p,1}\,c_{f,0} - \frac{1}{3}\,c^\dagger_{p,0}\,c^\dagger_{d,2}\,c_{p,-1}\,c_{f,3}
\\
&\qquad - \frac{2 \sqrt{3}}{9}\,c^\dagger_{p,0}\,c^\dagger_{d,2}\,c_{p,0}\,c_{f,2} + \frac{\sqrt{15}}{9}\,c^\dagger_{p,0}\,c^\dagger_{d,2}\,c_{p,1}\,c_{f,1} - \frac{\sqrt{2}}{3}\,c^\dagger_{p,1}\,c^\dagger_{d,1}\,c_{p,-1}\,c_{f,3}
\\
&\qquad - \frac{2 \sqrt{6}}{9}\,c^\dagger_{p,1}\,c^\dagger_{d,1}\,c_{p,0}\,c_{f,2} + \frac{\sqrt{30}}{9}\,c^\dagger_{p,1}\,c^\dagger_{d,1}\,c_{p,1}\,c_{f,1} - c^\dagger_{p,1}\,c^\dagger_{d,2}\,c_{p,0}\,c_{f,3}
\\
&\qquad + \frac{\sqrt{3}}{3}\,c^\dagger_{p,1}\,c^\dagger_{d,2}\,c_{p,1}\,c_{f,2} - \frac{\sqrt{3}}{3}\,c^\dagger_{p,-1}\,c^\dagger_{f,-2}\,c_{p,-1}\,c_{d,-2} + c^\dagger_{p,0}\,c^\dagger_{f,-3}\,c_{p,-1}\,c_{d,-2}
\\
&\qquad - \frac{\sqrt{30}}{9}\,c^\dagger_{p,-1}\,c^\dagger_{f,-1}\,c_{p,-1}\,c_{d,-1} + \frac{2 \sqrt{6}}{9}\,c^\dagger_{p,0}\,c^\dagger_{f,-2}\,c_{p,-1}\,c_{d,-1} + \frac{\sqrt{2}}{3}\,c^\dagger_{p,1}\,c^\dagger_{f,-3}\,c_{p,-1}\,c_{d,-1}
\\
&\qquad - \frac{\sqrt{15}}{9}\,c^\dagger_{p,-1}\,c^\dagger_{f,-1}\,c_{p,0}\,c_{d,-2} + \frac{2 \sqrt{3}}{9}\,c^\dagger_{p,0}\,c^\dagger_{f,-2}\,c_{p,0}\,c_{d,-2} + \frac{1}{3}\,c^\dagger_{p,1}\,c^\dagger_{f,-3}\,c_{p,0}\,c_{d,-2}
\\
&\qquad - \frac{2 \sqrt{15}}{15}\,c^\dagger_{p,-1}\,c^\dagger_{f,0}\,c_{p,-1}\,c_{d,0} + \frac{\sqrt{10}}{15}\,c^\dagger_{p,0}\,c^\dagger_{f,-1}\,c_{p,-1}\,c_{d,0} + \frac{\sqrt{2}}{3}\,c^\dagger_{p,1}\,c^\dagger_{f,-2}\,c_{p,-1}\,c_{d,0}
\\
&\qquad - \frac{4 \sqrt{5}}{15}\,c^\dagger_{p,-1}\,c^\dagger_{f,0}\,c_{p,0}\,c_{d,-1} + \frac{2 \sqrt{30}}{45}\,c^\dagger_{p,0}\,c^\dagger_{f,-1}\,c_{p,0}\,c_{d,-1} + \frac{2 \sqrt{6}}{9}\,c^\dagger_{p,1}\,c^\dagger_{f,-2}\,c_{p,0}\,c_{d,-1}
\\
&\qquad - \frac{\sqrt{10}}{15}\,c^\dagger_{p,-1}\,c^\dagger_{f,0}\,c_{p,1}\,c_{d,-2} + \frac{\sqrt{15}}{45}\,c^\dagger_{p,0}\,c^\dagger_{f,-1}\,c_{p,1}\,c_{d,-2} + \frac{\sqrt{3}}{9}\,c^\dagger_{p,1}\,c^\dagger_{f,-2}\,c_{p,1}\,c_{d,-2}
\\
&\qquad - \frac{\sqrt{30}}{15}\,c^\dagger_{p,-1}\,c^\dagger_{f,1}\,c_{p,-1}\,c_{d,1} + \frac{\sqrt{30}}{15}\,c^\dagger_{p,1}\,c^\dagger_{f,-1}\,c_{p,-1}\,c_{d,1} - \frac{\sqrt{10}}{5}\,c^\dagger_{p,-1}\,c^\dagger_{f,1}\,c_{p,0}\,c_{d,0}
\\
&\qquad + \frac{\sqrt{10}}{5}\,c^\dagger_{p,1}\,c^\dagger_{f,-1}\,c_{p,0}\,c_{d,0} - \frac{\sqrt{30}}{15}\,c^\dagger_{p,-1}\,c^\dagger_{f,1}\,c_{p,1}\,c_{d,-1} + \frac{\sqrt{30}}{15}\,c^\dagger_{p,1}\,c^\dagger_{f,-1}\,c_{p,1}\,c_{d,-1}
\\
&\qquad - \frac{\sqrt{3}}{9}\,c^\dagger_{p,-1}\,c^\dagger_{f,2}\,c_{p,-1}\,c_{d,2} - \frac{\sqrt{15}}{45}\,c^\dagger_{p,0}\,c^\dagger_{f,1}\,c_{p,-1}\,c_{d,2} + \frac{\sqrt{10}}{15}\,c^\dagger_{p,1}\,c^\dagger_{f,0}\,c_{p,-1}\,c_{d,2}
\\
&\qquad - \frac{2 \sqrt{6}}{9}\,c^\dagger_{p,-1}\,c^\dagger_{f,2}\,c_{p,0}\,c_{d,1} - \frac{2 \sqrt{30}}{45}\,c^\dagger_{p,0}\,c^\dagger_{f,1}\,c_{p,0}\,c_{d,1} + \frac{4 \sqrt{5}}{15}\,c^\dagger_{p,1}\,c^\dagger_{f,0}\,c_{p,0}\,c_{d,1}
\\
&\qquad - \frac{\sqrt{2}}{3}\,c^\dagger_{p,-1}\,c^\dagger_{f,2}\,c_{p,1}\,c_{d,0} - \frac{\sqrt{10}}{15}\,c^\dagger_{p,0}\,c^\dagger_{f,1}\,c_{p,1}\,c_{d,0} + \frac{2 \sqrt{15}}{15}\,c^\dagger_{p,1}\,c^\dagger_{f,0}\,c_{p,1}\,c_{d,0}
\\
&\qquad - \frac{1}{3}\,c^\dagger_{p,-1}\,c^\dagger_{f,3}\,c_{p,0}\,c_{d,2} - \frac{2 \sqrt{3}}{9}\,c^\dagger_{p,0}\,c^\dagger_{f,2}\,c_{p,0}\,c_{d,2} + \frac{\sqrt{15}}{9}\,c^\dagger_{p,1}\,c^\dagger_{f,1}\,c_{p,0}\,c_{d,2}
\\
&\qquad - \frac{\sqrt{2}}{3}\,c^\dagger_{p,-1}\,c^\dagger_{f,3}\,c_{p,1}\,c_{d,1} - \frac{2 \sqrt{6}}{9}\,c^\dagger_{p,0}\,c^\dagger_{f,2}\,c_{p,1}\,c_{d,1} + \frac{\sqrt{30}}{9}\,c^\dagger_{p,1}\,c^\dagger_{f,1}\,c_{p,1}\,c_{d,1}
\\
&\qquad - c^\dagger_{p,0}\,c^\dagger_{f,3}\,c_{p,1}\,c_{d,2} + \frac{\sqrt{3}}{3}\,c^\dagger_{p,1}\,c^\dagger_{f,2}\,c_{p,1}\,c_{d,2}\Big)
\\
{M_0^{(2)}}_{[2]} 
&=  
  \frac{i}{2} \left(\left[[c^\dagger_p \times c^\dagger_d]_{3} \times[\tilde c_f \times \tilde c_p]_{3}\right]_{0} - h.c. \right)\\
&=
\frac{i}{4} \left(\left[[c^\dagger_p \times c^\dagger_d]_{3} \times[\tilde c_f \times \tilde c_p]_{3}\right]_{0} 
+
\left[[c^\dagger_p \times c^\dagger_d]_{3} \times[\tilde c_p \times \tilde c_f]_{3}\right]_{0} 
- h.c. \right)\\
&=
  \frac{1}{2} \left\{  
    -\frac{2\sqrt{2}}{3} \left( 
    {\bm T}^{pf}_2 \cdot {\bm G}_2^{pd} - {\bm Q}^{pf}_2 \cdot {\bm M}_2^{pd} 
    \right)
    +\frac{1}{3} \left( 
    {\bm G}^{pf}_3 \cdot {\bm T}_3^{pd} - {\bm M}^{pf}_3 \cdot {\bm Q}_3^{pd} 
    \right)
    \right\}\\
    &\qquad 
    +\frac{1}{2} \left\{  
    -\frac{\sqrt{6}}{3} 
    {\bm M}^{p }_1 \cdot {\bm G}_1^{df} 
    +\frac{\sqrt{3}}{3}
     {\bm Q}^{p }_2 \cdot {\bm T}_2^{df} 
     \right\}
  \\
&=
\frac{\sqrt{21} i}{28}\Big(- \frac{\sqrt{3}}{3}\,c^\dagger_{p,-1}\,c^\dagger_{d,-2}\,c_{p,-1}\,c_{f,-2} + c^\dagger_{p,-1}\,c^\dagger_{d,-2}\,c_{p,0}\,c_{f,-3} - \frac{\sqrt{30}}{9}\,c^\dagger_{p,-1}\,c^\dagger_{d,-1}\,c_{p,-1}\,c_{f,-1}
\\
&\qquad + \frac{2 \sqrt{6}}{9}\,c^\dagger_{p,-1}\,c^\dagger_{d,-1}\,c_{p,0}\,c_{f,-2} + \frac{\sqrt{2}}{3}\,c^\dagger_{p,-1}\,c^\dagger_{d,-1}\,c_{p,1}\,c_{f,-3} - \frac{\sqrt{15}}{9}\,c^\dagger_{p,0}\,c^\dagger_{d,-2}\,c_{p,-1}\,c_{f,-1}
\\
&\qquad + \frac{2 \sqrt{3}}{9}\,c^\dagger_{p,0}\,c^\dagger_{d,-2}\,c_{p,0}\,c_{f,-2} + \frac{1}{3}\,c^\dagger_{p,0}\,c^\dagger_{d,-2}\,c_{p,1}\,c_{f,-3} - \frac{2 \sqrt{15}}{15}\,c^\dagger_{p,-1}\,c^\dagger_{d,0}\,c_{p,-1}\,c_{f,0}
\\
&\qquad + \frac{\sqrt{10}}{15}\,c^\dagger_{p,-1}\,c^\dagger_{d,0}\,c_{p,0}\,c_{f,-1} + \frac{\sqrt{2}}{3}\,c^\dagger_{p,-1}\,c^\dagger_{d,0}\,c_{p,1}\,c_{f,-2} - \frac{4 \sqrt{5}}{15}\,c^\dagger_{p,0}\,c^\dagger_{d,-1}\,c_{p,-1}\,c_{f,0}
\\
&\qquad + \frac{2 \sqrt{30}}{45}\,c^\dagger_{p,0}\,c^\dagger_{d,-1}\,c_{p,0}\,c_{f,-1} + \frac{2 \sqrt{6}}{9}\,c^\dagger_{p,0}\,c^\dagger_{d,-1}\,c_{p,1}\,c_{f,-2} - \frac{\sqrt{10}}{15}\,c^\dagger_{p,1}\,c^\dagger_{d,-2}\,c_{p,-1}\,c_{f,0}
\\
&\qquad + \frac{\sqrt{15}}{45}\,c^\dagger_{p,1}\,c^\dagger_{d,-2}\,c_{p,0}\,c_{f,-1} + \frac{\sqrt{3}}{9}\,c^\dagger_{p,1}\,c^\dagger_{d,-2}\,c_{p,1}\,c_{f,-2} - \frac{\sqrt{30}}{15}\,c^\dagger_{p,-1}\,c^\dagger_{d,1}\,c_{p,-1}\,c_{f,1}
\\
&\qquad + \frac{\sqrt{30}}{15}\,c^\dagger_{p,-1}\,c^\dagger_{d,1}\,c_{p,1}\,c_{f,-1} - \frac{\sqrt{10}}{5}\,c^\dagger_{p,0}\,c^\dagger_{d,0}\,c_{p,-1}\,c_{f,1} + \frac{\sqrt{10}}{5}\,c^\dagger_{p,0}\,c^\dagger_{d,0}\,c_{p,1}\,c_{f,-1}
\\
&\qquad - \frac{\sqrt{30}}{15}\,c^\dagger_{p,1}\,c^\dagger_{d,-1}\,c_{p,-1}\,c_{f,1} + \frac{\sqrt{30}}{15}\,c^\dagger_{p,1}\,c^\dagger_{d,-1}\,c_{p,1}\,c_{f,-1} - \frac{\sqrt{3}}{9}\,c^\dagger_{p,-1}\,c^\dagger_{d,2}\,c_{p,-1}\,c_{f,2}
\\
&\qquad - \frac{\sqrt{15}}{45}\,c^\dagger_{p,-1}\,c^\dagger_{d,2}\,c_{p,0}\,c_{f,1} + \frac{\sqrt{10}}{15}\,c^\dagger_{p,-1}\,c^\dagger_{d,2}\,c_{p,1}\,c_{f,0} - \frac{2 \sqrt{6}}{9}\,c^\dagger_{p,0}\,c^\dagger_{d,1}\,c_{p,-1}\,c_{f,2}
\\
&\qquad - \frac{2 \sqrt{30}}{45}\,c^\dagger_{p,0}\,c^\dagger_{d,1}\,c_{p,0}\,c_{f,1} + \frac{4 \sqrt{5}}{15}\,c^\dagger_{p,0}\,c^\dagger_{d,1}\,c_{p,1}\,c_{f,0} - \frac{\sqrt{2}}{3}\,c^\dagger_{p,1}\,c^\dagger_{d,0}\,c_{p,-1}\,c_{f,2}
\\
&\qquad - \frac{\sqrt{10}}{15}\,c^\dagger_{p,1}\,c^\dagger_{d,0}\,c_{p,0}\,c_{f,1} + \frac{2 \sqrt{15}}{15}\,c^\dagger_{p,1}\,c^\dagger_{d,0}\,c_{p,1}\,c_{f,0} - \frac{1}{3}\,c^\dagger_{p,0}\,c^\dagger_{d,2}\,c_{p,-1}\,c_{f,3}
\\
&\qquad - \frac{2 \sqrt{3}}{9}\,c^\dagger_{p,0}\,c^\dagger_{d,2}\,c_{p,0}\,c_{f,2} + \frac{\sqrt{15}}{9}\,c^\dagger_{p,0}\,c^\dagger_{d,2}\,c_{p,1}\,c_{f,1} - \frac{\sqrt{2}}{3}\,c^\dagger_{p,1}\,c^\dagger_{d,1}\,c_{p,-1}\,c_{f,3}
\\
&\qquad - \frac{2 \sqrt{6}}{9}\,c^\dagger_{p,1}\,c^\dagger_{d,1}\,c_{p,0}\,c_{f,2} + \frac{\sqrt{30}}{9}\,c^\dagger_{p,1}\,c^\dagger_{d,1}\,c_{p,1}\,c_{f,1} - c^\dagger_{p,1}\,c^\dagger_{d,2}\,c_{p,0}\,c_{f,3}
\\
&\qquad + \frac{\sqrt{3}}{3}\,c^\dagger_{p,1}\,c^\dagger_{d,2}\,c_{p,1}\,c_{f,2} + \frac{\sqrt{3}}{3}\,c^\dagger_{p,-1}\,c^\dagger_{f,-2}\,c_{p,-1}\,c_{d,-2} - c^\dagger_{p,0}\,c^\dagger_{f,-3}\,c_{p,-1}\,c_{d,-2}
\\
&\qquad + \frac{\sqrt{30}}{9}\,c^\dagger_{p,-1}\,c^\dagger_{f,-1}\,c_{p,-1}\,c_{d,-1} - \frac{2 \sqrt{6}}{9}\,c^\dagger_{p,0}\,c^\dagger_{f,-2}\,c_{p,-1}\,c_{d,-1} - \frac{\sqrt{2}}{3}\,c^\dagger_{p,1}\,c^\dagger_{f,-3}\,c_{p,-1}\,c_{d,-1}
\\
&\qquad + \frac{\sqrt{15}}{9}\,c^\dagger_{p,-1}\,c^\dagger_{f,-1}\,c_{p,0}\,c_{d,-2} - \frac{2 \sqrt{3}}{9}\,c^\dagger_{p,0}\,c^\dagger_{f,-2}\,c_{p,0}\,c_{d,-2} - \frac{1}{3}\,c^\dagger_{p,1}\,c^\dagger_{f,-3}\,c_{p,0}\,c_{d,-2}
\\
&\qquad + \frac{2 \sqrt{15}}{15}\,c^\dagger_{p,-1}\,c^\dagger_{f,0}\,c_{p,-1}\,c_{d,0} - \frac{\sqrt{10}}{15}\,c^\dagger_{p,0}\,c^\dagger_{f,-1}\,c_{p,-1}\,c_{d,0} - \frac{\sqrt{2}}{3}\,c^\dagger_{p,1}\,c^\dagger_{f,-2}\,c_{p,-1}\,c_{d,0}
\\
&\qquad + \frac{4 \sqrt{5}}{15}\,c^\dagger_{p,-1}\,c^\dagger_{f,0}\,c_{p,0}\,c_{d,-1} - \frac{2 \sqrt{30}}{45}\,c^\dagger_{p,0}\,c^\dagger_{f,-1}\,c_{p,0}\,c_{d,-1} - \frac{2 \sqrt{6}}{9}\,c^\dagger_{p,1}\,c^\dagger_{f,-2}\,c_{p,0}\,c_{d,-1}
\\
&\qquad + \frac{\sqrt{10}}{15}\,c^\dagger_{p,-1}\,c^\dagger_{f,0}\,c_{p,1}\,c_{d,-2} - \frac{\sqrt{15}}{45}\,c^\dagger_{p,0}\,c^\dagger_{f,-1}\,c_{p,1}\,c_{d,-2} - \frac{\sqrt{3}}{9}\,c^\dagger_{p,1}\,c^\dagger_{f,-2}\,c_{p,1}\,c_{d,-2}
\\
&\qquad + \frac{\sqrt{30}}{15}\,c^\dagger_{p,-1}\,c^\dagger_{f,1}\,c_{p,-1}\,c_{d,1} - \frac{\sqrt{30}}{15}\,c^\dagger_{p,1}\,c^\dagger_{f,-1}\,c_{p,-1}\,c_{d,1} + \frac{\sqrt{10}}{5}\,c^\dagger_{p,-1}\,c^\dagger_{f,1}\,c_{p,0}\,c_{d,0}
\\
&\qquad - \frac{\sqrt{10}}{5}\,c^\dagger_{p,1}\,c^\dagger_{f,-1}\,c_{p,0}\,c_{d,0} + \frac{\sqrt{30}}{15}\,c^\dagger_{p,-1}\,c^\dagger_{f,1}\,c_{p,1}\,c_{d,-1} - \frac{\sqrt{30}}{15}\,c^\dagger_{p,1}\,c^\dagger_{f,-1}\,c_{p,1}\,c_{d,-1}
\\
&\qquad + \frac{\sqrt{3}}{9}\,c^\dagger_{p,-1}\,c^\dagger_{f,2}\,c_{p,-1}\,c_{d,2} + \frac{\sqrt{15}}{45}\,c^\dagger_{p,0}\,c^\dagger_{f,1}\,c_{p,-1}\,c_{d,2} - \frac{\sqrt{10}}{15}\,c^\dagger_{p,1}\,c^\dagger_{f,0}\,c_{p,-1}\,c_{d,2}
\\
&\qquad + \frac{2 \sqrt{6}}{9}\,c^\dagger_{p,-1}\,c^\dagger_{f,2}\,c_{p,0}\,c_{d,1} + \frac{2 \sqrt{30}}{45}\,c^\dagger_{p,0}\,c^\dagger_{f,1}\,c_{p,0}\,c_{d,1} - \frac{4 \sqrt{5}}{15}\,c^\dagger_{p,1}\,c^\dagger_{f,0}\,c_{p,0}\,c_{d,1}
\\
&\qquad + \frac{\sqrt{2}}{3}\,c^\dagger_{p,-1}\,c^\dagger_{f,2}\,c_{p,1}\,c_{d,0} + \frac{\sqrt{10}}{15}\,c^\dagger_{p,0}\,c^\dagger_{f,1}\,c_{p,1}\,c_{d,0} - \frac{2 \sqrt{15}}{15}\,c^\dagger_{p,1}\,c^\dagger_{f,0}\,c_{p,1}\,c_{d,0}
\\
&\qquad + \frac{1}{3}\,c^\dagger_{p,-1}\,c^\dagger_{f,3}\,c_{p,0}\,c_{d,2} + \frac{2 \sqrt{3}}{9}\,c^\dagger_{p,0}\,c^\dagger_{f,2}\,c_{p,0}\,c_{d,2} - \frac{\sqrt{15}}{9}\,c^\dagger_{p,1}\,c^\dagger_{f,1}\,c_{p,0}\,c_{d,2}
\\
&\qquad + \frac{\sqrt{2}}{3}\,c^\dagger_{p,-1}\,c^\dagger_{f,3}\,c_{p,1}\,c_{d,1} + \frac{2 \sqrt{6}}{9}\,c^\dagger_{p,0}\,c^\dagger_{f,2}\,c_{p,1}\,c_{d,1} - \frac{\sqrt{30}}{9}\,c^\dagger_{p,1}\,c^\dagger_{f,1}\,c_{p,1}\,c_{d,1}
\\
&\qquad + c^\dagger_{p,0}\,c^\dagger_{f,3}\,c_{p,1}\,c_{d,2} - \frac{\sqrt{3}}{3}\,c^\dagger_{p,1}\,c^\dagger_{f,2}\,c_{p,1}\,c_{d,2}\Big)
\end{align*}

% pdfp : 3×3
% j13,j24,A wigner_9j： 2 , 2 , -2*sqrt(2)/3
% j13,j24,A wigner_9j： 3 , 3 , 1/3
% pdpf : 3×3
% j13,j24,A wigner_9j： 1 , 1 , sqrt(6)/3
% j13,j24,A wigner_9j： 2 , 2 , -sqrt(3)/3

\begin{align*}
{G_0^{(2)}}_{[3]} 
&=  
  \frac{1}{2} \left(\left[[c^\dagger_p \times c^\dagger_p]_{1} \times[\tilde c_d \times \tilde c_f]_{1}\right]_{0} + h.c. \right)\\
&=
    - \frac{\sqrt{3}}{3} \left( 
    {\bm Q}^{pd}_2 \cdot {\bm G}_2^{pf} - {\bm T}^{pd}_2 \cdot {\bm M}_2^{pf} 
    \right)
    - \frac{\sqrt{6}}{3} \left( 
    {\bm M}^{pd}_3 \cdot {\bm T}_3^{pf} - {\bm G}^{pd}_3 \cdot {\bm Q}_3^{pf} 
    \right)
  \\
&=
\frac{\sqrt{14}}{14}\Big(c^\dagger_{p,0}\,c^\dagger_{p,-1}\,c_{d,2}\,c_{f,-3} - \frac{\sqrt{6}}{3}\,c^\dagger_{p,0}\,c^\dagger_{p,-1}\,c_{d,1}\,c_{f,-2} + \frac{\sqrt{10}}{5}\,c^\dagger_{p,0}\,c^\dagger_{p,-1}\,c_{d,0}\,c_{f,-1}
\\
&\qquad - \frac{\sqrt{5}}{5}\,c^\dagger_{p,0}\,c^\dagger_{p,-1}\,c_{d,-1}\,c_{f,0} + \frac{\sqrt{15}}{15}\,c^\dagger_{p,0}\,c^\dagger_{p,-1}\,c_{d,-2}\,c_{f,1} + \frac{\sqrt{3}}{3}\,c^\dagger_{p,1}\,c^\dagger_{p,-1}\,c_{d,2}\,c_{f,-2}
\\
&\qquad - \frac{2 \sqrt{30}}{15}\,c^\dagger_{p,1}\,c^\dagger_{p,-1}\,c_{d,1}\,c_{f,-1} + \frac{\sqrt{15}}{5}\,c^\dagger_{p,1}\,c^\dagger_{p,-1}\,c_{d,0}\,c_{f,0} - \frac{2 \sqrt{30}}{15}\,c^\dagger_{p,1}\,c^\dagger_{p,-1}\,c_{d,-1}\,c_{f,1}
\\
&\qquad + \frac{\sqrt{3}}{3}\,c^\dagger_{p,1}\,c^\dagger_{p,-1}\,c_{d,-2}\,c_{f,2} + \frac{\sqrt{15}}{15}\,c^\dagger_{p,1}\,c^\dagger_{p,0}\,c_{d,2}\,c_{f,-1} - \frac{\sqrt{5}}{5}\,c^\dagger_{p,1}\,c^\dagger_{p,0}\,c_{d,1}\,c_{f,0}
\\
&\qquad + \frac{\sqrt{10}}{5}\,c^\dagger_{p,1}\,c^\dagger_{p,0}\,c_{d,0}\,c_{f,1} - \frac{\sqrt{6}}{3}\,c^\dagger_{p,1}\,c^\dagger_{p,0}\,c_{d,-1}\,c_{f,2} + c^\dagger_{p,1}\,c^\dagger_{p,0}\,c_{d,-2}\,c_{f,3}
\\
&\qquad + c^\dagger_{d,2}\,c^\dagger_{f,-3}\,c_{p,0}\,c_{p,-1} - \frac{\sqrt{6}}{3}\,c^\dagger_{d,1}\,c^\dagger_{f,-2}\,c_{p,0}\,c_{p,-1} + \frac{\sqrt{10}}{5}\,c^\dagger_{d,0}\,c^\dagger_{f,-1}\,c_{p,0}\,c_{p,-1}
\\
&\qquad - \frac{\sqrt{5}}{5}\,c^\dagger_{d,-1}\,c^\dagger_{f,0}\,c_{p,0}\,c_{p,-1} + \frac{\sqrt{15}}{15}\,c^\dagger_{d,-2}\,c^\dagger_{f,1}\,c_{p,0}\,c_{p,-1} + \frac{\sqrt{3}}{3}\,c^\dagger_{d,2}\,c^\dagger_{f,-2}\,c_{p,1}\,c_{p,-1}
\\
&\qquad - \frac{2 \sqrt{30}}{15}\,c^\dagger_{d,1}\,c^\dagger_{f,-1}\,c_{p,1}\,c_{p,-1} + \frac{\sqrt{15}}{5}\,c^\dagger_{d,0}\,c^\dagger_{f,0}\,c_{p,1}\,c_{p,-1} - \frac{2 \sqrt{30}}{15}\,c^\dagger_{d,-1}\,c^\dagger_{f,1}\,c_{p,1}\,c_{p,-1}
\\
&\qquad + \frac{\sqrt{3}}{3}\,c^\dagger_{d,-2}\,c^\dagger_{f,2}\,c_{p,1}\,c_{p,-1} + \frac{\sqrt{15}}{15}\,c^\dagger_{d,2}\,c^\dagger_{f,-1}\,c_{p,1}\,c_{p,0} - \frac{\sqrt{5}}{5}\,c^\dagger_{d,1}\,c^\dagger_{f,0}\,c_{p,1}\,c_{p,0}
\\
&\qquad + \frac{\sqrt{10}}{5}\,c^\dagger_{d,0}\,c^\dagger_{f,1}\,c_{p,1}\,c_{p,0} - \frac{\sqrt{6}}{3}\,c^\dagger_{d,-1}\,c^\dagger_{f,2}\,c_{p,1}\,c_{p,0} + c^\dagger_{d,-2}\,c^\dagger_{f,3}\,c_{p,1}\,c_{p,0}\Big)
\\
{M_0^{(2)}}_{[3]} 
&=  
  \frac{1}{2} \left(\left[[c^\dagger_p \times c^\dagger_p]_{1} \times[\tilde c_d \times \tilde c_f]_{1}\right]_{0} + h.c. \right)\\
&=
    - \frac{\sqrt{3}}{3} \left( 
   {\bm T}^{pd}_2 \cdot {\bm G}_2^{pf}  +{\bm Q}^{pd}_2 \cdot {\bm M}_2^{pf}  
    \right)
    - \frac{\sqrt{6}}{3} \left( 
   {\bm G}^{pd}_3 \cdot {\bm T}_3^{pf}  +{\bm M}^{pd}_3 \cdot {\bm Q}_3^{pf}  
    \right)
  \\
&=
\frac{\sqrt{14} i}{14}\Big(c^\dagger_{p,0}\,c^\dagger_{p,-1}\,c_{d,2}\,c_{f,-3} - \frac{\sqrt{6}}{3}\,c^\dagger_{p,0}\,c^\dagger_{p,-1}\,c_{d,1}\,c_{f,-2} + \frac{\sqrt{10}}{5}\,c^\dagger_{p,0}\,c^\dagger_{p,-1}\,c_{d,0}\,c_{f,-1}
\\
&\qquad - \frac{\sqrt{5}}{5}\,c^\dagger_{p,0}\,c^\dagger_{p,-1}\,c_{d,-1}\,c_{f,0} + \frac{\sqrt{15}}{15}\,c^\dagger_{p,0}\,c^\dagger_{p,-1}\,c_{d,-2}\,c_{f,1} + \frac{\sqrt{3}}{3}\,c^\dagger_{p,1}\,c^\dagger_{p,-1}\,c_{d,2}\,c_{f,-2}
\\
&\qquad - \frac{2 \sqrt{30}}{15}\,c^\dagger_{p,1}\,c^\dagger_{p,-1}\,c_{d,1}\,c_{f,-1} + \frac{\sqrt{15}}{5}\,c^\dagger_{p,1}\,c^\dagger_{p,-1}\,c_{d,0}\,c_{f,0} - \frac{2 \sqrt{30}}{15}\,c^\dagger_{p,1}\,c^\dagger_{p,-1}\,c_{d,-1}\,c_{f,1}
\\
&\qquad + \frac{\sqrt{3}}{3}\,c^\dagger_{p,1}\,c^\dagger_{p,-1}\,c_{d,-2}\,c_{f,2} + \frac{\sqrt{15}}{15}\,c^\dagger_{p,1}\,c^\dagger_{p,0}\,c_{d,2}\,c_{f,-1} - \frac{\sqrt{5}}{5}\,c^\dagger_{p,1}\,c^\dagger_{p,0}\,c_{d,1}\,c_{f,0}
\\
&\qquad + \frac{\sqrt{10}}{5}\,c^\dagger_{p,1}\,c^\dagger_{p,0}\,c_{d,0}\,c_{f,1} - \frac{\sqrt{6}}{3}\,c^\dagger_{p,1}\,c^\dagger_{p,0}\,c_{d,-1}\,c_{f,2} + c^\dagger_{p,1}\,c^\dagger_{p,0}\,c_{d,-2}\,c_{f,3}
\\
&\qquad - c^\dagger_{d,2}\,c^\dagger_{f,-3}\,c_{p,0}\,c_{p,-1} + \frac{\sqrt{6}}{3}\,c^\dagger_{d,1}\,c^\dagger_{f,-2}\,c_{p,0}\,c_{p,-1} - \frac{\sqrt{10}}{5}\,c^\dagger_{d,0}\,c^\dagger_{f,-1}\,c_{p,0}\,c_{p,-1}
\\
&\qquad + \frac{\sqrt{5}}{5}\,c^\dagger_{d,-1}\,c^\dagger_{f,0}\,c_{p,0}\,c_{p,-1} - \frac{\sqrt{15}}{15}\,c^\dagger_{d,-2}\,c^\dagger_{f,1}\,c_{p,0}\,c_{p,-1} - \frac{\sqrt{3}}{3}\,c^\dagger_{d,2}\,c^\dagger_{f,-2}\,c_{p,1}\,c_{p,-1}
\\
&\qquad + \frac{2 \sqrt{30}}{15}\,c^\dagger_{d,1}\,c^\dagger_{f,-1}\,c_{p,1}\,c_{p,-1} - \frac{\sqrt{15}}{5}\,c^\dagger_{d,0}\,c^\dagger_{f,0}\,c_{p,1}\,c_{p,-1} + \frac{2 \sqrt{30}}{15}\,c^\dagger_{d,-1}\,c^\dagger_{f,1}\,c_{p,1}\,c_{p,-1}
\\
&\qquad - \frac{\sqrt{3}}{3}\,c^\dagger_{d,-2}\,c^\dagger_{f,2}\,c_{p,1}\,c_{p,-1} - \frac{\sqrt{15}}{15}\,c^\dagger_{d,2}\,c^\dagger_{f,-1}\,c_{p,1}\,c_{p,0} + \frac{\sqrt{5}}{5}\,c^\dagger_{d,1}\,c^\dagger_{f,0}\,c_{p,1}\,c_{p,0}
\\
&\qquad - \frac{\sqrt{10}}{5}\,c^\dagger_{d,0}\,c^\dagger_{f,1}\,c_{p,1}\,c_{p,0} + \frac{\sqrt{6}}{3}\,c^\dagger_{d,-1}\,c^\dagger_{f,2}\,c_{p,1}\,c_{p,0} - c^\dagger_{d,-2}\,c^\dagger_{f,3}\,c_{p,1}\,c_{p,0}\Big)
\end{align*}

% ppdf : 1×1
% j13,j24,A wigner_9j： 2 , 2 , sqrt(3)/3
% j13,j24,A wigner_9j： 3 , 3 , sqrt(6)/3

\begin{align*}
{G_0^{(2)}}_{[4]} 
&=  
  \frac{1}{2} \left(\left[[c^\dagger_p \times c^\dagger_d]_{1} \times[\tilde c_f \times \tilde c_f]_{1}\right]_{0} + h.c. \right)\\
&=
    \frac{\sqrt{7}}{7} \left( 
    {\bm Q}^{pf}_2 \cdot {\bm G}_2^{df} - {\bm T}^{pf}_2 \cdot {\bm M}_2^{df} 
    \right)
    +\frac{\sqrt{6}}{4} \left( 
    {\bm M}^{pf}_3 \cdot {\bm T}_3^{df} - {\bm G}^{pf}_3 \cdot {\bm Q}_3^{df} 
    \right)
    \\&\qquad    
    + \frac{3\sqrt{42}}{28} \left( 
    {\bm Q}^{pf}_4 \cdot {\bm G}_4^{df} - {\bm T}^{pf}_4 \cdot {\bm M}_4^{df} 
    \right)
  \\
&=
\frac{\sqrt{210}}{70}\Big(\frac{\sqrt{3}}{6}\,c^\dagger_{p,-1}\,c^\dagger_{d,0}\,c_{f,2}\,c_{f,-3} - \frac{\sqrt{5}}{6}\,c^\dagger_{p,-1}\,c^\dagger_{d,0}\,c_{f,1}\,c_{f,-2} + \frac{\sqrt{6}}{6}\,c^\dagger_{p,-1}\,c^\dagger_{d,0}\,c_{f,0}\,c_{f,-1}
\\
&\qquad - \frac{1}{2}\,c^\dagger_{p,0}\,c^\dagger_{d,-1}\,c_{f,2}\,c_{f,-3} + \frac{\sqrt{15}}{6}\,c^\dagger_{p,0}\,c^\dagger_{d,-1}\,c_{f,1}\,c_{f,-2} - \frac{\sqrt{2}}{2}\,c^\dagger_{p,0}\,c^\dagger_{d,-1}\,c_{f,0}\,c_{f,-1}
\\
&\qquad + \frac{\sqrt{2}}{2}\,c^\dagger_{p,1}\,c^\dagger_{d,-2}\,c_{f,2}\,c_{f,-3} - \frac{\sqrt{30}}{6}\,c^\dagger_{p,1}\,c^\dagger_{d,-2}\,c_{f,1}\,c_{f,-2} + c^\dagger_{p,1}\,c^\dagger_{d,-2}\,c_{f,0}\,c_{f,-1}
\\
&\qquad + \frac{\sqrt{3}}{2}\,c^\dagger_{p,-1}\,c^\dagger_{d,1}\,c_{f,3}\,c_{f,-3} - \frac{\sqrt{3}}{3}\,c^\dagger_{p,-1}\,c^\dagger_{d,1}\,c_{f,2}\,c_{f,-2} + \frac{\sqrt{3}}{6}\,c^\dagger_{p,-1}\,c^\dagger_{d,1}\,c_{f,1}\,c_{f,-1}
\\
&\qquad - c^\dagger_{p,0}\,c^\dagger_{d,0}\,c_{f,3}\,c_{f,-3} + \frac{2}{3}\,c^\dagger_{p,0}\,c^\dagger_{d,0}\,c_{f,2}\,c_{f,-2} - \frac{1}{3}\,c^\dagger_{p,0}\,c^\dagger_{d,0}\,c_{f,1}\,c_{f,-1}
\\
&\qquad + \frac{\sqrt{3}}{2}\,c^\dagger_{p,1}\,c^\dagger_{d,-1}\,c_{f,3}\,c_{f,-3} - \frac{\sqrt{3}}{3}\,c^\dagger_{p,1}\,c^\dagger_{d,-1}\,c_{f,2}\,c_{f,-2} + \frac{\sqrt{3}}{6}\,c^\dagger_{p,1}\,c^\dagger_{d,-1}\,c_{f,1}\,c_{f,-1}
\\
&\qquad + \frac{\sqrt{2}}{2}\,c^\dagger_{p,-1}\,c^\dagger_{d,2}\,c_{f,3}\,c_{f,-2} - \frac{\sqrt{30}}{6}\,c^\dagger_{p,-1}\,c^\dagger_{d,2}\,c_{f,2}\,c_{f,-1} + c^\dagger_{p,-1}\,c^\dagger_{d,2}\,c_{f,1}\,c_{f,0}
\\
&\qquad - \frac{1}{2}\,c^\dagger_{p,0}\,c^\dagger_{d,1}\,c_{f,3}\,c_{f,-2} + \frac{\sqrt{15}}{6}\,c^\dagger_{p,0}\,c^\dagger_{d,1}\,c_{f,2}\,c_{f,-1} - \frac{\sqrt{2}}{2}\,c^\dagger_{p,0}\,c^\dagger_{d,1}\,c_{f,1}\,c_{f,0}
\\
&\qquad + \frac{\sqrt{3}}{6}\,c^\dagger_{p,1}\,c^\dagger_{d,0}\,c_{f,3}\,c_{f,-2} - \frac{\sqrt{5}}{6}\,c^\dagger_{p,1}\,c^\dagger_{d,0}\,c_{f,2}\,c_{f,-1} + \frac{\sqrt{6}}{6}\,c^\dagger_{p,1}\,c^\dagger_{d,0}\,c_{f,1}\,c_{f,0}
\\
&\qquad + \frac{\sqrt{3}}{6}\,c^\dagger_{f,2}\,c^\dagger_{f,-3}\,c_{p,-1}\,c_{d,0} - \frac{\sqrt{5}}{6}\,c^\dagger_{f,1}\,c^\dagger_{f,-2}\,c_{p,-1}\,c_{d,0} + \frac{\sqrt{6}}{6}\,c^\dagger_{f,0}\,c^\dagger_{f,-1}\,c_{p,-1}\,c_{d,0}
\\
&\qquad - \frac{1}{2}\,c^\dagger_{f,2}\,c^\dagger_{f,-3}\,c_{p,0}\,c_{d,-1} + \frac{\sqrt{15}}{6}\,c^\dagger_{f,1}\,c^\dagger_{f,-2}\,c_{p,0}\,c_{d,-1} - \frac{\sqrt{2}}{2}\,c^\dagger_{f,0}\,c^\dagger_{f,-1}\,c_{p,0}\,c_{d,-1}
\\
&\qquad + \frac{\sqrt{2}}{2}\,c^\dagger_{f,2}\,c^\dagger_{f,-3}\,c_{p,1}\,c_{d,-2} - \frac{\sqrt{30}}{6}\,c^\dagger_{f,1}\,c^\dagger_{f,-2}\,c_{p,1}\,c_{d,-2} + c^\dagger_{f,0}\,c^\dagger_{f,-1}\,c_{p,1}\,c_{d,-2}
\\
&\qquad + \frac{\sqrt{3}}{2}\,c^\dagger_{f,3}\,c^\dagger_{f,-3}\,c_{p,-1}\,c_{d,1} - \frac{\sqrt{3}}{3}\,c^\dagger_{f,2}\,c^\dagger_{f,-2}\,c_{p,-1}\,c_{d,1} + \frac{\sqrt{3}}{6}\,c^\dagger_{f,1}\,c^\dagger_{f,-1}\,c_{p,-1}\,c_{d,1}
\\
&\qquad - c^\dagger_{f,3}\,c^\dagger_{f,-3}\,c_{p,0}\,c_{d,0} + \frac{2}{3}\,c^\dagger_{f,2}\,c^\dagger_{f,-2}\,c_{p,0}\,c_{d,0} - \frac{1}{3}\,c^\dagger_{f,1}\,c^\dagger_{f,-1}\,c_{p,0}\,c_{d,0}
\\
&\qquad + \frac{\sqrt{3}}{2}\,c^\dagger_{f,3}\,c^\dagger_{f,-3}\,c_{p,1}\,c_{d,-1} - \frac{\sqrt{3}}{3}\,c^\dagger_{f,2}\,c^\dagger_{f,-2}\,c_{p,1}\,c_{d,-1} + \frac{\sqrt{3}}{6}\,c^\dagger_{f,1}\,c^\dagger_{f,-1}\,c_{p,1}\,c_{d,-1}
\\
&\qquad + \frac{\sqrt{2}}{2}\,c^\dagger_{f,3}\,c^\dagger_{f,-2}\,c_{p,-1}\,c_{d,2} - \frac{\sqrt{30}}{6}\,c^\dagger_{f,2}\,c^\dagger_{f,-1}\,c_{p,-1}\,c_{d,2} + c^\dagger_{f,1}\,c^\dagger_{f,0}\,c_{p,-1}\,c_{d,2}
\\
&\qquad - \frac{1}{2}\,c^\dagger_{f,3}\,c^\dagger_{f,-2}\,c_{p,0}\,c_{d,1} + \frac{\sqrt{15}}{6}\,c^\dagger_{f,2}\,c^\dagger_{f,-1}\,c_{p,0}\,c_{d,1} - \frac{\sqrt{2}}{2}\,c^\dagger_{f,1}\,c^\dagger_{f,0}\,c_{p,0}\,c_{d,1}
\\
&\qquad + \frac{\sqrt{3}}{6}\,c^\dagger_{f,3}\,c^\dagger_{f,-2}\,c_{p,1}\,c_{d,0} - \frac{\sqrt{5}}{6}\,c^\dagger_{f,2}\,c^\dagger_{f,-1}\,c_{p,1}\,c_{d,0} + \frac{\sqrt{6}}{6}\,c^\dagger_{f,1}\,c^\dagger_{f,0}\,c_{p,1}\,c_{d,0}\Big)
\\
{M_0^{(2)}}_{[4]} 
&=  
  \frac{i}{2} \left(\left[[c^\dagger_p \times c^\dagger_d]_{1} \times[\tilde c_f \times \tilde c_f]_{1}\right]_{0} - h.c. \right)\\
&=
    \frac{\sqrt{7}}{7} \left( 
   {\bm T}^{pf}_2 \cdot {\bm G}_2^{df}  +{\bm Q}^{pf}_2 \cdot {\bm M}_2^{df}  
    \right)
    +\frac{\sqrt{6}}{4} \left( 
   {\bm G}^{pf}_3 \cdot {\bm T}_3^{df}  +{\bm M}^{pf}_3 \cdot {\bm Q}_3^{df}  
    \right)
    \\&\qquad    
    + \frac{3\sqrt{42}}{28} \left( 
   {\bm T}^{pf}_4 \cdot {\bm G}_4^{df}  +{\bm Q}^{pf}_4 \cdot {\bm M}_4^{df}  
    \right)
  \\
&=
\frac{\sqrt{210} i}{70}\Big(\frac{\sqrt{3}}{6}\,c^\dagger_{p,-1}\,c^\dagger_{d,0}\,c_{f,2}\,c_{f,-3} - \frac{\sqrt{5}}{6}\,c^\dagger_{p,-1}\,c^\dagger_{d,0}\,c_{f,1}\,c_{f,-2} + \frac{\sqrt{6}}{6}\,c^\dagger_{p,-1}\,c^\dagger_{d,0}\,c_{f,0}\,c_{f,-1}
\\
&\qquad - \frac{1}{2}\,c^\dagger_{p,0}\,c^\dagger_{d,-1}\,c_{f,2}\,c_{f,-3} + \frac{\sqrt{15}}{6}\,c^\dagger_{p,0}\,c^\dagger_{d,-1}\,c_{f,1}\,c_{f,-2} - \frac{\sqrt{2}}{2}\,c^\dagger_{p,0}\,c^\dagger_{d,-1}\,c_{f,0}\,c_{f,-1}
\\
&\qquad + \frac{\sqrt{2}}{2}\,c^\dagger_{p,1}\,c^\dagger_{d,-2}\,c_{f,2}\,c_{f,-3} - \frac{\sqrt{30}}{6}\,c^\dagger_{p,1}\,c^\dagger_{d,-2}\,c_{f,1}\,c_{f,-2} + c^\dagger_{p,1}\,c^\dagger_{d,-2}\,c_{f,0}\,c_{f,-1}
\\
&\qquad + \frac{\sqrt{3}}{2}\,c^\dagger_{p,-1}\,c^\dagger_{d,1}\,c_{f,3}\,c_{f,-3} - \frac{\sqrt{3}}{3}\,c^\dagger_{p,-1}\,c^\dagger_{d,1}\,c_{f,2}\,c_{f,-2} + \frac{\sqrt{3}}{6}\,c^\dagger_{p,-1}\,c^\dagger_{d,1}\,c_{f,1}\,c_{f,-1}
\\
&\qquad - c^\dagger_{p,0}\,c^\dagger_{d,0}\,c_{f,3}\,c_{f,-3} + \frac{2}{3}\,c^\dagger_{p,0}\,c^\dagger_{d,0}\,c_{f,2}\,c_{f,-2} - \frac{1}{3}\,c^\dagger_{p,0}\,c^\dagger_{d,0}\,c_{f,1}\,c_{f,-1}
\\
&\qquad + \frac{\sqrt{3}}{2}\,c^\dagger_{p,1}\,c^\dagger_{d,-1}\,c_{f,3}\,c_{f,-3} - \frac{\sqrt{3}}{3}\,c^\dagger_{p,1}\,c^\dagger_{d,-1}\,c_{f,2}\,c_{f,-2} + \frac{\sqrt{3}}{6}\,c^\dagger_{p,1}\,c^\dagger_{d,-1}\,c_{f,1}\,c_{f,-1}
\\
&\qquad + \frac{\sqrt{2}}{2}\,c^\dagger_{p,-1}\,c^\dagger_{d,2}\,c_{f,3}\,c_{f,-2} - \frac{\sqrt{30}}{6}\,c^\dagger_{p,-1}\,c^\dagger_{d,2}\,c_{f,2}\,c_{f,-1} + c^\dagger_{p,-1}\,c^\dagger_{d,2}\,c_{f,1}\,c_{f,0}
\\
&\qquad - \frac{1}{2}\,c^\dagger_{p,0}\,c^\dagger_{d,1}\,c_{f,3}\,c_{f,-2} + \frac{\sqrt{15}}{6}\,c^\dagger_{p,0}\,c^\dagger_{d,1}\,c_{f,2}\,c_{f,-1} - \frac{\sqrt{2}}{2}\,c^\dagger_{p,0}\,c^\dagger_{d,1}\,c_{f,1}\,c_{f,0}
\\
&\qquad + \frac{\sqrt{3}}{6}\,c^\dagger_{p,1}\,c^\dagger_{d,0}\,c_{f,3}\,c_{f,-2} - \frac{\sqrt{5}}{6}\,c^\dagger_{p,1}\,c^\dagger_{d,0}\,c_{f,2}\,c_{f,-1} + \frac{\sqrt{6}}{6}\,c^\dagger_{p,1}\,c^\dagger_{d,0}\,c_{f,1}\,c_{f,0}
\\
&\qquad - \frac{\sqrt{3}}{6}\,c^\dagger_{f,2}\,c^\dagger_{f,-3}\,c_{p,-1}\,c_{d,0} + \frac{\sqrt{5}}{6}\,c^\dagger_{f,1}\,c^\dagger_{f,-2}\,c_{p,-1}\,c_{d,0} - \frac{\sqrt{6}}{6}\,c^\dagger_{f,0}\,c^\dagger_{f,-1}\,c_{p,-1}\,c_{d,0}
\\
&\qquad + \frac{1}{2}\,c^\dagger_{f,2}\,c^\dagger_{f,-3}\,c_{p,0}\,c_{d,-1} - \frac{\sqrt{15}}{6}\,c^\dagger_{f,1}\,c^\dagger_{f,-2}\,c_{p,0}\,c_{d,-1} + \frac{\sqrt{2}}{2}\,c^\dagger_{f,0}\,c^\dagger_{f,-1}\,c_{p,0}\,c_{d,-1}
\\
&\qquad - \frac{\sqrt{2}}{2}\,c^\dagger_{f,2}\,c^\dagger_{f,-3}\,c_{p,1}\,c_{d,-2} + \frac{\sqrt{30}}{6}\,c^\dagger_{f,1}\,c^\dagger_{f,-2}\,c_{p,1}\,c_{d,-2} - c^\dagger_{f,0}\,c^\dagger_{f,-1}\,c_{p,1}\,c_{d,-2}
\\
&\qquad - \frac{\sqrt{3}}{2}\,c^\dagger_{f,3}\,c^\dagger_{f,-3}\,c_{p,-1}\,c_{d,1} + \frac{\sqrt{3}}{3}\,c^\dagger_{f,2}\,c^\dagger_{f,-2}\,c_{p,-1}\,c_{d,1} - \frac{\sqrt{3}}{6}\,c^\dagger_{f,1}\,c^\dagger_{f,-1}\,c_{p,-1}\,c_{d,1}
\\
&\qquad + c^\dagger_{f,3}\,c^\dagger_{f,-3}\,c_{p,0}\,c_{d,0} - \frac{2}{3}\,c^\dagger_{f,2}\,c^\dagger_{f,-2}\,c_{p,0}\,c_{d,0} + \frac{1}{3}\,c^\dagger_{f,1}\,c^\dagger_{f,-1}\,c_{p,0}\,c_{d,0}
\\
&\qquad - \frac{\sqrt{3}}{2}\,c^\dagger_{f,3}\,c^\dagger_{f,-3}\,c_{p,1}\,c_{d,-1} + \frac{\sqrt{3}}{3}\,c^\dagger_{f,2}\,c^\dagger_{f,-2}\,c_{p,1}\,c_{d,-1} - \frac{\sqrt{3}}{6}\,c^\dagger_{f,1}\,c^\dagger_{f,-1}\,c_{p,1}\,c_{d,-1}
\\
&\qquad - \frac{\sqrt{2}}{2}\,c^\dagger_{f,3}\,c^\dagger_{f,-2}\,c_{p,-1}\,c_{d,2} + \frac{\sqrt{30}}{6}\,c^\dagger_{f,2}\,c^\dagger_{f,-1}\,c_{p,-1}\,c_{d,2} - c^\dagger_{f,1}\,c^\dagger_{f,0}\,c_{p,-1}\,c_{d,2}
\\
&\qquad + \frac{1}{2}\,c^\dagger_{f,3}\,c^\dagger_{f,-2}\,c_{p,0}\,c_{d,1} - \frac{\sqrt{15}}{6}\,c^\dagger_{f,2}\,c^\dagger_{f,-1}\,c_{p,0}\,c_{d,1} + \frac{\sqrt{2}}{2}\,c^\dagger_{f,1}\,c^\dagger_{f,0}\,c_{p,0}\,c_{d,1}
\\
&\qquad - \frac{\sqrt{3}}{6}\,c^\dagger_{f,3}\,c^\dagger_{f,-2}\,c_{p,1}\,c_{d,0} + \frac{\sqrt{5}}{6}\,c^\dagger_{f,2}\,c^\dagger_{f,-1}\,c_{p,1}\,c_{d,0} - \frac{\sqrt{6}}{6}\,c^\dagger_{f,1}\,c^\dagger_{f,0}\,c_{p,1}\,c_{d,0}\Big)
\end{align*}
% pdff : 1×1
% j13,j24,A wigner_9j： 2 , 2 , -sqrt(7)/7
% j13,j24,A wigner_9j： 3 , 3 , -sqrt(6)/4
% j13,j24,A wigner_9j： 4 , 4 , -3*sqrt(42)/28

\begin{align*}
{G_0^{(2)}}_{[5]} 
&=  
  \frac{1}{2} \left(\left[[c^\dagger_p \times c^\dagger_d]_{3} \times[\tilde c_f \times \tilde c_f]_{3}\right]_{0} + h.c. \right)\\
&=
    \frac{\sqrt{21}}{7} \left( 
   {\bm Q}^{pf}_2 \cdot {\bm G}_2^{df} -{\bm T}^{pf}_2 \cdot {\bm M}_2^{df} 
    \right)
    -\frac{\sqrt{2}}{2} \left( 
   {\bm M}^{pf}_3 \cdot {\bm T}_3^{df} -{\bm G}^{pf}_3 \cdot {\bm Q}_3^{df} 
    \right)
    \\&\qquad    
    + \frac{\sqrt{14}}{14} \left( 
   {\bm Q}^{pf}_4 \cdot {\bm G}_4^{df} -{\bm T}^{pf}_4 \cdot {\bm M}_4^{df} 
    \right)
  \\
&=
- \frac{\sqrt{21}}{21}\Big(- \frac{\sqrt{2}}{2}\,c^\dagger_{p,-1}\,c^\dagger_{d,-2}\,c_{f,0}\,c_{f,-3} + c^\dagger_{p,-1}\,c^\dagger_{d,-2}\,c_{f,-1}\,c_{f,-2} - \frac{\sqrt{6}}{3}\,c^\dagger_{p,-1}\,c^\dagger_{d,-1}\,c_{f,1}\,c_{f,-3}
\\
&\qquad + \frac{\sqrt{3}}{3}\,c^\dagger_{p,-1}\,c^\dagger_{d,-1}\,c_{f,0}\,c_{f,-2} - \frac{\sqrt{3}}{3}\,c^\dagger_{p,0}\,c^\dagger_{d,-2}\,c_{f,1}\,c_{f,-3} + \frac{\sqrt{6}}{6}\,c^\dagger_{p,0}\,c^\dagger_{d,-2}\,c_{f,0}\,c_{f,-2}
\\
&\qquad - \frac{\sqrt{10}}{5}\,c^\dagger_{p,-1}\,c^\dagger_{d,0}\,c_{f,2}\,c_{f,-3} + \frac{\sqrt{5}}{5}\,c^\dagger_{p,-1}\,c^\dagger_{d,0}\,c_{f,0}\,c_{f,-1} - \frac{2 \sqrt{30}}{15}\,c^\dagger_{p,0}\,c^\dagger_{d,-1}\,c_{f,2}\,c_{f,-3}
\\
&\qquad + \frac{2 \sqrt{15}}{15}\,c^\dagger_{p,0}\,c^\dagger_{d,-1}\,c_{f,0}\,c_{f,-1} - \frac{\sqrt{15}}{15}\,c^\dagger_{p,1}\,c^\dagger_{d,-2}\,c_{f,2}\,c_{f,-3} + \frac{\sqrt{30}}{30}\,c^\dagger_{p,1}\,c^\dagger_{d,-2}\,c_{f,0}\,c_{f,-1}
\\
&\qquad - \frac{\sqrt{10}}{10}\,c^\dagger_{p,-1}\,c^\dagger_{d,1}\,c_{f,3}\,c_{f,-3} - \frac{\sqrt{10}}{10}\,c^\dagger_{p,-1}\,c^\dagger_{d,1}\,c_{f,2}\,c_{f,-2} + \frac{\sqrt{10}}{10}\,c^\dagger_{p,-1}\,c^\dagger_{d,1}\,c_{f,1}\,c_{f,-1}
\\
&\qquad - \frac{\sqrt{30}}{10}\,c^\dagger_{p,0}\,c^\dagger_{d,0}\,c_{f,3}\,c_{f,-3} - \frac{\sqrt{30}}{10}\,c^\dagger_{p,0}\,c^\dagger_{d,0}\,c_{f,2}\,c_{f,-2} + \frac{\sqrt{30}}{10}\,c^\dagger_{p,0}\,c^\dagger_{d,0}\,c_{f,1}\,c_{f,-1}
\\
&\qquad - \frac{\sqrt{10}}{10}\,c^\dagger_{p,1}\,c^\dagger_{d,-1}\,c_{f,3}\,c_{f,-3} - \frac{\sqrt{10}}{10}\,c^\dagger_{p,1}\,c^\dagger_{d,-1}\,c_{f,2}\,c_{f,-2} + \frac{\sqrt{10}}{10}\,c^\dagger_{p,1}\,c^\dagger_{d,-1}\,c_{f,1}\,c_{f,-1}
\\
&\qquad - \frac{\sqrt{15}}{15}\,c^\dagger_{p,-1}\,c^\dagger_{d,2}\,c_{f,3}\,c_{f,-2} + \frac{\sqrt{30}}{30}\,c^\dagger_{p,-1}\,c^\dagger_{d,2}\,c_{f,1}\,c_{f,0} - \frac{2 \sqrt{30}}{15}\,c^\dagger_{p,0}\,c^\dagger_{d,1}\,c_{f,3}\,c_{f,-2}
\\
&\qquad + \frac{2 \sqrt{15}}{15}\,c^\dagger_{p,0}\,c^\dagger_{d,1}\,c_{f,1}\,c_{f,0} - \frac{\sqrt{10}}{5}\,c^\dagger_{p,1}\,c^\dagger_{d,0}\,c_{f,3}\,c_{f,-2} + \frac{\sqrt{5}}{5}\,c^\dagger_{p,1}\,c^\dagger_{d,0}\,c_{f,1}\,c_{f,0}
\\
&\qquad - \frac{\sqrt{3}}{3}\,c^\dagger_{p,0}\,c^\dagger_{d,2}\,c_{f,3}\,c_{f,-1} + \frac{\sqrt{6}}{6}\,c^\dagger_{p,0}\,c^\dagger_{d,2}\,c_{f,2}\,c_{f,0} - \frac{\sqrt{6}}{3}\,c^\dagger_{p,1}\,c^\dagger_{d,1}\,c_{f,3}\,c_{f,-1}
\\
&\qquad + \frac{\sqrt{3}}{3}\,c^\dagger_{p,1}\,c^\dagger_{d,1}\,c_{f,2}\,c_{f,0} - \frac{\sqrt{2}}{2}\,c^\dagger_{p,1}\,c^\dagger_{d,2}\,c_{f,3}\,c_{f,0} + c^\dagger_{p,1}\,c^\dagger_{d,2}\,c_{f,2}\,c_{f,1}
\\
&\qquad - \frac{\sqrt{2}}{2}\,c^\dagger_{f,0}\,c^\dagger_{f,-3}\,c_{p,-1}\,c_{d,-2} + c^\dagger_{f,-1}\,c^\dagger_{f,-2}\,c_{p,-1}\,c_{d,-2} - \frac{\sqrt{6}}{3}\,c^\dagger_{f,1}\,c^\dagger_{f,-3}\,c_{p,-1}\,c_{d,-1}
\\
&\qquad + \frac{\sqrt{3}}{3}\,c^\dagger_{f,0}\,c^\dagger_{f,-2}\,c_{p,-1}\,c_{d,-1} - \frac{\sqrt{3}}{3}\,c^\dagger_{f,1}\,c^\dagger_{f,-3}\,c_{p,0}\,c_{d,-2} + \frac{\sqrt{6}}{6}\,c^\dagger_{f,0}\,c^\dagger_{f,-2}\,c_{p,0}\,c_{d,-2}
\\
&\qquad - \frac{\sqrt{10}}{5}\,c^\dagger_{f,2}\,c^\dagger_{f,-3}\,c_{p,-1}\,c_{d,0} + \frac{\sqrt{5}}{5}\,c^\dagger_{f,0}\,c^\dagger_{f,-1}\,c_{p,-1}\,c_{d,0} - \frac{2 \sqrt{30}}{15}\,c^\dagger_{f,2}\,c^\dagger_{f,-3}\,c_{p,0}\,c_{d,-1}
\\
&\qquad + \frac{2 \sqrt{15}}{15}\,c^\dagger_{f,0}\,c^\dagger_{f,-1}\,c_{p,0}\,c_{d,-1} - \frac{\sqrt{15}}{15}\,c^\dagger_{f,2}\,c^\dagger_{f,-3}\,c_{p,1}\,c_{d,-2} + \frac{\sqrt{30}}{30}\,c^\dagger_{f,0}\,c^\dagger_{f,-1}\,c_{p,1}\,c_{d,-2}
\\
&\qquad - \frac{\sqrt{10}}{10}\,c^\dagger_{f,3}\,c^\dagger_{f,-3}\,c_{p,-1}\,c_{d,1} - \frac{\sqrt{10}}{10}\,c^\dagger_{f,2}\,c^\dagger_{f,-2}\,c_{p,-1}\,c_{d,1} + \frac{\sqrt{10}}{10}\,c^\dagger_{f,1}\,c^\dagger_{f,-1}\,c_{p,-1}\,c_{d,1}
\\
&\qquad - \frac{\sqrt{30}}{10}\,c^\dagger_{f,3}\,c^\dagger_{f,-3}\,c_{p,0}\,c_{d,0} - \frac{\sqrt{30}}{10}\,c^\dagger_{f,2}\,c^\dagger_{f,-2}\,c_{p,0}\,c_{d,0} + \frac{\sqrt{30}}{10}\,c^\dagger_{f,1}\,c^\dagger_{f,-1}\,c_{p,0}\,c_{d,0}
\\
&\qquad - \frac{\sqrt{10}}{10}\,c^\dagger_{f,3}\,c^\dagger_{f,-3}\,c_{p,1}\,c_{d,-1} - \frac{\sqrt{10}}{10}\,c^\dagger_{f,2}\,c^\dagger_{f,-2}\,c_{p,1}\,c_{d,-1} + \frac{\sqrt{10}}{10}\,c^\dagger_{f,1}\,c^\dagger_{f,-1}\,c_{p,1}\,c_{d,-1}
\\
&\qquad - \frac{\sqrt{15}}{15}\,c^\dagger_{f,3}\,c^\dagger_{f,-2}\,c_{p,-1}\,c_{d,2} + \frac{\sqrt{30}}{30}\,c^\dagger_{f,1}\,c^\dagger_{f,0}\,c_{p,-1}\,c_{d,2} - \frac{2 \sqrt{30}}{15}\,c^\dagger_{f,3}\,c^\dagger_{f,-2}\,c_{p,0}\,c_{d,1}
\\
&\qquad + \frac{2 \sqrt{15}}{15}\,c^\dagger_{f,1}\,c^\dagger_{f,0}\,c_{p,0}\,c_{d,1} - \frac{\sqrt{10}}{5}\,c^\dagger_{f,3}\,c^\dagger_{f,-2}\,c_{p,1}\,c_{d,0} + \frac{\sqrt{5}}{5}\,c^\dagger_{f,1}\,c^\dagger_{f,0}\,c_{p,1}\,c_{d,0}
\\
&\qquad - \frac{\sqrt{3}}{3}\,c^\dagger_{f,3}\,c^\dagger_{f,-1}\,c_{p,0}\,c_{d,2} + \frac{\sqrt{6}}{6}\,c^\dagger_{f,2}\,c^\dagger_{f,0}\,c_{p,0}\,c_{d,2} - \frac{\sqrt{6}}{3}\,c^\dagger_{f,3}\,c^\dagger_{f,-1}\,c_{p,1}\,c_{d,1}
\\
&\qquad + \frac{\sqrt{3}}{3}\,c^\dagger_{f,2}\,c^\dagger_{f,0}\,c_{p,1}\,c_{d,1} - \frac{\sqrt{2}}{2}\,c^\dagger_{f,3}\,c^\dagger_{f,0}\,c_{p,1}\,c_{d,2} + c^\dagger_{f,2}\,c^\dagger_{f,1}\,c_{p,1}\,c_{d,2}\Big)
\\
{M_0^{(2)}}_{[4]} 
&=  
  \frac{i}{2} \left(\left[[c^\dagger_p \times c^\dagger_d]_{3} \times[\tilde c_f \times \tilde c_f]_{3}\right]_{0} - h.c. \right)\\
&=
    \frac{\sqrt{21}}{7} \left( 
   {\bm T}^{pf}_2 \cdot {\bm G}_2^{df} +{\bm Q}^{pf}_2 \cdot {\bm M}_2^{df} 
    \right)
     -\frac{\sqrt{2}}{2} \left( 
   {\bm G}^{pf}_3 \cdot {\bm T}_3^{df} +{\bm M}^{pf}_3 \cdot {\bm Q}_3^{df} 
    \right)
    \\&\qquad    
    + \frac{\sqrt{14}}{14} \left( 
   {\bm T}^{pf}_4 \cdot {\bm G}_4^{df} +{\bm Q}^{pf}_4 \cdot {\bm M}_4^{df} 
    \right)
  \\
&=
- \frac{\sqrt{21} i}{21}\Big(- \frac{\sqrt{2}}{2}\,c^\dagger_{p,-1}\,c^\dagger_{d,-2}\,c_{f,0}\,c_{f,-3} + c^\dagger_{p,-1}\,c^\dagger_{d,-2}\,c_{f,-1}\,c_{f,-2} - \frac{\sqrt{6}}{3}\,c^\dagger_{p,-1}\,c^\dagger_{d,-1}\,c_{f,1}\,c_{f,-3}
\\
&\qquad + \frac{\sqrt{3}}{3}\,c^\dagger_{p,-1}\,c^\dagger_{d,-1}\,c_{f,0}\,c_{f,-2} - \frac{\sqrt{3}}{3}\,c^\dagger_{p,0}\,c^\dagger_{d,-2}\,c_{f,1}\,c_{f,-3} + \frac{\sqrt{6}}{6}\,c^\dagger_{p,0}\,c^\dagger_{d,-2}\,c_{f,0}\,c_{f,-2}
\\
&\qquad - \frac{\sqrt{10}}{5}\,c^\dagger_{p,-1}\,c^\dagger_{d,0}\,c_{f,2}\,c_{f,-3} + \frac{\sqrt{5}}{5}\,c^\dagger_{p,-1}\,c^\dagger_{d,0}\,c_{f,0}\,c_{f,-1} - \frac{2 \sqrt{30}}{15}\,c^\dagger_{p,0}\,c^\dagger_{d,-1}\,c_{f,2}\,c_{f,-3}
\\
&\qquad + \frac{2 \sqrt{15}}{15}\,c^\dagger_{p,0}\,c^\dagger_{d,-1}\,c_{f,0}\,c_{f,-1} - \frac{\sqrt{15}}{15}\,c^\dagger_{p,1}\,c^\dagger_{d,-2}\,c_{f,2}\,c_{f,-3} + \frac{\sqrt{30}}{30}\,c^\dagger_{p,1}\,c^\dagger_{d,-2}\,c_{f,0}\,c_{f,-1}
\\
&\qquad - \frac{\sqrt{10}}{10}\,c^\dagger_{p,-1}\,c^\dagger_{d,1}\,c_{f,3}\,c_{f,-3} - \frac{\sqrt{10}}{10}\,c^\dagger_{p,-1}\,c^\dagger_{d,1}\,c_{f,2}\,c_{f,-2} + \frac{\sqrt{10}}{10}\,c^\dagger_{p,-1}\,c^\dagger_{d,1}\,c_{f,1}\,c_{f,-1}
\\
&\qquad - \frac{\sqrt{30}}{10}\,c^\dagger_{p,0}\,c^\dagger_{d,0}\,c_{f,3}\,c_{f,-3} - \frac{\sqrt{30}}{10}\,c^\dagger_{p,0}\,c^\dagger_{d,0}\,c_{f,2}\,c_{f,-2} + \frac{\sqrt{30}}{10}\,c^\dagger_{p,0}\,c^\dagger_{d,0}\,c_{f,1}\,c_{f,-1}
\\
&\qquad - \frac{\sqrt{10}}{10}\,c^\dagger_{p,1}\,c^\dagger_{d,-1}\,c_{f,3}\,c_{f,-3} - \frac{\sqrt{10}}{10}\,c^\dagger_{p,1}\,c^\dagger_{d,-1}\,c_{f,2}\,c_{f,-2} + \frac{\sqrt{10}}{10}\,c^\dagger_{p,1}\,c^\dagger_{d,-1}\,c_{f,1}\,c_{f,-1}
\\
&\qquad - \frac{\sqrt{15}}{15}\,c^\dagger_{p,-1}\,c^\dagger_{d,2}\,c_{f,3}\,c_{f,-2} + \frac{\sqrt{30}}{30}\,c^\dagger_{p,-1}\,c^\dagger_{d,2}\,c_{f,1}\,c_{f,0} - \frac{2 \sqrt{30}}{15}\,c^\dagger_{p,0}\,c^\dagger_{d,1}\,c_{f,3}\,c_{f,-2}
\\
&\qquad + \frac{2 \sqrt{15}}{15}\,c^\dagger_{p,0}\,c^\dagger_{d,1}\,c_{f,1}\,c_{f,0} - \frac{\sqrt{10}}{5}\,c^\dagger_{p,1}\,c^\dagger_{d,0}\,c_{f,3}\,c_{f,-2} + \frac{\sqrt{5}}{5}\,c^\dagger_{p,1}\,c^\dagger_{d,0}\,c_{f,1}\,c_{f,0}
\\
&\qquad - \frac{\sqrt{3}}{3}\,c^\dagger_{p,0}\,c^\dagger_{d,2}\,c_{f,3}\,c_{f,-1} + \frac{\sqrt{6}}{6}\,c^\dagger_{p,0}\,c^\dagger_{d,2}\,c_{f,2}\,c_{f,0} - \frac{\sqrt{6}}{3}\,c^\dagger_{p,1}\,c^\dagger_{d,1}\,c_{f,3}\,c_{f,-1}
\\
&\qquad + \frac{\sqrt{3}}{3}\,c^\dagger_{p,1}\,c^\dagger_{d,1}\,c_{f,2}\,c_{f,0} - \frac{\sqrt{2}}{2}\,c^\dagger_{p,1}\,c^\dagger_{d,2}\,c_{f,3}\,c_{f,0} + c^\dagger_{p,1}\,c^\dagger_{d,2}\,c_{f,2}\,c_{f,1}
\\
&\qquad + \frac{\sqrt{2}}{2}\,c^\dagger_{f,0}\,c^\dagger_{f,-3}\,c_{p,-1}\,c_{d,-2} - c^\dagger_{f,-1}\,c^\dagger_{f,-2}\,c_{p,-1}\,c_{d,-2} + \frac{\sqrt{6}}{3}\,c^\dagger_{f,1}\,c^\dagger_{f,-3}\,c_{p,-1}\,c_{d,-1}
\\
&\qquad - \frac{\sqrt{3}}{3}\,c^\dagger_{f,0}\,c^\dagger_{f,-2}\,c_{p,-1}\,c_{d,-1} + \frac{\sqrt{3}}{3}\,c^\dagger_{f,1}\,c^\dagger_{f,-3}\,c_{p,0}\,c_{d,-2} - \frac{\sqrt{6}}{6}\,c^\dagger_{f,0}\,c^\dagger_{f,-2}\,c_{p,0}\,c_{d,-2}
\\
&\qquad + \frac{\sqrt{10}}{5}\,c^\dagger_{f,2}\,c^\dagger_{f,-3}\,c_{p,-1}\,c_{d,0} - \frac{\sqrt{5}}{5}\,c^\dagger_{f,0}\,c^\dagger_{f,-1}\,c_{p,-1}\,c_{d,0} + \frac{2 \sqrt{30}}{15}\,c^\dagger_{f,2}\,c^\dagger_{f,-3}\,c_{p,0}\,c_{d,-1}
\\
&\qquad - \frac{2 \sqrt{15}}{15}\,c^\dagger_{f,0}\,c^\dagger_{f,-1}\,c_{p,0}\,c_{d,-1} + \frac{\sqrt{15}}{15}\,c^\dagger_{f,2}\,c^\dagger_{f,-3}\,c_{p,1}\,c_{d,-2} - \frac{\sqrt{30}}{30}\,c^\dagger_{f,0}\,c^\dagger_{f,-1}\,c_{p,1}\,c_{d,-2}
\\
&\qquad + \frac{\sqrt{10}}{10}\,c^\dagger_{f,3}\,c^\dagger_{f,-3}\,c_{p,-1}\,c_{d,1} + \frac{\sqrt{10}}{10}\,c^\dagger_{f,2}\,c^\dagger_{f,-2}\,c_{p,-1}\,c_{d,1} - \frac{\sqrt{10}}{10}\,c^\dagger_{f,1}\,c^\dagger_{f,-1}\,c_{p,-1}\,c_{d,1}
\\
&\qquad + \frac{\sqrt{30}}{10}\,c^\dagger_{f,3}\,c^\dagger_{f,-3}\,c_{p,0}\,c_{d,0} + \frac{\sqrt{30}}{10}\,c^\dagger_{f,2}\,c^\dagger_{f,-2}\,c_{p,0}\,c_{d,0} - \frac{\sqrt{30}}{10}\,c^\dagger_{f,1}\,c^\dagger_{f,-1}\,c_{p,0}\,c_{d,0}
\\
&\qquad + \frac{\sqrt{10}}{10}\,c^\dagger_{f,3}\,c^\dagger_{f,-3}\,c_{p,1}\,c_{d,-1} + \frac{\sqrt{10}}{10}\,c^\dagger_{f,2}\,c^\dagger_{f,-2}\,c_{p,1}\,c_{d,-1} - \frac{\sqrt{10}}{10}\,c^\dagger_{f,1}\,c^\dagger_{f,-1}\,c_{p,1}\,c_{d,-1}
\\
&\qquad + \frac{\sqrt{15}}{15}\,c^\dagger_{f,3}\,c^\dagger_{f,-2}\,c_{p,-1}\,c_{d,2} - \frac{\sqrt{30}}{30}\,c^\dagger_{f,1}\,c^\dagger_{f,0}\,c_{p,-1}\,c_{d,2} + \frac{2 \sqrt{30}}{15}\,c^\dagger_{f,3}\,c^\dagger_{f,-2}\,c_{p,0}\,c_{d,1}
\\
&\qquad - \frac{2 \sqrt{15}}{15}\,c^\dagger_{f,1}\,c^\dagger_{f,0}\,c_{p,0}\,c_{d,1} + \frac{\sqrt{10}}{5}\,c^\dagger_{f,3}\,c^\dagger_{f,-2}\,c_{p,1}\,c_{d,0} - \frac{\sqrt{5}}{5}\,c^\dagger_{f,1}\,c^\dagger_{f,0}\,c_{p,1}\,c_{d,0}
\\
&\qquad + \frac{\sqrt{3}}{3}\,c^\dagger_{f,3}\,c^\dagger_{f,-1}\,c_{p,0}\,c_{d,2} - \frac{\sqrt{6}}{6}\,c^\dagger_{f,2}\,c^\dagger_{f,0}\,c_{p,0}\,c_{d,2} + \frac{\sqrt{6}}{3}\,c^\dagger_{f,3}\,c^\dagger_{f,-1}\,c_{p,1}\,c_{d,1}
\\
&\qquad - \frac{\sqrt{3}}{3}\,c^\dagger_{f,2}\,c^\dagger_{f,0}\,c_{p,1}\,c_{d,1} + \frac{\sqrt{2}}{2}\,c^\dagger_{f,3}\,c^\dagger_{f,0}\,c_{p,1}\,c_{d,2} - c^\dagger_{f,2}\,c^\dagger_{f,1}\,c_{p,1}\,c_{d,2}\Big)
\end{align*}

% pdff : 3×3
% j13,j24,A wigner_9j： 2 , 2 , -sqrt(21)/7
% j13,j24,A wigner_9j： 3 , 3 , sqrt(2)/2
% j13,j24,A wigner_9j： 4 , 4 , -sqrt(14)/14



\subsection*{$spdf$ system}

\begin{align*}
{Q_0^{(2)}}_{[1]}
&=  
  \frac{1}{2} \left(
\left[[c^\dagger_s \times c^\dagger_p]_{1} \times[\tilde c_d \times \tilde c_f]_{1}\right]_{0} 
    + h.c. 
    \right)\\
&=  
- {\bm Q}^{sd}_2 \cdot {\bm Q}^{pf}_2  
+ {\bm T}^{sd}_2 \cdot {\bm T}^{pf}_2  
\\
&=
\frac{\sqrt{7}}{14}\Big(c^\dagger_{s,0}\,c^\dagger_{p,-1}\,c_{d,2}\,c_{f,-3} - \frac{\sqrt{6}}{3}\,c^\dagger_{s,0}\,c^\dagger_{p,-1}\,c_{d,1}\,c_{f,-2} + \frac{\sqrt{10}}{5}\,c^\dagger_{s,0}\,c^\dagger_{p,-1}\,c_{d,0}\,c_{f,-1}
\\
&\qquad - \frac{\sqrt{5}}{5}\,c^\dagger_{s,0}\,c^\dagger_{p,-1}\,c_{d,-1}\,c_{f,0} + \frac{\sqrt{15}}{15}\,c^\dagger_{s,0}\,c^\dagger_{p,-1}\,c_{d,-2}\,c_{f,1} + \frac{\sqrt{3}}{3}\,c^\dagger_{s,0}\,c^\dagger_{p,0}\,c_{d,2}\,c_{f,-2}
\\
&\qquad - \frac{2 \sqrt{30}}{15}\,c^\dagger_{s,0}\,c^\dagger_{p,0}\,c_{d,1}\,c_{f,-1} + \frac{\sqrt{15}}{5}\,c^\dagger_{s,0}\,c^\dagger_{p,0}\,c_{d,0}\,c_{f,0} - \frac{2 \sqrt{30}}{15}\,c^\dagger_{s,0}\,c^\dagger_{p,0}\,c_{d,-1}\,c_{f,1}
\\
&\qquad + \frac{\sqrt{3}}{3}\,c^\dagger_{s,0}\,c^\dagger_{p,0}\,c_{d,-2}\,c_{f,2} + \frac{\sqrt{15}}{15}\,c^\dagger_{s,0}\,c^\dagger_{p,1}\,c_{d,2}\,c_{f,-1} - \frac{\sqrt{5}}{5}\,c^\dagger_{s,0}\,c^\dagger_{p,1}\,c_{d,1}\,c_{f,0}
\\
&\qquad + \frac{\sqrt{10}}{5}\,c^\dagger_{s,0}\,c^\dagger_{p,1}\,c_{d,0}\,c_{f,1} - \frac{\sqrt{6}}{3}\,c^\dagger_{s,0}\,c^\dagger_{p,1}\,c_{d,-1}\,c_{f,2} + c^\dagger_{s,0}\,c^\dagger_{p,1}\,c_{d,-2}\,c_{f,3}
\\
&\qquad + c^\dagger_{d,2}\,c^\dagger_{f,-3}\,c_{s,0}\,c_{p,-1} - \frac{\sqrt{6}}{3}\,c^\dagger_{d,1}\,c^\dagger_{f,-2}\,c_{s,0}\,c_{p,-1} + \frac{\sqrt{10}}{5}\,c^\dagger_{d,0}\,c^\dagger_{f,-1}\,c_{s,0}\,c_{p,-1}
\\
&\qquad - \frac{\sqrt{5}}{5}\,c^\dagger_{d,-1}\,c^\dagger_{f,0}\,c_{s,0}\,c_{p,-1} + \frac{\sqrt{15}}{15}\,c^\dagger_{d,-2}\,c^\dagger_{f,1}\,c_{s,0}\,c_{p,-1} + \frac{\sqrt{3}}{3}\,c^\dagger_{d,2}\,c^\dagger_{f,-2}\,c_{s,0}\,c_{p,0}
\\
&\qquad - \frac{2 \sqrt{30}}{15}\,c^\dagger_{d,1}\,c^\dagger_{f,-1}\,c_{s,0}\,c_{p,0} + \frac{\sqrt{15}}{5}\,c^\dagger_{d,0}\,c^\dagger_{f,0}\,c_{s,0}\,c_{p,0} - \frac{2 \sqrt{30}}{15}\,c^\dagger_{d,-1}\,c^\dagger_{f,1}\,c_{s,0}\,c_{p,0}
\\
&\qquad + \frac{\sqrt{3}}{3}\,c^\dagger_{d,-2}\,c^\dagger_{f,2}\,c_{s,0}\,c_{p,0} + \frac{\sqrt{15}}{15}\,c^\dagger_{d,2}\,c^\dagger_{f,-1}\,c_{s,0}\,c_{p,1} - \frac{\sqrt{5}}{5}\,c^\dagger_{d,1}\,c^\dagger_{f,0}\,c_{s,0}\,c_{p,1}
\\
&\qquad + \frac{\sqrt{10}}{5}\,c^\dagger_{d,0}\,c^\dagger_{f,1}\,c_{s,0}\,c_{p,1} - \frac{\sqrt{6}}{3}\,c^\dagger_{d,-1}\,c^\dagger_{f,2}\,c_{s,0}\,c_{p,1} + c^\dagger_{d,-2}\,c^\dagger_{f,3}\,c_{s,0}\,c_{p,1}\Big)
\\
{T_0^{(2)}}_{[1]}
&=  
  \frac{i}{2} \left(
\left[[c^\dagger_s \times c^\dagger_p]_{1} \times[\tilde c_d \times \tilde c_f]_{1}\right]_{0} 
    - h.c. 
    \right)\\
&=  
- {\bm T}^{sd}_2 \cdot {\bm Q}^{pf}_2  
- {\bm Q}^{sd}_2 \cdot {\bm T}^{pf}_2  
\\
&=
\frac{\sqrt{7} i}{14}\Big(c^\dagger_{s,0}\,c^\dagger_{p,-1}\,c_{d,2}\,c_{f,-3} - \frac{\sqrt{6}}{3}\,c^\dagger_{s,0}\,c^\dagger_{p,-1}\,c_{d,1}\,c_{f,-2} + \frac{\sqrt{10}}{5}\,c^\dagger_{s,0}\,c^\dagger_{p,-1}\,c_{d,0}\,c_{f,-1}
\\
&\qquad - \frac{\sqrt{5}}{5}\,c^\dagger_{s,0}\,c^\dagger_{p,-1}\,c_{d,-1}\,c_{f,0} + \frac{\sqrt{15}}{15}\,c^\dagger_{s,0}\,c^\dagger_{p,-1}\,c_{d,-2}\,c_{f,1} + \frac{\sqrt{3}}{3}\,c^\dagger_{s,0}\,c^\dagger_{p,0}\,c_{d,2}\,c_{f,-2}
\\
&\qquad - \frac{2 \sqrt{30}}{15}\,c^\dagger_{s,0}\,c^\dagger_{p,0}\,c_{d,1}\,c_{f,-1} + \frac{\sqrt{15}}{5}\,c^\dagger_{s,0}\,c^\dagger_{p,0}\,c_{d,0}\,c_{f,0} - \frac{2 \sqrt{30}}{15}\,c^\dagger_{s,0}\,c^\dagger_{p,0}\,c_{d,-1}\,c_{f,1}
\\
&\qquad + \frac{\sqrt{3}}{3}\,c^\dagger_{s,0}\,c^\dagger_{p,0}\,c_{d,-2}\,c_{f,2} + \frac{\sqrt{15}}{15}\,c^\dagger_{s,0}\,c^\dagger_{p,1}\,c_{d,2}\,c_{f,-1} - \frac{\sqrt{5}}{5}\,c^\dagger_{s,0}\,c^\dagger_{p,1}\,c_{d,1}\,c_{f,0}
\\
&\qquad + \frac{\sqrt{10}}{5}\,c^\dagger_{s,0}\,c^\dagger_{p,1}\,c_{d,0}\,c_{f,1} - \frac{\sqrt{6}}{3}\,c^\dagger_{s,0}\,c^\dagger_{p,1}\,c_{d,-1}\,c_{f,2} + c^\dagger_{s,0}\,c^\dagger_{p,1}\,c_{d,-2}\,c_{f,3}
\\
&\qquad - c^\dagger_{d,2}\,c^\dagger_{f,-3}\,c_{s,0}\,c_{p,-1} + \frac{\sqrt{6}}{3}\,c^\dagger_{d,1}\,c^\dagger_{f,-2}\,c_{s,0}\,c_{p,-1} - \frac{\sqrt{10}}{5}\,c^\dagger_{d,0}\,c^\dagger_{f,-1}\,c_{s,0}\,c_{p,-1}
\\
&\qquad + \frac{\sqrt{5}}{5}\,c^\dagger_{d,-1}\,c^\dagger_{f,0}\,c_{s,0}\,c_{p,-1} - \frac{\sqrt{15}}{15}\,c^\dagger_{d,-2}\,c^\dagger_{f,1}\,c_{s,0}\,c_{p,-1} - \frac{\sqrt{3}}{3}\,c^\dagger_{d,2}\,c^\dagger_{f,-2}\,c_{s,0}\,c_{p,0}
\\
&\qquad + \frac{2 \sqrt{30}}{15}\,c^\dagger_{d,1}\,c^\dagger_{f,-1}\,c_{s,0}\,c_{p,0} - \frac{\sqrt{15}}{5}\,c^\dagger_{d,0}\,c^\dagger_{f,0}\,c_{s,0}\,c_{p,0} + \frac{2 \sqrt{30}}{15}\,c^\dagger_{d,-1}\,c^\dagger_{f,1}\,c_{s,0}\,c_{p,0}
\\
&\qquad - \frac{\sqrt{3}}{3}\,c^\dagger_{d,-2}\,c^\dagger_{f,2}\,c_{s,0}\,c_{p,0} - \frac{\sqrt{15}}{15}\,c^\dagger_{d,2}\,c^\dagger_{f,-1}\,c_{s,0}\,c_{p,1} + \frac{\sqrt{5}}{5}\,c^\dagger_{d,1}\,c^\dagger_{f,0}\,c_{s,0}\,c_{p,1}
\\
&\qquad - \frac{\sqrt{10}}{5}\,c^\dagger_{d,0}\,c^\dagger_{f,1}\,c_{s,0}\,c_{p,1} + \frac{\sqrt{6}}{3}\,c^\dagger_{d,-1}\,c^\dagger_{f,2}\,c_{s,0}\,c_{p,1} - c^\dagger_{d,-2}\,c^\dagger_{f,3}\,c_{s,0}\,c_{p,1}\Big)
\end{align*}

% sdpf : 2×2
% j13,j24,A wigner_9j： 1 , 1 , 1

\begin{align*}
{Q_0^{(2)}}_{[2]}
&=  
  \frac{1}{2} \left(
\left[[c^\dagger_s \times c^\dagger_d]_{2} \times[\tilde c_p \times \tilde c_f]_{2}\right]_{0} 
    + h.c. 
    \right)\\
&=  
  {\bm Q}^{sp}_1 \cdot {\bm Q}^{df}_1  
- {\bm T}^{sp}_1 \cdot {\bm T}^{df}_1  
\\
&=
\frac{\sqrt{7}}{14}\Big(c^\dagger_{s,0}\,c^\dagger_{d,-2}\,c_{p,1}\,c_{f,-3} - \frac{\sqrt{3}}{3}\,c^\dagger_{s,0}\,c^\dagger_{d,-2}\,c_{p,0}\,c_{f,-2} + \frac{\sqrt{15}}{15}\,c^\dagger_{s,0}\,c^\dagger_{d,-2}\,c_{p,-1}\,c_{f,-1}
\\
&\qquad + \frac{\sqrt{6}}{3}\,c^\dagger_{s,0}\,c^\dagger_{d,-1}\,c_{p,1}\,c_{f,-2} - \frac{2 \sqrt{30}}{15}\,c^\dagger_{s,0}\,c^\dagger_{d,-1}\,c_{p,0}\,c_{f,-1} + \frac{\sqrt{5}}{5}\,c^\dagger_{s,0}\,c^\dagger_{d,-1}\,c_{p,-1}\,c_{f,0}
\\
&\qquad + \frac{\sqrt{10}}{5}\,c^\dagger_{s,0}\,c^\dagger_{d,0}\,c_{p,1}\,c_{f,-1} - \frac{\sqrt{15}}{5}\,c^\dagger_{s,0}\,c^\dagger_{d,0}\,c_{p,0}\,c_{f,0} + \frac{\sqrt{10}}{5}\,c^\dagger_{s,0}\,c^\dagger_{d,0}\,c_{p,-1}\,c_{f,1}
\\
&\qquad + \frac{\sqrt{5}}{5}\,c^\dagger_{s,0}\,c^\dagger_{d,1}\,c_{p,1}\,c_{f,0} - \frac{2 \sqrt{30}}{15}\,c^\dagger_{s,0}\,c^\dagger_{d,1}\,c_{p,0}\,c_{f,1} + \frac{\sqrt{6}}{3}\,c^\dagger_{s,0}\,c^\dagger_{d,1}\,c_{p,-1}\,c_{f,2}
\\
&\qquad + \frac{\sqrt{15}}{15}\,c^\dagger_{s,0}\,c^\dagger_{d,2}\,c_{p,1}\,c_{f,1} - \frac{\sqrt{3}}{3}\,c^\dagger_{s,0}\,c^\dagger_{d,2}\,c_{p,0}\,c_{f,2} + c^\dagger_{s,0}\,c^\dagger_{d,2}\,c_{p,-1}\,c_{f,3}
\\
&\qquad + c^\dagger_{p,1}\,c^\dagger_{f,-3}\,c_{s,0}\,c_{d,-2} - \frac{\sqrt{3}}{3}\,c^\dagger_{p,0}\,c^\dagger_{f,-2}\,c_{s,0}\,c_{d,-2} + \frac{\sqrt{15}}{15}\,c^\dagger_{p,-1}\,c^\dagger_{f,-1}\,c_{s,0}\,c_{d,-2}
\\
&\qquad + \frac{\sqrt{6}}{3}\,c^\dagger_{p,1}\,c^\dagger_{f,-2}\,c_{s,0}\,c_{d,-1} - \frac{2 \sqrt{30}}{15}\,c^\dagger_{p,0}\,c^\dagger_{f,-1}\,c_{s,0}\,c_{d,-1} + \frac{\sqrt{5}}{5}\,c^\dagger_{p,-1}\,c^\dagger_{f,0}\,c_{s,0}\,c_{d,-1}
\\
&\qquad + \frac{\sqrt{10}}{5}\,c^\dagger_{p,1}\,c^\dagger_{f,-1}\,c_{s,0}\,c_{d,0} - \frac{\sqrt{15}}{5}\,c^\dagger_{p,0}\,c^\dagger_{f,0}\,c_{s,0}\,c_{d,0} + \frac{\sqrt{10}}{5}\,c^\dagger_{p,-1}\,c^\dagger_{f,1}\,c_{s,0}\,c_{d,0}
\\
&\qquad + \frac{\sqrt{5}}{5}\,c^\dagger_{p,1}\,c^\dagger_{f,0}\,c_{s,0}\,c_{d,1} - \frac{2 \sqrt{30}}{15}\,c^\dagger_{p,0}\,c^\dagger_{f,1}\,c_{s,0}\,c_{d,1} + \frac{\sqrt{6}}{3}\,c^\dagger_{p,-1}\,c^\dagger_{f,2}\,c_{s,0}\,c_{d,1}
\\
&\qquad + \frac{\sqrt{15}}{15}\,c^\dagger_{p,1}\,c^\dagger_{f,1}\,c_{s,0}\,c_{d,2} - \frac{\sqrt{3}}{3}\,c^\dagger_{p,0}\,c^\dagger_{f,2}\,c_{s,0}\,c_{d,2} + c^\dagger_{p,-1}\,c^\dagger_{f,3}\,c_{s,0}\,c_{d,2}\Big)
\\
{T_0^{(2)}}_{[2]}
&=  
  \frac{i}{2} \left(
\left[[c^\dagger_s \times c^\dagger_d]_{2} \times[\tilde c_p \times \tilde c_f]_{2}\right]_{0} 
    - h.c. 
    \right)\\
&=  
- {\bm T}^{sp}_1 \cdot {\bm Q}^{df}_1  
- {\bm Q}^{sp}_1 \cdot {\bm T}^{df}_1  
\\
&=
\frac{\sqrt{7} i}{14}\Big(c^\dagger_{s,0}\,c^\dagger_{d,-2}\,c_{p,1}\,c_{f,-3} - \frac{\sqrt{3}}{3}\,c^\dagger_{s,0}\,c^\dagger_{d,-2}\,c_{p,0}\,c_{f,-2} + \frac{\sqrt{15}}{15}\,c^\dagger_{s,0}\,c^\dagger_{d,-2}\,c_{p,-1}\,c_{f,-1}
\\
&\qquad + \frac{\sqrt{6}}{3}\,c^\dagger_{s,0}\,c^\dagger_{d,-1}\,c_{p,1}\,c_{f,-2} - \frac{2 \sqrt{30}}{15}\,c^\dagger_{s,0}\,c^\dagger_{d,-1}\,c_{p,0}\,c_{f,-1} + \frac{\sqrt{5}}{5}\,c^\dagger_{s,0}\,c^\dagger_{d,-1}\,c_{p,-1}\,c_{f,0}
\\
&\qquad + \frac{\sqrt{10}}{5}\,c^\dagger_{s,0}\,c^\dagger_{d,0}\,c_{p,1}\,c_{f,-1} - \frac{\sqrt{15}}{5}\,c^\dagger_{s,0}\,c^\dagger_{d,0}\,c_{p,0}\,c_{f,0} + \frac{\sqrt{10}}{5}\,c^\dagger_{s,0}\,c^\dagger_{d,0}\,c_{p,-1}\,c_{f,1}
\\
&\qquad + \frac{\sqrt{5}}{5}\,c^\dagger_{s,0}\,c^\dagger_{d,1}\,c_{p,1}\,c_{f,0} - \frac{2 \sqrt{30}}{15}\,c^\dagger_{s,0}\,c^\dagger_{d,1}\,c_{p,0}\,c_{f,1} + \frac{\sqrt{6}}{3}\,c^\dagger_{s,0}\,c^\dagger_{d,1}\,c_{p,-1}\,c_{f,2}
\\
&\qquad + \frac{\sqrt{15}}{15}\,c^\dagger_{s,0}\,c^\dagger_{d,2}\,c_{p,1}\,c_{f,1} - \frac{\sqrt{3}}{3}\,c^\dagger_{s,0}\,c^\dagger_{d,2}\,c_{p,0}\,c_{f,2} + c^\dagger_{s,0}\,c^\dagger_{d,2}\,c_{p,-1}\,c_{f,3}
\\
&\qquad - c^\dagger_{p,1}\,c^\dagger_{f,-3}\,c_{s,0}\,c_{d,-2} + \frac{\sqrt{3}}{3}\,c^\dagger_{p,0}\,c^\dagger_{f,-2}\,c_{s,0}\,c_{d,-2} - \frac{\sqrt{15}}{15}\,c^\dagger_{p,-1}\,c^\dagger_{f,-1}\,c_{s,0}\,c_{d,-2}
\\
&\qquad - \frac{\sqrt{6}}{3}\,c^\dagger_{p,1}\,c^\dagger_{f,-2}\,c_{s,0}\,c_{d,-1} + \frac{2 \sqrt{30}}{15}\,c^\dagger_{p,0}\,c^\dagger_{f,-1}\,c_{s,0}\,c_{d,-1} - \frac{\sqrt{5}}{5}\,c^\dagger_{p,-1}\,c^\dagger_{f,0}\,c_{s,0}\,c_{d,-1}
\\
&\qquad - \frac{\sqrt{10}}{5}\,c^\dagger_{p,1}\,c^\dagger_{f,-1}\,c_{s,0}\,c_{d,0} + \frac{\sqrt{15}}{5}\,c^\dagger_{p,0}\,c^\dagger_{f,0}\,c_{s,0}\,c_{d,0} - \frac{\sqrt{10}}{5}\,c^\dagger_{p,-1}\,c^\dagger_{f,1}\,c_{s,0}\,c_{d,0}
\\
&\qquad - \frac{\sqrt{5}}{5}\,c^\dagger_{p,1}\,c^\dagger_{f,0}\,c_{s,0}\,c_{d,1} + \frac{2 \sqrt{30}}{15}\,c^\dagger_{p,0}\,c^\dagger_{f,1}\,c_{s,0}\,c_{d,1} - \frac{\sqrt{6}}{3}\,c^\dagger_{p,-1}\,c^\dagger_{f,2}\,c_{s,0}\,c_{d,1}
\\
&\qquad - \frac{\sqrt{15}}{15}\,c^\dagger_{p,1}\,c^\dagger_{f,1}\,c_{s,0}\,c_{d,2} + \frac{\sqrt{3}}{3}\,c^\dagger_{p,0}\,c^\dagger_{f,2}\,c_{s,0}\,c_{d,2} - c^\dagger_{p,-1}\,c^\dagger_{f,3}\,c_{s,0}\,c_{d,2}\Big)
\end{align*}

% sfdp : 3×3
% j13,j24,A wigner_9j： 1 , 1 , 1
\begin{align*}
{Q_0^{(2)}}_{[3]}
&=  
  \frac{1}{2} \left(
\left[[c^\dagger_s \times c^\dagger_f]_{3} \times[\tilde c_d \times \tilde c_p]_{3}\right]_{0} 
    + h.c. 
    \right)\\
&=  
- {\bm Q}^{sd}_2 \cdot {\bm Q}^{pf}_2  
- {\bm T}^{sd}_2 \cdot {\bm T}^{pf}_2  
\\
&=
- \frac{\sqrt{7}}{14}\Big(c^\dagger_{s,0}\,c^\dagger_{f,-3}\,c_{p,-1}\,c_{d,-2} + \frac{\sqrt{6}}{3}\,c^\dagger_{s,0}\,c^\dagger_{f,-2}\,c_{p,-1}\,c_{d,-1} + \frac{\sqrt{3}}{3}\,c^\dagger_{s,0}\,c^\dagger_{f,-2}\,c_{p,0}\,c_{d,-2}
\\
&\qquad + \frac{\sqrt{10}}{5}\,c^\dagger_{s,0}\,c^\dagger_{f,-1}\,c_{p,-1}\,c_{d,0} + \frac{2 \sqrt{30}}{15}\,c^\dagger_{s,0}\,c^\dagger_{f,-1}\,c_{p,0}\,c_{d,-1} + \frac{\sqrt{15}}{15}\,c^\dagger_{s,0}\,c^\dagger_{f,-1}\,c_{p,1}\,c_{d,-2}
\\
&\qquad + \frac{\sqrt{5}}{5}\,c^\dagger_{s,0}\,c^\dagger_{f,0}\,c_{p,-1}\,c_{d,1} + \frac{\sqrt{15}}{5}\,c^\dagger_{s,0}\,c^\dagger_{f,0}\,c_{p,0}\,c_{d,0} + \frac{\sqrt{5}}{5}\,c^\dagger_{s,0}\,c^\dagger_{f,0}\,c_{p,1}\,c_{d,-1}
\\
&\qquad + \frac{\sqrt{15}}{15}\,c^\dagger_{s,0}\,c^\dagger_{f,1}\,c_{p,-1}\,c_{d,2} + \frac{2 \sqrt{30}}{15}\,c^\dagger_{s,0}\,c^\dagger_{f,1}\,c_{p,0}\,c_{d,1} + \frac{\sqrt{10}}{5}\,c^\dagger_{s,0}\,c^\dagger_{f,1}\,c_{p,1}\,c_{d,0}
\\
&\qquad + \frac{\sqrt{3}}{3}\,c^\dagger_{s,0}\,c^\dagger_{f,2}\,c_{p,0}\,c_{d,2} + \frac{\sqrt{6}}{3}\,c^\dagger_{s,0}\,c^\dagger_{f,2}\,c_{p,1}\,c_{d,1} + c^\dagger_{s,0}\,c^\dagger_{f,3}\,c_{p,1}\,c_{d,2}
\\
&\qquad + c^\dagger_{p,-1}\,c^\dagger_{d,-2}\,c_{s,0}\,c_{f,-3} + \frac{\sqrt{6}}{3}\,c^\dagger_{p,-1}\,c^\dagger_{d,-1}\,c_{s,0}\,c_{f,-2} + \frac{\sqrt{3}}{3}\,c^\dagger_{p,0}\,c^\dagger_{d,-2}\,c_{s,0}\,c_{f,-2}
\\
&\qquad + \frac{\sqrt{10}}{5}\,c^\dagger_{p,-1}\,c^\dagger_{d,0}\,c_{s,0}\,c_{f,-1} + \frac{2 \sqrt{30}}{15}\,c^\dagger_{p,0}\,c^\dagger_{d,-1}\,c_{s,0}\,c_{f,-1} + \frac{\sqrt{15}}{15}\,c^\dagger_{p,1}\,c^\dagger_{d,-2}\,c_{s,0}\,c_{f,-1}
\\
&\qquad + \frac{\sqrt{5}}{5}\,c^\dagger_{p,-1}\,c^\dagger_{d,1}\,c_{s,0}\,c_{f,0} + \frac{\sqrt{15}}{5}\,c^\dagger_{p,0}\,c^\dagger_{d,0}\,c_{s,0}\,c_{f,0} + \frac{\sqrt{5}}{5}\,c^\dagger_{p,1}\,c^\dagger_{d,-1}\,c_{s,0}\,c_{f,0}
\\
&\qquad + \frac{\sqrt{15}}{15}\,c^\dagger_{p,-1}\,c^\dagger_{d,2}\,c_{s,0}\,c_{f,1} + \frac{2 \sqrt{30}}{15}\,c^\dagger_{p,0}\,c^\dagger_{d,1}\,c_{s,0}\,c_{f,1} + \frac{\sqrt{10}}{5}\,c^\dagger_{p,1}\,c^\dagger_{d,0}\,c_{s,0}\,c_{f,1}
\\
&\qquad + \frac{\sqrt{3}}{3}\,c^\dagger_{p,0}\,c^\dagger_{d,2}\,c_{s,0}\,c_{f,2} + \frac{\sqrt{6}}{3}\,c^\dagger_{p,1}\,c^\dagger_{d,1}\,c_{s,0}\,c_{f,2} + c^\dagger_{p,1}\,c^\dagger_{d,2}\,c_{s,0}\,c_{f,3}\Big)
\\
{T_0^{(2)}}_{[3]}
&=  
  \frac{i}{2} \left(
\left[[c^\dagger_s \times c^\dagger_f]_{3} \times[\tilde c_d \times \tilde c_p]_{3}\right]_{0} 
    - h.c. 
    \right)\\
&=  
- {\bm T}^{sd}_2 \cdot {\bm Q}^{pf}_2  
+ {\bm Q}^{sd}_2 \cdot {\bm T}^{pf}_2  
\\
&=
- \frac{\sqrt{7} i}{14}\Big(c^\dagger_{s,0}\,c^\dagger_{f,-3}\,c_{p,-1}\,c_{d,-2} + \frac{\sqrt{6}}{3}\,c^\dagger_{s,0}\,c^\dagger_{f,-2}\,c_{p,-1}\,c_{d,-1} + \frac{\sqrt{3}}{3}\,c^\dagger_{s,0}\,c^\dagger_{f,-2}\,c_{p,0}\,c_{d,-2}
\\
&\qquad + \frac{\sqrt{10}}{5}\,c^\dagger_{s,0}\,c^\dagger_{f,-1}\,c_{p,-1}\,c_{d,0} + \frac{2 \sqrt{30}}{15}\,c^\dagger_{s,0}\,c^\dagger_{f,-1}\,c_{p,0}\,c_{d,-1} + \frac{\sqrt{15}}{15}\,c^\dagger_{s,0}\,c^\dagger_{f,-1}\,c_{p,1}\,c_{d,-2}
\\
&\qquad + \frac{\sqrt{5}}{5}\,c^\dagger_{s,0}\,c^\dagger_{f,0}\,c_{p,-1}\,c_{d,1} + \frac{\sqrt{15}}{5}\,c^\dagger_{s,0}\,c^\dagger_{f,0}\,c_{p,0}\,c_{d,0} + \frac{\sqrt{5}}{5}\,c^\dagger_{s,0}\,c^\dagger_{f,0}\,c_{p,1}\,c_{d,-1}
\\
&\qquad + \frac{\sqrt{15}}{15}\,c^\dagger_{s,0}\,c^\dagger_{f,1}\,c_{p,-1}\,c_{d,2} + \frac{2 \sqrt{30}}{15}\,c^\dagger_{s,0}\,c^\dagger_{f,1}\,c_{p,0}\,c_{d,1} + \frac{\sqrt{10}}{5}\,c^\dagger_{s,0}\,c^\dagger_{f,1}\,c_{p,1}\,c_{d,0}
\\
&\qquad + \frac{\sqrt{3}}{3}\,c^\dagger_{s,0}\,c^\dagger_{f,2}\,c_{p,0}\,c_{d,2} + \frac{\sqrt{6}}{3}\,c^\dagger_{s,0}\,c^\dagger_{f,2}\,c_{p,1}\,c_{d,1} + c^\dagger_{s,0}\,c^\dagger_{f,3}\,c_{p,1}\,c_{d,2}
\\
&\qquad - c^\dagger_{p,-1}\,c^\dagger_{d,-2}\,c_{s,0}\,c_{f,-3} - \frac{\sqrt{6}}{3}\,c^\dagger_{p,-1}\,c^\dagger_{d,-1}\,c_{s,0}\,c_{f,-2} - \frac{\sqrt{3}}{3}\,c^\dagger_{p,0}\,c^\dagger_{d,-2}\,c_{s,0}\,c_{f,-2}
\\
&\qquad - \frac{\sqrt{10}}{5}\,c^\dagger_{p,-1}\,c^\dagger_{d,0}\,c_{s,0}\,c_{f,-1} - \frac{2 \sqrt{30}}{15}\,c^\dagger_{p,0}\,c^\dagger_{d,-1}\,c_{s,0}\,c_{f,-1} - \frac{\sqrt{15}}{15}\,c^\dagger_{p,1}\,c^\dagger_{d,-2}\,c_{s,0}\,c_{f,-1}
\\
&\qquad - \frac{\sqrt{5}}{5}\,c^\dagger_{p,-1}\,c^\dagger_{d,1}\,c_{s,0}\,c_{f,0} - \frac{\sqrt{15}}{5}\,c^\dagger_{p,0}\,c^\dagger_{d,0}\,c_{s,0}\,c_{f,0} - \frac{\sqrt{5}}{5}\,c^\dagger_{p,1}\,c^\dagger_{d,-1}\,c_{s,0}\,c_{f,0}
\\
&\qquad - \frac{\sqrt{15}}{15}\,c^\dagger_{p,-1}\,c^\dagger_{d,2}\,c_{s,0}\,c_{f,1} - \frac{2 \sqrt{30}}{15}\,c^\dagger_{p,0}\,c^\dagger_{d,1}\,c_{s,0}\,c_{f,1} - \frac{\sqrt{10}}{5}\,c^\dagger_{p,1}\,c^\dagger_{d,0}\,c_{s,0}\,c_{f,1}
\\
&\qquad - \frac{\sqrt{3}}{3}\,c^\dagger_{p,0}\,c^\dagger_{d,2}\,c_{s,0}\,c_{f,2} - \frac{\sqrt{6}}{3}\,c^\dagger_{p,1}\,c^\dagger_{d,1}\,c_{s,0}\,c_{f,2} - c^\dagger_{p,1}\,c^\dagger_{d,2}\,c_{s,0}\,c_{f,3}\Big)
\end{align*}

{
  % \color{red}
\section{Minimal two-electron model realizing a finite two-body ET monopole}

In this section, we present a minimal spinless two-electron $sp$ model in which the two-body ET monopole $G_0^{(2)}$ acquires a finite expectation value without spin--orbit coupling. The purpose is to demonstrate explicitly that the $G_0^{(2)}$ channel is not only symmetry-allowed, but can be activated microscopically by interactions.

We consider the $N=2$ subspace spanned by the following three states,
\begin{align}
|P\rangle
&=
c_{p,-1}^{\dagger}c_{p,1}^{\dagger}|0\rangle,
\\
|S_{-}\rangle
&=
c_{s,0}^{\dagger}c_{p,-1}^{\dagger}|0\rangle,
\\
|S_{0}\rangle
&=
c_{s,0}^{\dagger}c_{p,0}^{\dagger}|0\rangle.
\end{align}
Other two-electron states are decoupled in the minimal model below. Within this subspace, the relevant component of the two-body ET monopole may be written as
\begin{align}
G_0^{(2)} 
&=
\frac{1}{\sqrt{6}}\Big(c^\dagger_{s,0}\,c^\dagger_{p,-1}\,c_{p,0}\,c_{p,-1} + c^\dagger_{s,0}\,c^\dagger_{p,0}\,c_{p,1}\,c_{p,-1} + c^\dagger_{s,0}\,c^\dagger_{p,1}\,c_{p,1}\,c_{p,0} \nonumber\\
&\qquad + c^\dagger_{p,0}\,c^\dagger_{p,-1}\,c_{s,0}\,c_{p,-1} + c^\dagger_{p,1}\,c^\dagger_{p,-1}\,c_{s,0}\,c_{p,0} + c^\dagger_{p,1}\,c^\dagger_{p,0}\,c_{s,0}\,c_{p,1}\Big)\blue{,}\nonumber\\
&= 
\frac{1}{\sqrt{6}}
\left(
|S_0\rangle\langle P|
+
|P\rangle\langle S_0|
\right).
\label{eq:G0_minimal_matrix}
\end{align}

We introduce %the
% \blue{a} minimal Hamiltonian \blue{given by}
a minimal Hamiltonian given by
\begin{align}
H_{\rm min}
=&\,
\Delta
\,\Big(
|S_{-}\rangle\langle S_{-}|
+
|S_{0}\rangle\langle S_{0}|
\Big)
+
v_{sp}
\,\Big(
|S_{-}\rangle\langle P|
+
|P\rangle\langle S_{-}|
\Big)
+
u
\,\Big(
|S_{0}\rangle\langle S_{-}|
+
|S_{-}\rangle\langle S_{0}|
\Big),
\label{eq:H_minimal_sp}
\end{align}
% where $\Delta>0$ is the energy separation between the $\blue{\ket{P}}$ and 
% $\blue{(}\blue{\ket{S_-}},\blue{\ket{S_0}}\blue{)}$ sectors%.
% %The parameter 
% \blue{,} $v_{sp}$ denotes an inversion-odd $s$-$p$ hybridization%. The parameter 
% \blue{, and} $u$ is introduced as an electric quadrupolar interaction. 
where $\Delta>0$ is the energy separation between the $\ket{P}$ and 
$(\ket{S_-},\ket{S_0})$ sectors, $v_{sp}$ denotes an inversion-odd $s$-$p$ hybridization, and $u$ is introduced as an electric quadrupolar interaction. 
Schematically, such an interaction may be written as
\begin{align}
H_Q^{(2)}
\sim
n_s Q_{2\mu}^{p},
\qquad
n_s=c_{s,0}^\dagger c_{s,0},
\label{eq:density_assisted_Q}
\end{align}
where $Q_{2\mu}^{p}$ is a time-reversal-even electric quadrupole operator in the $p$-orbital sector. Since $n_s$ is the electric monopole density in the $s$ sector, the product $n_s Q_{2\mu}^{p}$ represents a two-body electric multipolar degree of freedom. 

In the basis
$
\left(
|P\rangle,\ |S_{-}\rangle,\ |S_0\rangle
\right),
$
Eq.~\eqref{eq:H_minimal_sp} is represented by
\begin{align}
H_{\rm min}
=
\begin{pmatrix}
0 & v_{sp} & 0 \\
v_{sp} & \Delta & u \\
0 & u & \Delta
\end{pmatrix},
\label{eq:H_minimal_matrix}
\end{align}
while Eq.~\eqref{eq:G0_minimal_matrix} is represented by
\begin{align}
G_0^{(2)}
=
\frac{1}{\sqrt{6}}
\begin{pmatrix}
0 & 0 & 1 \\
0 & 0 & 0 \\
1 & 0 & 0
\end{pmatrix}.
\label{eq:G0_minimal_matrix_rep}
\end{align}

For $|v_{sp}|,|u|\ll \Delta$, the ground state is obtained perturbatively as
\begin{align}
|\psi_0\rangle
\simeq
|P\rangle
-
\frac{v_{sp}}{\Delta}|S_{-}\rangle
+
\frac{v_{sp}u}{\Delta^2}|S_0\rangle
+\cdots .
\label{eq:psi0_perturbative}
\end{align}
Substituting Eq.~\eqref{eq:psi0_perturbative} into
Eq.~\eqref{eq:G0_minimal_matrix}, we obtain
\begin{align}
\langle G_0^{(2)}\rangle
&=
\langle \psi_0|G_0^{(2)}|\psi_0\rangle
\simeq
\frac{2}{\sqrt{6}}
\frac{v_{sp}u}{\Delta^2}
+
O\left[
\frac{1}{\Delta^4}
\right].
\label{eq:G0_expectation_minimal}
\end{align}
Thus, the activation of $G_0^{(2)}$ requires both inversion breaking and the two-body electric quadrupolar interaction.

}